\documentclass[a4paper,11pt]{article}
\usepackage{graphicx}
\pdfoutput=1 

\usepackage{mathrsfs}
\usepackage[T1]{fontenc} 
\usepackage{bbold}
\usepackage{url}
\usepackage[english]{babel}
\usepackage[utf8]{inputenc}
\usepackage{amsmath}

\usepackage{amsfonts}
\usepackage{amssymb}
\usepackage{mathtools}
\usepackage{placeins}
\usepackage{float}
\usepackage{tabularx}
\usepackage{adjustbox}
\usepackage{caption}
\usepackage{subcaption}
\DeclareUnicodeCharacter{2212}{-}
\usepackage{jcappub}

\title{\boldmath Constraining an Early Dark Energy Motivated Quintessential \texorpdfstring{$\alpha$}{a}-Attractor Inflaton Potential}
\author{Arunoday Sarkar and}
\author{Buddhadeb Ghosh}

\affiliation{Centre of Advanced Studies, Department of Physics, The University of Burdwan,\\Burdwan 713 104, India}

\emailAdd{adsarkar@scholar.buruniv.ac.in}
\emailAdd{bghosh@phys.buruniv.ac.in}
\abstract{We construct a new model of quintessential $\alpha$ attractor inflation in conjunction with the features of non-oscillating early dark energy (EDE). Slow-roll plateau of this model is obtained, and analyzed in $k$-space, through the inflaton field and its first-order perturbation over a quasi de-Sitter metric fluctuation in the range $k=0.001-0.009$ Mpc$^{-1}$. The estimated cosmological parameters are found to obey Planck+BICEP2/Keck bounds with $68\%$ CL with the required trend of spectral tilts in the $n_s-r$ parametric space. We verify that, the inclusion of the EDE does not significantly affect the observed parameters. Its presence manifests in obtaining \textit{improved values} of the energy scale of inflation ($M$) and the present-day vacuum density ($V_{\Lambda}$). They are found to be $M=5.58\times 10^{-4}-4.57\times 10^{-3} M_P$ and $V_{\Lambda}=1.042\times 10^{-119}-4.688\times 10^{-116} M_P^4$. 
   However, the $\alpha$-parameter is drastically constrained in two ways. Its lower end is fixed by the consistency analysis of the $k$-mode equations, while the upper end is evaluated as a derived expression of $\alpha$-cut-off through the aspects of EDE \textit{viz.,} the effects of \textit{Enhanced Symmetry Point} (ESP) in the potential during inflation. Improvised range of $\alpha$ is found to be $0.001\leq\alpha<0.1$ for the model parameters $\gamma$ and $n$ lying within $0.01\leq\gamma\leq 0.09$ and $8\leq n\leq 10$ respectively. These ranges are shown to be essential for satisfying the COBE/Planck normalized energy scale of inflation and the Planck-value of present-day vacuum density. If we choose $\gamma=0.0818$ and $n=8$, then we get $0.001\leq\alpha\leq 0.0186$. Thus, the lower and upper limits of $\alpha$ are diminished substantially, compared with those in the earlier studies. We also argue that, the lower values of $\alpha$ can be instrumental in resolving the $H_0$-discrepancy, whereas its higher values are capable of explaining both the early- and the late-time expansions of the universe, as discussed in some current literature.}

\begin{document}
\maketitle
\flushbottom






\section{Introduction}
\label{sec:intro}
One of the great achievements of $\Lambda$-cold dark matter ($\Lambda$CDM) model is to identify the major component \cite{WMAP:2012nax} of our universe, \textit{viz.,} the \textit{`dark energy'} (DE) \cite{Huterer:1998qv}. The current accelerating expansion of the universe, which is confirmed by several cosmological surveys, based on redshift observations \cite{SupernovaSearchTeam:1998fmf,SupernovaCosmologyProject:1998vns}, cosmic microwave background (CMB) anisotropies plus polarization sky-maps \cite{WMAP:2012nax,Planck:2015fie,Planck:2015sxf,Planck:2018vyg,Planck:2018jri} and various other astrophysical explorations \cite{KiDS:2020suj,DES:2021wwk} supports the existence of DE, having barotropic parameter $\omega=-1.03\pm0.03$ \cite{Planck:2018vyg}. One way of elucidating this exotic element is to introduce a new scalar field, the \textit{`quintessence'}\footnote{considered to be the fifth element after normal baryonic matter, dark matter, photon and neutrino.} \cite{Caldwell:1997ii,Peebles:2002gy,Ratra:1987rm,Copeland:2006wr} in order to solve the \textit{cosmological constant problem} \cite{Weinberg:1988cp,Dimopoulos:2022wzo}. The quintessential scalar field is unique in so far as other models of DE are concerned, because it is characterised by a variable barotropic parameter in the range $-1\leq \omega\leq -0.95$ \cite{Planck:2018vyg}. However, the quintessence proposition has its own internal inconsistencies, especially regarding its initial conditions. This is known as the \textit{coincidence problem} \cite{Dimopoulos:2022wzo,delCampo:2008jx,Huey:2004qv,Velten:2014nra,Wang:2016lxa,Bolotin:2013jpa,Zlatev:1998tr}. It is required to merge coherently with the inflaton field to address both the early and the late-time expansions of the universe, along with the period of \textit{kination} \cite{Spokoiny:1993kt,Pallis:2005hm,Pallis:2005bb,Gomez:2008js} within a single unified framework, \textit{viz.,} the \textit{quintessential inflation} \cite{Peebles:1998qn,Peloso:1999dm,Sen:2000ym,Kaganovich:2000fc,Yahiro:2001uh,Martin:2004ba,Barenboim:2005np,Rosenfeld:2005mt,Cardenas:2006py,BuenoSanchez:2006fhh,Membiela:2006rj,Rosenfeld:2006hs,Neupane:2007mu,Bastero-Gil:2009wdy,Piedipalumbo:2011bj,Wetterich:2014gaa,Hossain:2014xha,Hossain:2014coa,Hossain:2014ova,Geng:2015fla,WaliHossain:2014usl,Haro:2015ljc,deHaro:2016ftq,deHaro:2016cdm,Guendelman:2016kwj,Rubio:2017gty,Ahmad:2017itq,Haro:2018zdb,Bettoni:2018pbl,Selvaganapathy:2019bpm,Lima:2019yyv,Kleidis:2019ywv,Haro:2019peq,Benisty:2020xqm,Benisty:2020vvm,deHaro:2021swo,Dimopoulos:2021xld,Tian:2021cqq,AresteSalo:2021wgb,Akrami:2020zxw,Garcia-Garcia:2019ees,Garcia-Garcia:2018hlc,Akrami:2017cir,Kepuladze:2021tsb,Dimopoulos:2017zvq,Geng:2017mic,Agarwal:2017wxo,AresteSalo:2017lkv,DeHaro:2017abf,Haro:2019gsv,deHaro:2019oki,Benisty:2020qta,Shokri:2021zqw,Bettoni:2021qfs,Jaman:2022bho,Jesus:2021bxq,Fujikura:2022udt,Karciauskas:2021fdu,Basak:2021cgk,AresteSalo:2020yxl,Hung:2023pby,Alho:2023pkl}.\par Despite being so successful, the $\Lambda$CDM model, which is a standard theoretical tool for understanding the present universe from first few seconds until today, is not yet complete. There are two missing links \textit{viz.,} the \textit{Hubble tension} \cite{Riess:2019cxk,Riess:2020fzl,Dainotti:2022bzg,Dainotti:2021pqg,Vagnozzi:2019ezj,DiValentino:2017iww,DiValentino:2019ffd,DiValentino:2021izs,Cyr-Racine:2021aqp,Nunes:2022bhn,Yang:2018euj,Wang:2021kxc,Gariazzo:2021qtg,Jiang:2023bsz,Adil:2023exv,Ben-Dayan:2023htq,Pan:2023mie} and the \textit{$\sigma_8$ discrepancy}\footnote{related to the dark matter-dark energy interactions and the matter clustering processes.} \cite{Lucca:2021dxo,Nunes:2021ipq,DiValentino:2020vvd,Gariazzo:2021qtg,Wang:2021kxc,Yang:2018euj,Ben-Dayan:2023rgt,vanderWesthuizen:2023hcl,deSa:2022hsh,Vagnozzi:2021gjh,Ben-Dayan:2023htq}.\par Cosmological parameters, specifically the observed expansion rate and the current age of the universe are estimated by the precise measurement of Hubble constant $H_0$. Various processes to determine $H_0$ can be broadly categorized into two major classes \cite{Brissenden:2023yko}: (i) Analysis of the fluctuations in cosmic microwave background radiation (CMBR) at redshift $z=1089.80\pm0.21$ \cite{Planck:2018vyg} (at the last scattering surface) (or from the Baryon Acoustic Oscillations (BAO)) by the Planck satellite \cite{Planck:2018vyg}; (ii) The cosmic distance ladder (CDL) methods such as redshift observations \cite{Riess:2019cxk,Riess:2020fzl} of distant type-1a supernovae or Cepheid stars at redshift $z=\mathcal{O}(1)$. However, there is a $5\sigma$ level disagreement\footnote{Actually it is in between $4.56\sigma$ and $6.36\sigma$ \cite{Brissenden:2023cne}.} between the values of $H_0$, obtained by these two methods. The value given by Planck \cite{Planck:2018vyg} is  
\begin{equation}
    H_0^{\mathrm{Planck}}=67.44\pm 0.58 \quad \mathrm{Km~ s}^{-1}\mathrm{Mpc}^{-1}
    \label{eq:Planck_Hubble}
\end{equation} whereas the CDL measurement from Cepheid-SN-1a by SH0ES collaboration \cite{Riess:2021jrx} gives
\begin{equation}
    H_0^{\mathrm{SN}}=73.04\pm 1.04  \quad \mathrm{Km~ s}^{-1}\mathrm{Mpc}^{-1}. 
    \label{eq:Planck_SN}
\end{equation}
Many possible resolutions of the discrepancy between these two values have been discussed in Refs. \cite{Escudero:2022rbq,Haridasu:2022dyp,Cai:2021weh,Cai:2022dkh,Knox:2019rjx,Gomez-Valent:2022bku}. Most of them rely on early universe phenomenology. \par In many contemporary literature \cite{Karwal:2016vyq,Pettorino:2013ia,Calabrese:2011hg,Doran:2006kp,Sabla:2021nfy,Smith:2019ihp,Murai:2022zur,Capparelli:2019rtn,Berghaus:2022cwf,Berghaus:2019cls,Sakstein:2019fmf,Karwal:2021vpk,Sabla:2022xzj,Lin:2019qug,McDonough:2022pku,Poulin:2018dzj,Niedermann:2020dwg,Hill:2020osr,Smith:2020rxx,Nojiri:2021dze,Poulin:2018cxd,Freese:2021rjq,Agrawal:2019lmo,Braglia:2020bym,Moshafi:2022mva,Guendelman:2022cop,Seto:2021xua,Vagnozzi:2019ezj,Reeves:2022aoi,MohseniSadjadi:2022pfz,Goldstein:2023gnw,Poulin:2023lkg,Reboucas:2023rjm,Cruz:2023cxy,CarrilloGonzalez:2023lma,Brissenden:2023yko,Eskilt:2023nxm,Odintsov:2023cli,Copeland:2023zqz,Raveri:2023zmr,Sharma:2023kzr,Benaoum:2023ekz,Niedermann:2023ssr,Kodama:2023fjq,Vagnozzi:2023nrq,Cicoli:2023qri} it has been suggested that a viable option to alleviate the $5\sigma$ tension is to propose a new version of dark energy, the \textit{Early Dark Energy} (EDE)\footnote{The connotation `\textit{early}' is used to differentiate it from the usual one which is responsible for the present accelerated expansion of the universe.} during matter-radiation equality without any substantial modification of $\Lambda$CDM. The cosmological parameters which are tightly constrained by the Planck observations remain unaltered by the EDE. Actually, the EDE is formulated in such a way \cite{Escudero:2022rbq,Knox:2019rjx,Gomez-Valent:2022bku} that its decay rate is faster than the background energy density (particularly the radiation\footnote{The decay is so rapid that the EDE fades away before the decoupling of CMB photon from the last scattering surface.}) \cite{Pettorino:2013ia}, so that its effects can not be tested in CMB observations. However, according to Friedmann equations the value of $H_0$ can be lifted slightly by the increase in the overall density of the universe, induced by the EDE. This has been shown to be possible by considering either a first order phase transition \cite{Niedermann:2020dwg} or an EDE oscillation \cite{Knox:2019rjx,Karwal:2016vyq,Pettorino:2013ia,Niedermann:2020dwg,Hill:2020osr,Smith:2020rxx,Nojiri:2021dze,Poulin:2018cxd,Freese:2021rjq,Agrawal:2019lmo,Braglia:2020bym} near the potential minimum.\par Recently, the EDE has been considered \cite{Brissenden:2023yko} to be a promising candidate for resolving the $H_0$ tension in the quintessential form of the $\alpha$-attractor potential. In this model, the EDE is identified with a scalar field of non-oscillating (NO) type near the matter-radiation equality at the Enhanced Symmetry Point (ESP) \cite{Kofman:2004yc,Dimopoulos:2019ogl}, present in the potential profile, which can decay faster than the oscillating one \cite{Braglia:2020bym} and is believed to be the fastest, that is possible \cite{Brissenden:2023yko}. In this work \cite{Brissenden:2023yko}, the authors perform a simulation, engaging the Friedmann equations, the Klein-Gordon equation and the continuity equations for radiation and matter, with the conditions from matter-radiation equality, decoupling and the present values of various cosmological parameters. The results of the simulation process reveal that within reasonable choices of the parameter space, the model yields required constraints for post-inflationary as well as for the current DE-related observations. We feel, in order to make the outcome of this survey more concrete, the scenario should be connected with the inflationary paradigm \textit{vis-\`{a}-vis} the microscopic mode behaviour of the inflaton field and the relevant cosmological parameters. \par In a previous work \cite{Sarkar:2023cpd}, we made a quite similar kind of study. There, the inflationary slow-roll regime was probed for a specific class of quintessential $\alpha$-attractor models through a sub-Planckian $k$-space first order quantum mode analysis in the background of a quasi de-Sitter metric fluctuation, developed in Ref. \cite{Sarkar:2021ird}, under spatially flat gauge. The self-consistent solutions for $k=0.001-0.009$ Mpc$^{-1}$ revealed that the relevant cosmological parameters agree with the Planck-2018 data very well within a given range and accuracy. Specifically, the cumulative mode responses of the scalar spectral index ($n_s$) and the tensor-to-scalar ratio ($r$) showed that the spectral tilts obeyed the Planck bounds with $68\%$ CL, which eventually constrain the model parameters as $n=122$ and $\frac{1}{10}\leq \alpha \leq 4.3$ continuously. It was found that, below $\alpha =\frac{1}{10}$ the evolution equations are insensitive to $\alpha$ and beyond $\alpha=4.3$ no convergence is observed for obtaining a consistent solution. Four values of $\alpha$ \textit{viz.,} $\alpha=1/10,~ 1/6,~ 1$ and $4.3$ were chosen for which the slow-roll plateau with the inflationary energy scales $M=4.28\times 10^{15} - 1.11\times 10 ^{16}$ GeV was found to satisfy COBE/Planck normalisation and the quintessential tail part yielded dynamically the present day vacuum density with the amplitude $V_{\Lambda}\approx 10^{-115} - 10^{-117}$ $M_P^4$. We also compared our results with ordinary $\alpha$ attractor $E$ and $T$ models and discovered that the quintessence always restricts the potential to be single-field concave type and prevents its mutation into the simple chaotic one which is ruled out by Planck, thereby making the concerned model just appropriate for quintessential inflation.\par In the present paper, we make a connection of our earlier work \cite{Sarkar:2023cpd} to the EDE proposal. Our primary focus will be on further constraining the important parameter $\alpha$ using the EDE as a tool in the light of Planck-2018 data. We therefore extend the formalism developed in Ref. \cite{Sarkar:2023cpd} keeping in mind the aspects of EDE in the inflaton potential.\par The paper is organised in this way. In Section \ref{sec:our model}, we set up our model in non-canonical and canonical field spaces, describe its pole behaviour, study the EDE dynamics and display the variations of the obtained potential with respect to the model parameters. Section \ref{sec:formalism} outlines the quasi de-Sitter Hubble-exit of the quantum inflaton modes and their evolutions with a brief introduction of the relevant cosmological parameters. The corresponding boundary conditions and the mode responses of the parameters are illustrated in Section \ref{sec: result}. Then in the same section we constrain the parameter $\alpha$ by the mode equations in conjunction with the EDE. In this course, we find that the lower values of $\alpha$ fit the Planck data in improved way. We use these results to make a comparative study with the Planck+BICEP2/Keck bounds and then write a brief discussion here regarding further constraining the ranges of the model parameters, which could be helpful in resolution of the Hubble tension. Finally Section \ref{sec:conclusion} contains some concluding remarks.
\section{EDE-motivated quintessential \texorpdfstring{$\alpha$}{a}-attractor (EMQA) model}
\label{sec:our model}
We consider the action,
\begin{equation}
    S_{\mathrm{minimal}}[x]=\int_{\Omega} d^4x \sqrt{-g(x)}\left[\frac{1}{2}R(x)+\mathbf{\mathcal{L}}_{\mathrm{non-canonical}}\left(\theta (x),\partial\theta (x)\right)\right]
    \label{eq:the_action}
\end{equation} of a non-canonical inflaton field $\theta$ measured in the reduced Planck scale, $M_P=(8\pi G)^{-1/2}=2.43\times 10^{18}$ GeV, minimally coupled with gravity, whose symmetry and dynamics are controlled by the Lagrangian\footnote{The Lagrangian is written following the metric signature $(-,+,+,+)$.} \cite{Brissenden:2023yko}
\begin{equation}
    \mathbf{\mathcal{L}}_{\mathrm{non-canonical}}(\theta,\partial\theta)=-\left[1-\frac{\theta^2}{(\sqrt{6\alpha}M_P)^2}\right]^{-2}\frac{g^{\mu\nu}\partial_\mu\theta\partial_\nu\theta}{2}-V(\theta).
    \label{eq:the_lagrangian}
\end{equation}
The kinetic part of Eq. (\ref{eq:the_lagrangian}) contains the same non-trivial quadratic pole structure of quintessential $\alpha$-attractor inflation as described in detail in Ref. \cite{Sarkar:2023cpd}. The poles at $\theta=\pm \sqrt{6\alpha}M_P$ save the model from the issues arising from the super-Planckian effects like the \textit{`fifth force problem'} \cite{Wetterich:2004ff,Dimopoulos:2017zvq,Dimopoulos:2022wzo} and the radiative corrections \cite{Linde:2017pwt,Kallosh:2016gqp,Chojnacki:2021fag} due to infinite field excursion of the non-canonical scalar field $\theta$, which therefore ensure that the inflaton field behaviour near the pole boundary is genuinely sub-Planckian. In Ref. \cite{Sarkar:2023cpd} it was discussed that the presence of these non-canonical poles is in fact a generic feature of $\alpha$-attractors \cite{Kallosh:2013yoa,Ferrara:2013rsa,Kallosh:2013pby,Kallosh:2013maa,Kallosh:2015lwa,Carrasco:2015pla,Kallosh:2013hoa,Maeda:2018sje,Kallosh:2014rga,Galante:2014ifa,Kallosh:2013tua,Ferrara:2013kca,Linde:2018hmx,Cecotti:2014ipa,Ferrara:2013eqa,Ferrara:2014rya} in the context of \textit{`pole inflation'} in $\mathcal{N}=1$ minimal supergravity and exclusive as well, because such poles are not generally found even in string theory \cite{Dasgupta:2018rtp,Let:2022fmu,Let:2023dtb}.\par So far as the potential part\footnote{The structure of the potential is inspired by that in Ref. \cite{Brissenden:2023yko}. Because it shows that such a functional dependence is efficient in resolving the Hubble tension. In this paper, the structure of the potential is slightly customized in comparison with that in \cite{Brissenden:2023yko} to incorporate both inflation and quintessence. Our aim is to construct a model which can accommodate inflation, quintessence as well as EDE.} is concerned, we take,
\begin{equation}
    V(\theta)=\exp{\left(\gamma e^{-\sigma\sqrt{6\alpha}}\right)}M^4 \exp{\left(-\gamma e^{-\frac{\sigma\theta}{M_P}}\right)}.
    \label{eq:non_can_pot}
\end{equation}
Eq. (\ref{eq:non_can_pot}) is radically different from what is considered in \cite{Sarkar:2023cpd}. The double exponential structure is a modification over the simple `$\exp$' type of potential in quintessential inflation such that it can embed the EDE smoothly in the field dynamics. $\gamma$ and $\sigma$ are the dimensionless positive constants which can deform the shape of the potential substantially. This will be seen shortly. However, the most valuable parameter here is $\alpha$, which is the characteristic constant of the well known $\alpha$-attractor formulation, signifying the reciprocal curvature of the underlying $SU(1,1)/U(1)$ K\"{a}hler manifold,
\begin{equation}
    \alpha=-\frac{2}{3\mathscr{R}_\mathscr{K}}.
\end{equation} Optimising the value of $\alpha$ is indispensable in the assimilation of several theoretical \cite{Carrasco:2015rva,Kallosh:2015zsa,Carrasco:2015uma,Kallosh:2017ced,Odintsov:2016vzz,Scalisi:2018eaz,Kallosh:2017wku,Kallosh:2021vcf} as well as experimental aspects \cite{Kallosh:2019hzo,Kallosh:2019eeu,Ferrara:2016fwe,Planck:2018vyg,Planck:2018jri} in conjunction with the various cosmological observations regarding CMB $B$-mode polarization \cite{Kallosh:2019hzo,Kallosh:2019eeu,Ferrara:2016fwe,Planck:2018vyg,Planck:2018jri}, gravitational waves \cite{Dimopoulos:2017zvq} and the stage III/IV DE and large scale structure (LSS) surveys \cite{Akrami:2017cir,Garcia-Garcia:2018hlc,Garcia-Garcia:2019ees,Akrami:2020zxw}.\par In Ref. \cite{Sarkar:2021ird}, we analysed the quantum modes of the linear inflaton perturbation in $k\rightarrow 0$ limit incorporating the primitive forms of the $\alpha$-attractor $E$ and $T$ models and the results showed that for $\alpha_E=1,~5,~10$; $\alpha_T=1,~6,~10$ (with $68\%$ CL) and $\alpha_{E,T}=15$ (with $95\%$ CL) for the exponent $n=2$; and $\alpha_E=1,~6,~11$ and $\alpha_T=1,~4,~9$ for $n=4$ (with $68\%$ CL), the $E/T$ models can explain all the necessary constraints of the Planck-2018 data. Similarly in Ref. \cite{Sarkar:2023cpd} it was discovered that attaching quintessence with the $\alpha$-attractor framework, the $\alpha$ values are diminished considerably into the closed interval $0.1\leq\alpha\leq 4.3$ for a large power $n=122$ to accommodate both the early and the late-time expansions of the universe. Now, it will be worthwhile to explore the behaviour of $\alpha$ when we equip the potential with EDE in addition to DE. This might bear some clues regarding further understanding of EDE. With this objective, we present below some related derivations.\par 
The potential in Eq. (\ref{eq:non_can_pot}) has two asymptotic limits, as the field $\theta$ approaches the poles of the inflaton-kinetic term in Eq. (\ref{eq:the_lagrangian}):
\begin{equation}
\begin{split}
    \lim_{\theta\rightarrow +\sqrt{6\alpha}M_P}V(\theta)&=\exp{\left(\gamma e^{-\sigma\sqrt{6\alpha}}\right)}M^4 \exp{\left(-\gamma e^{-\sigma\sqrt{6\alpha}}\right)}\\
    &=\exp{\left[-\gamma\left(e^{-\sigma\sqrt{6\alpha}}-e^{-\sigma\sqrt{6\alpha}}\right)\right]}M^4\\
    &= M^4
\end{split}
\label{eq:non_can_limit1}
\end{equation} and
\begin{equation}
\begin{split}
    \lim_{\theta\rightarrow -\sqrt{6\alpha}M_P}V(\theta)&=\exp{\left(\gamma e^{-\sigma\sqrt{6\alpha}}\right)}M^4 \exp{\left(-\gamma e^{\sigma\sqrt{6\alpha}}\right)}\\
    &=\exp{\left[-\gamma\left(e^{\sigma\sqrt{6\alpha}}-e^{-\sigma\sqrt{6\alpha}}\right)\right]}M^4\\
    &=\exp{\left[-2\gamma\sinh{\left(\sigma\sqrt{6\alpha}\right)}\right]}M^4\\
    &=V_{\Lambda}.
\end{split}
\label{eq:non_can_limit2}
\end{equation}
These two extreme limits indicate the energy scales of inflation $M$ ($\sim 10^{15}$ GeV) and the present day vacuum density $V_{\Lambda}$ of the universe ($\sqrt[4]{V_{\Lambda}}\sim 10^{-12}$ GeV) \cite{Planck:2018jri,Planck:2018vyg}  respectively.\par
Now, we redefine the inflaton field as,
\begin{equation}
    \xi = \sqrt{6\alpha}M_P \tanh^{-1}{\left(\frac{\theta}{\sqrt{6\alpha}M_P}\right)}
    \label{eq:NonCanToCanTransform}
\end{equation} 
in such a way that the non-canonical kinetic term transforms into a canonical one as
\begin{equation}
    -\left[1-\frac{\theta^2}{(\sqrt{6\alpha}M_P)^2}\right]^{-2}\frac{g^{\mu\nu}\partial_\mu\theta\partial_\nu\theta}{2}=-\frac{1}{2}\partial_{\mu}\xi\partial^{\mu}\xi
\end{equation}
from which we obtain a new form of the quintessential $\alpha$-attractor inflaton potential
\begin{equation}
    V(\xi)=\exp{\left(\gamma e^{-\sigma\sqrt{6\alpha}}\right)}M^4 \exp{\left(-\gamma e^{-\sigma\sqrt{6\alpha}\tanh{\left(\frac{\xi}{\sqrt{6\alpha}M_P}\right)}}\right)}.
\end{equation}
We can express this potential in a more convenient form by defining a power index $n=\sigma\sqrt{6\alpha}$
\begin{equation}
    V(\xi)=\exp{\left(\gamma e^{-n}\right)}M^4 \exp{\left(-\gamma e^{-n\tanh{\left(\frac{\xi}{\sqrt{6\alpha}M_P}\right)}}\right)}.
    \label{eq:new_can_pot}
\end{equation}
This newly obtained potential features identical pole behaviours as that of Eq. (\ref{eq:non_can_pot}) with the correspondences
\begin{equation}
    \lim_{\xi\rightarrow +\infty}V(\xi)\implies \lim_{\theta\rightarrow +\sqrt{6\alpha}M_P}V(\theta) = M^4
    \end{equation} and
    \begin{equation}
    \lim_{\xi\rightarrow -\infty}V(\xi)\implies \lim_{\theta\rightarrow -\sqrt{6\alpha}M_P}V(\theta) = V_{\Lambda},
    \end{equation}
    following Eq. (\ref{eq:NonCanToCanTransform}).
 In this new version of the potential, the poles at $\theta=\pm\sqrt{6\alpha} M_P$ are stretched to $\xi=\pm\infty$ and the boundaries of these poles manifest as the inflationary plateau and the quintessential tail in the positive and negative limits, respectively.   
\begin{figure}[H]
	\centering
	\includegraphics[width=0.8\linewidth]{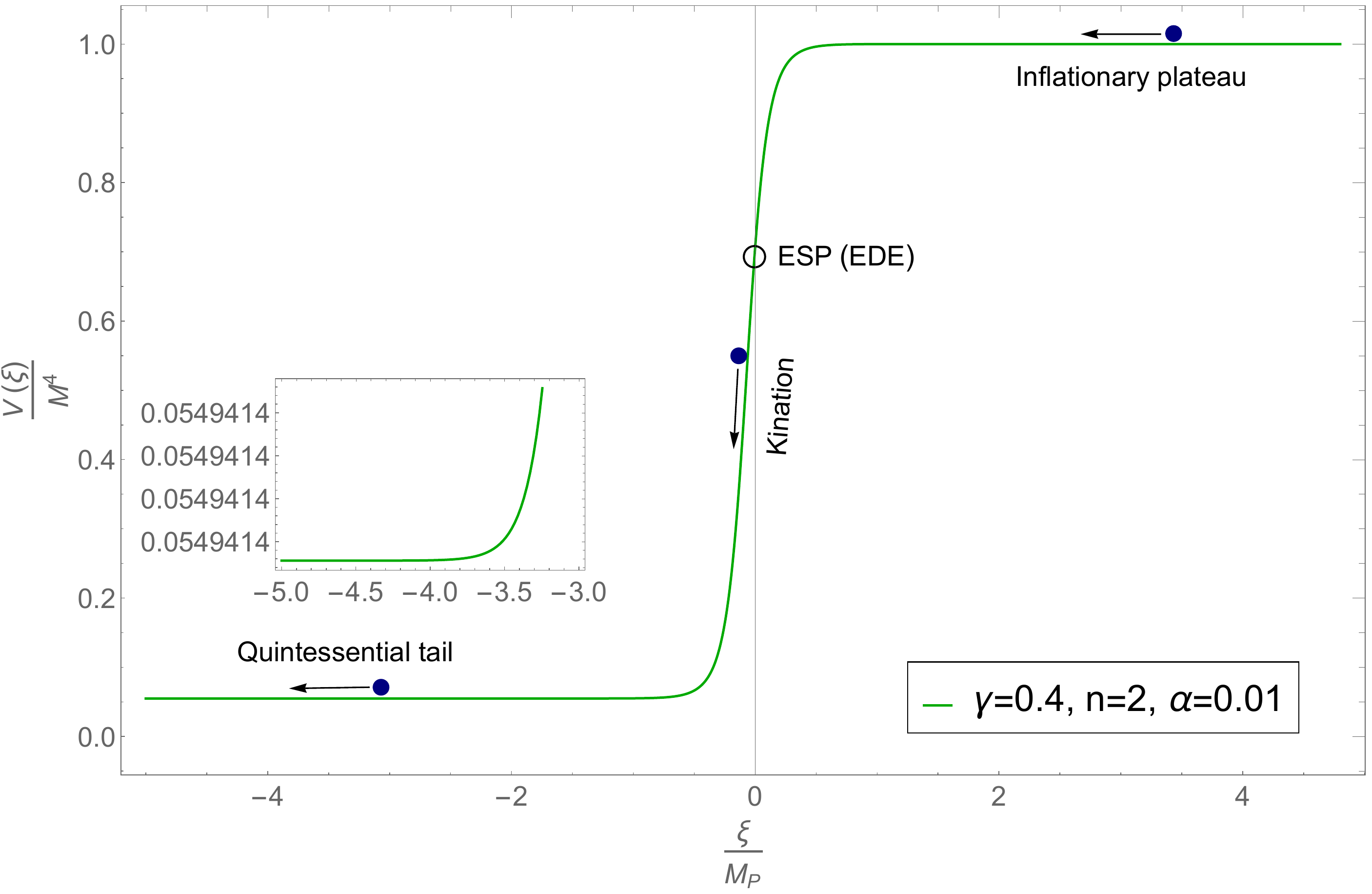}
	\caption{Graphical representation of the potential of Eq. (\ref{eq:new_can_pot}) for $\gamma=0.4$, $n=2$ and $\alpha=0.01$, as an example. These values are chosen to show the important phases of the quintessential inflation with EDE. The inflaton field starts its journey over the slow-roll plateau from infinity (or from the pole in non-canonical field space). When it reaches the ESP at origin, indicated by the ring with a positive energy $V(0)$, it gets stuck for a while due to an interaction with some other heavy fields. It remains frozen there until matter-radiation equality when its density parameter reaches a maximum value before decoupling. The time of freezing is actually fixed by the interaction time plus the time required for the decay of the additional particles. Then it experiences the sharp gradient of the potential and the field energy drains away to a negative value at the throat of the quintessential runaway. As long as the field remains frozen at the ESP, the field acts as an early version of the dark energy without disturbing the $\Lambda$CDM parameters at all. It doesn't even leave any traces in CMBR, such that the observations can not detect its presence in the background. When the period of kination terminates, the field enters into the final part of the potential to serve the late-time expansion of the universe in the form of DE with energy much smaller than that of the inflationary one.}
	\label{fig:Fig1}
\end{figure}
In figure \ref{fig:Fig1} we depict the potential of Eq. (\ref{eq:new_can_pot}) normalized by $M^4$ against the canonical field $\xi$ in unit of $M_P$ for three sample model parameters \textit{viz.,} $\gamma = 0.4$, $n=2$ and $\alpha = 0.01$. The inflaton field, denoted by the blue ball starts rolling slowly from $\xi>0$ region over the inflationary plateau with a significant amount of energy ($\sim 10^{15}$ GeV) against the Hubble drag. As a result, the potential energy dominates over the kinetic one until the field reaches a point where the first slow-roll parameter $\epsilon_V$ is unity. It is the place where the inflation ends and after that, the field enters into a new region, called `\textit{kination}'. Generically, this phase is characterized by a fast decay rate (possibly fastest \cite{Brissenden:2023yko}) of the energy density of the field $\rho_{\xi}\propto \frac{1}{a^6}$ \cite{Dimopoulos:2022wzo} with scale factor $a$ and eventually the field undergoes a kinetically driven free-fall until it moves towards $\xi<0$ region and resurrects as DE in the form of quintessence \cite{Dimopoulos:2021xld} with the energy $\sim 10^{-12}$ GeV, as we observe today. But, here the scenario is strikingly different, rather novel.\par 
The potential considered in Eq. (\ref{eq:non_can_pot}) or (\ref{eq:new_can_pot}) has a new feature, called the `\textit{Enhanced Symmetry Point (ESP)}' \cite{Brissenden:2023yko,Kofman:2004yc,Dimopoulos:2019ogl} at $\theta=0$ (or, $\xi=0$) (shown by a ring in figure \ref{fig:Fig1}). Thus, when the field hits the origin, it gets trapped due to a quadratic interaction with some other heavy fields \textit{viz.,} the $\Theta$ particles of type $\delta V_{\mathrm{int}}(\theta,\Theta)=\frac{1}{2}p^2\theta^2\Theta^2$ for the coupling strength $p<1$ \cite{Brissenden:2023yko,Kofman:2004yc,Dimopoulos:2019ogl,Dimopoulos:2022wzo,Dimopoulos:2017tud}. As a result,  all the kinetic energy of the $\xi$-field (or $\theta$ field) developed during traversing from the end of inflation to the origin, is transferred to $\Theta$ particles, and the $\theta$ field (or the $\xi$ field) sticks at the origin with constant potential energy density $V(0)$ \textit{vis-\`{a}-vis} a fixed (total) energy density $\rho_{\xi}(0)$. On the other hand, in the background, the $\Theta$ particles being massive decay into matter and radiation in the thermal bath of the hot big bang (HBB) \cite{Brissenden:2023yko,Dimopoulos:2019ogl,Dimopoulos:2022wzo}. The densities of matter ($\rho_{\mathrm{matter}}$) and radiation ($\rho_{\mathrm{rad}}$) then start to decrease with scale factor as $\rho_{\mathrm{matter}}\propto a^{-3}$ and $\rho_{\mathrm{rad}}\propto a^{-4}$ \cite{Baumann:2009ds,Baumann:2022mni,Dimopoulos:2022wzo}, diminishing the total background energy density ($\rho_{\mathrm{matter}}+\rho_{\mathrm{rad}}$) of the universe. As radiation density decreases faster than the matter density, at a particular scale factor, the two densities will be equal, called the \textit{matter-radiation equality} at redshift $z_{\mathrm{equal}}=3387.4$ \cite{Planck:2018vyg}. \par Now, from the starting of production of matter and radiation to the matter-radiation equality, as the total density of the universe in the background decreases, the density parameter of the $\xi$-field at origin (\textit{i.e.} at the ESP) $\Omega (\xi=0)=\frac{\rho_{\xi}(0)}{\rho_{\mathrm{matter}}+\rho_{\mathrm{rad}}}$ increases and attains a maximum value $\Omega_{\mathrm{equal}}$ at the instant of equality. According to Refs. \cite{Copeland:1997et,Copeland:2006wr,Brissenden:2023yko,Dimopoulos:2022wzo}, when the density parameter of $\xi$-field becomes maximum, the frozen scalar field unfreezes before dominating. Afterward, the $\xi$-field faces the steepest section of the potential due to the $\exp{(\exp)}$ structure of the potential and freely falls to a negative value $-\frac{\sqrt{48\alpha}}{n\gamma}M_P=-0.87M_P$ (derived from an expression given in Ref. \cite{Brissenden:2023yko} for the dynamics of the field during free-fall and the model parameters chosen in figure \ref{fig:Fig1}) in the quintessential runaway. Then, it again refreezes at the potential minimum at a constant potential energy density comparable to the present-day density of matter until it becomes dominant in the present universe as dark energy, causing the observed late-time acceleration of the universe.\par 
As long as the $\xi$-field stays frozen at the ESP ($\xi=0$), its kinetic energy density is zero and potential energy density is constant at $V(0)$. Therefore, during this period of freezing the barotropic parameter of $\xi$-field is effectively $(-1)$, which is the same as that of dark energy. This indicates that starting from being trapped at the ESP to the matter-radiation equality, the $\xi$-field behaves like an early version of dark energy and hence the name: \textit{`Early Dark Energy (EDE)'}. In this course, it contributes additional energy to the dark energy sector, which slightly hikes the expansion rate of the universe \cite{Smith:2020rxx} in such a way that the value of $H_0^{\mathrm{Planck}}$ can be increased by $8.3\%$ in order to match with the value of $H_0^{\mathrm{SN}}$\footnote{Because $\frac{H_0^{\mathrm{SN}}-H_0^{\mathrm{Planck}}}{H_0^{\mathrm{Planck}}}\times 100 \approx 8.3$.}. Therefore, the scalar field $\xi$, which plays the roles of inflaton during slow-roll and DE at the present day, can also play another role, called EDE at ESP near matter-radiation equality before CMB decoupling. For more details see figure 9 of Ref. \cite{Brissenden:2023yko}. In this way, the idea of EDE shows an effective resolution of the Hubble tension. However, this change of the Hubble expansion does not calibrate the cosmological parameters of $\Lambda$CDM model constrained by Planck because the decay rate of the EDE field energy is so fast that its effect does not appear in the CMB anisotropies and polarizations \cite{Pettorino:2013ia}. Consequently, the EDE remains a hypothetical entity \cite{Brissenden:2023yko} whose apparent role is to tune the expansion to alleviate the $H_0$ discrepancy. \par In the present work, we employ the above-mentioned aspects in constraining the model parameters, specifically the $\alpha$, through a quantum dynamical mode analysis of the inflationary perturbations and the resulting cosmological parameter estimations, following the path of earlier studies \cite{Sarkar:2021ird,Sarkar:2023cpd}. But, at first, we would like to study how the potential responds to the variations of its model parameters, which will guide us to pick up right choices of the parameter space for the subsequent analysis.
\begin{figure}[H]
    \begin{subfigure}{0.5\linewidth}
  \centering
   \includegraphics[width=70mm,height=55mm]{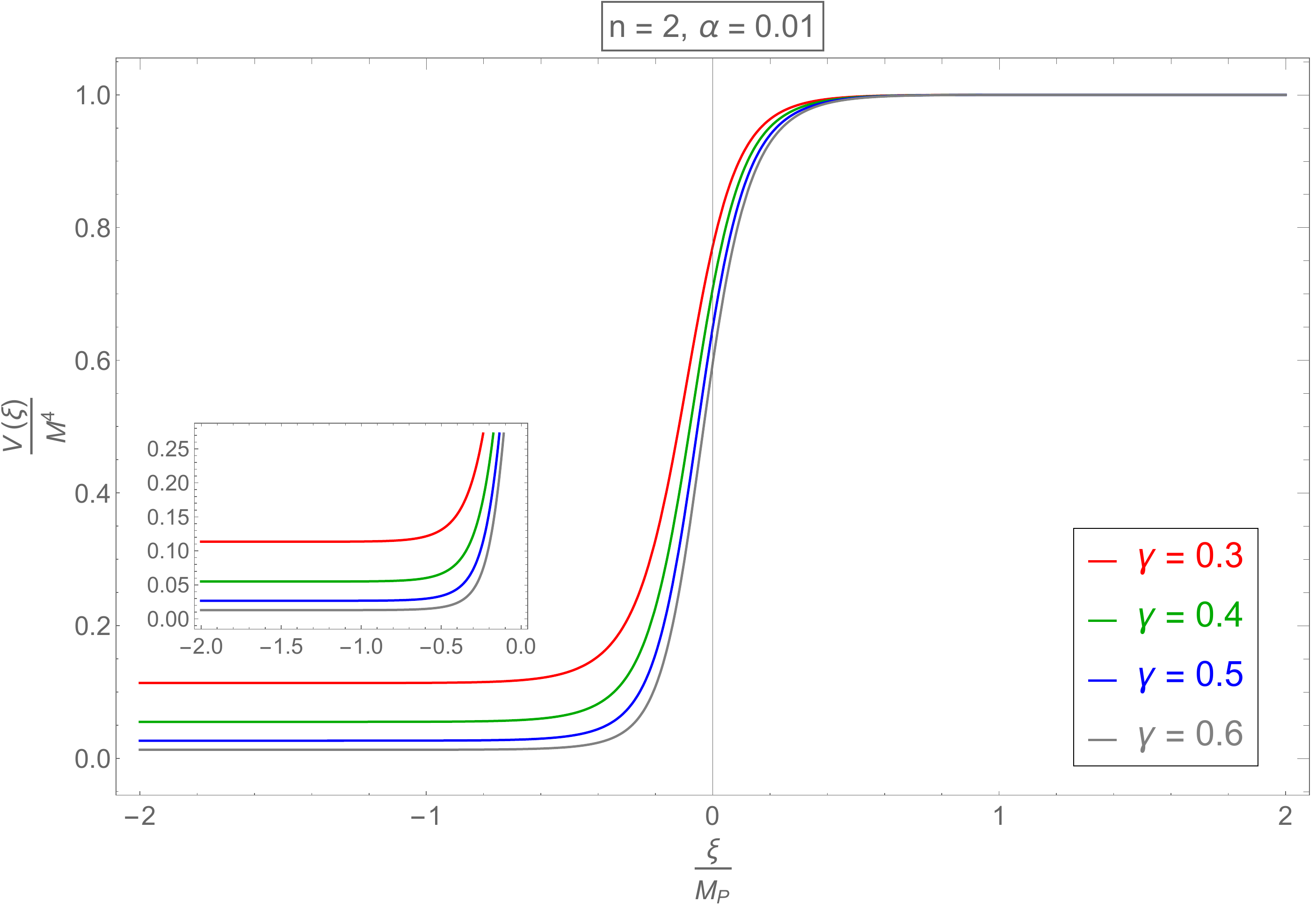}
   \subcaption{}
   \label{fig:Fig2a}
\end{subfigure}
\begin{subfigure}{0.5\linewidth}
  \centering
   \includegraphics[width=70mm,height=55mm]{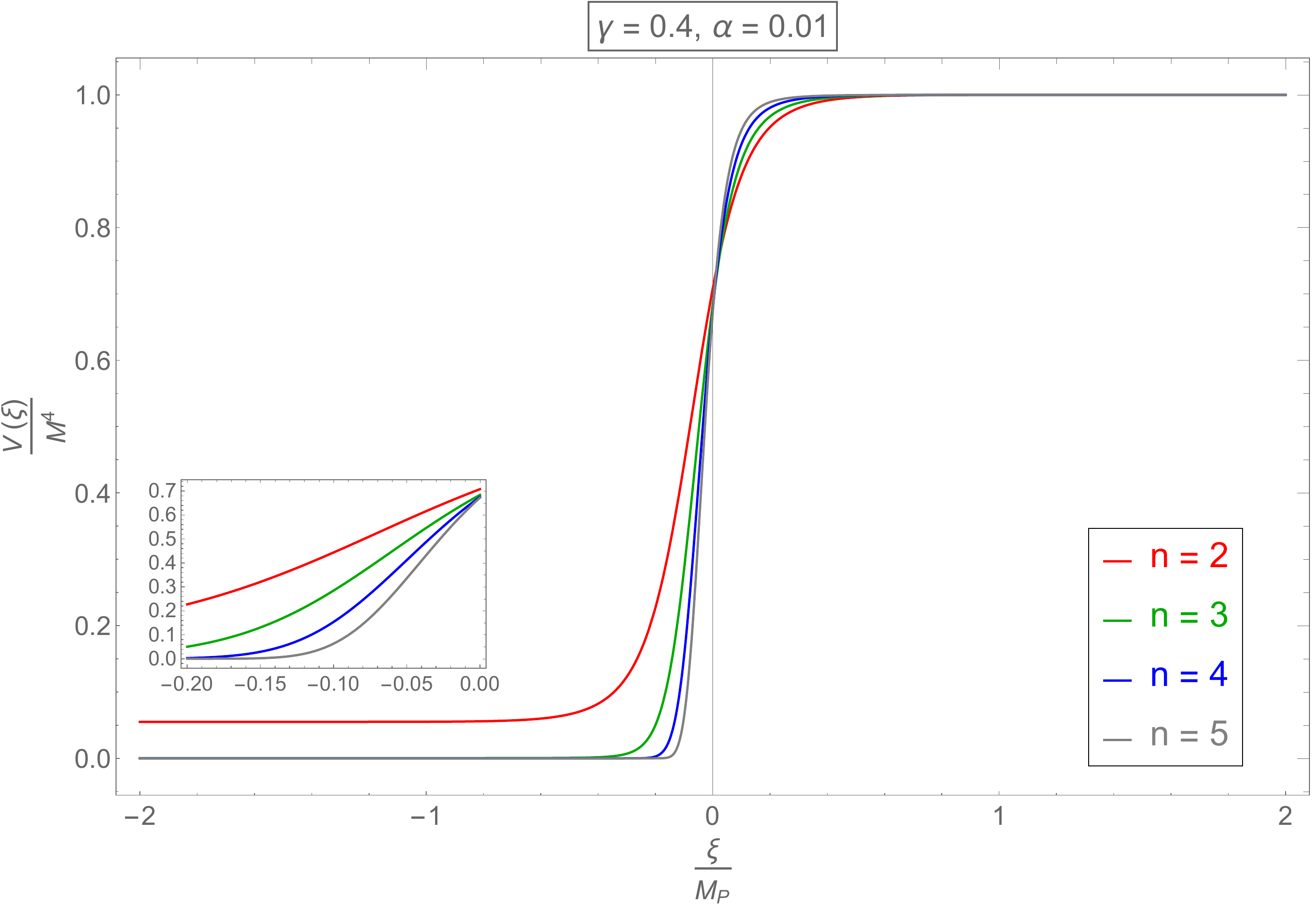}
   \subcaption{}
    \label{fig:Fig2b}
\end{subfigure}
     \caption{Figure \ref{fig:Fig2a} shows the dependence of the ESP with $\gamma$ for $n=2$ and $\alpha=0.01$. As $\gamma$ varies from $\gamma=0.6$ to $\gamma=0.3$ the point of freezing of EDE is uplifted, while in figure \ref{fig:Fig2b} it is almost unaltered with the change of $n$ from $n=2$ to $n=5$ for $\gamma=0.4$ and $\alpha=0.01$. The high values of $n$ play the major role to determine the correct slope of the potential compatible with the EDE-expansion of the universe. Thus small $\gamma (\ll 1)$ and large $n (>1)$ together govern the overall shape of the required potential.}
    \label{fig:Fig2}
\end{figure}
Figures (\ref{fig:Fig2a}) and (\ref{fig:Fig2b}) illustrate the changes in geometry and energy scales of the potential for variations in $\gamma$ and $n$ respectively, for a fixed value of $\alpha$ (here, $\alpha=0.01$). The energy scale of inflation $M$, being independent of $\gamma$ and $n$ (see Eq. (\ref{eq:non_can_limit1})), remains the same while that of quintessence, the $V_{\Lambda}=\exp{\left(-2\gamma\sinh{\left(n\right)}\right)}M^4$ varies with $n$ and $\gamma$ (see Eq. (\ref{eq:non_can_limit2})). Here, the noticeable fact is that when $\gamma$ decreases from $\gamma=0.6$ to $\gamma=0.3$ in figure \ref{fig:Fig2a}, the ESP, $V(0)=\exp{\left[-\gamma\left(1-e^{-n}\right)\right]}M^4$ increases slowly. In figure \ref{fig:Fig2b}, the ESP is almost fixed by the variation of $n$ for a given value of $\gamma$ (here, $\gamma=0.4$), because $\left[-\gamma (1-e^{-n})\right]\approx (-\gamma)$ for $n>1$. Actually the index $n$ controls the steepness of the potential. As $n$ increases from $n=2$ to $n=5$ the gradient increases, signifying faster and faster decay rate of the EDE field energy which is also an important criterion for the EDE dynamics. Therefore, slightly higher values of $n$ are preferable. Now, if the potential is approximated near equality at origin, then we get,
\begin{equation}
    \lim_{\xi\rightarrow 0} V(\xi)=\exp{\left[-\gamma\left(1-e^{-n}\right)\right]}M^4 \exp{\left(\frac{n\gamma\xi}{\sqrt{6\alpha}M_P}\right)}.
\end{equation}
Refs. \cite{Copeland:1997et,Copeland:2006wr,Brissenden:2023yko} show that for such an approximate exponential dependence of the potential near equality (called the \textit{exponential attractor} or \textit{scaling attractor}), the density parameter of the field at equality takes the form
\begin{equation}
    \Omega_{\mathrm{equal}}\simeq \frac{18\alpha}{(n\gamma)^2}<1.
    \label{eq:EDEomega}
\end{equation}
Therefore, for a fixed value of $\Omega_{\mathrm{equal}}$, slightly high value of $n$ corresponds to a small value of $\gamma$, which are the necessary conditions for the successful model building of quintessential inflation with EDE. In fact, such choices of $n$ and $\gamma$ are consistent with the experimental bounds of energy scales of inflation and present-day vacuum density, which we shall explain in Section \ref{sec: result}.
\begin{figure}[H]
	\centering
	\includegraphics[width=0.8\linewidth]{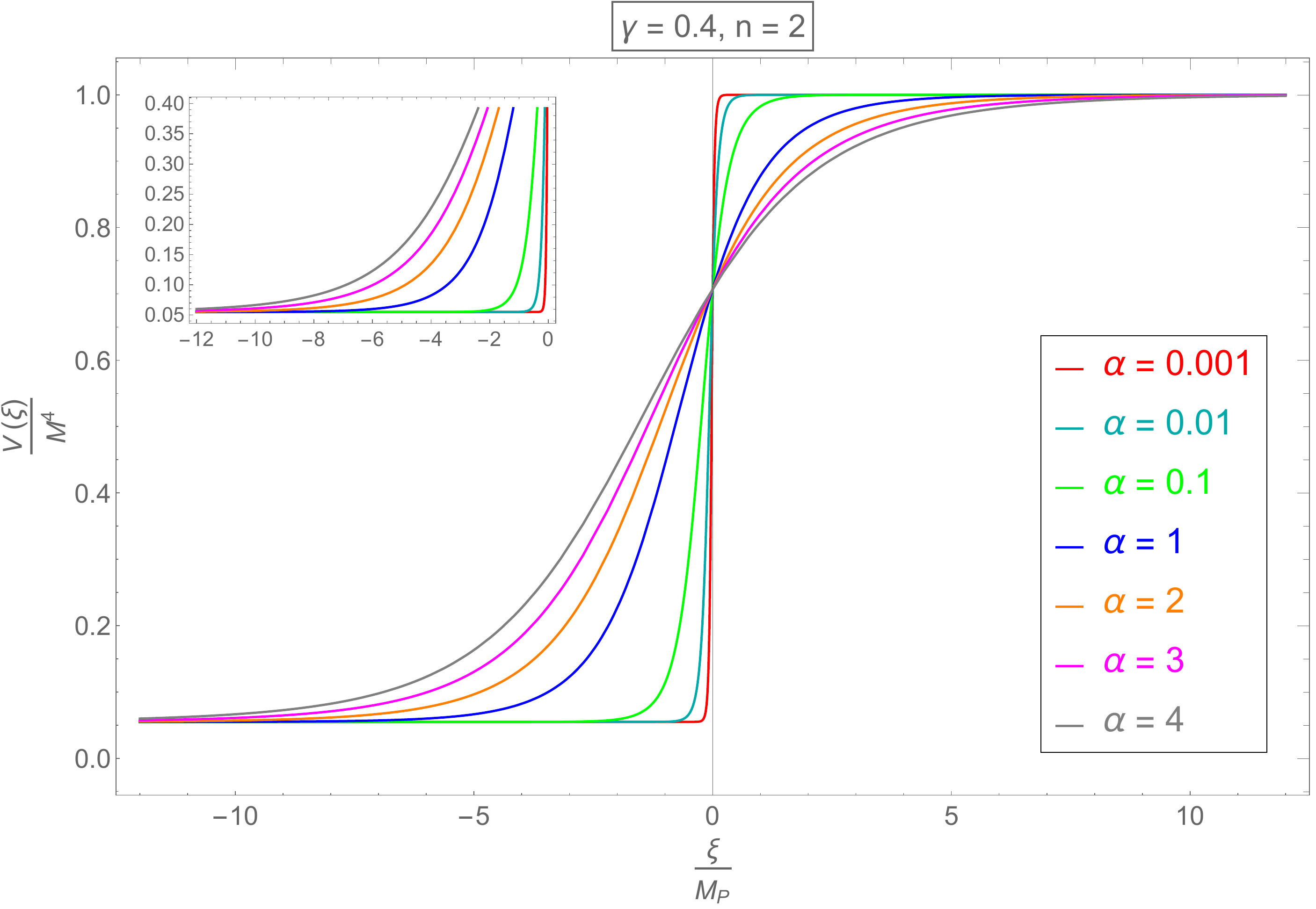}
	\caption{As $\alpha$ increases from $\alpha=0.001$ to $\alpha=4$ the potential is deformed from slow-roll to polynomial type. Low values of $\alpha$ ($\alpha<1$) give steeper free-fall of the field compared to the high values of $\alpha (\alpha >1)$. Therefore the fractional values of $\alpha$ are suitable for the framework of EDE in quintessential inflation.}
	\label{fig:Fig3}
\end{figure}
The parameter set $(n,\gamma)$, is not adequate for fixing the energy densities $M$ and $V_{\Lambda}$ completely. An appropriate value of $\alpha$ is equally important here, not only for fitting the scales, but also for achieving a single-field  model of concave type with a flat direction during inflation. In this context, figure \ref{fig:Fig3} depicts the variation of the potential with $\alpha$ for $n=2$ and $\gamma =0.4$, for example. As $\alpha$ grows from $\alpha=0.001$ to $\alpha=4$, the potential shows the same double pole behaviour as that of ordinary $\alpha$-attractor and quintessential $\alpha$-attractor models, for which mode analyses were done in Refs. \cite{Sarkar:2021ird,Sarkar:2023cpd} respectively. For the smaller values of $\alpha$, the potential is highly concave resulting in an infinitely extended region over which the inflaton field rolls very slowly. As $\alpha$ crosses the value $\alpha\approx 0.1$, the potential loses its concave nature and starts to become convex of the simple polynomial-type chaotic inflation. This is one way of looking at the picture from the angle of inflation. We can also interpret the phenomenon in the perspective of EDE. When $\alpha$ is fractional, the potential shows a steeper slope, which is beneficial for the fast decay of the EDE energy, which does not appear in the CMBR \cite{Brissenden:2023yko}. Therefore, small fractional values of $\alpha$ are quite favourable for the EDE field dynamics in quintessential formulation of inflation. We shall further analyze these points in the sub-Planckian $k$-space analysis of the inflationary modes in Section \ref{sec: result} and understand that the ESP will restrict the allowed range of $\alpha$ more precisely. \par It would be convenient to redesign the complicated potential of Eq. (\ref{eq:new_can_pot}) into two simplified forms by successive approximations for two extreme limits of $\xi$ ($\textit{i.e.}~ \xi =\pm \infty$) corresponding to the inflationary plateau and quintessential tail. From Eq. (\ref{eq:new_can_pot}) in $\xi\rightarrow +\infty$ we get,
\begin{equation}
    \begin{split}
        V(\xi)&= \exp{\left(\gamma e^{-n}\right)}M^4 \exp{\left[-\gamma e^{-n\tanh{\left(\frac{\xi}{\sqrt{6\alpha}M_P}\right)}}\right]}\\
        &\approx \exp{\left(\gamma e^{-n}\right)}M^4 \exp{\left[-\gamma e^{-n\left(1-2e^{-\sqrt{\frac{2}{3\alpha}}\frac{\xi}{M_P}}\right)}\right]}\\
        &\approx \exp{\left(\gamma e^{-n}\right)}M^4\exp{\left[-\gamma e^{-n}\left(1+2n e^{-\sqrt{\frac{2}{3\alpha}}\frac{\xi}{M_P}}\right)\right]}\\
        &\approx M^4\left[1-e^{-n}2n\gamma\exp{\left(-\sqrt{\frac{2}{3\alpha}}\frac{\xi}{M_P}\right)}\right]\\
        &= V_{\mathrm{inf}}(\xi).
    \end{split}
\end{equation}
Similarly for $\xi\rightarrow -\infty$ gives,
\begin{equation}
    \begin{split}
        V(\xi)&= \exp{\left(\gamma e^{-n}\right)}M^4 \exp{\left[-\gamma e^{-n\tanh{\left(\frac{\xi}{\sqrt{6\alpha}M_P}\right)}}\right]}\\
        &\approx \exp{\left(\gamma e^{-n}\right)}M^4 \exp{\left[-\gamma e^{-n\left(-1+2e^{\sqrt{\frac{2}{3\alpha}}\frac{\xi}{M_P}}\right)}\right]}\\
         &\approx \exp{\left(\gamma e^{-n}\right)}M^4\exp{\left[-\gamma e^{n}\left(1-2n e^{\sqrt{\frac{2}{3\alpha}}\frac{\xi}{M_P}}\right)\right]}\\
         &\approx \exp{\left[-\gamma\left(e^{n}-e^{-n}\right)\right]}M^4\left[1+e^{n}2n\gamma \exp{\left(\sqrt{\frac{2}{3\alpha}}\frac{\xi}{M_P}\right)}\right]\\
         &=\exp{\left[-2\gamma\sinh{(n)}\right]}M^4\left[1+e^{n}2n\gamma \exp{\left(\sqrt{\frac{2}{3\alpha}}\frac{\xi}{M_P}\right)}\right]\\
         &= V_{\Lambda}\left[1+e^{n}2n\gamma \exp{\left(\sqrt{\frac{2}{3\alpha}}\frac{\xi}{M_P}\right)}\right]\\
         &= V_{\mathrm{quint}}(\xi).
    \end{split}
    \label{eq:def_CC}
\end{equation}
\begin{figure}[H]
    \begin{subfigure}{0.5\linewidth}
  \centering
   \includegraphics[width=70mm,height=55mm]{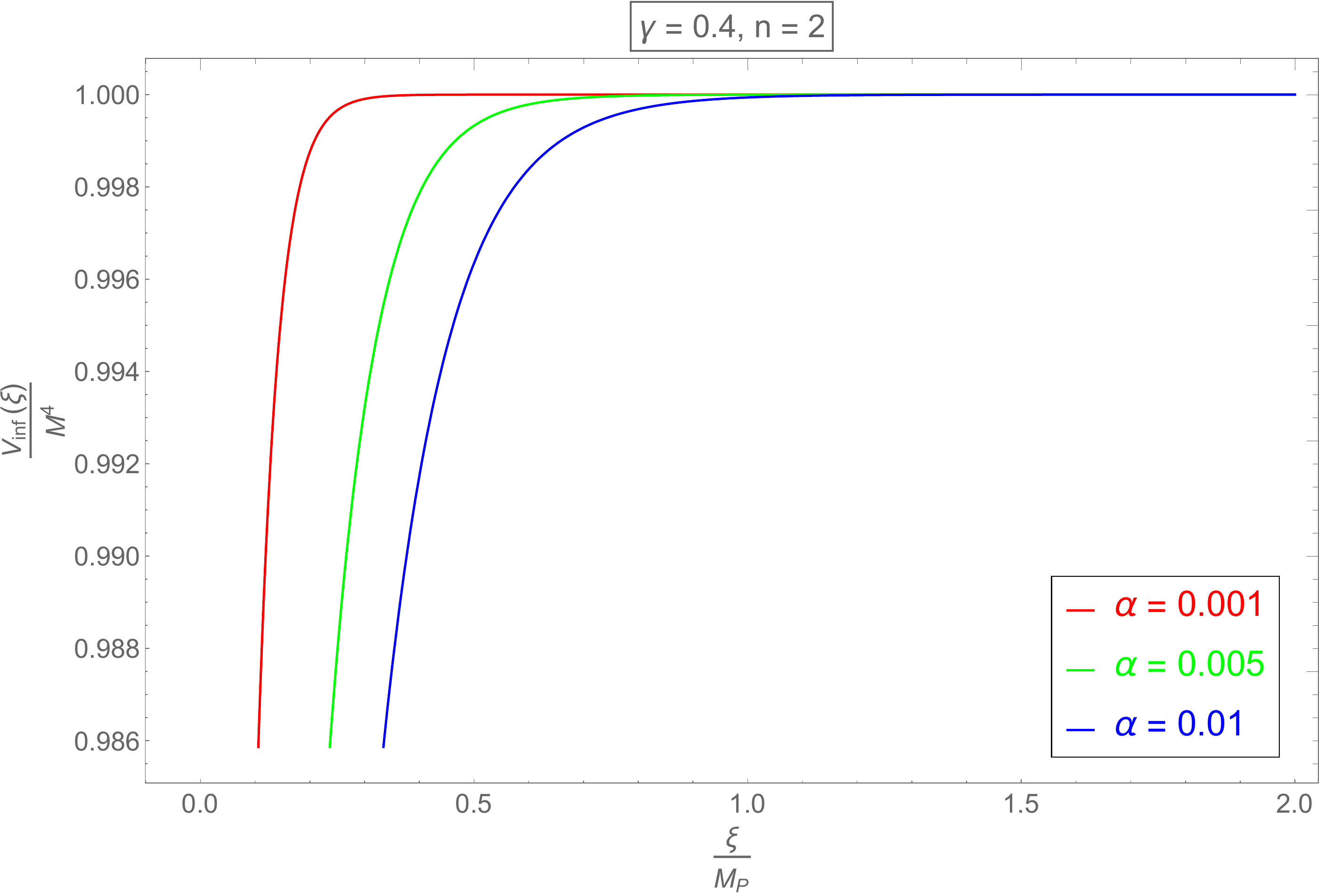}
   \subcaption{}
   \label{fig:Fig4a}
\end{subfigure}
\begin{subfigure}{0.5\linewidth}
  \centering
   \includegraphics[width=70mm,height=55mm]{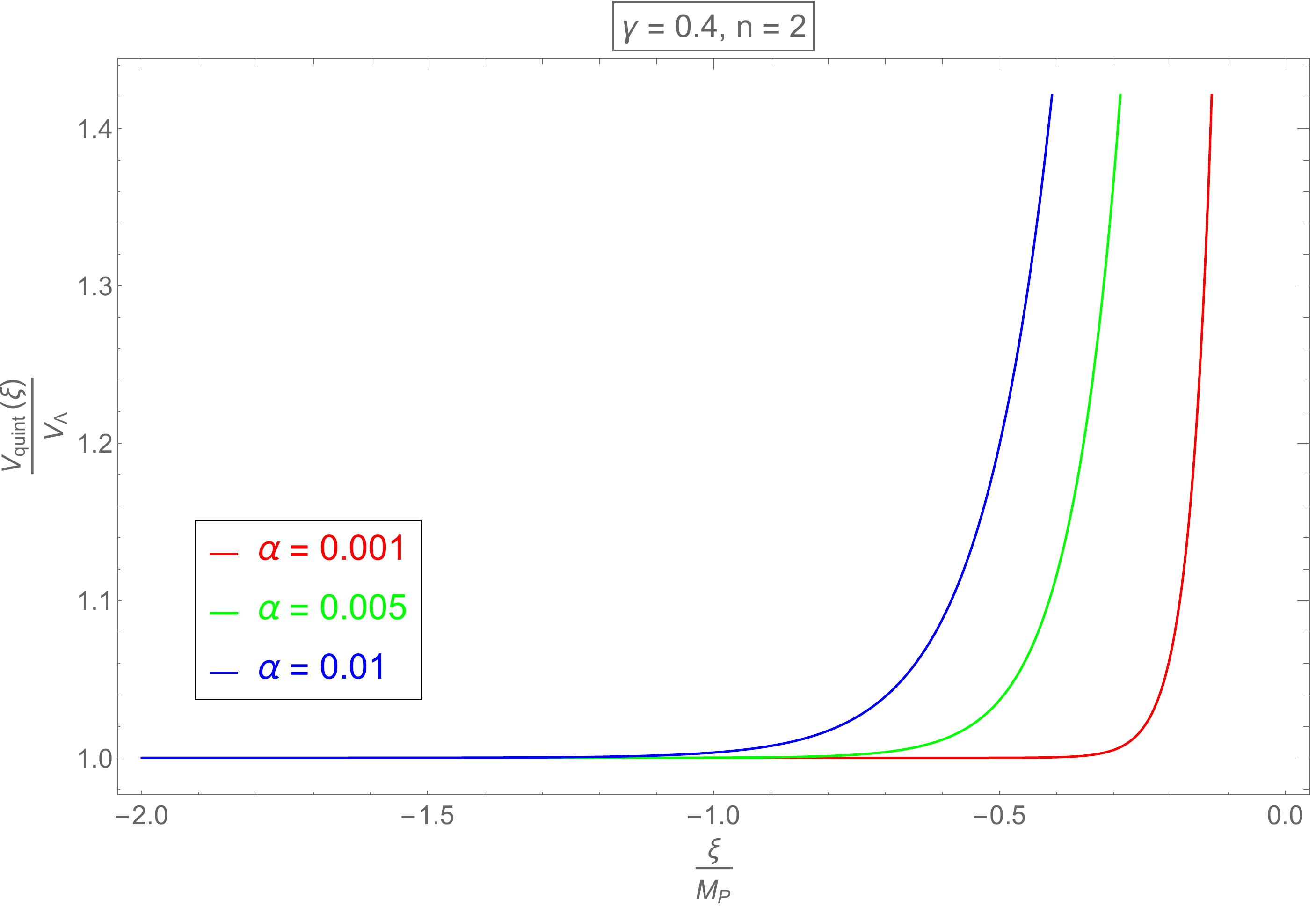}
   \subcaption{}
    \label{fig:Fig4b}
\end{subfigure}
     \caption{Approximated versions of Eq. (\ref{eq:new_can_pot}) in $\xi\gg 0$ and $\xi\ll 0$ limits for $\gamma =0.4$, $n=2$ and $\alpha=0.001, ~0.005$ and $0.01$. Figure \ref{fig:Fig4a} corresponding to the former limit describes the simplified inflaton potential and likewise the latter limit signifies the quintessential potential, shown in figure \ref{fig:Fig4b}. Both the potentials end with their respective normalization constants asymptotically.}
    \label{fig:Fig4}
\end{figure}
Thus, we finally obtain the required limiting potentials for inflation and quintessence, respectively, as
\begin{equation}
    V_{\mathrm{inf}}(\xi)=M^4\left[1-e^{-n}2n\gamma\exp{\left(-\sqrt{\frac{2}{3\alpha}}\frac{\xi}{M_P}\right)}\right]
    \label{eq:final_inf_pot}
\end{equation} and 
\begin{equation}
    V_{\mathrm{quint}}(\xi)=V_{\Lambda}\left[1+e^{n}2n\gamma \exp{\left(\sqrt{\frac{2}{3\alpha}}\frac{\xi}{M_P}\right)}\right].
    \label{eq:final_quint_pot}
\end{equation}
In figures \ref{fig:Fig4a} and \ref{fig:Fig4b} we have plotted these two potentials normalized by $M^4$ and $V_{\Lambda}$ respectively, against $\xi/M_P$ for $n=2$, $\gamma = 0.4$ and $\alpha =0.001,~ 0.005$ and $0.01$. $V_{\mathrm{inf}}(\xi)$ reaches $M^4$ asymptotically in positive side of $\xi$ and the $V_{\mathrm{quint}}(\xi)$ dies down to $V_{\Lambda}$ in the same way for the negative $\xi$-values. \par Now, we shall proceed with the simplified version of quintesssential $\alpha$-attractor potential with EDE of Eq. (\ref{eq:final_inf_pot}) with the concept that $\xi$ is sub-Planckian\footnote{that means non-canonically it remains always within the pole boundaries.} in entire field space, which is an essential requirement for the dynamical mode analysis, described in the next section\footnote{We do not directly deal with the quintessential potential of Eq. (\ref{eq:final_quint_pot}), as the main work of the present paper is the microscopic mode analysis of the inflaton perturbation in $k$-space.}.
\section{Quantum treatment of perturbation during inflation and associated parameters --- a very brief overview}
\label{sec:formalism}
Now, we shall describe, in a nutshell, the theories applied for the quantum $k$-mode analysis to explore the aspects of EDE during inflation in parameter estimation process. This calculational framework has been developed in Ref. \cite{Sarkar:2021ird} for $\alpha$-atractor and further elaborated in Ref. \cite{Sarkar:2023cpd} in the context of quintessential $\alpha$-attractor. The reader is referred to Refs. \cite{Sarkar:2021ird,Sarkar:2023cpd} for details.
\subsection{Quasi de-Sitter Hubble-exit of the dynamical inflaton modes and their evolutions}
\label{sec:k space}
The primordial quantum nuggets are found to be nearly scale-invariant, non-Gaussian and adiabatic, which implies that all the components of the universe including the dark energy/quintessence have emerged from the undulations of one single scalar field - the inflaton field \cite{Baumann:2009ds,Baumann:2022mni}. In this process, the inflaton modes play the pivotal role. During the contraction of the Hubble sphere along with the exponential expansion of spacetime, they exit the horizon and become frozen. After these modes, encoded with the initial conditions, re-enter the causal sphere, they evolve as classical density perturbations \cite{Dodelson:2003ip}.\par The inherent uncertainties of the quantum modes ascribe a dynamical behaviour to the exiting mechanism, over a quasi de-Sitter metric background following $k=aH$ for the scale factor $a$ and Hubble parameter $H$, which we term as \textit{dynamical horizon exit} (DHE) \cite{Sarkar:2023cpd}. This acts like a one-to-one mapping $\varphi:\mathbb{R}\rightarrow \mathbb{R}$ between $t$ and $k$ spaces as $k=\varphi (t)=a(t) H(t)$. We can consider the mapping $\varphi$ as an operator $\hat{\varphi}$ in $\mathbb{R}$ and then DHE implies a non-standard derivative identity
\begin{equation}
    \hat{\varphi}\left(\Ddot{a}^{-1}\partial_t\right)\equiv \partial_k \quad \mathrm{for}\quad k\in \mathbb{K}\subset\mathbb{R}
    \label{eq:derivative_identity}
\end{equation}
where $\mathbb{K}$ is a subspace of all the \textit{dynamical modes} crossing the horizon. This relation converts the temporal evolution of the inflaton field into its mode-dependent evolution in the inflationary metric space.\par A particular statistical correlation always exists  between two quantum nuggets over the expanding background, which carries the information about the specific initial condition buried indside the quantum fluctuations. It can be of scalar or tensor type depending upon the nature of perturbation in the metric. This correlation acts like a cosmic code of the present observable universe involving the CMBR through various parameters like power spectra, spectral indices and tensor-to-scalar ratio \textit{etc.} We shall briefly describe them for the model concerned, in the next subsection.\par Here, we follow the linear perturbative framework which is the simplest and the most effective way for decryption of CMB angular power spectra, as suggested by Planck data \cite{Planck:2018jri,Planck:2018vyg}. Thus, we express the quantum mode function $\xi(t,k)$ as a linear combination of its zerorth order ($\xi^{(0)}(t)$) and first order ($\delta\xi(t,k)$) parts as
\begin{equation}
    \xi(t,k)=\xi^{(0)}(t)+\delta\xi(t,k).
\end{equation} The dynamics of $\xi^{(0)}(t)$ and $\delta\xi(t,k)$ are dictated by their respective time domain evolution equations derived in Ref. \cite{Sarkar:2021ird} from Friedmann equations and continuity equation of Einstein's general theory of relativity.
\par Now, when we map the temporal dependencies of the evolution equations into the mode dependencies by the DHE condition $k=a(t)H(t)$ using the derivative transformation of Eq. (\ref{eq:derivative_identity}), the modes become dynamical which cross the Hubble horizon at random times. This randomness is microscopically induced in the initial conditions, manifesting in the statistical correlations of the perturbations through mode-dependent power spectra and spectral indices. In order to measure these parameters precisely in momentum space, we need to solve a set of $k$-mode evolution equations of the inflaton field ($\xi^{(0)}(k)$) and its first order perturbation ($\delta\xi(k)$) over the fluctuating metric background, which is a certain gauge invariant combination of metric perturbation, called the Badeen ptential \cite{Baumann:2009ds}. Such equations have been derived in \cite{Sarkar:2021ird} using the DHE and the attributes for slow-roll inflation.\par The perturbative evolution equations in $k$-space thus constitute a system of non-linearly coupled ordinary differential equations of three independent variables $\xi^{(0)}$, $\delta\xi$ and the Bardeen potential. But, according to the adiabaticity condition\footnote{The condition says that the amplitude of the adiabatic spectrum of the perturbation is characterized by co-moving curvature perturbation $\mathcal{R}$ only. In spatially flat gauge, $\mathcal{R}$ is measured in terms of the inflaton perturbation (which is $\delta\xi$, here) as $\mathcal{R}=H\frac{\delta\xi}{\Dot{\xi}^{(0)}}$ \cite{Baumann:2009ds}. Therefore, in our context, $\delta\xi$ is the only degree of freedom for calculating the cosmological parameters.} \cite{Bartolo:2001rt,Gordon:2000hv}, only the self-consistent solution of $\delta\xi$ under appropriate boundary values will be utilised to monitor the mode responses of the cosmological parameters, described below.
\subsection{Inflationary parameters for EMQA model}
\label{subsec:cosmoparam}
The statistical correlations of scalar and tensor perturbations connected with the quantum fluctuations as initial conditions are quantified by various cosmological parameters constrained by Planck \cite{Planck:2018vyg,Planck:2018jri}. These parameters are dynamical and hence momentum dependent. In this section, we derive some of the required parameters for the potential considered in Eq. (\ref{eq:final_inf_pot}) using all the formulae mentioned in earlier works \cite{Sarkar:2021ird,Sarkar:2023cpd}.\par
The number of remaining e-folds of an inflaton mode (sometimes called, \textit{scale}) at the moment of leaving the horizon is obtained as,
\begin{equation} N(\xi)=\frac{3\alpha}{4n\gamma e^{-n}}\left[e^{\sqrt{\frac{2}{3\alpha}}\xi}-e^{\sqrt{\frac{2}{3\alpha}}\xi_{\mathrm{end}}}\right]-\sqrt{\frac{3\alpha}{2}}\left(\xi-\xi_{\mathrm{end}}\right)
    \label{eq:e_folds_1}
\end{equation}
where, the end-value of the scalar field,
\begin{equation}
    \xi_{\mathrm{end}}=\sqrt{\frac{3\alpha}{2}}\ln{\left[\gamma e^{-n}\left(\frac{2n}{\sqrt{3\alpha}}+2n\right)\right]}
    \label{eq:e_folds_3}
\end{equation} corresponds to unit value of the first potential slow-roll parameter. This is the point where the inflation ends. For a successful model building, this point should be as far as possible from the starting point\footnote{This is actually an important requirement for large field model \cite{Baumann:2009ds} \textit{e.g.} slow-roll inflation, which is found to be effective in parameter estimation as per Planck data.} \textit{i.e.} $\xi_{\mathrm{end}}\ll\xi$. In fact, we can verify it by putting, for example, $n=2$, $\alpha=0.001$ and $\gamma=0.4$ in Eq. (\ref{eq:e_folds_3}) which yields $\xi_{\mathrm{end}}=0.055$, which is significantly small.\par Putting the expression of $\xi_{\mathrm{end}}$ of Eq. (\ref{eq:e_folds_3}) in Eq. (\ref{eq:e_folds_1}) and keeping only the dominating exponential terms we obtain,
\begin{equation}
    N(\xi)\approx \frac{3\alpha}{4n\gamma e^{-n}}\left[e^{\sqrt{\frac{2}{3\alpha}}\xi}-\gamma e^{-n}\left(\frac{2n}{\sqrt{3\alpha}}+2n\right)\right].
\end{equation}
For $n>>1$ and $\gamma\ll 1$ (needed for EDE model building) we can further approximate $N(\xi)$ as,
\begin{equation}
     N(\xi)\approx \frac{3\alpha}{4n\gamma e^{-n}}e^{\sqrt{\frac{2}{3\alpha}}\xi}.
     \label{eq:Preefolds}
\end{equation}
Now, we can write the $k$-dependent number of remaining e-folds as
\begin{equation}
    N(k)\approx \frac{3\alpha}{4n\gamma e^{-n}}e^{\sqrt{\frac{2}{3\alpha}}\xi(k)}.
     \label{eq:e_folds_4}
\end{equation}
Here the factor $\gamma e^{-n}$ in the denominator is a new term, which was not present in the earlier model \cite{Sarkar:2023cpd}. This term carries the signature of EDE. The expression of $N(k)$ in $k$-space will be exploited to fix the proper boundary values for solving the dynamical mode equations of perturbations.\par
A two-point correlation \cite{Baumann:2009ds} is found to exist due to quantum fluctuations during inflation  between two scalar perturbations of different momenta. All the crucial information regarding this correlation are encoded in a statistical measure, called the \textit{scalar power spectrum} $\Delta_s(k)$. In the simplest version of linear cosmological perturbation the $\Delta_s(k)$ is Gaussian. Primordial non-Gaussianity \cite{Bartolo:2004if} is found to be negligible in first order single field slow-roll perturbation \cite{Acquaviva:2002ud,Maldacena:2002vr} and is significant mainly in higher order correlations like three-point correlation or \textit{cosmic bispectrum} \cite{Komatsu:2001rj} and multi-field models \cite{Kaiser:2012ak}. Therefore, we do not attempt to compute higher order correlations and proceed with two-point correlation which is effectively Gaussian. In the present model, we get the $\Delta_s(k)$ in dimensionless format in terms of $N(k)$ using Eqs. (\ref{eq:final_inf_pot}) and 
 (\ref{eq:e_folds_4}) as,
\begin{equation}
     \Delta_s(k) = \left(\frac{M^2}{3\pi\sqrt{2\alpha}}\right)^2\left(\frac{\left(N(k)+\sqrt{\frac{3\alpha}{4}}\right)}{\left(N(k)+\sqrt{\frac{3\alpha}{4}}+\frac{3\alpha}{2}\right)^{1/3}}\right)^3.
    \label{eq:modeScalarPower}
\end{equation}\par Apart from probing the statistical information of primordial perturbations, the power spectra specifically $\Delta_s(k)$ has another crucial role to play. It actually determines the scale of inflation, $M$, through the amplitude of scalar perturbation $A_s(k)$ by the technique described in Ref. \cite{Sarkar:2023cpd} with respect to a pivot scale $k_{*}$\footnote{The order of this scale is actually same as that of the modes of the inflaton field, which indicates the range of $k$ at which the horizon exit occurs. This aspect will be more clear in next section.} for which $\Delta_s(k_{*})=A_s(k_{*})$. Following this method we obtain
\begin{equation}
     M^4(A_s(k_{*}),N(k_{*}),\alpha) = 18\pi^2 \alpha A_s(k_{*})\left(N(k_{*})+\sqrt{\frac{3\alpha}{4}}\right)^{-2}e^{\left(\frac{3\alpha}{2}\left(N(k_{*})+\sqrt{\frac{3\alpha}{4}}\right)^{-1}\right)}.
     \label{eq:COBEnormalization}
\end{equation}
This is widely known as the `\textit{COBE/Planck normalization}' for inflationary energy scale or the scale of inflaton potential.   
\begin{figure}[H]
	\centering
	\includegraphics[width=0.8\linewidth]{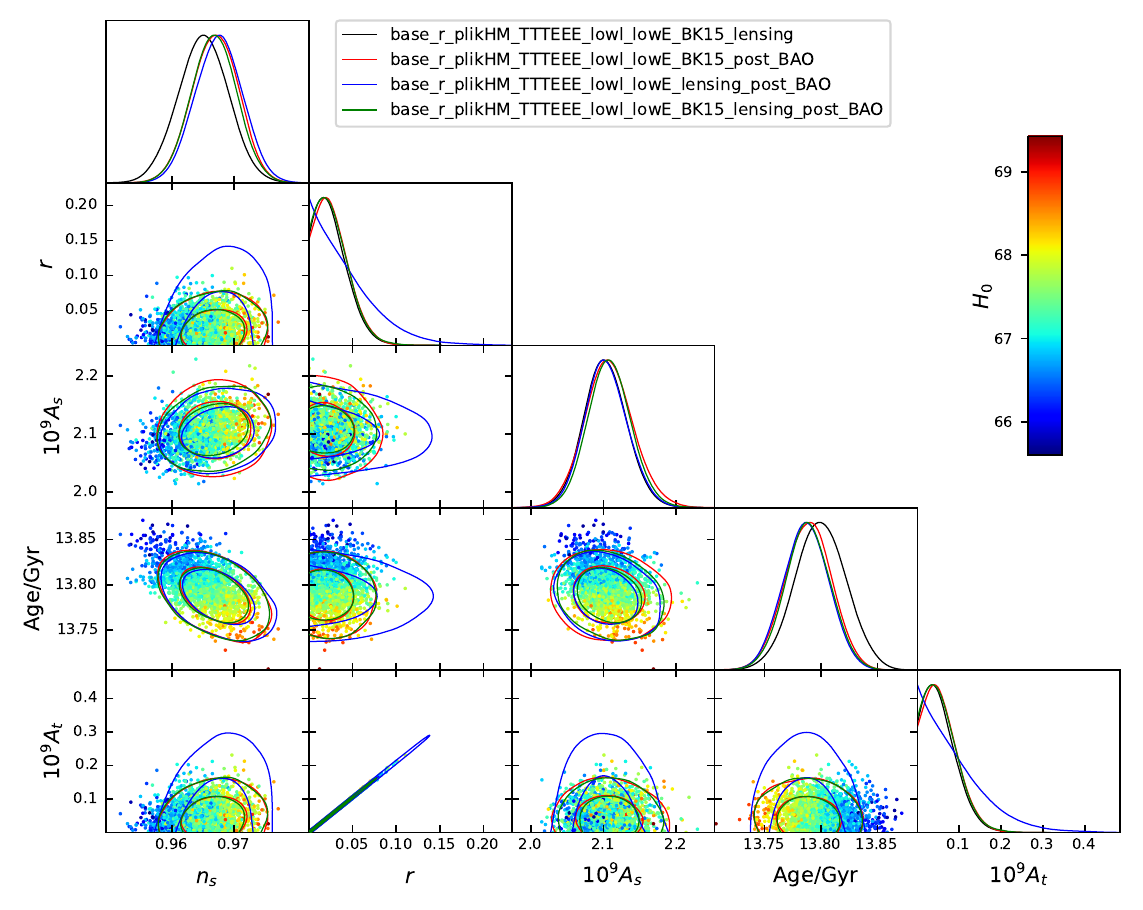}
	\caption{Planck constraints for various parameters to satisfy the present day Hubble parameter indicated in Eq. (\ref{eq:Planck_Hubble}). This plot has been generated within GetDist (\url{https://getdist.readthedocs.io/en/latest/}) plotting utility by running the simulation data available in Planck Legacy Archive (\url{https://pla.esac.esa.int/}).}
	\label{fig:5}
\end{figure}
In the same way, the two-point correlation function for two mutually perpendicular tensor perturbations is determined by a dimensionless tensor power spectrum $\Delta_t(k)$. Using Eqs. (\ref{eq:final_inf_pot}) and (\ref{eq:e_folds_4}) we get
\begin{equation}
    \Delta_t(k)=\left(\sqrt{\frac{2}{3}}\frac{M^2}{\pi}\right)^2\left(\frac{\frac{2N(k)}{3\alpha}+\sqrt{\frac{1}{3\alpha}}}{\frac{2N(k)}{3\alpha}+\sqrt{\frac{1}{3\alpha}}+1}\right).
    \label{eq:modeTensorPower}
\end{equation}
\par
The first order logarithmic  scale dependencies of scalar ($\Delta_s(k)$) and tensor ($\Delta_t(k)$) power spectra are determined by two spectral indices \textit{viz.,} the scalar spectral index $n_s(k)$ and tensor spectral index $n_t(k)$ and the relative change of $\Delta_t(k)$ with respect to $\Delta_s(k)$ is given by the tensor-to-scalar ratio $r(k)$. \par It is customary to measure the above-mentioned parameters in terms of number of remaning e-folds $N(k)$. Therefore, using Eqs. (\ref{eq:final_inf_pot}) and (\ref{eq:e_folds_4}) for large $N(k)$ (so that $\mathcal{O}(\frac{1}{N^2})\ll1$) we obtain,
\begin{equation}
    n_s(k)\approx 1-\frac{2}{N(k)},
    \label{eq:final_ns}
\end{equation}
\begin{equation}
    n_t(k)\approx -\frac{3\alpha}{2N(k)^2}
    \label{eq:final_nt}
\end{equation} and 
\begin{equation}
    r(k)\approx \frac{12\alpha}{N(k)^2}.
    \label{eq:final_r}
\end{equation}
Eqs. (\ref{eq:final_ns})-(\ref{eq:final_r}) are of the same forms as those of the famous attractor equations of $\alpha$-attractors, which are also identical to the predictions of quintessential $\alpha$-attractors (see \cite{Sarkar:2023cpd}, \cite{Sarkar:2021ird}). The reason behind the latter resemblance is that all these models consist of the same type of non-canonical quadratic pole structures in kinetic parts of their respective Lagrangians. It is indeed a fundamental property of pole inflation that the cosmological predictions depend upon the nature of kinetic poles and are independent of the form and origin of the potential.\par
Now, Eq. (\ref{eq:final_ns}) implies that $n_s\lesssim 1$, signifying that the primordial cosmological perturbation is \textit{nearly scale invariant}. The tiny mode dependency can be measured in terms of its logarithmic derivatives of various orders. In linear perturbative formalism, it is sufficient to measure the first order one, which is called the running of spectral index $\alpha_s(k)$. \par The rate of inflationary spacetime expansion in Friedmann universe is measured by the corresponding Hubble parameter $H_{\mathrm{inf}}(k)$. Using Eqs. (\ref{eq:final_inf_pot}) and (\ref{eq:e_folds_4}) $H_{\mathrm{inf}}(k)$ can be expressed in terms of $N(k)$ as,
\begin{equation}
    H_{\mathrm{inf}}(k)=M^2\sqrt{\frac{\frac{2N(k)}{\sqrt{3\alpha}}+1}{\frac{6N(k)}{\sqrt{3\alpha}}+3+3\sqrt{3\alpha}}}.
    \label{eq:Inf_Hubble_parameter}
\end{equation}
 \par
According to  Planck-2018 \cite{Planck:2018jri,Planck:2018vyg}, in order to obtain the correct amount of scalar perturbation of amplitude $A_s(k_{*})=2.1\pm 0.03\times 10^{-9}$, tensor perturbation of amplitude $A_t(k_{*})$\footnote{The $A_t(k_{*})$ is calculated by the empirical formula \cite{Baumann:2009ds} $A_t(k_{*})=\Delta_t(k)\left(\frac{k}{k_{*}}\right)^{-n_t(k_{*})}$.} $<0.1\times 10^{-9}$, present day vacuum density $V_{\Lambda}\sim 10^{-120} M_P^4$, scalar spectral index $n_s=0.9649~\pm~ 0.0042$ with tensor-to-scalar ratio $r_{0.002}<0.064$ (see figure \ref{fig:5}\footnote{Based on observations obtained with Planck (http://www.esa.int/Planck), an ESA science mission with instruments and contributions directly funded by ESA Member States, NASA, and Canada.}), running of spectral index $\alpha_s(k_{*})=-0.0045\pm 0.0067$ within $68\%$ CL and the inflationary Hubble parameter $H_{\mathrm{inf}}(k_{*})<6.07\times 10^{13}$ GeV or $H_{\mathrm{inf}}(k_{*})<2.5\times 10^{-5}$ in reduced Planck unit within $95\%$ CL at $k_{*}=0.002$ Mpc$^{-1}$ with $TT+TE+EE+\mathrm{low}l+\mathrm{low}E+\mathrm{lensing}+\mathrm{BAO}$, the scale of the inflaton potential should be $V_{*}^{1/4}=M<6.99\times 10^{-3}$ in reduced Planck unit or $M<1.7\times 10^{16}$ GeV. This requirement is equally important for satisfying the present day Hubble parameter $H_0^{\mathrm{Planck}}=67.44\pm 0.58$ Km$~$s$^{-1}$Mpc$^{-1}$ which determines that the current age of the universe is roughly 13.8 billion years.
\par In the next section, we shall continue our discussion regarding estimation of all necessary cosmological parameters described above in the context of the EMQA model. Our plan is first
to make ready the model for calculations by fitting the inflationary and the quintessential energy scales with COBE/Planck normalisation condition derived in Eq. (\ref{eq:COBEnormalization}). Then, we numerically solve the dynamical mode equations of perturbations as in Refs. \cite{Sarkar:2021ird,Sarkar:2023cpd} by Wolfram Mathematica $12.0$, for the simplified inflaton potential of Eq. (\ref{eq:final_inf_pot}) under appropriate boundary conditions within a certain range of $k$-values. The results are then employed to compute and plot the required parameters using Eqs. (\ref{eq:e_folds_4}), (\ref{eq:modeScalarPower}) and (\ref{eq:modeTensorPower})-(\ref{eq:Inf_Hubble_parameter}) within the same specified $k$-range for a set of predetermined model parameters. In this way, we explore the effects of EDE in parameter estimation process and thereby verify some of its interesting features. At the end, we shall make use of the Planck bounds to constrain the $\alpha$ factor, which is the primary motivation of the present work.
\section{Results and discussions}
\label{sec: result}
\subsection{Prerequisites}
\label{subsec:basic_consideration}
\begin{enumerate}
    \item In the present formalism, the mechanism underlying inflation is the \textit{`dynamical horizon crossing'} of cosmological scales or the inflaton modes. Now, which mode (of momentum $k$) will exit first, depends upon the corresponding number of remaining e-folds $N(k)$. This means that those modes will cross the horizon first for which the number of remaining e-folds is maximum, which is roughly $60$ e-folds as required by Planck observations. The order of these modes/scales can be estimated by that of the pivot scale $k_{*}$ which is around $0.002$ Mpc$^{-1}$. This is the reference scale with respect to which Planck constrains the cosmological parameters. Therefore we can expect that the exiting modes should be of the order of $10^{-3}$ Mpc$^{-1}$. \par Motivated by these ideas we select $N=63.49$\footnote{which is the same value as taken in the quintessential model \cite{Sarkar:2023cpd}.} as the starting point of dynamical crossing of quantum modes. We also choose the same initial conditions as considered in Refs \cite{Sarkar:2021ird,Sarkar:2023cpd}. The outputs are then allowed to plot in the $k$-range $0.001-0.009$ Mpc$^{-1}$. We assume, here, that all the required conditions for the selection of Bunch-Davies vacuum state \cite{Bunch:1978yq,Birrell:1982ix,Kundu:2011sg,Jiang:2016nok} during DHE are fulfilled, as the self-consistency for the solutions of the mode equations is concerned. 
    \item In Ref. \cite{Sarkar:2023cpd}, it was shown that $0.1\leq \alpha\leq 4.3$ is a valid range for $\alpha$ in quintessential $\alpha$-attractor model to explain both the inflationary and the DE expansions successfully. A recent study \cite{Brissenden:2023yko} on non-oscillating EDE model of $\alpha$-attractor reveals that the parameter $\alpha$ should be $\sim 10^{-3}$ in order to solve the $H_0$-tension. Therefore, in the present analysis we consider three hierarchies of $\alpha$-values. In the low end of range we choose $\alpha =0.001,~ 0.005,~ 0.010$, in the mid range $\alpha=0.05,~ 0.10,~ 0.50$ and in high end of range $\alpha=1.0,~ 2.5,~ 4.3$.
    \item We determine the amplitudes of power spectra through COBE/Planck normalisation condition of Eq. (\ref{eq:COBEnormalization}) and the nine value of $\alpha$ described earlier by evaluating the factor $M$ with the considerations $N(k_{*})=63.49$ and $A_s(k_{*})=2.1\times 10^{-9}$ as enlisted in table \ref{tab:Table1}. All the $M$ values are less than $M^{\mathrm{Planck}}$ and therefore conform to the Planck requirements.
    \begin{table}[H]
    \captionsetup{justification=centering,width=0.7\textwidth}
    \caption{COBE/Planck normalised $M$ for nine values of $\alpha$ with $M_P=2.43\times 10^{18}$ GeV.}
    \begin{center}
        \begin{adjustbox}{width=0.6\textwidth}
        \begin{tabular}{|c|c|c|c|c|}
    \hline
     $\alpha$ & $M (M_P)$ & $M$ (GeV) & $M^{\mathrm{Planck}} (M_P)$ & $M^{\mathrm{Planck}}$ (GeV)\\
    \hline\hline
    $0.001$ & $5.58\times 10^{-4}$ & $1.35\times 10^{15}$ & & \\
    $0.005$ & $8.35\times 10^{-4}$ & $2.03\times 10^{15}$ & & \\
    $0.010$ & $9.92\times 10^{-4}$ & $2.41\times 10^{15}$ & & \\
    $0.050$ & $1.48\times 10^{-3}$ & $3.60\times 10^{15}$ & $6.99\times 10^{-3}$ & $1.70\times 10^{16}$\\
    $0.100$ & $1.76\times 10^{-3}$ & $4.28\times 10^{15}$ & & \\
    $0.500$ & $2.64\times 10^{-3}$ & $6.41\times 10^{15}$ & & \\
    $1.000$ & $3.14\times 10^{-3}$ & $7.63\times 10^{15}$ & & \\
    $2.500$ & $3.96\times 10^{-3}$ & $9.62\times 10^{15}$ & & \\
    $4.300$ & $4.57\times 10^{-3}$ & $1.11\times 10^{16}$ & & \\
    \hline
    \end{tabular}
    \end{adjustbox}
    \end{center}
         \label{tab:Table1}
    \end{table}
    \item The other model parameters \textit{viz.,} $\gamma$ and $n$ are picked up on the basis of the model requirements for EDE mentioned in Section \ref{sec:our model}. We choose a small fractional value of $\gamma$, $\gamma=0.0818$ and a little large value of $n$, $n=8$. With this choice of the parameter space, the responses of the simplified potentials for inflation of Eq. (\ref{eq:final_inf_pot}) and for quintessence of Eq.  (\ref{eq:final_quint_pot}) are represented in figures \ref{fig:Fig6a} and \ref{fig:Fig6b}, respectively, for the first family of $\alpha$-values.
\begin{figure}[H]
    \begin{subfigure}{0.5\linewidth}
  \centering
   \includegraphics[width=70mm,height=55mm]{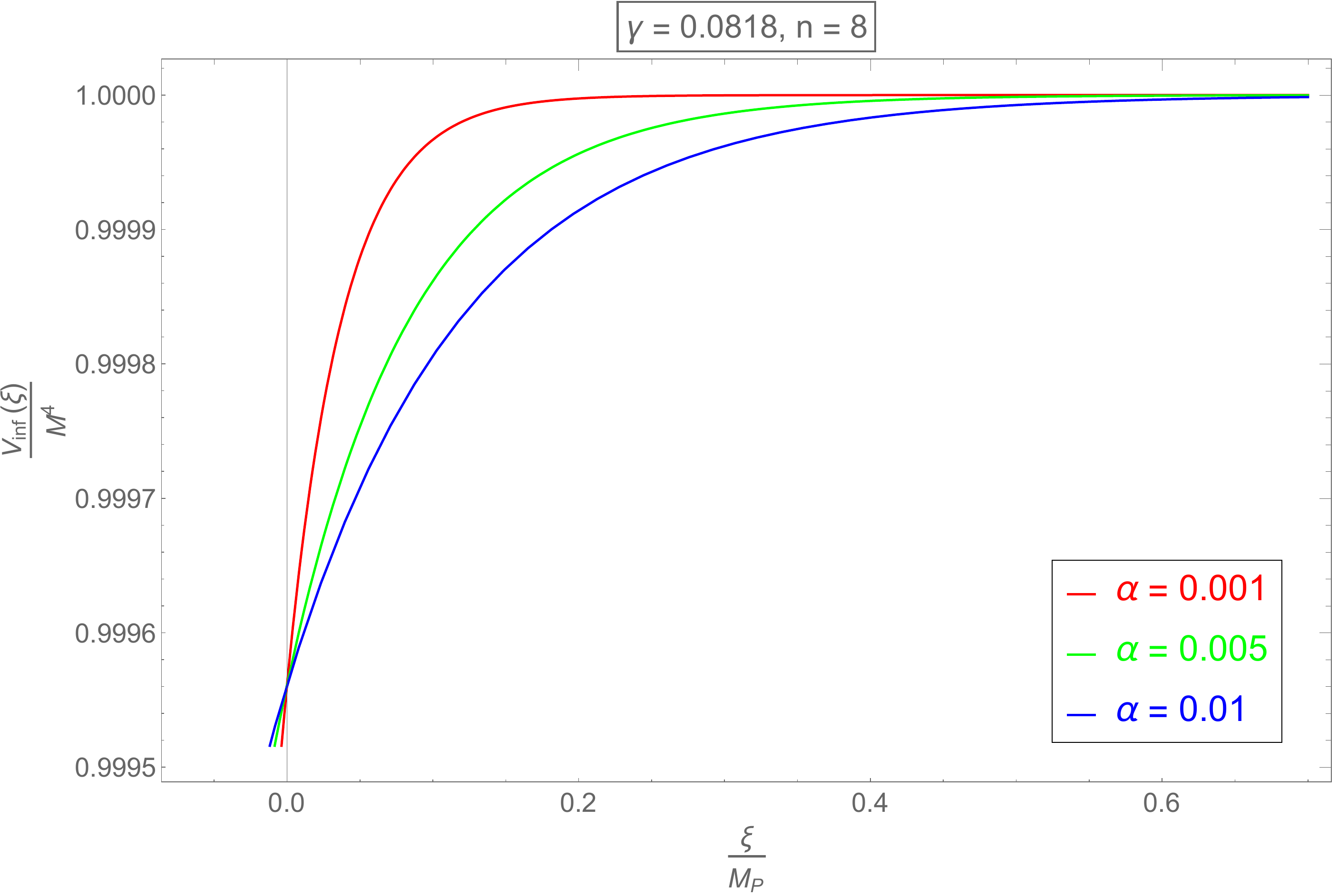}
   \subcaption{}
   \label{fig:Fig6a}
\end{subfigure}
\begin{subfigure}{0.5\linewidth}
  \centering
   \includegraphics[width=70mm,height=55mm]{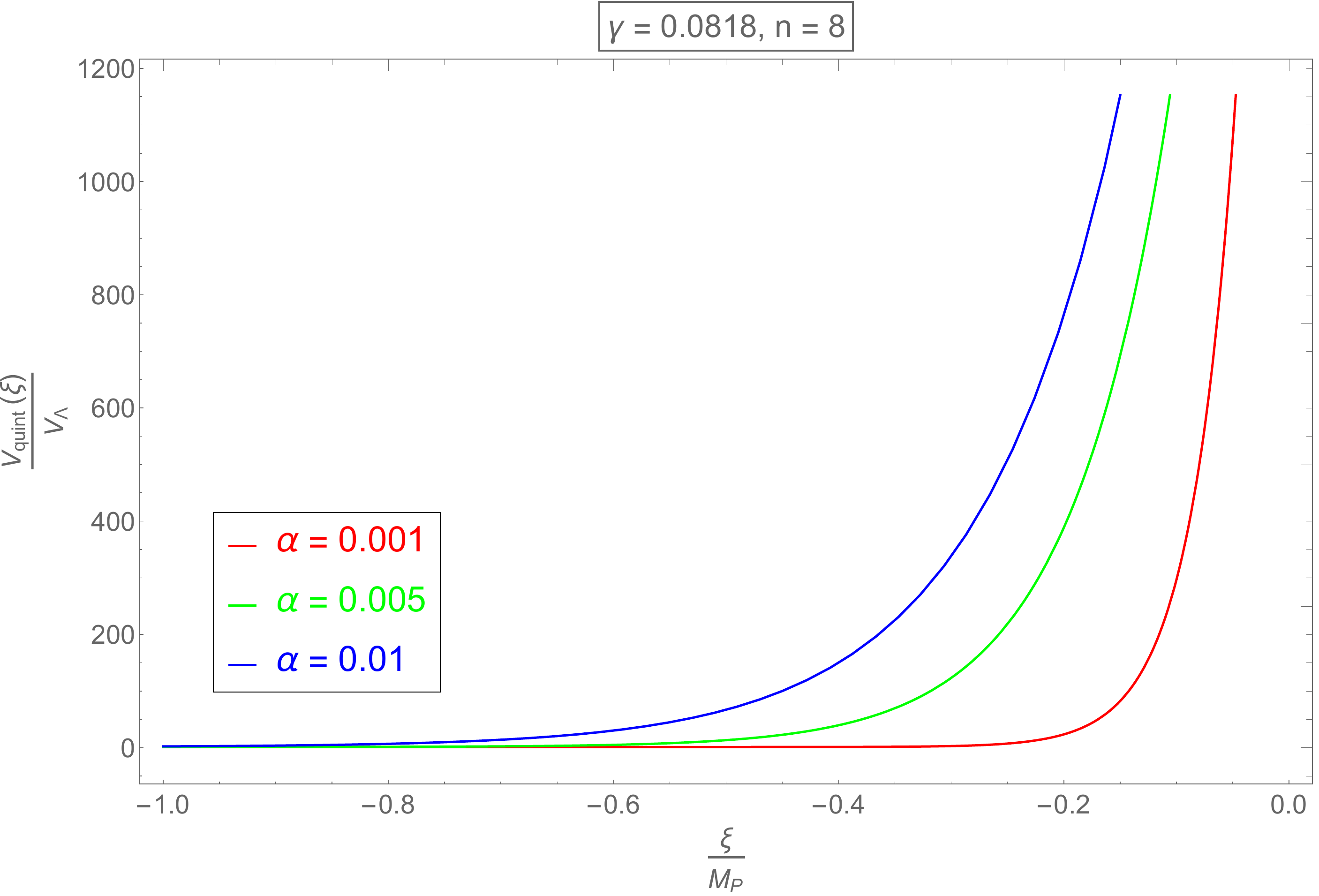}
   \subcaption{}
    \label{fig:Fig6b}
\end{subfigure}
     \caption{Inflaton and quintessential potentials for $\gamma=0.0818$, $n=8$ and three values of $\alpha$.}
    \label{fig:Fig6}
\end{figure}
\item The choice of the parameter space would be satisfactory if the corresponding vacuum density matches with that of the Planck bound. Therefore, let us calculate the amplitude of the quintessential potential of Eq. (\ref{eq:final_quint_pot}) \textit{i.e.} the present vacuum density $V_{\Lambda}$ using the constraints discussed above. We first use the expression of $V_{\Lambda}$ defined in Eq. (\ref{eq:def_CC}),
\begin{equation}
    V_{\Lambda}=\exp{\left[-2\gamma\sinh{(n)}\right]}M^4=10^{-u}\quad (\mathrm{say}).
    \label{eq:final_V_lambda_1}
\end{equation}
Then 
\begin{equation}
    u=\frac{2\gamma\sinh{(n)}-4\ln{M}}{\ln{10}},
    \label{eq:final_V_lambda_2}
\end{equation}
where $M$ is the COBE/Planck normalisation constant measured in $M_P$ unit and $V_{\Lambda}$ is obtained in $M_P^4$ unit. Now, we convert $V_{\Lambda}$ into a new expression given in \cite{Brissenden:2023yko} in presence of EDE
\begin{equation}
\begin{split}
    &V^{\mathrm{exact}}_{\Lambda}=\left(\frac{H_0^{\mathrm{Planck}}}{H_0^{\mathrm{SN}}}\right)^2V_{\Lambda}=0.8525 V_{\Lambda}\\
    \mathrm{or,}&\quad 10^{-v}=0.8525\times 10^{-u}\quad (\mathrm{say}).\\
\end{split}
    \label{eq:final_V_lambda_3}
\end{equation}
Here, $H_0^{\mathrm{Planck}}$ and $H_0^{\mathrm{SN}}$ are the two values of the Hubble parameters from Planck and supernovae measurements, given in Eqs. (\ref{eq:Planck_Hubble}) and (\ref{eq:Planck_SN}) respectively. Ref. \cite{Brissenden:2023yko} shows that, this modified form of $V_{\Lambda}$ is suitable for solving the Hubble tension. In this way, we encode the aspects of resolution of $H_0$ tension by EDE in the chosen set of parameters for the model, discussed in this paper.\par Eq. (\ref{eq:final_V_lambda_3}) gives
\begin{equation}
   v=u-\frac{\ln{0.8525}}{\ln{10}}.
    \label{eq:final_V_lambda_4}
\end{equation}
Putting the expression of $u$ from Eq. (\ref{eq:final_V_lambda_2}) in Eq. (\ref{eq:final_V_lambda_4}) we get,
\begin{equation}
    v=\frac{2\gamma\sinh{(n)}-4\ln{M}}{\ln{10}}-\frac{\ln{0.8525}}{\ln{10}}=\frac{2\gamma\sinh{(n)}-4\ln{M}-\ln{0.8525}}{\ln{10}}.
     \label{eq:final_V_lambda_5}
\end{equation}
\begin{figure}[H]
	\centering
	\includegraphics[width=90mm,height=75mm]{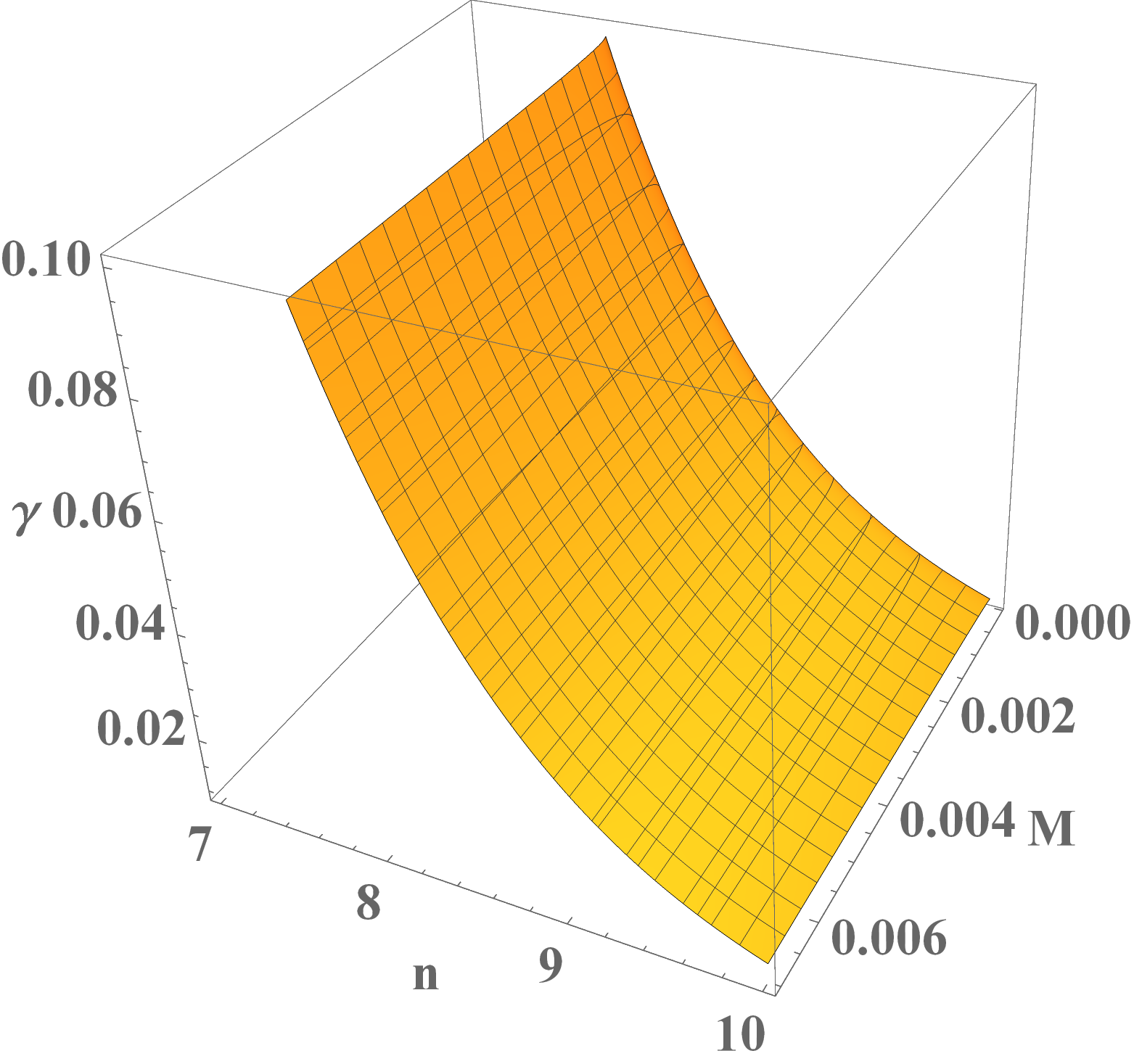}
	\caption{Three dimensional contour plot of $n$, $\gamma$ and $M$ for $v=120$ according to Eq. (\ref{eq:final_V_lambda_5}). The figure shows that $\gamma$ and $n$ should belong in the closed intervals $[0.01,0.09]$ and $[8,10]$ respectively in order to have the $M$ values within the COBE/Planck limit $M<6.99\times 10^{-3} M_P$ and vacuum density $V_{\Lambda}\sim 10^{-120} M_P^4$.}
	\label{fig:3D}
\end{figure}
The order of the vacuum density, $v$, depends on the model parameters, $n$, $\gamma$ and $M$. It also depends on $\alpha$ through $M$ according to Eq. (\ref{eq:COBEnormalization}). If this order matches with the experimental value, then the model parameters automatically conform to the potential suitable for inflation, quintessence and EDE. Figure \ref{fig:3D} shows the necessary ranges of model parameters in order to obtain the experimentally favoured value of vacuum density \textit{i.e.} $V_{\Lambda}\sim 10^{-120} M_P^4$. Our chosen parameter space described by $n=8$ and $\gamma =0.0818$ lies within these ranges, so we can fairly expect that it will give satisfactory amount of vacuum density.\par Now, using Eq. (\ref{eq:final_V_lambda_5}) the calculated values of the present-day vacuum density for considered values of $\alpha$ in the EDE-influenced quintessential $\alpha$-attractor model are given in table \ref{tab:Table2}.
\begin{table}[H]
    \captionsetup{justification=centering,width=0.5\textwidth}
    \caption{The EDE-modified values of $V_{\Lambda}$ in the present model.}
    \begin{center}
        \begin{adjustbox}{width=0.6\textwidth}
        \begin{tabular}{|c|c|c|c|c|c|c|}
    \hline
     $n$ & $\gamma$ & $\alpha$ & $M (M_P)$ & $v$ & $V^{\mathrm{exact}}_{\Lambda}(M_P^4)$ & $V_{\Lambda}^{\mathrm{Planck}} (M_P^4)$\\
    \hline\hline
  & & $0.001$ & $5.58\times 10^{-4}$ & $118.982$ & $1.042\times 10^{-119}$ &\\
  & & $0.005$ & $8.35\times 10^{-4}$ & $118.282$ & $5.224\times 10^{-119}$ &\\
  & & $0.010$ & $9.92\times 10^{-4}$ & $117.983$ & $1.039\times 10^{-118}$ &\\
  $8$ & $0.0818$ & $0.050$ & $1.48\times 10^{-3}$ & $117.288$ & $5.152\times 10^{-118}$ & $10^{-120}$\\
  & & $0.100$ & $1.76\times 10^{-3}$ & $116.987$ & $1.030\times 10^{-117}$ &\\
  & & $0.500$ & $2.64\times 10^{-3}$ & $116.281$ & $5.240\times 10^{-117}$ &\\
  & & $1.000$ & $3.14\times 10^{-3}$ & $115.981$ & $1.045\times 10^{-116}$ &\\
  & & $2.500$ & $3.96\times 10^{-3}$ & $115.578$ & $2.642\times 10^{-116}$ &\\
  & & $4.300$ & $4.57\times 10^{-3}$ & $115.329$ & $4.688\times 10^{-116}$ &\\
    \hline
    \end{tabular}
    \end{adjustbox}
    \end{center}
         \label{tab:Table2}
    \end{table}
    Tables \ref{tab:Table1} and \ref{tab:Table2} clearly show that the $M$ and $V_{\Lambda}$ for the chosen values of model parameters $n$, $\gamma$, $\alpha$ are in excellent agreement with Planck data.\par
\end{enumerate}
In the above, we have set all the prerequisites and can proceed to analyse the solutions of the dynamical mode equations and estimate the cosmological parameters within the model constraints. 
\subsection{Inflaton field and its first order perturbation}
\label{subsec:Validity}
\begin{figure}[H]
\begin{subfigure}{0.33\linewidth}
  \centering
   \includegraphics[width=46mm,height=40mm]{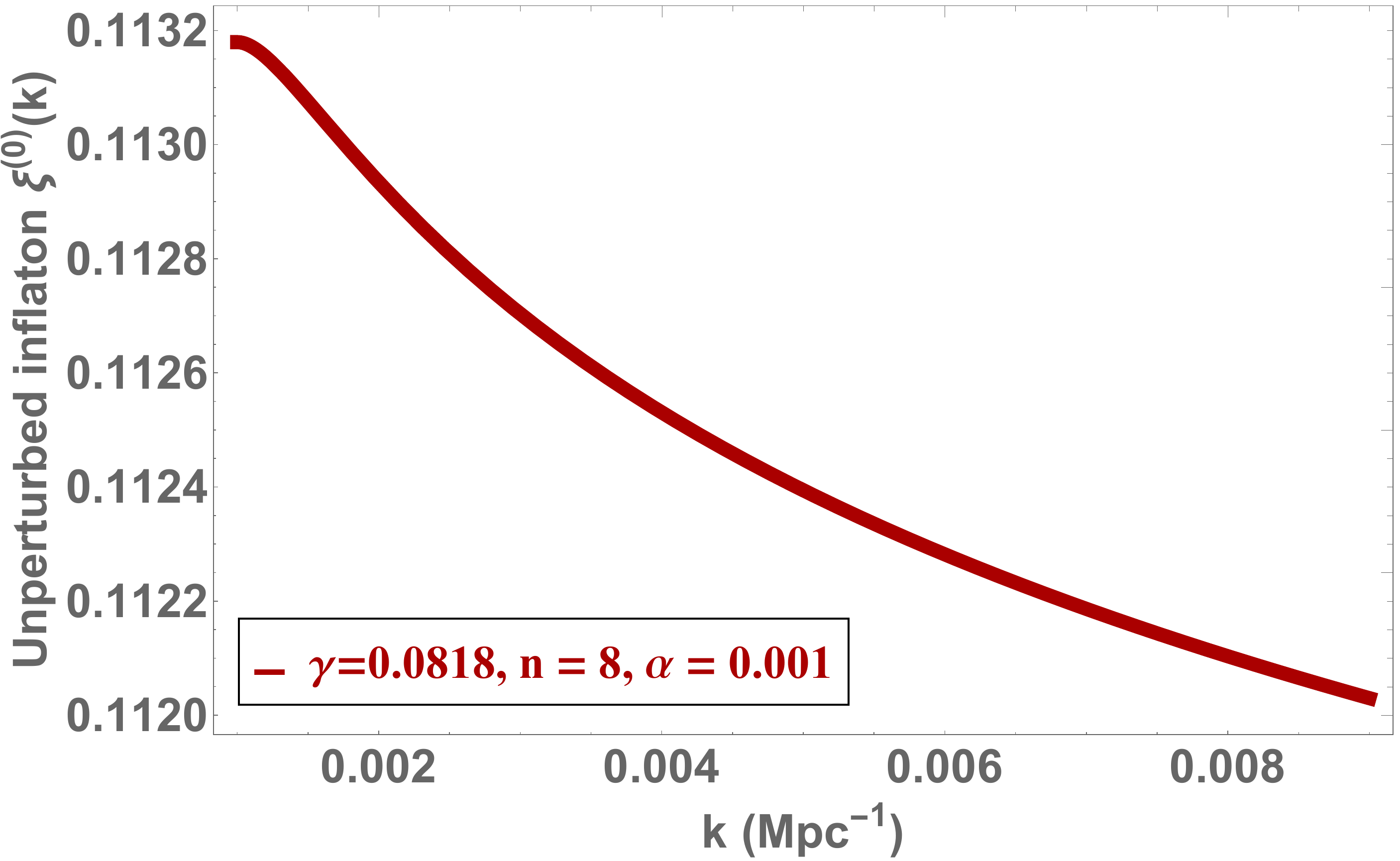} 
   \subcaption{}
   \label{fig:unperturbedINF_1}
\end{subfigure}%
\begin{subfigure}{0.33\linewidth}
  \centering
   \includegraphics[width=46mm,height=40mm]{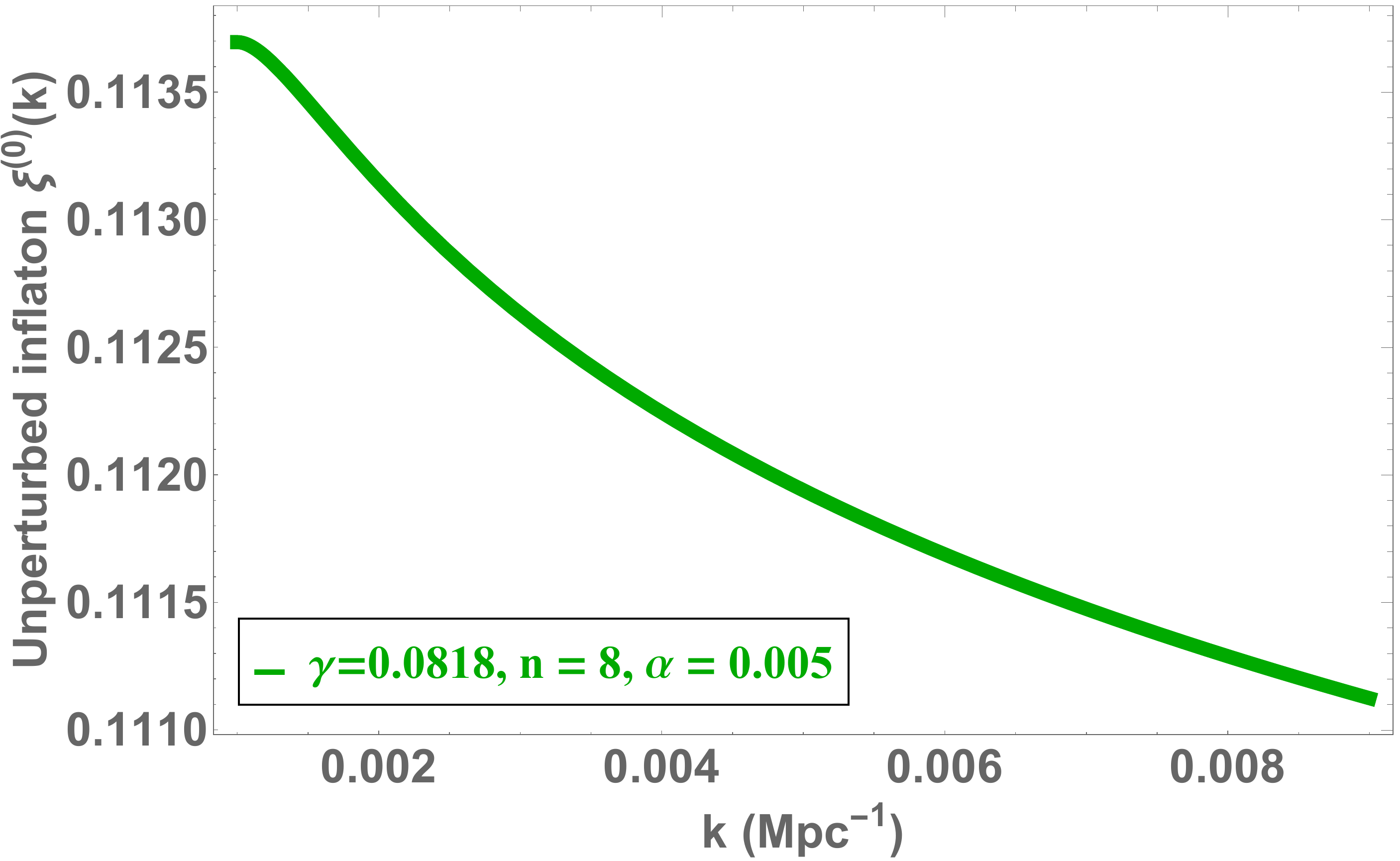}
   \subcaption{}
   \label{fig:unperturbedINF_2}
\end{subfigure}%
\begin{subfigure}{0.33\linewidth}
  \centering
   \includegraphics[width=46mm,height=40mm]{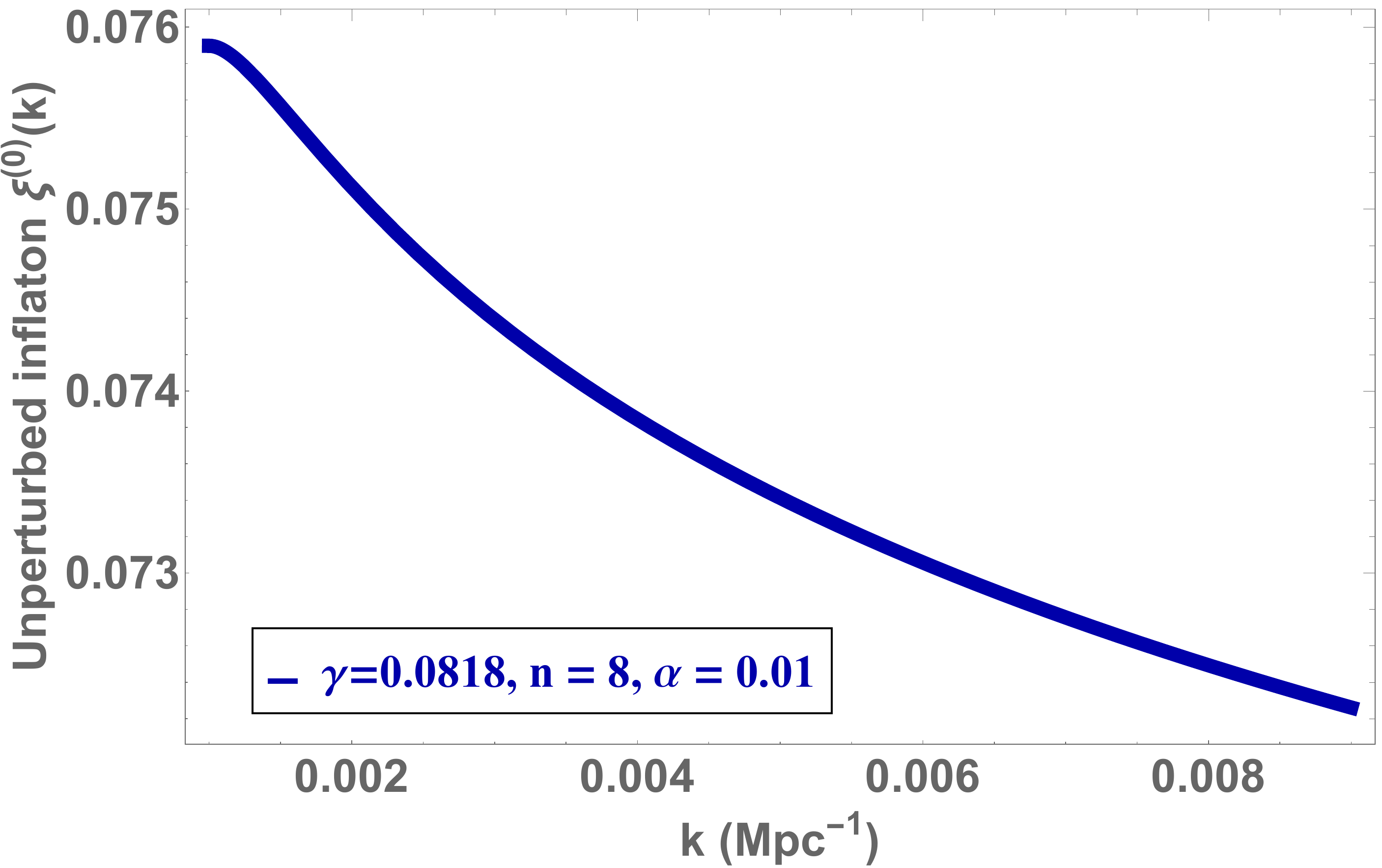}
   \subcaption{}
   \label{fig:unperturbedINF_3}
\end{subfigure}%
\vspace{0.05\linewidth}
\begin{subfigure}{0.33\linewidth}
  \centering
   \includegraphics[width=46mm,height=40mm]{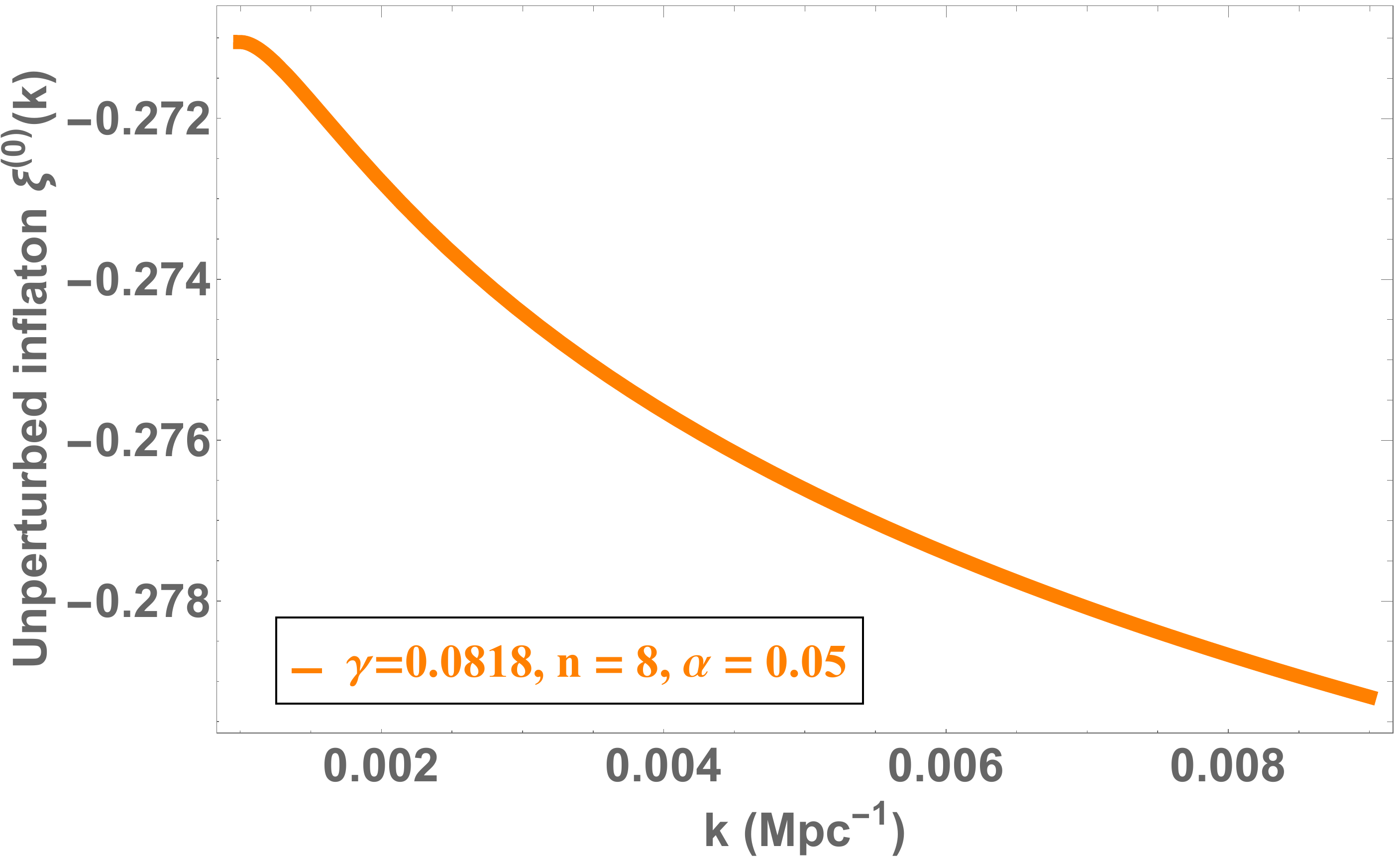}
   \subcaption{}
    \label{fig:unperturbedINF_4}
\end{subfigure}%
\begin{subfigure}{0.33\linewidth}
  \centering
   \includegraphics[width=46mm,height=40mm]{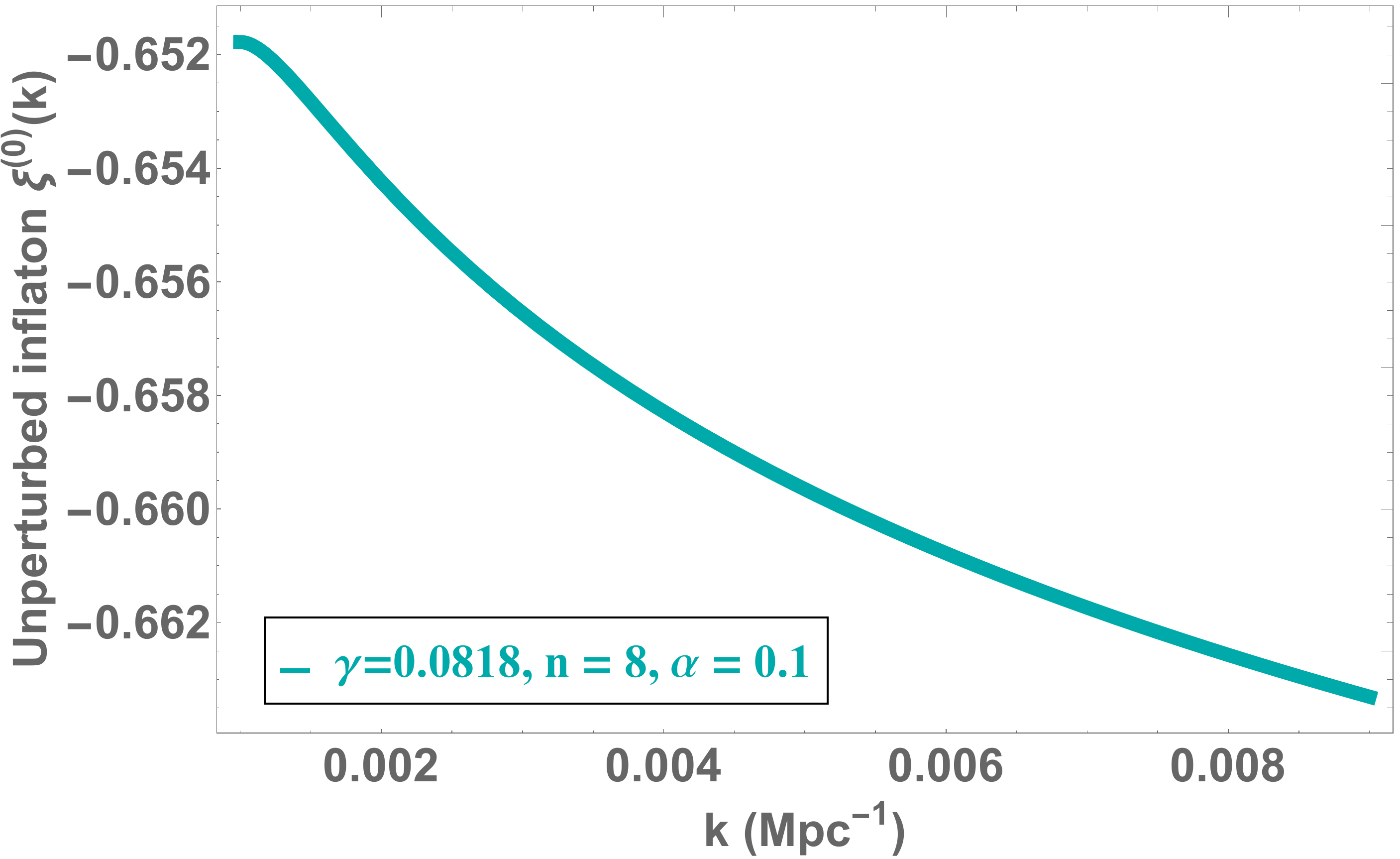}
   \subcaption{}
    \label{fig:unperturbedINF_5}
\end{subfigure}%
\begin{subfigure}{0.33\linewidth}
  \centering
   \includegraphics[width=46mm,height=40mm]{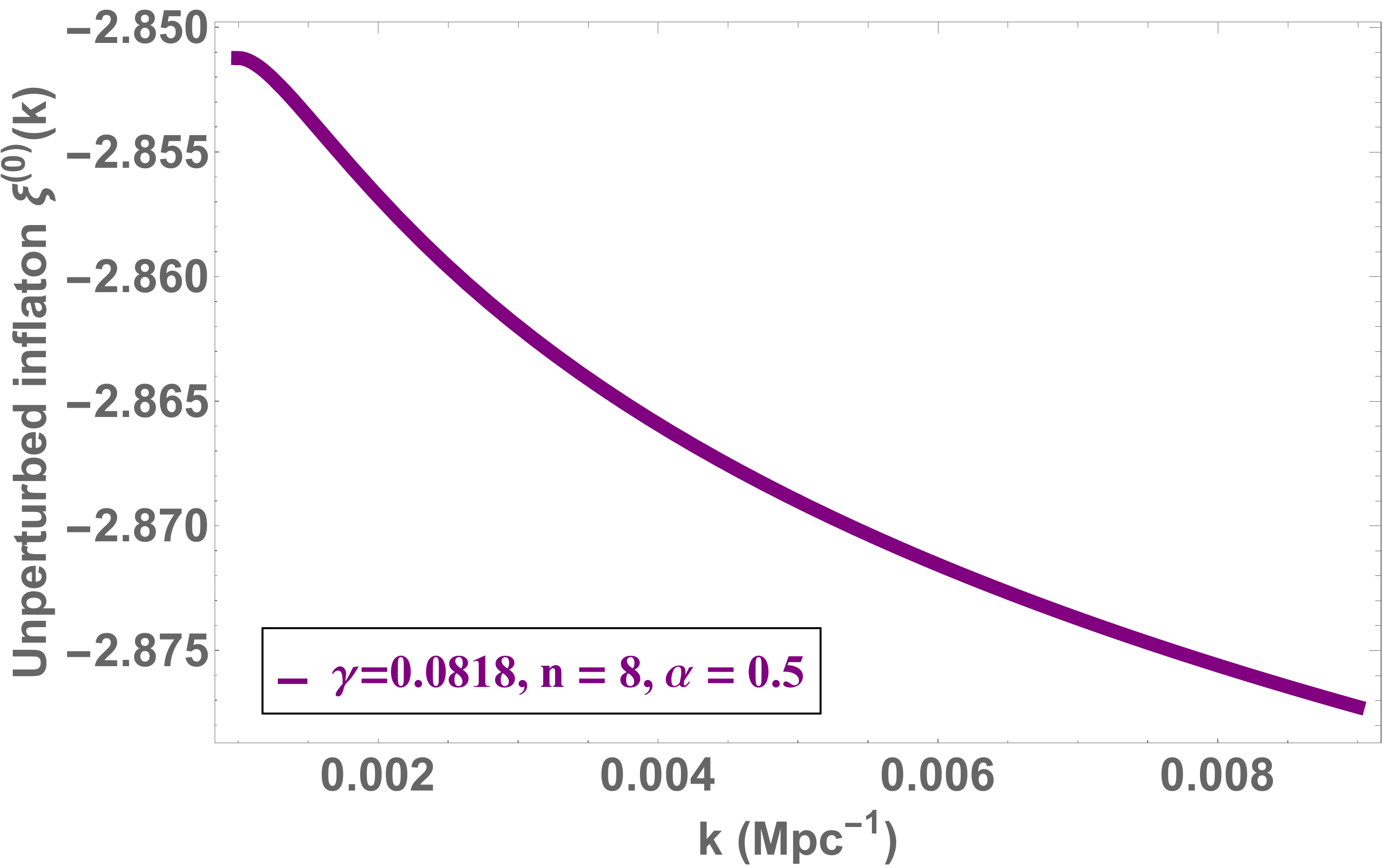}
   \subcaption{}
    \label{fig:unperturbedINF_6}
\end{subfigure}%
\vspace{0.05\linewidth}
\begin{subfigure}{0.33\linewidth}
  \centering
   \includegraphics[width=46mm,height=40mm]{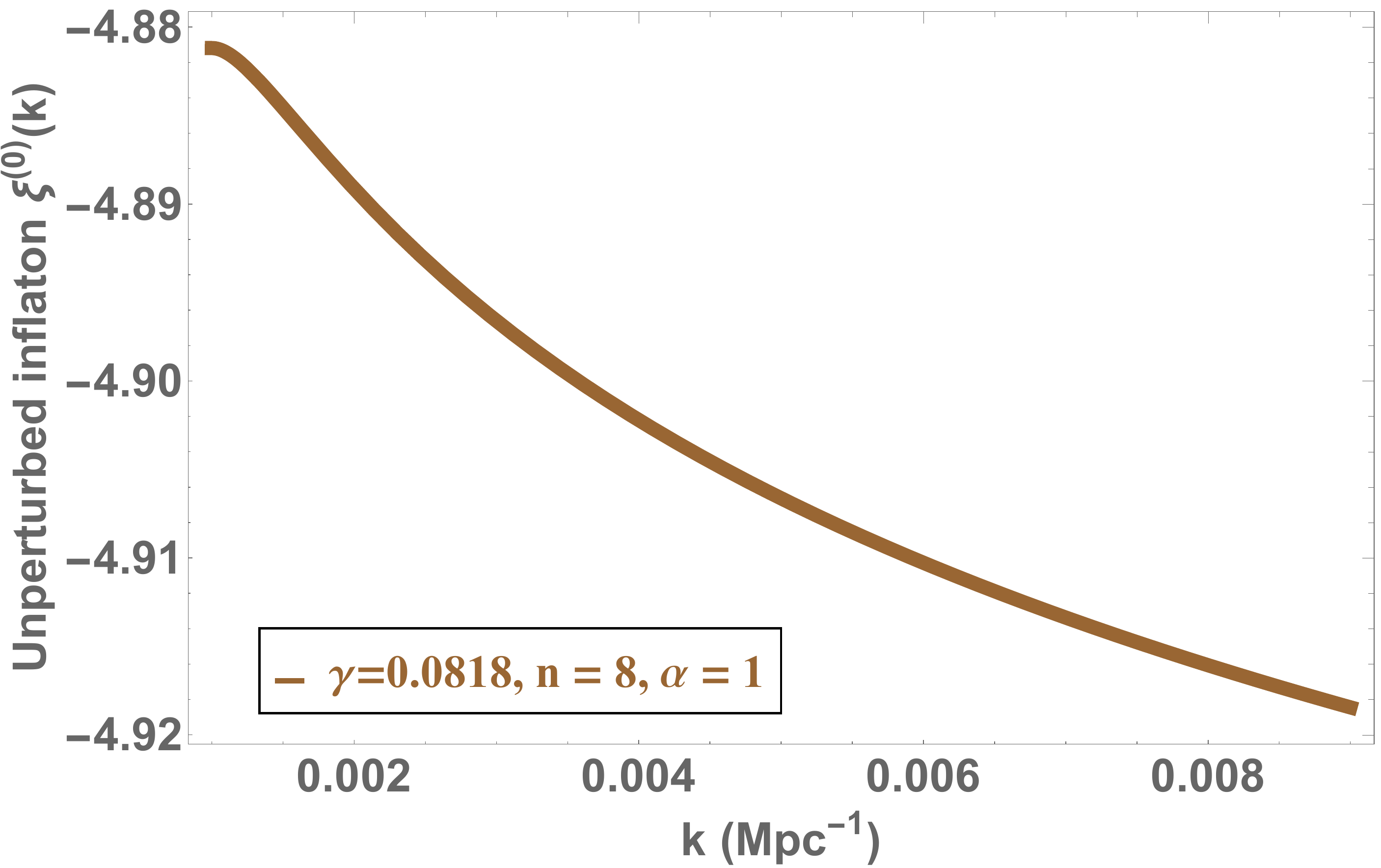}
   \subcaption{}
    \label{fig:unperturbedINF_7}
\end{subfigure}%
\begin{subfigure}{0.33\linewidth}
  \centering
   \includegraphics[width=46mm,height=40mm]{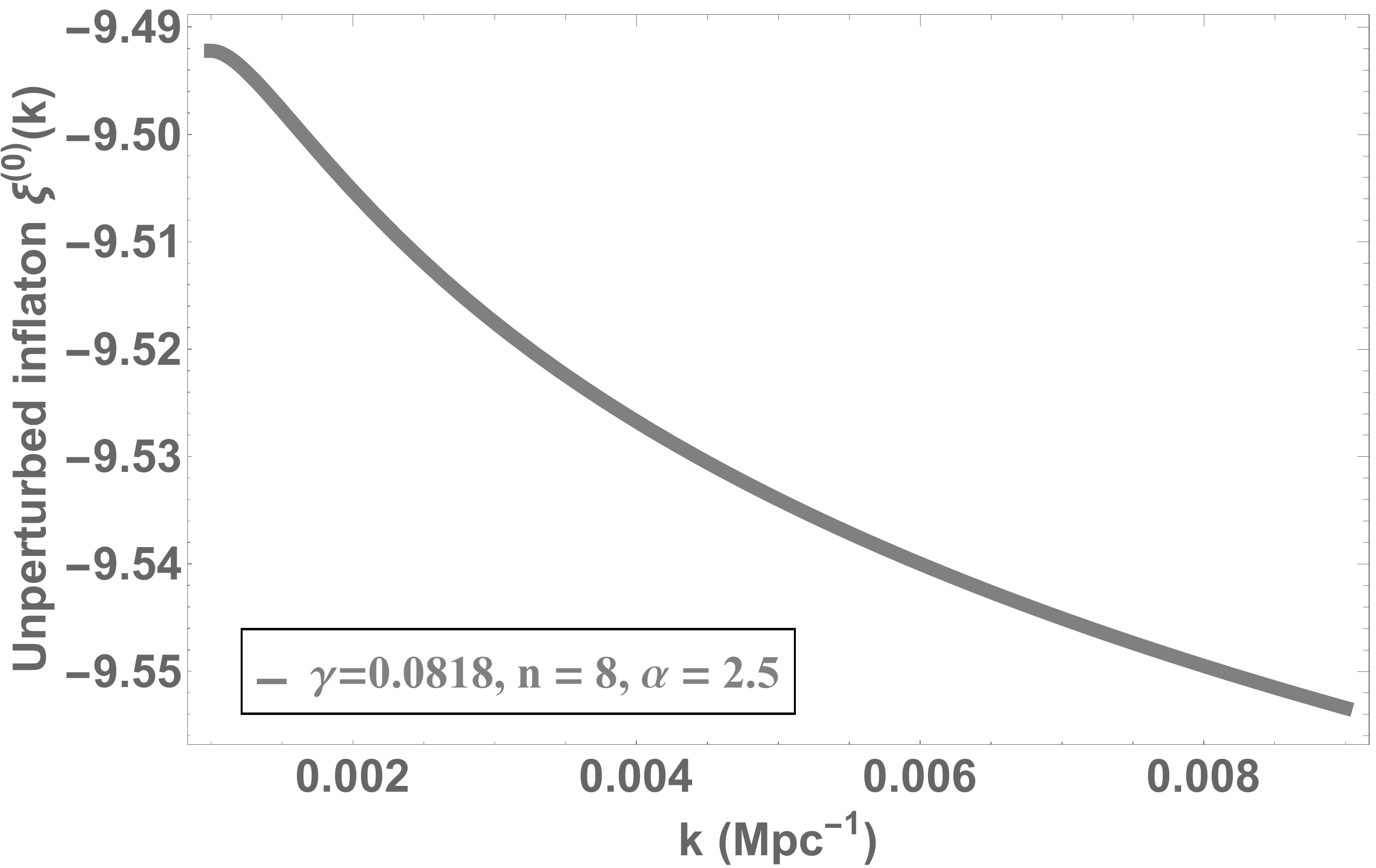}
   \subcaption{}
    \label{fig:unperturbedINF_8}
\end{subfigure}%
\begin{subfigure}{0.33\linewidth}
  \centering
   \includegraphics[width=46mm,height=40mm]{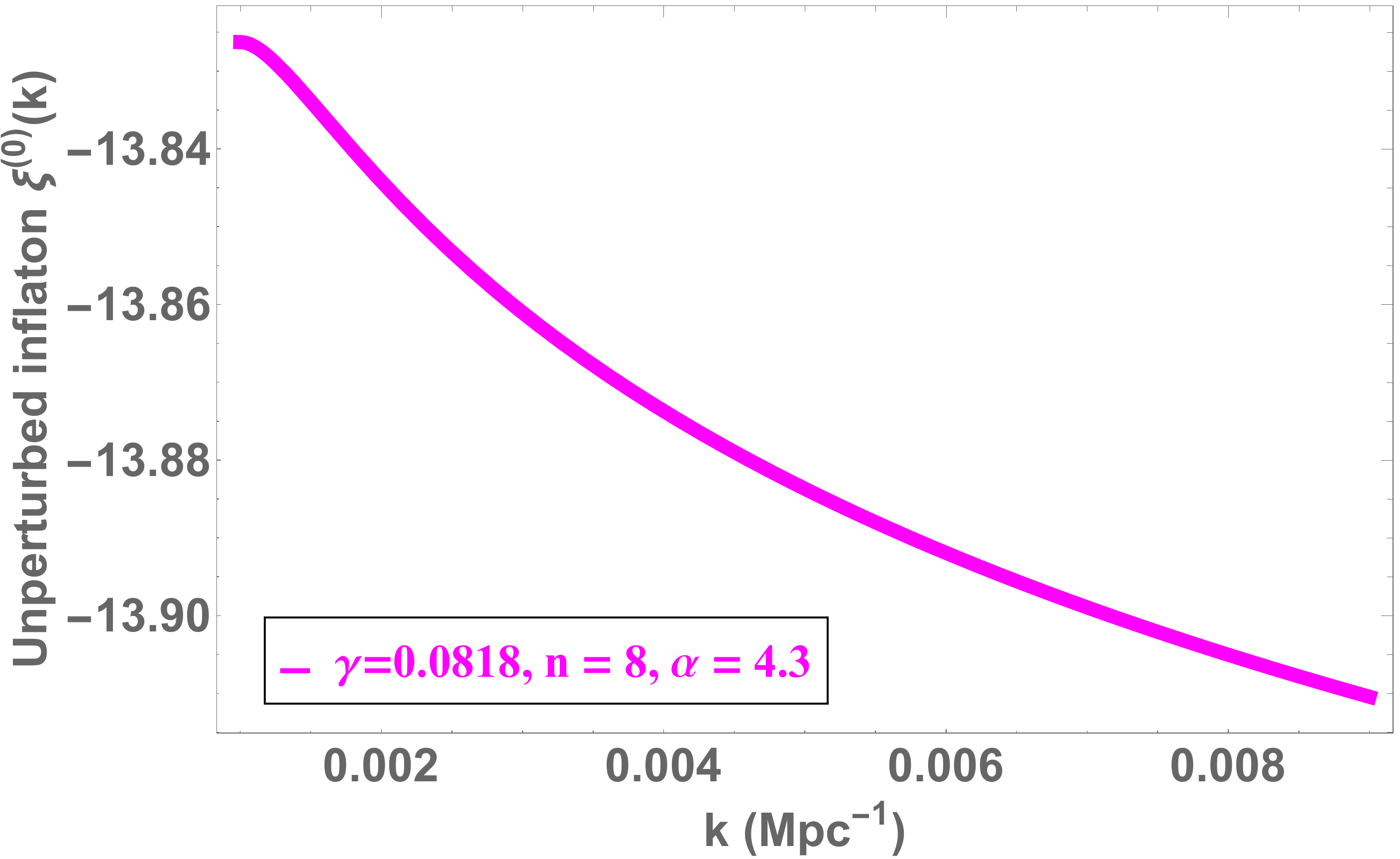}
   \subcaption{}
    \label{fig:unperturbedINF_9}
\end{subfigure}
\caption{ Zeroth order parts of the inflaton field $\xi(k)$ for nine values of $\alpha$ for $\gamma=0.0818$ and $n=8$. The values of $\xi^{(0)}(k)$ become increasingly negative from $\alpha=0.05$ at a particular $k$ value.}
\label{fig:unperturbedINF}
\end{figure}

\begin{figure}[H]
\begin{subfigure}{0.33\linewidth}
  \centering
   \includegraphics[width=46mm,height=40mm]{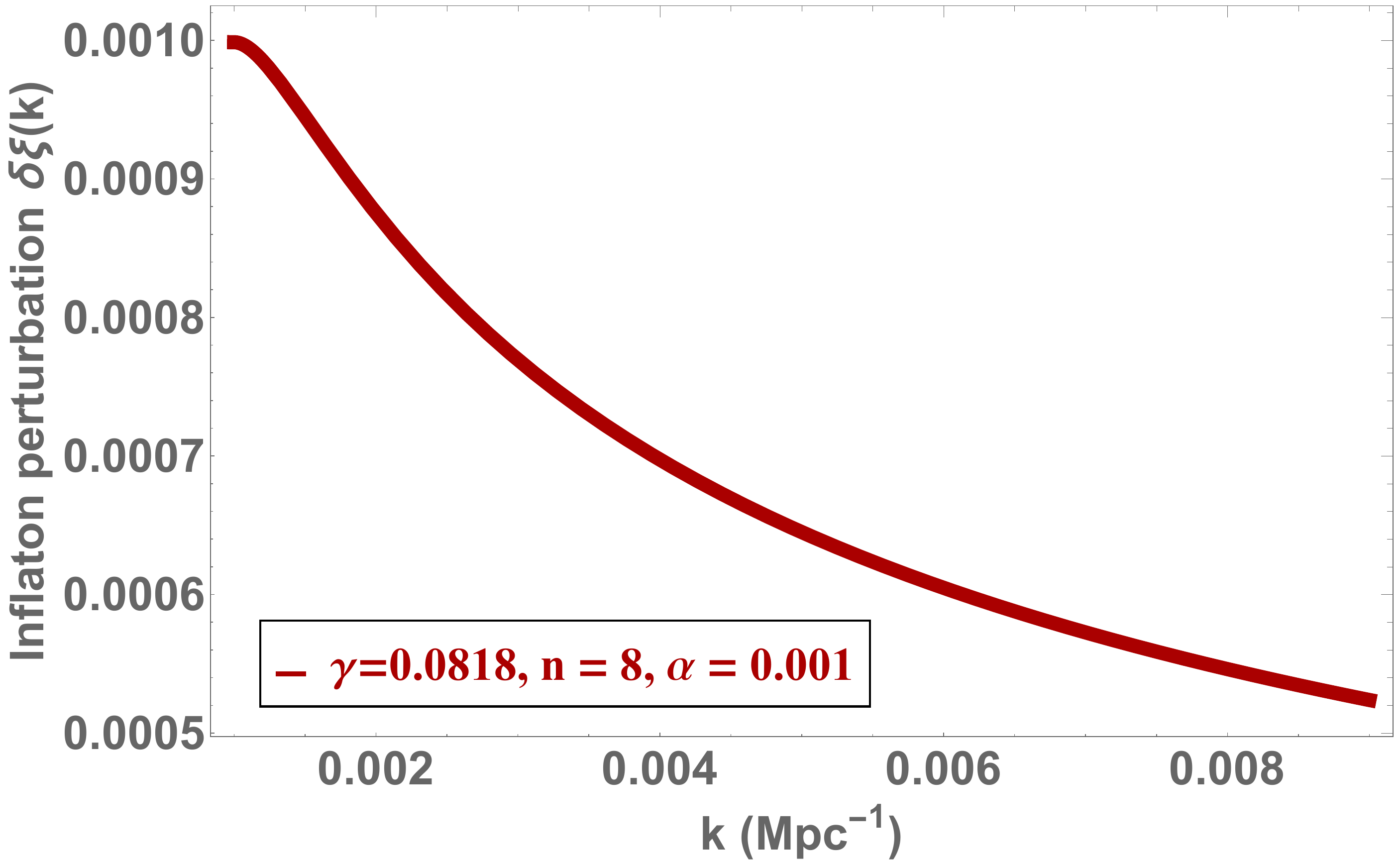} 
   \subcaption{}
   \label{fig:perturbedINF_1}
\end{subfigure}%
\begin{subfigure}{0.33\linewidth}
  \centering
   \includegraphics[width=46mm,height=40mm]{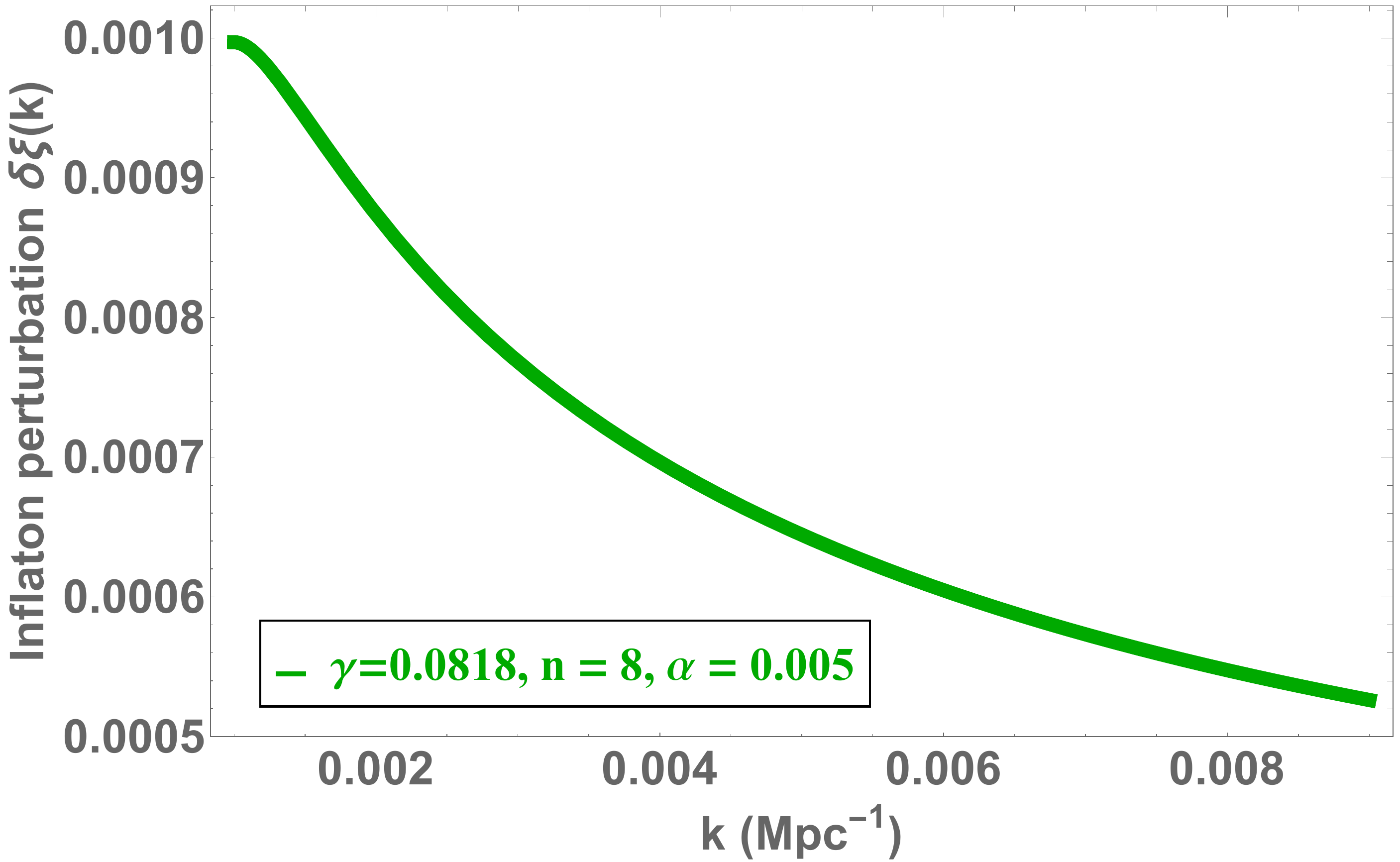}
   \subcaption{}
   \label{fig:perturbedINF_2}
\end{subfigure}%
\begin{subfigure}{0.33\linewidth}
  \centering
   \includegraphics[width=46mm,height=40mm]{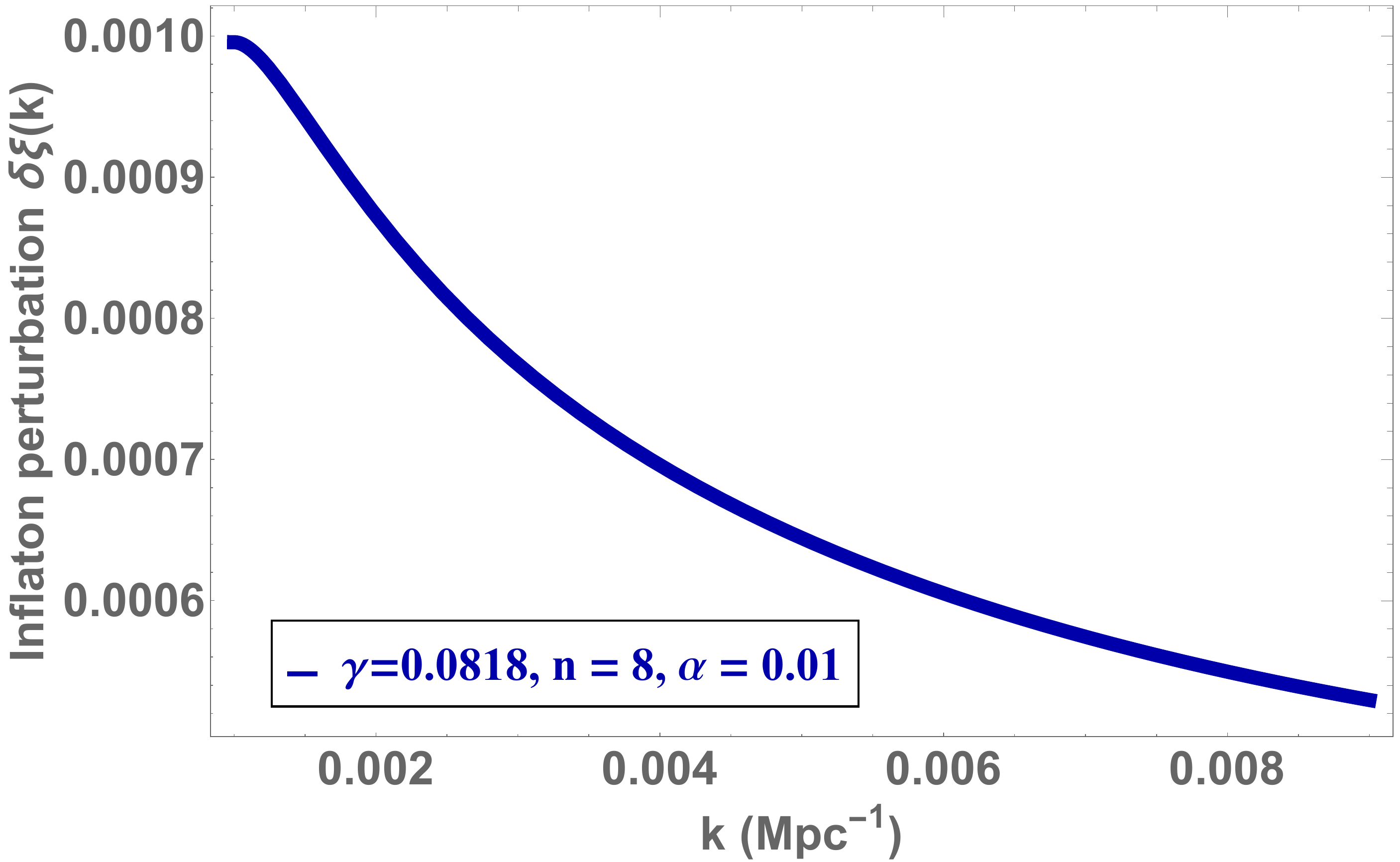}
   \subcaption{}
   \label{fig:perturbedINF_3}
\end{subfigure}%
\vspace{0.05\linewidth}
\begin{subfigure}{0.33\linewidth}
  \centering
   \includegraphics[width=46mm,height=40mm]{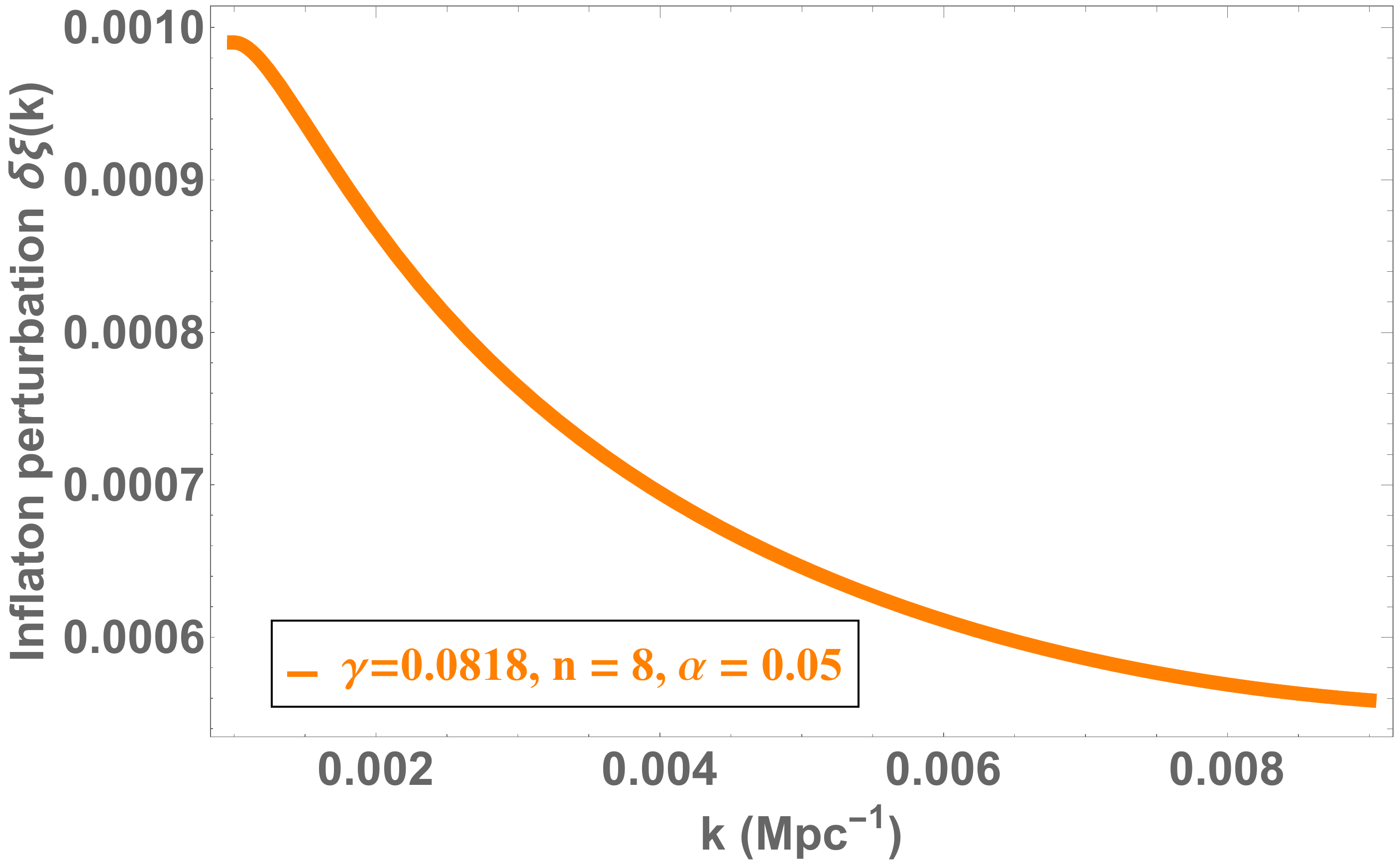}
   \subcaption{}
    \label{fig:perturbedINF_4}
\end{subfigure}%
\begin{subfigure}{0.33\linewidth}
  \centering
   \includegraphics[width=46mm,height=40mm]{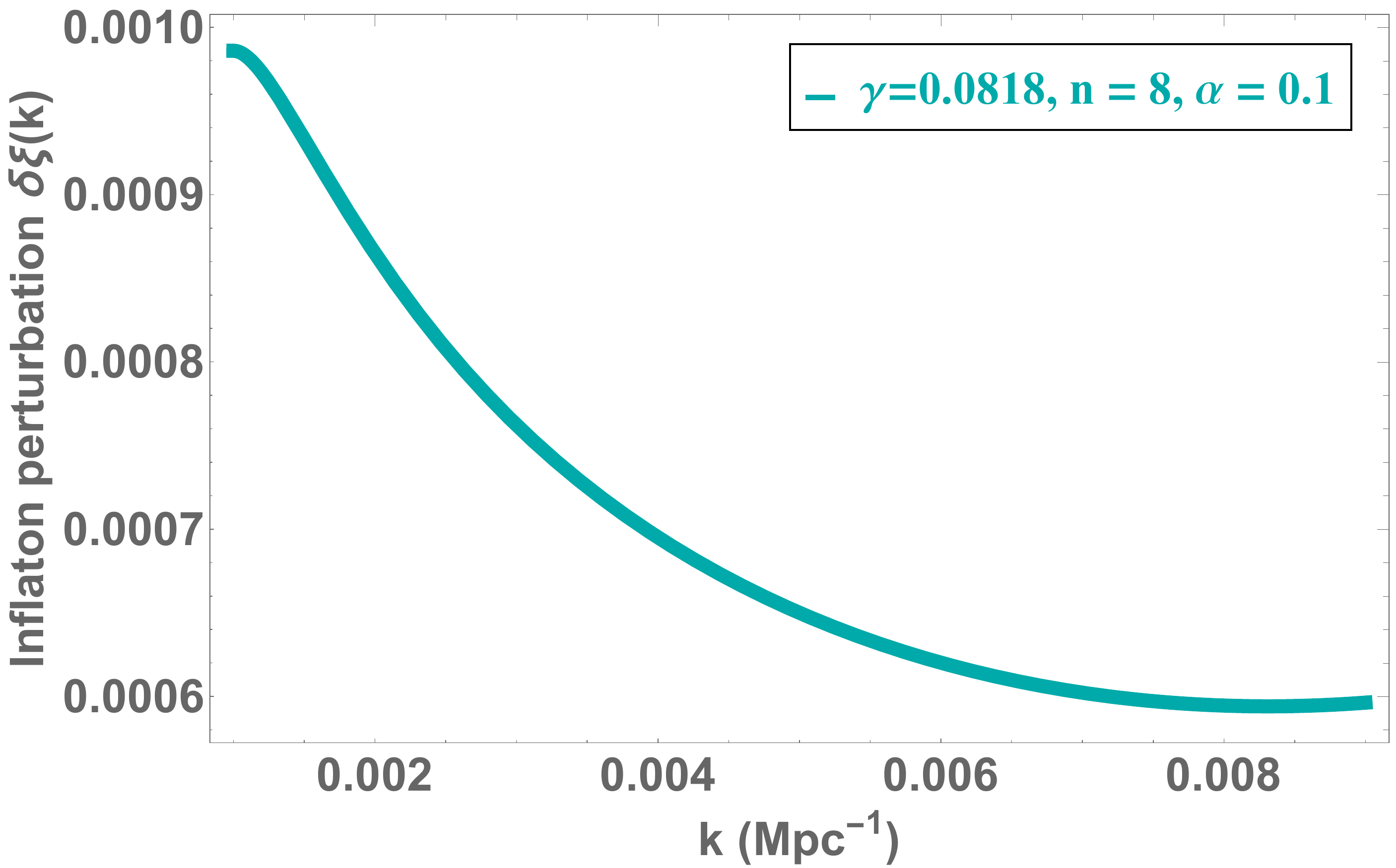}
   \subcaption{}
    \label{fig:perturbedINF_5}
\end{subfigure}%
\begin{subfigure}{0.33\linewidth}
  \centering
   \includegraphics[width=46mm,height=40mm]{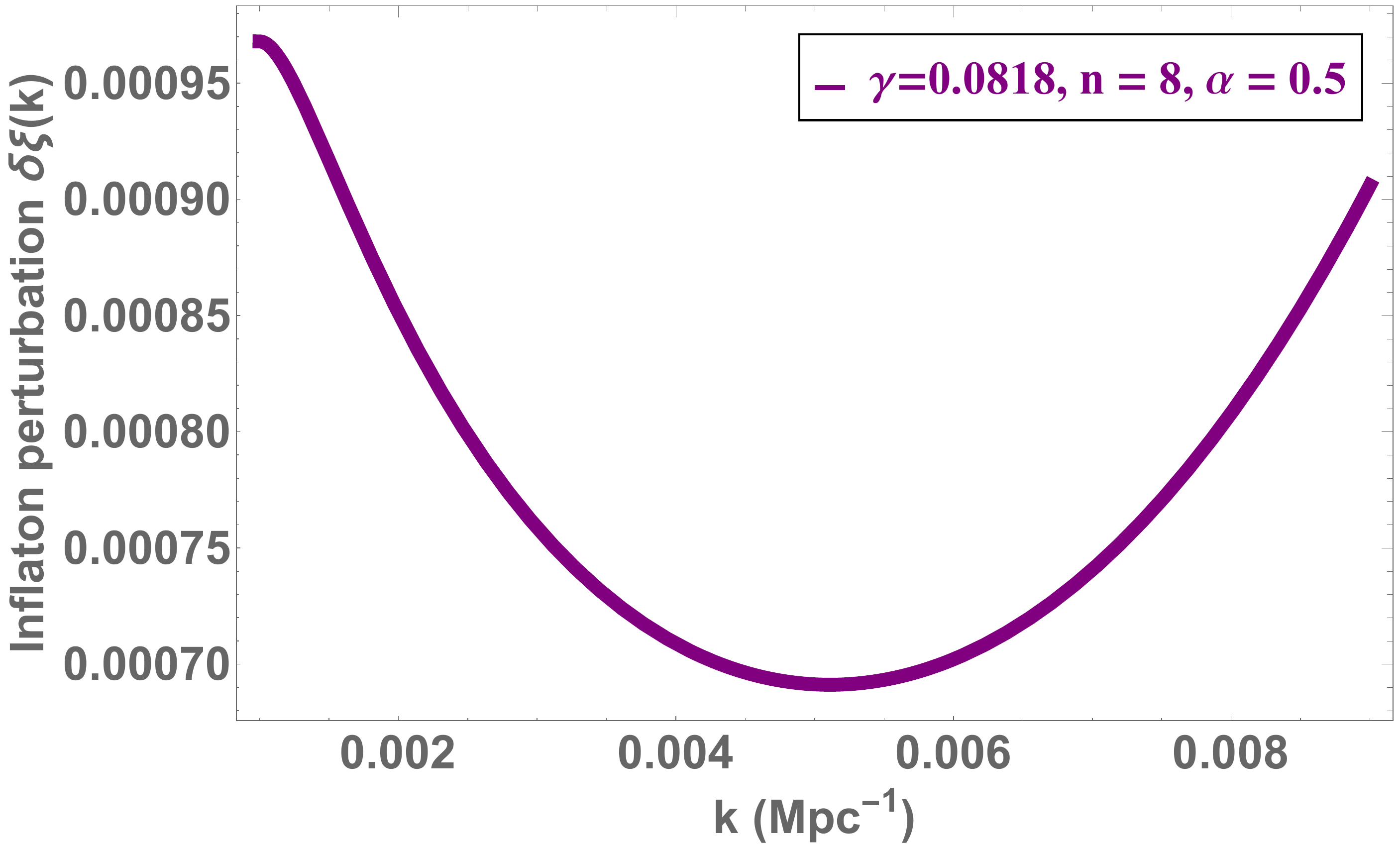}
   \subcaption{}
    \label{fig:perturbedINF_6}
\end{subfigure}%
\vspace{0.05\linewidth}
\begin{subfigure}{0.33\linewidth}
  \centering
   \includegraphics[width=46mm,height=40mm]{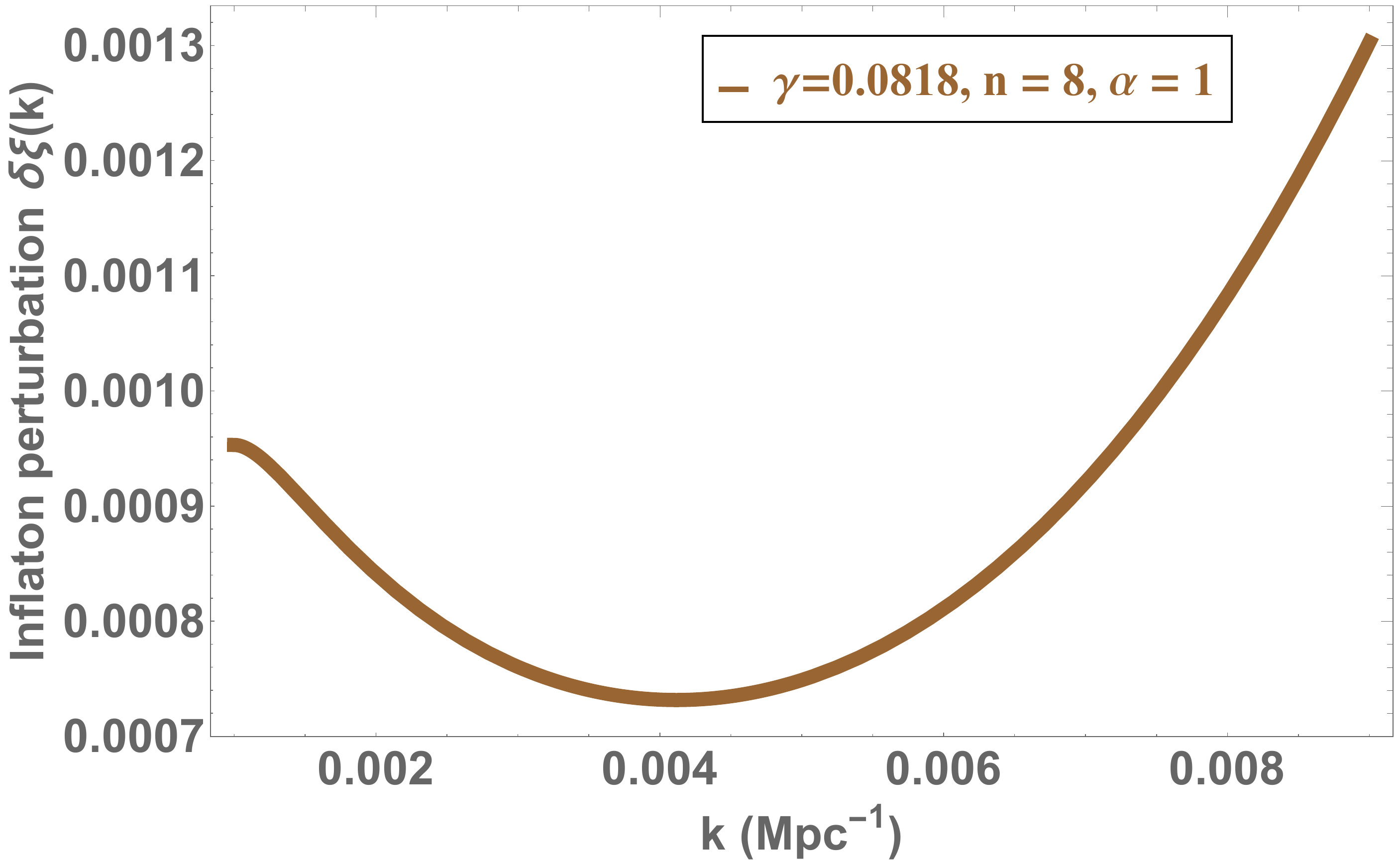}
   \subcaption{}
    \label{fig:perturbedINF_7}
\end{subfigure}%
\begin{subfigure}{0.33\linewidth}
  \centering
   \includegraphics[width=46mm,height=40mm]{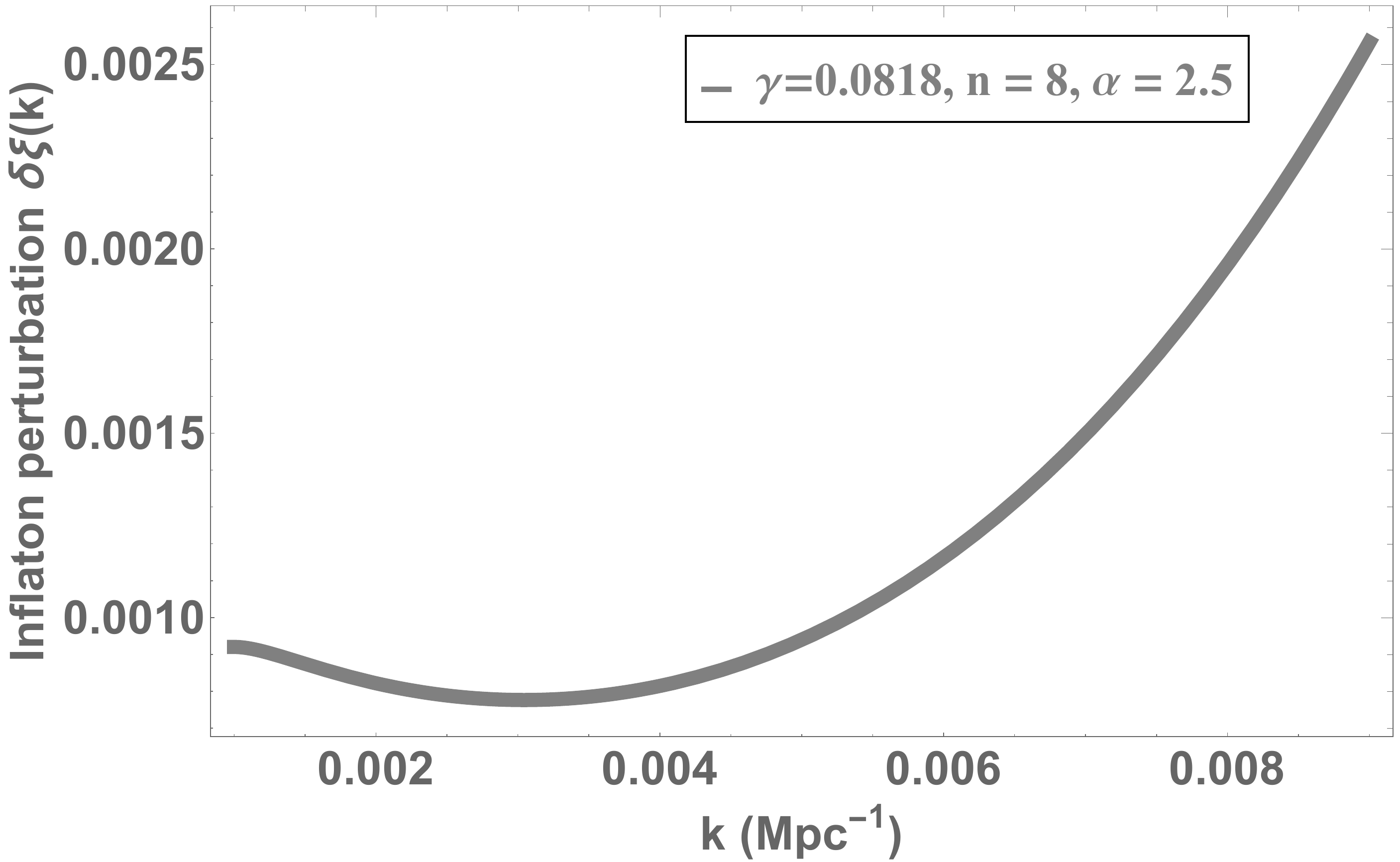}
   \subcaption{}
    \label{fig:perturbedINF_8}
\end{subfigure}%
\begin{subfigure}{0.33\linewidth}
  \centering
   \includegraphics[width=46mm,height=40mm]{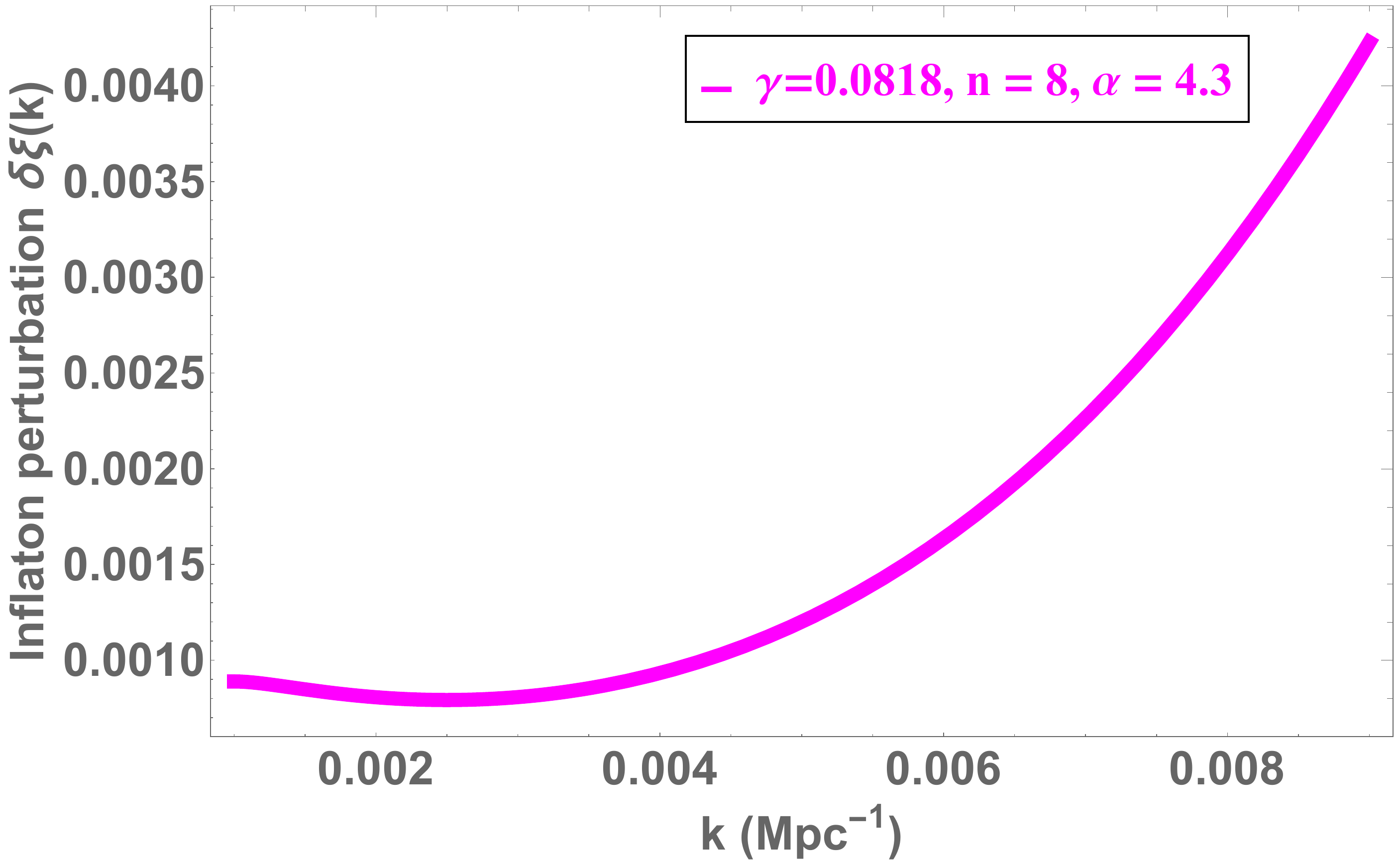}
   \subcaption{}
    \label{fig:perturbedINF_9}
\end{subfigure}
\caption{First order perturbed parts of the inflaton field for nine values of $\alpha$ for $\gamma=0.0818$ and $n=8$. Up to $\alpha=0.1$ the $\delta\xi(k)$ is almost insensitive to $\alpha$. After that, it tends to increase with the increase in $\alpha$ at a particular $k$-value.  For $\alpha\geq 0.1$ the perturbation becomes stronger for $k\geq 0.005$ Mpc$^{-1}$ than $k<0.005$ Mpc$^{-1}$.}
\label{fig:perturbedINF}
\end{figure}

In figures \ref{fig:unperturbedINF} and \ref{fig:perturbedINF} we show the solutions of the inflaton field $\xi^{(0)}(k)$ and its first order perturbation $\delta\xi(k)$ for $\alpha=0.001-4.3$ within $k=0.001-0.009$ Mpc$^{-1}$. The values of $\xi^{(0)}(k)$ are positive ranging from $0.072$ to $0.1136$ from $\alpha=0.001$ to $\alpha=0.01$ and afterwards it become negative from $\alpha=0.05$ to $\alpha=4.3$. The corresponding $\delta\xi(k)$ values range from $0.0005$ to $0.004$ in the entire range of $\alpha$. The negative values of $\xi$ beyond $\alpha=0.01$ (\textit{i.e.} $0.05\leq\alpha\leq 4.3$) seem to be inconsistent. It should be positive for quintessential model of inflation since in that case the inflationary plateau remains at positive values of $\xi$. Actually there is a deep connection lying behind this odd behaviour with the EDE scenario and it necessitates further constraining of the parameter $\alpha$. We shall explain this crucial aspect in detail in subsection \ref{subsec:Role_of_ESP}.\par  So far as the values of $\delta\xi(k)$ are concerned, they are roughly $10-100$ times smaller than $\xi^{(0)}(k)$\footnote{which can be easily realised by computing the ratio $\frac{\delta\xi(k)}{|\xi^{(0)}(k)|}$ at a particular $k$ value.}. Therefore it confirms that the linear perturbative framework is equally consistent in $k$-space as in $t$-space. Actually the $k$-space formulation appears to be an alternative to the usual formulation in time domain (see \cite{Brissenden:2023yko} for example), especially when the mode responses and estimations of the cosmological parameters are required during inflation. It is indeed a great advantage. However it has some limitations also. It can not be extended to post inflationary regime. Therefore all the ideas of reheating, preheating and particle production are outside the scope of the present framework. \par
\begin{figure}[H]
\begin{subfigure}{0.33\linewidth}
  \centering
   \includegraphics[width=46mm,height=40mm]{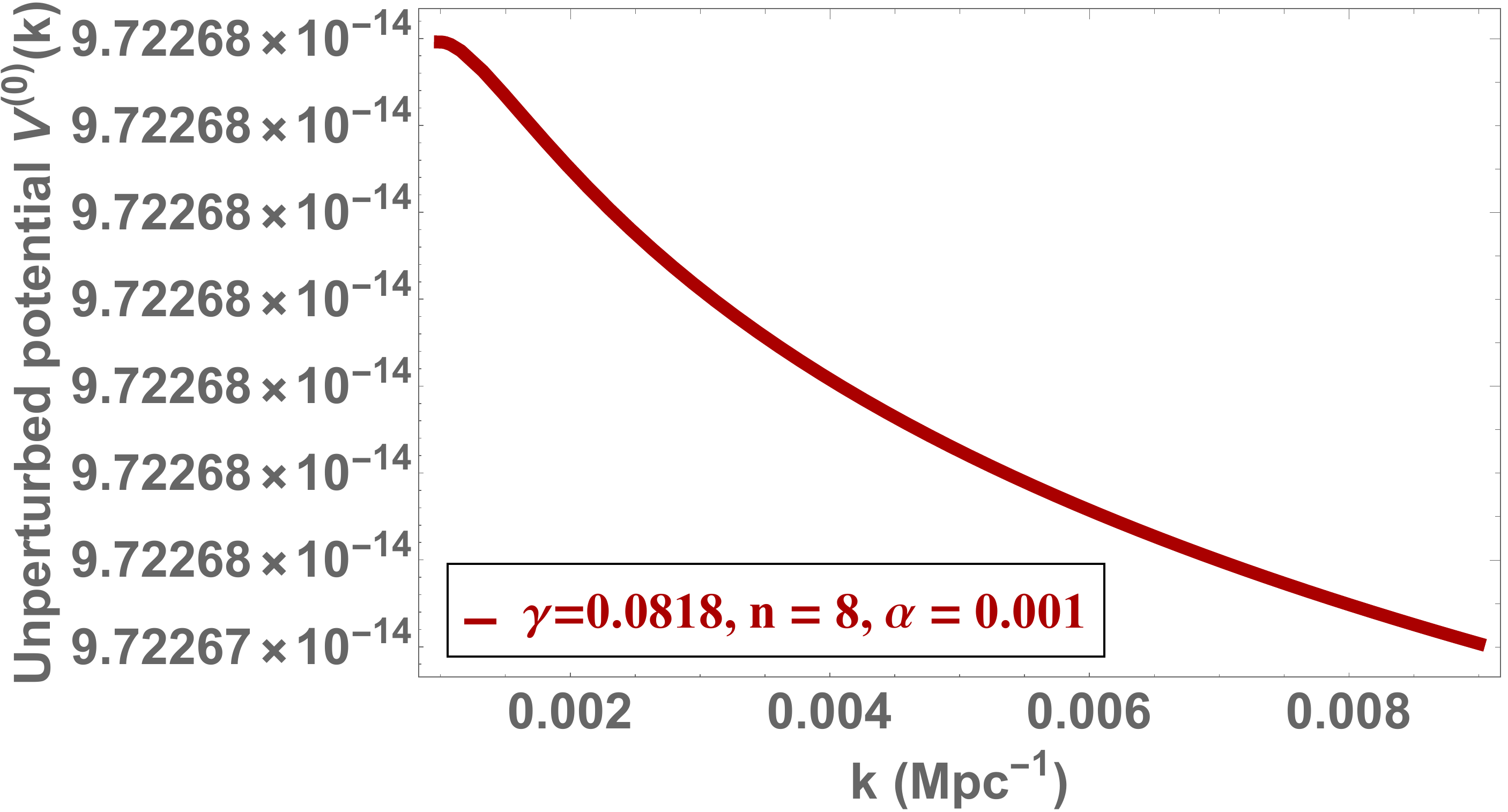} 
   \subcaption{}
   \label{fig:unperturbedPOT_1}
\end{subfigure}%
\begin{subfigure}{0.33\linewidth}
  \centering
   \includegraphics[width=46mm,height=40mm]{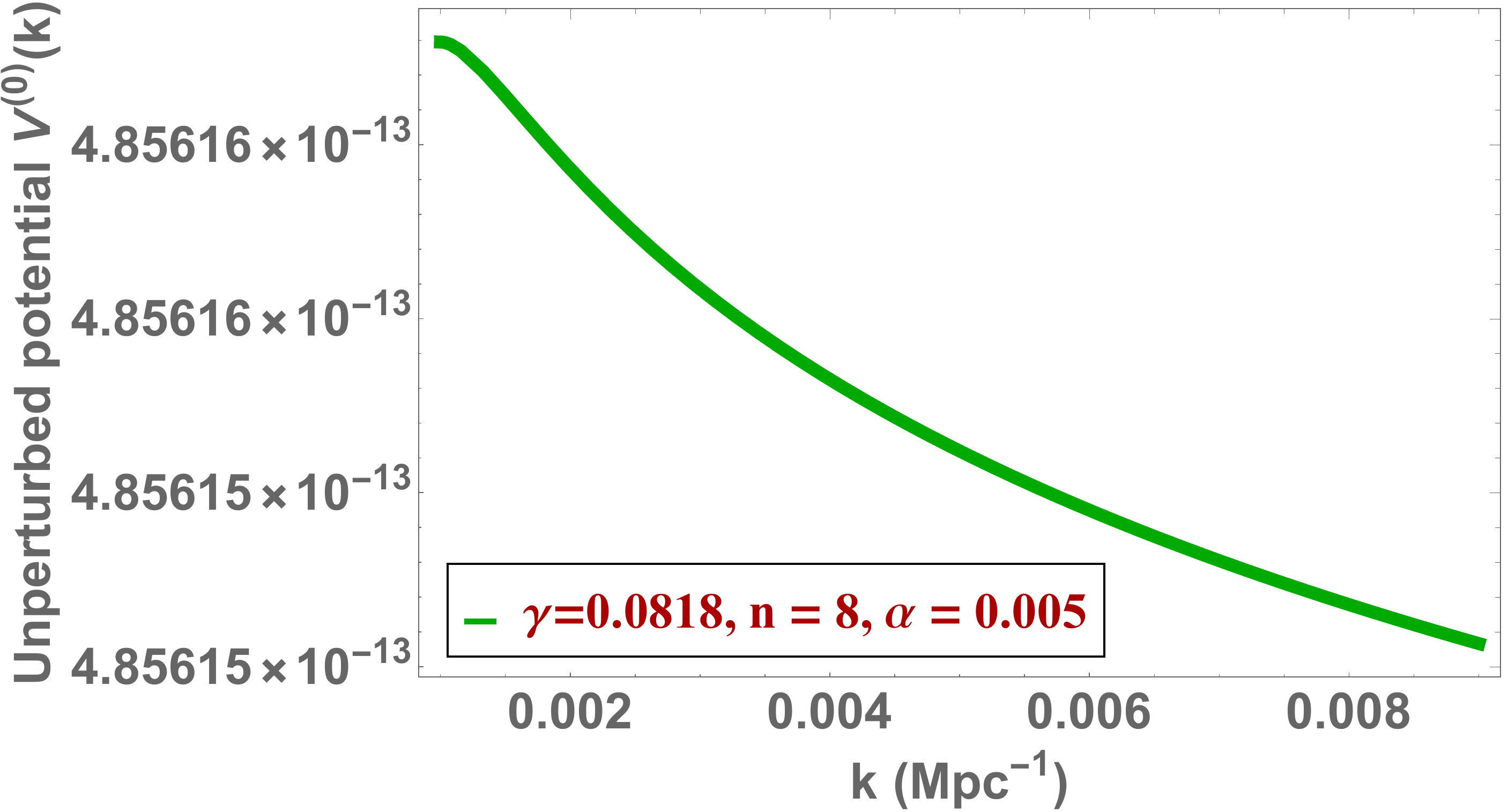}
   \subcaption{}
   \label{fig:unperturbedPOT_2}
\end{subfigure}%
\begin{subfigure}{0.33\linewidth}
  \centering
   \includegraphics[width=46mm,height=40mm]{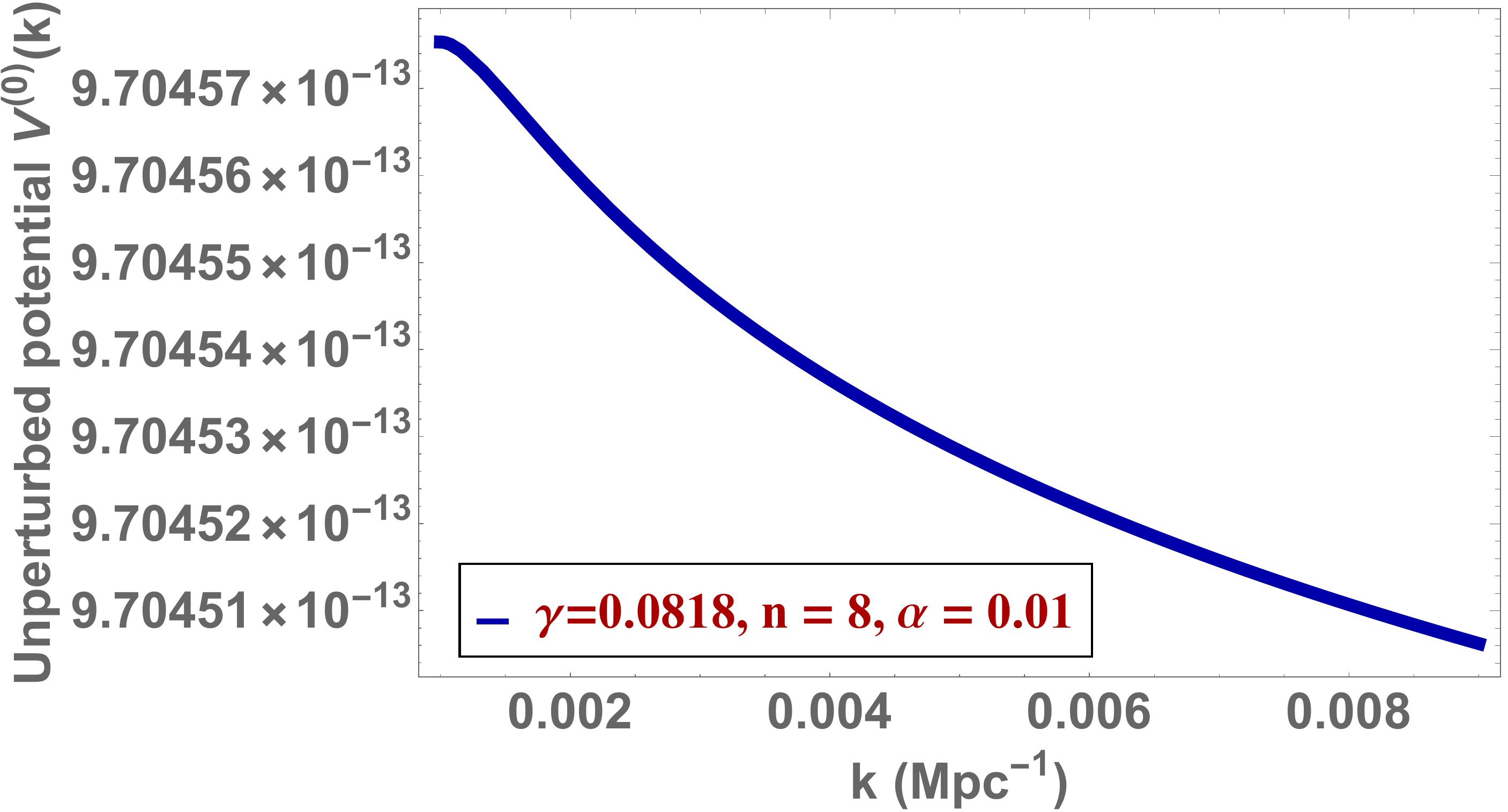}
   \subcaption{}
   \label{fig:unperturbedPOT_3}
\end{subfigure}%
\vspace{0.05\linewidth}
\begin{subfigure}{0.33\linewidth}
  \centering
   \includegraphics[width=46mm,height=40mm]{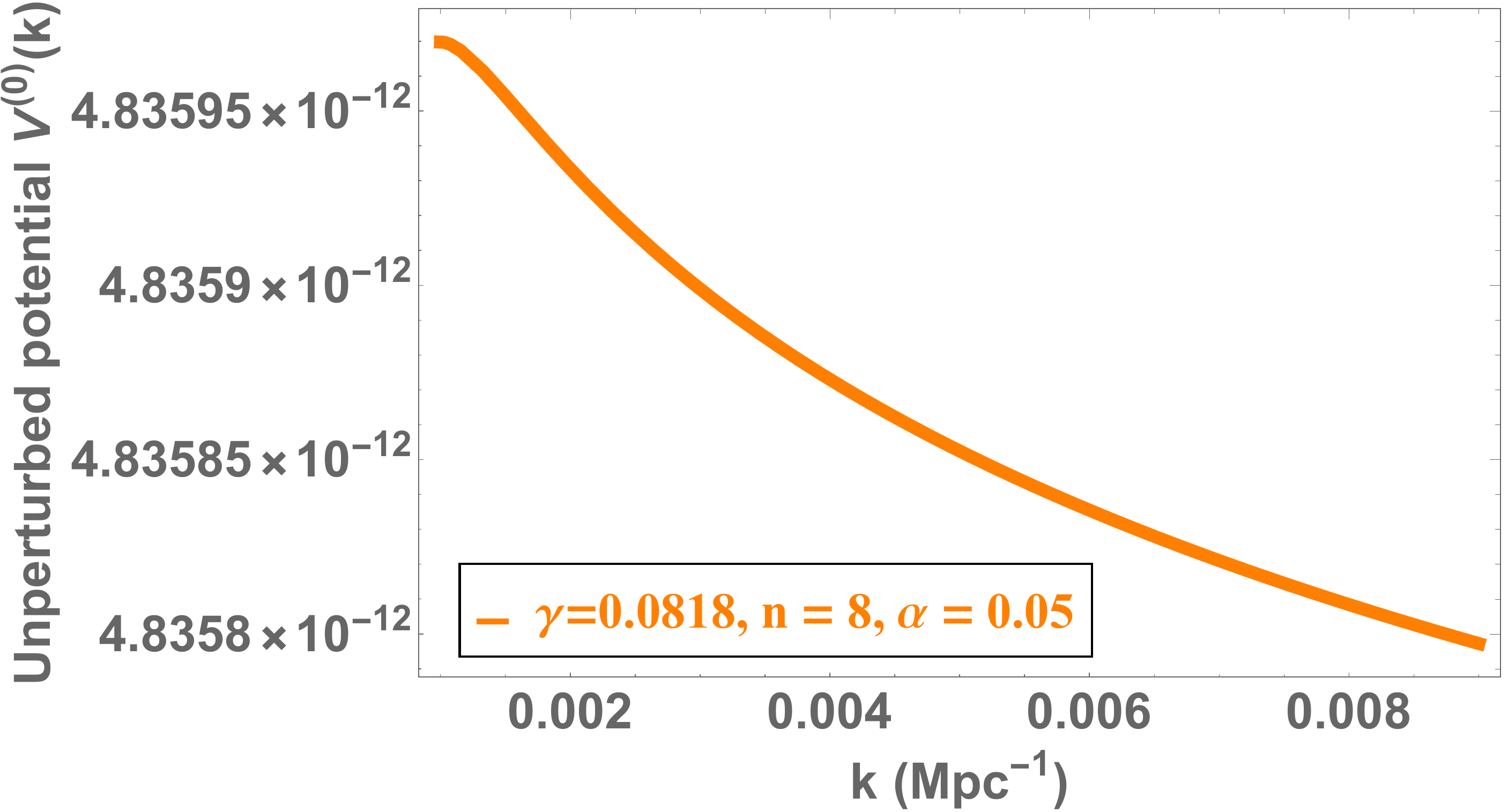}
   \subcaption{}
    \label{fig:unperturbedPOT_4}
\end{subfigure}%
\begin{subfigure}{0.33\linewidth}
  \centering
   \includegraphics[width=46mm,height=40mm]{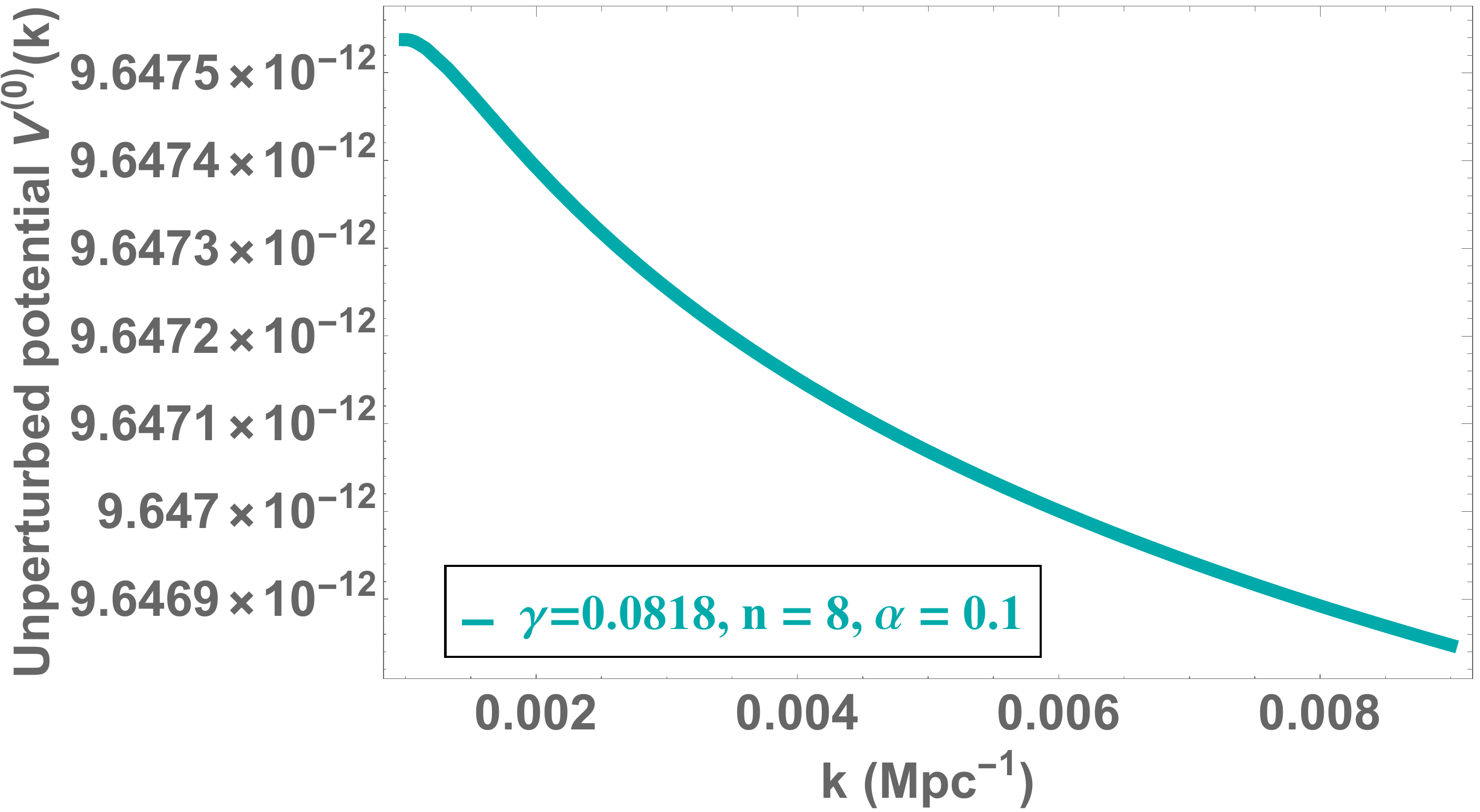}
   \subcaption{}
    \label{fig:unperturbedPOT_5}
\end{subfigure}%
\begin{subfigure}{0.33\linewidth}
  \centering
   \includegraphics[width=46mm,height=40mm]{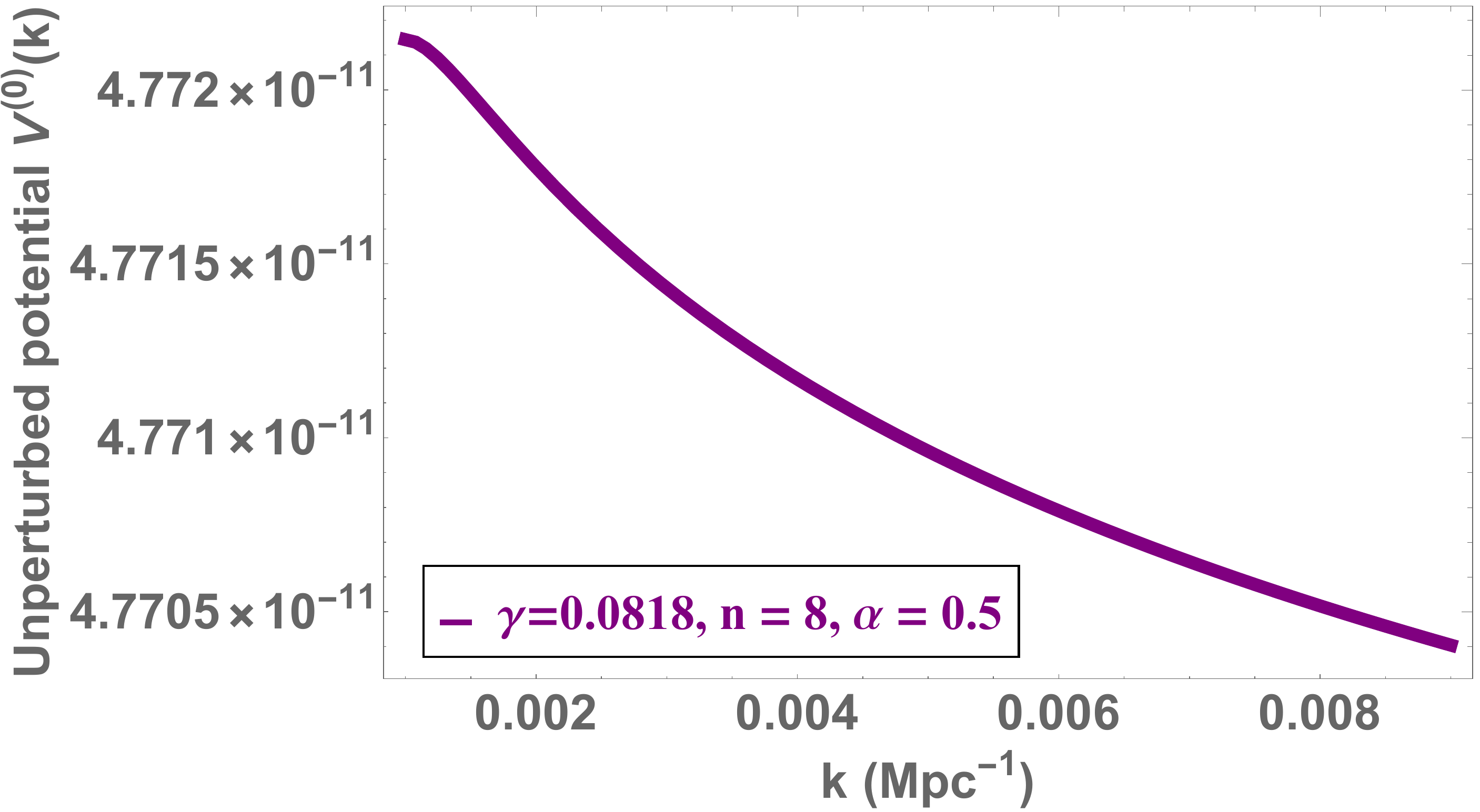}
   \subcaption{}
    \label{fig:unperturbedPOT_6}
\end{subfigure}%
\vspace{0.05\linewidth}
\begin{subfigure}{0.33\linewidth}
  \centering
   \includegraphics[width=46mm,height=40mm]{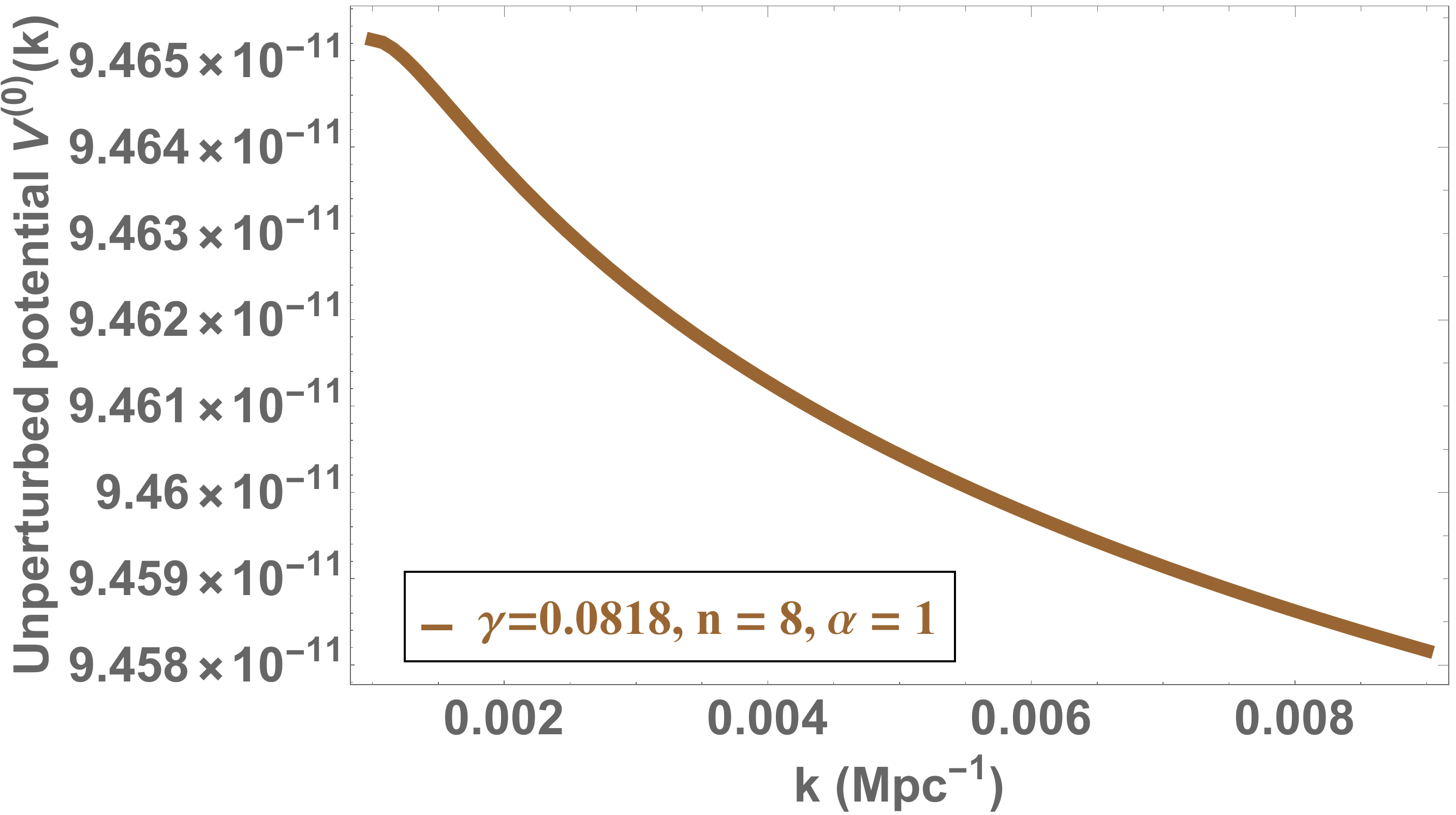}
   \subcaption{}
    \label{fig:unperturbedPOT_7}
\end{subfigure}%
\begin{subfigure}{0.33\linewidth}
  \centering
   \includegraphics[width=46mm,height=40mm]{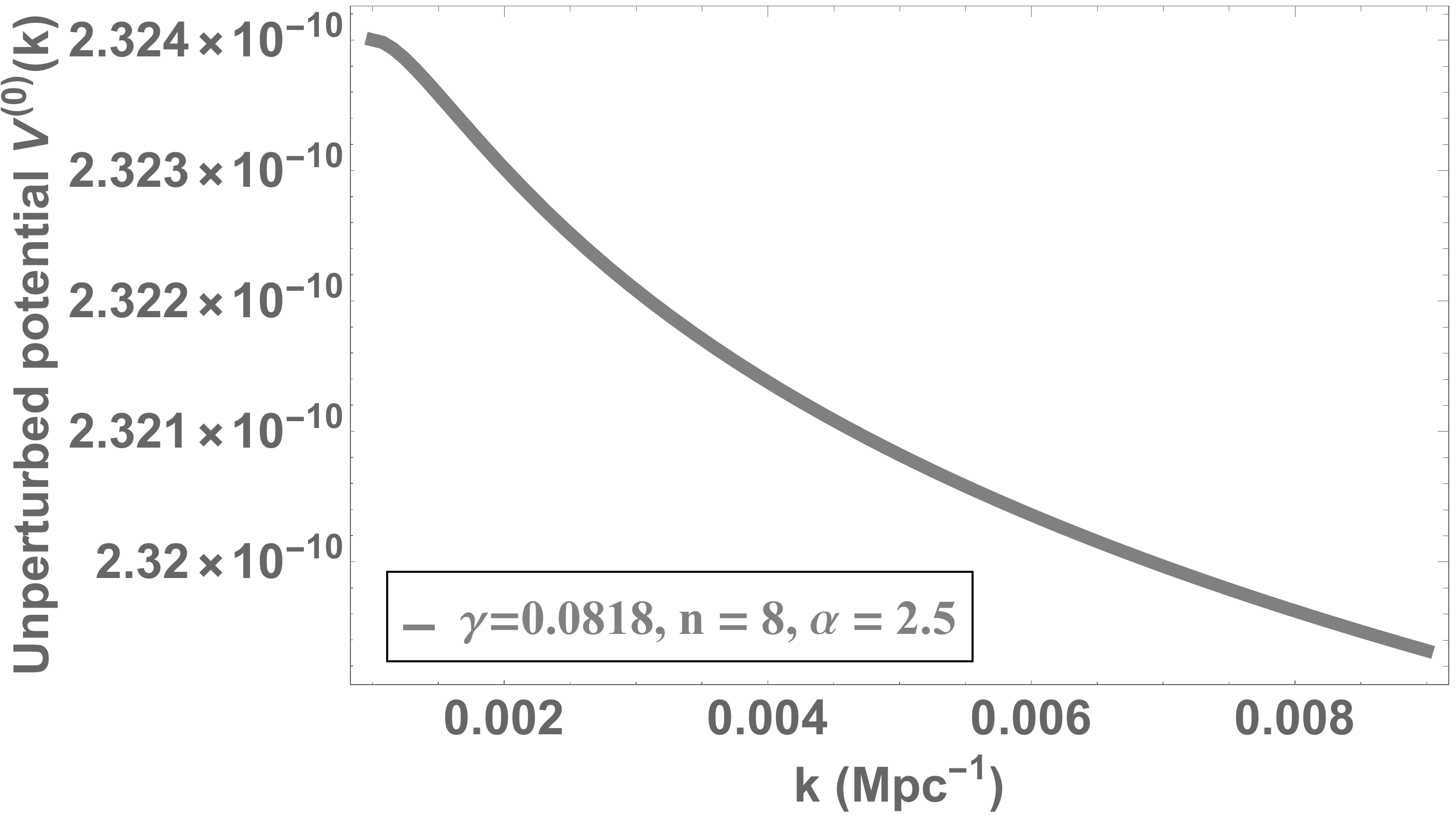}
   \subcaption{}
    \label{fig:unperturbedPOT_8}
\end{subfigure}%
\begin{subfigure}{0.33\linewidth}
  \centering
   \includegraphics[width=46mm,height=40mm]{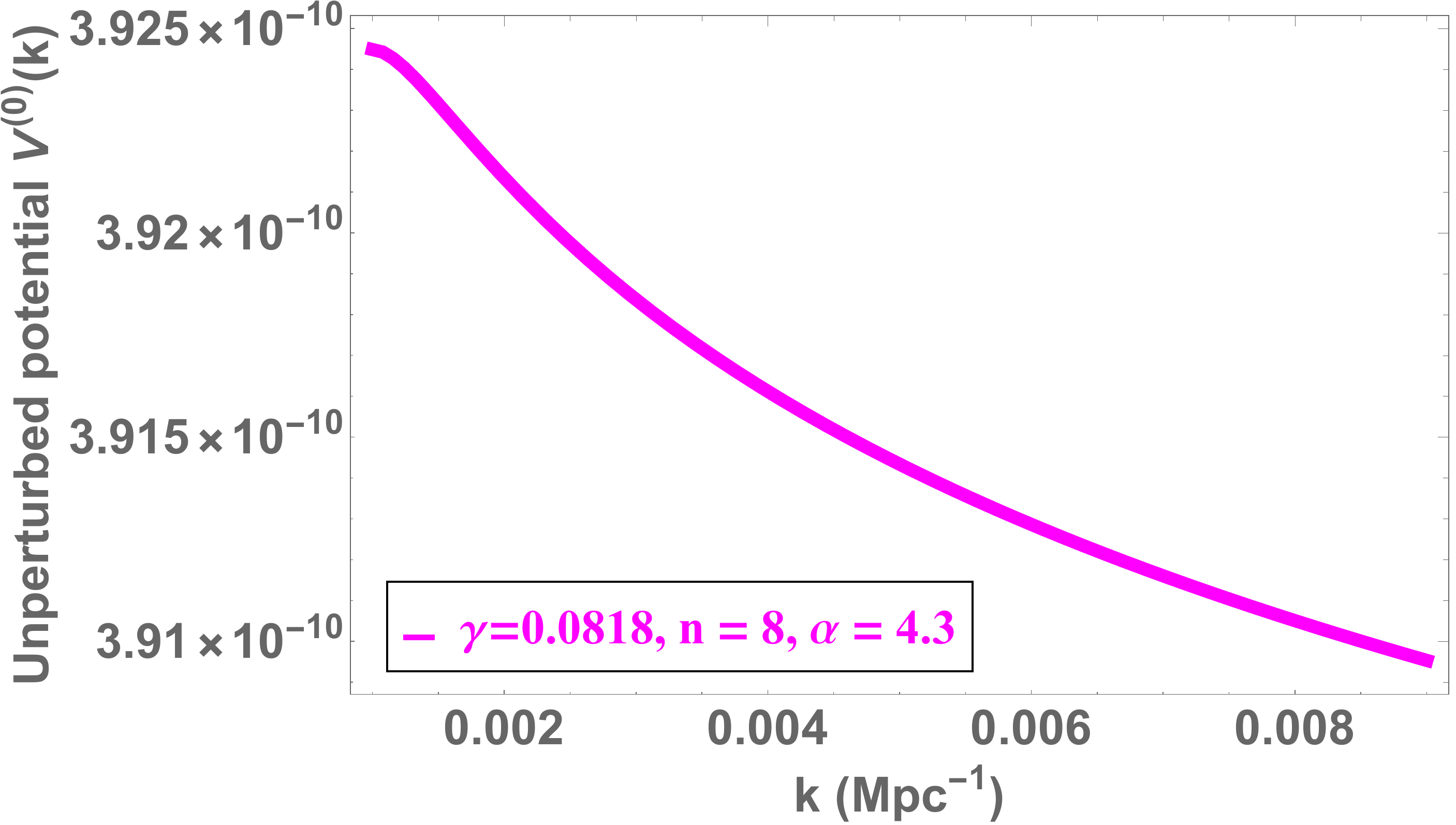}
   \subcaption{}
    \label{fig:unperturbedPOT_9}
\end{subfigure}
\caption{ Zeroth order parts of the potential of Eq. (\ref{eq:final_inf_pot}) for nine values of $\alpha$ for $\gamma=0.0818$ and $n=8$. The values of $V^{(0)}(k)$ tend to increase with the increase in $\alpha$ at a particular $k$ value.}
\label{fig:unperturbedPOT}
\end{figure}

\begin{figure}[H]
\begin{subfigure}{0.33\linewidth}
  \centering
   \includegraphics[width=46mm,height=40mm]{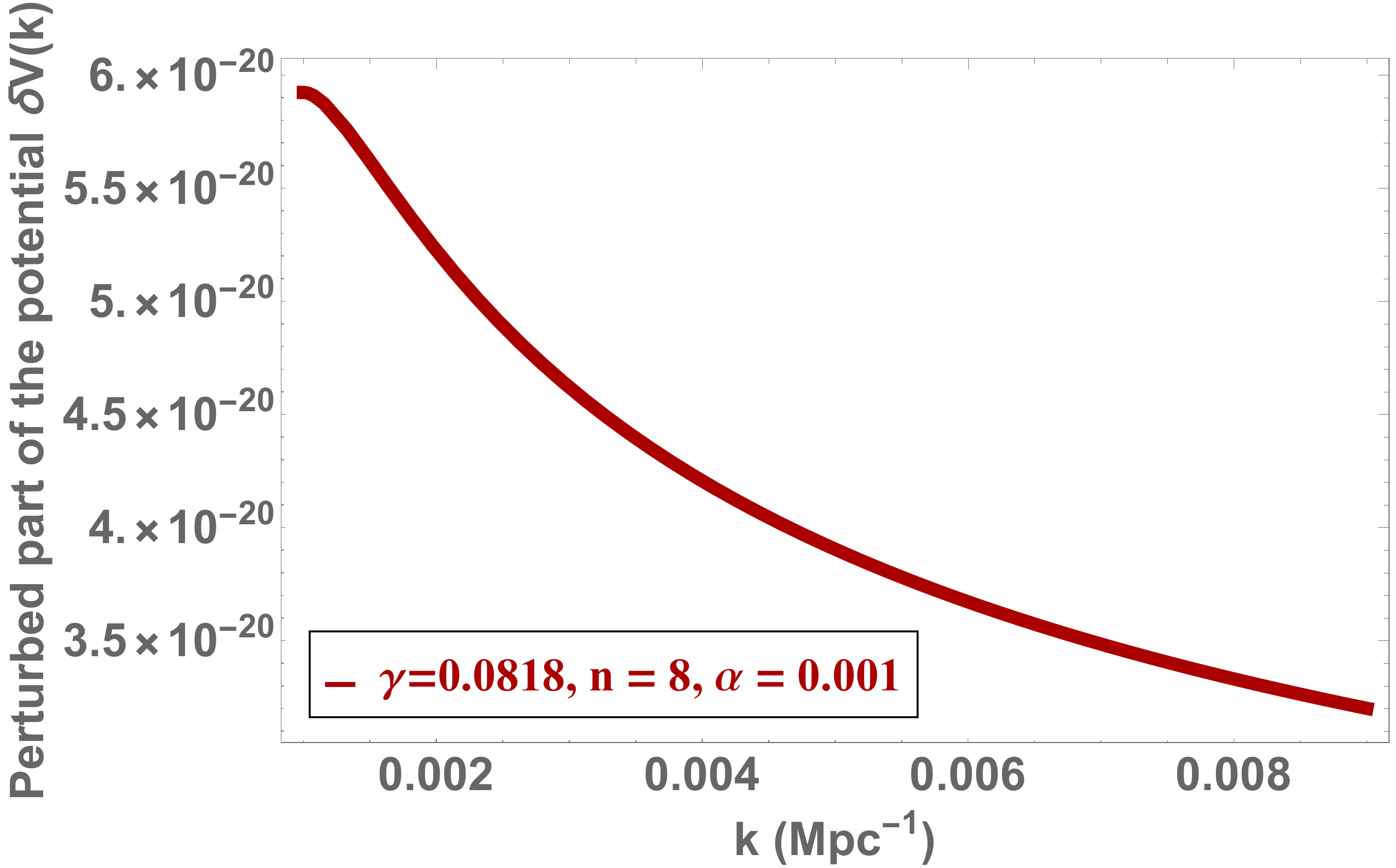} 
   \subcaption{}
   \label{fig:perturbedPOT_1}
\end{subfigure}%
\begin{subfigure}{0.33\linewidth}
  \centering
   \includegraphics[width=46mm,height=40mm]{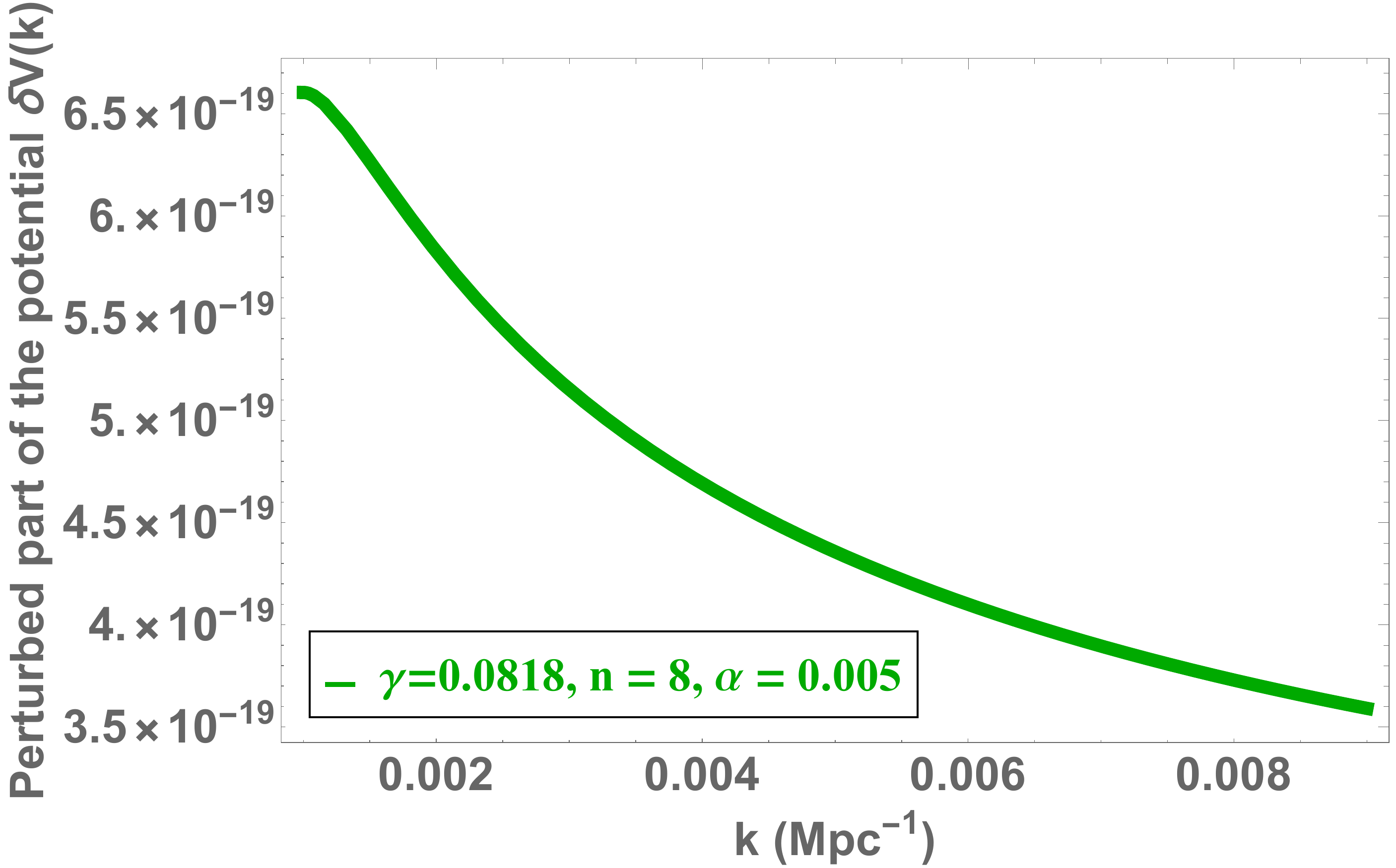}
   \subcaption{}
   \label{fig:perturbedPOT_2}
\end{subfigure}%
\begin{subfigure}{0.33\linewidth}
  \centering
   \includegraphics[width=46mm,height=40mm]{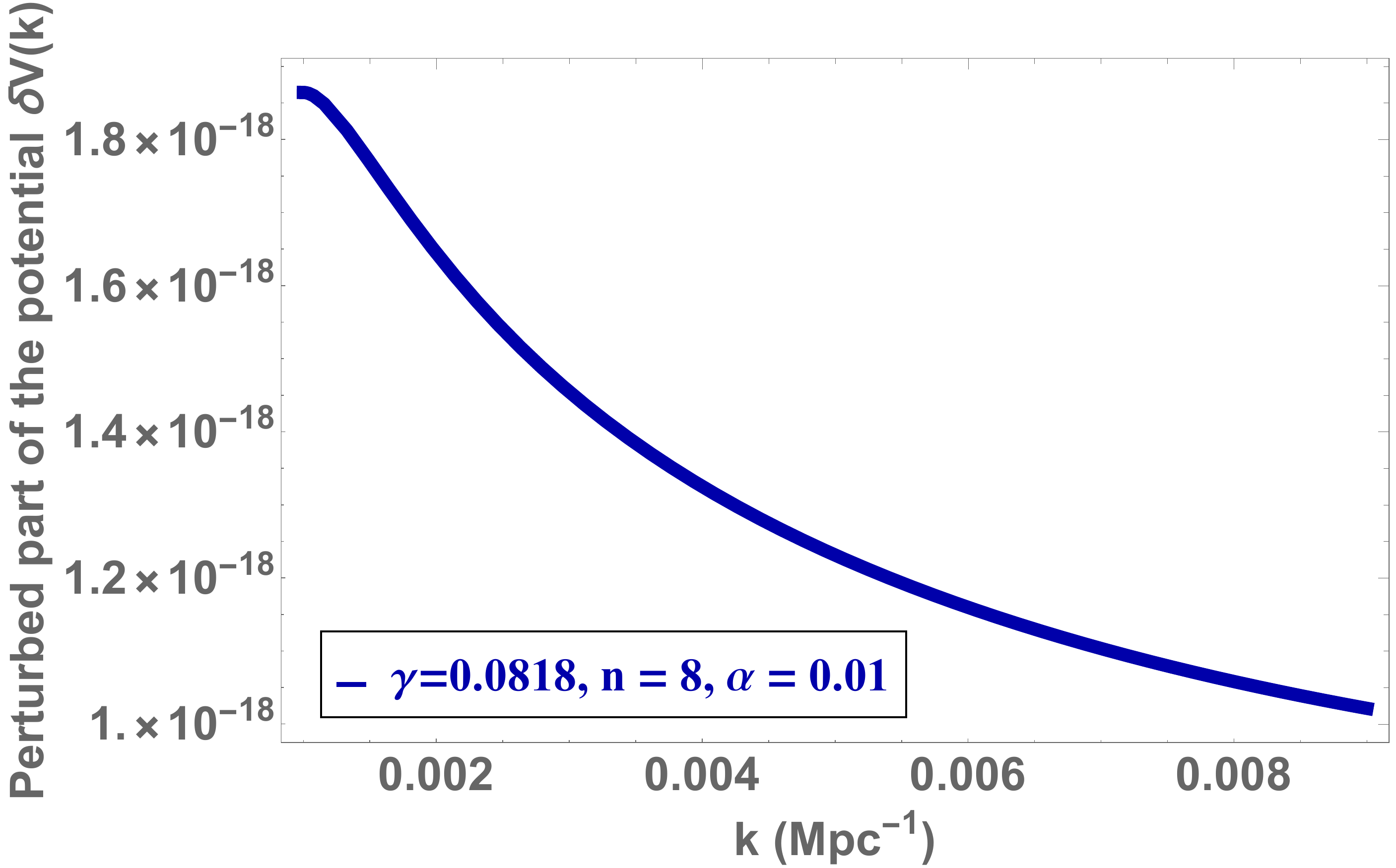}
   \subcaption{}
   \label{fig:perturbedPOT_3}
\end{subfigure}%
\vspace{0.05\linewidth}
\begin{subfigure}{0.33\linewidth}
  \centering
   \includegraphics[width=46mm,height=40mm]{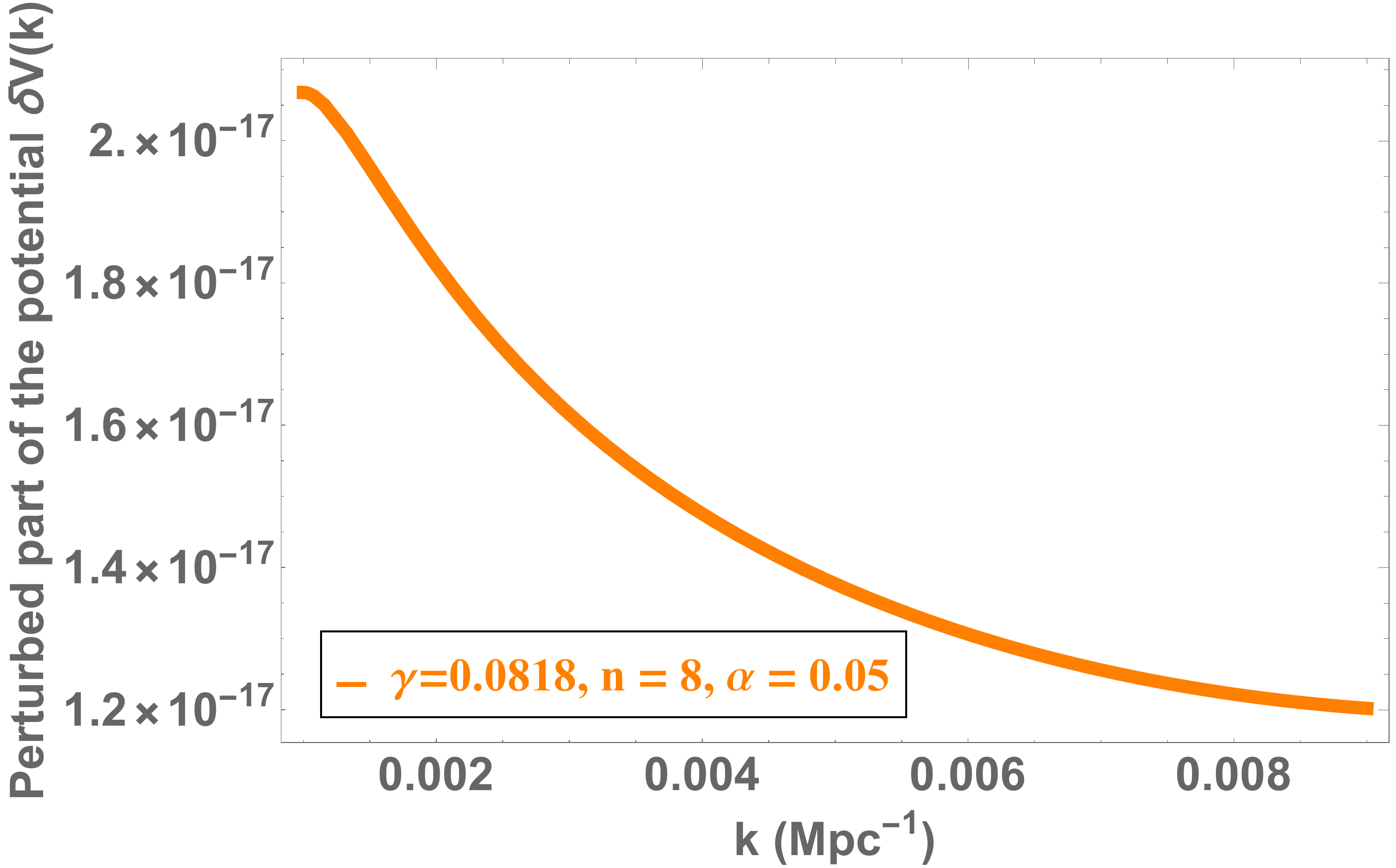}
   \subcaption{}
    \label{fig:perturbedPOT_4}
\end{subfigure}%
\begin{subfigure}{0.33\linewidth}
  \centering
   \includegraphics[width=46mm,height=40mm]{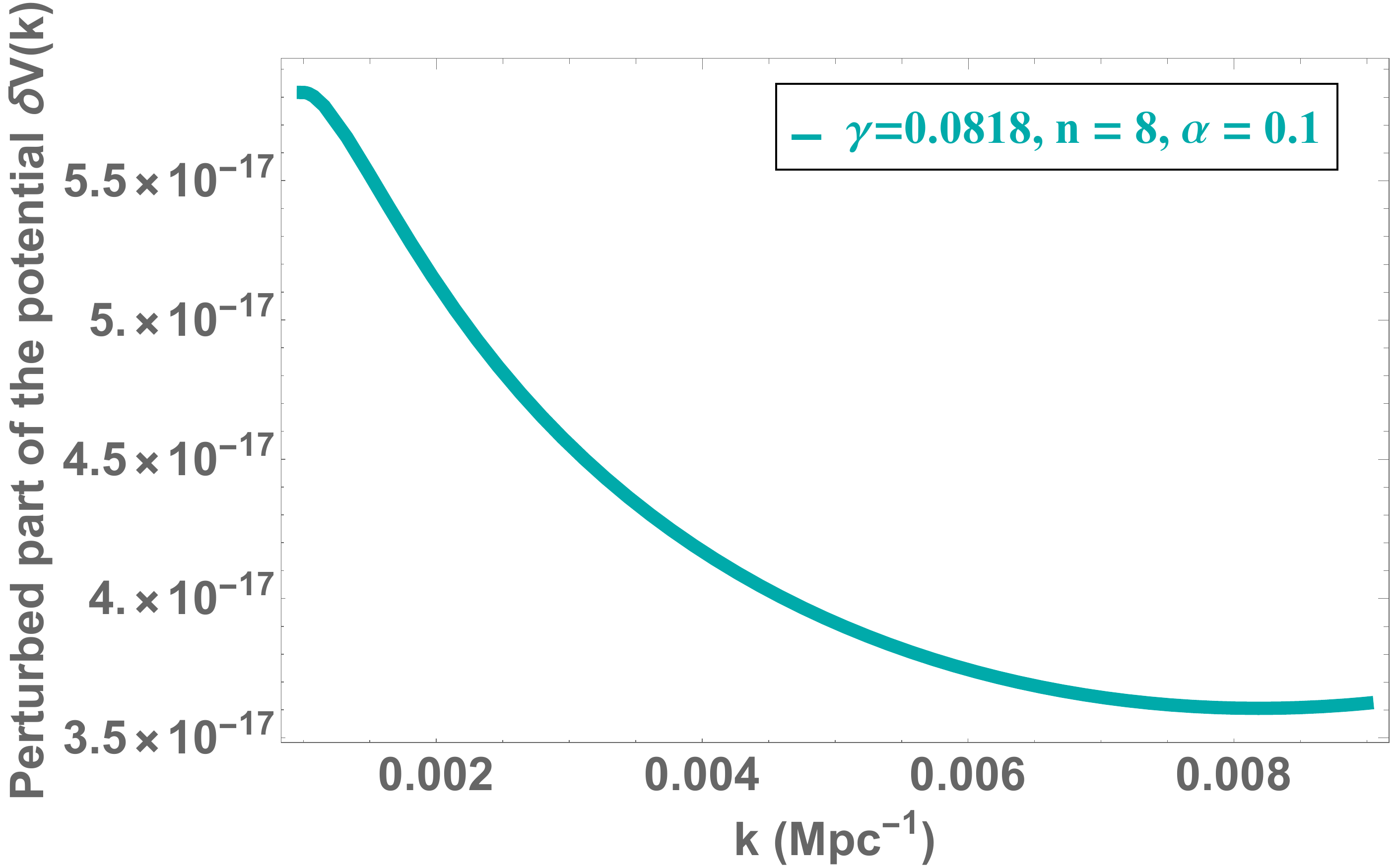}
   \subcaption{}
    \label{fig:perturbedPOT_5}
\end{subfigure}%
\begin{subfigure}{0.33\linewidth}
  \centering
   \includegraphics[width=46mm,height=40mm]{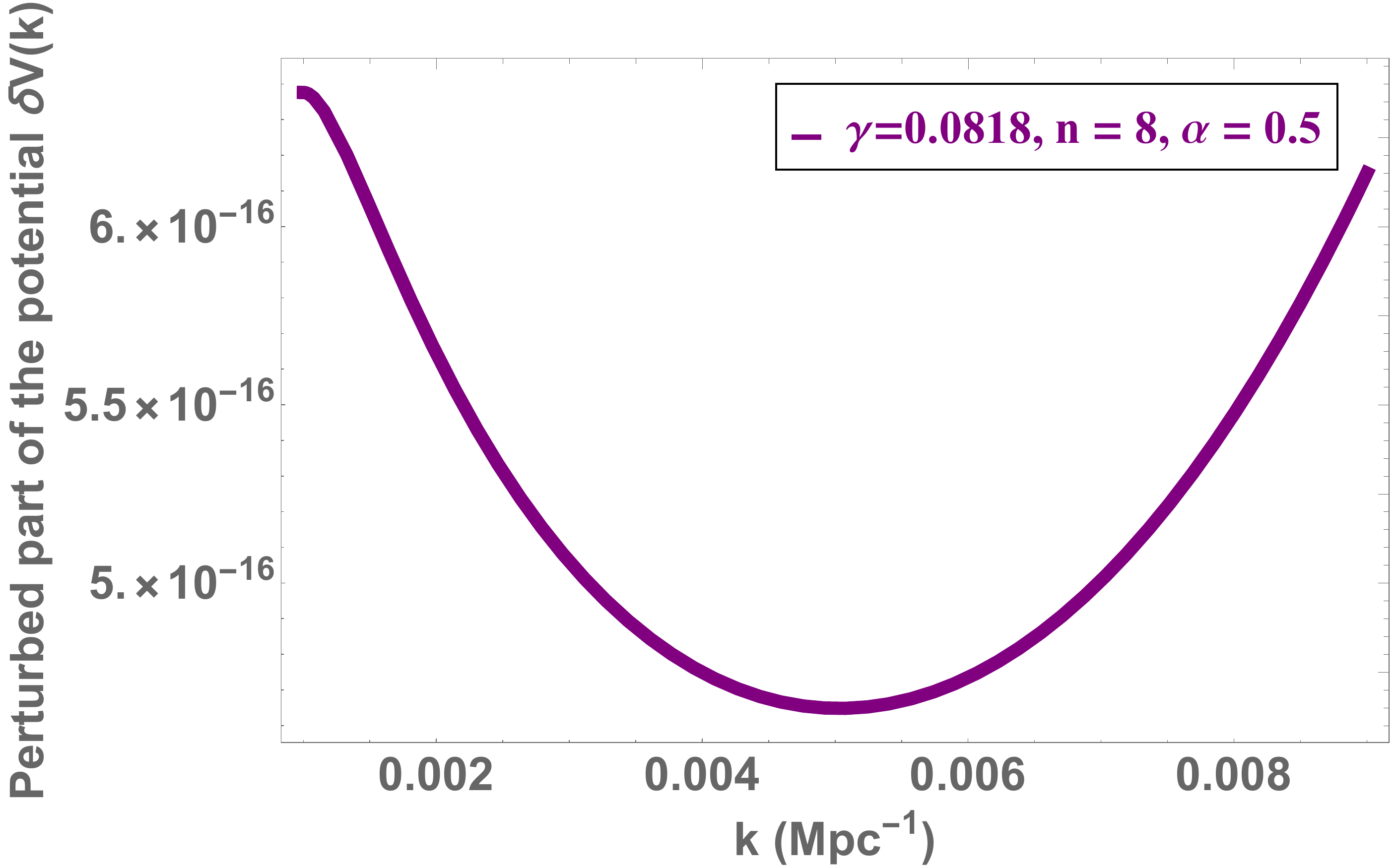}
   \subcaption{}
    \label{fig:perturbedPOT_6}
\end{subfigure}%
\vspace{0.05\linewidth}
\begin{subfigure}{0.33\linewidth}
  \centering
   \includegraphics[width=46mm,height=40mm]{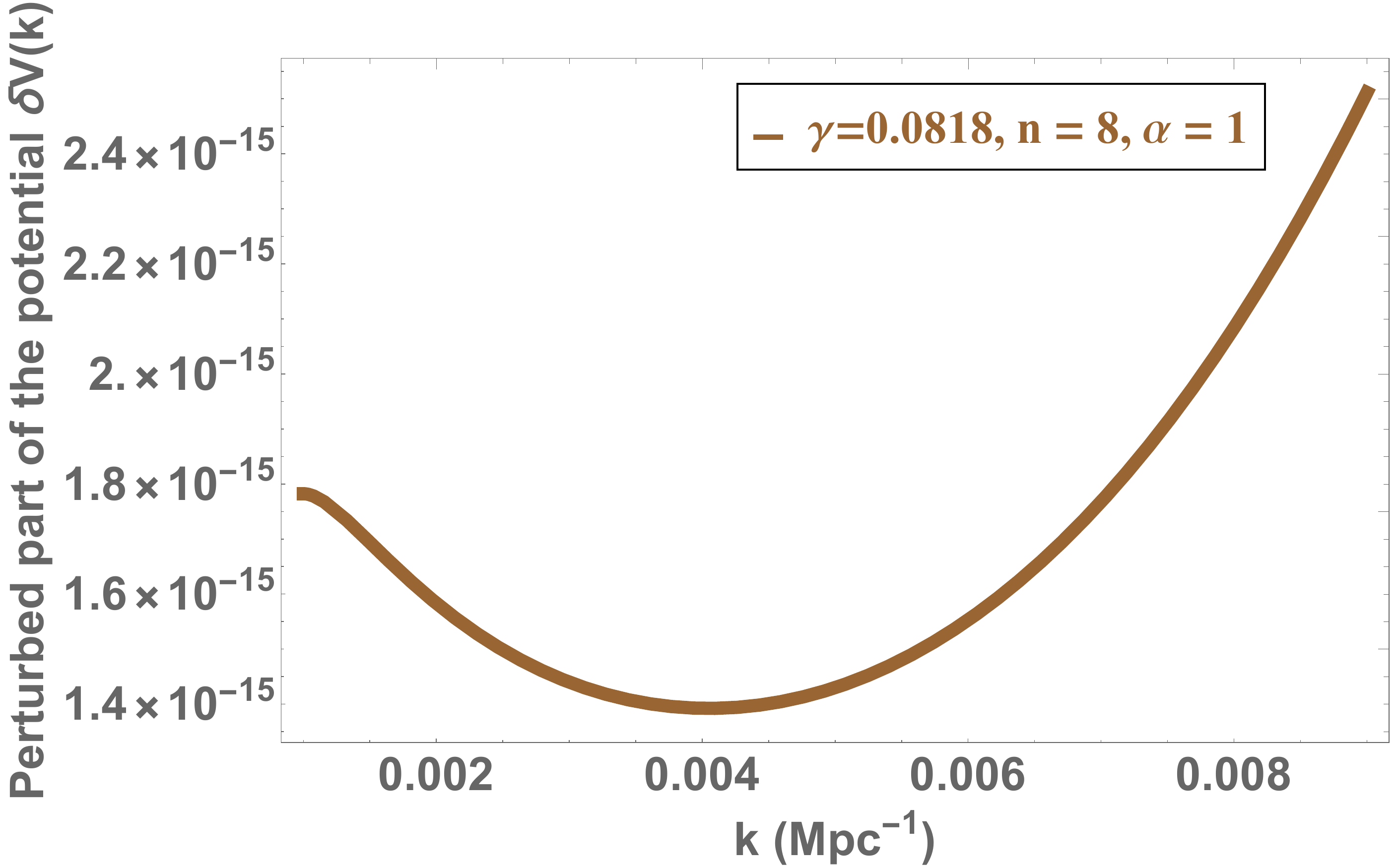}
   \subcaption{}
    \label{fig:perturbedPOT_7}
\end{subfigure}%
\begin{subfigure}{0.33\linewidth}
  \centering
   \includegraphics[width=46mm,height=40mm]{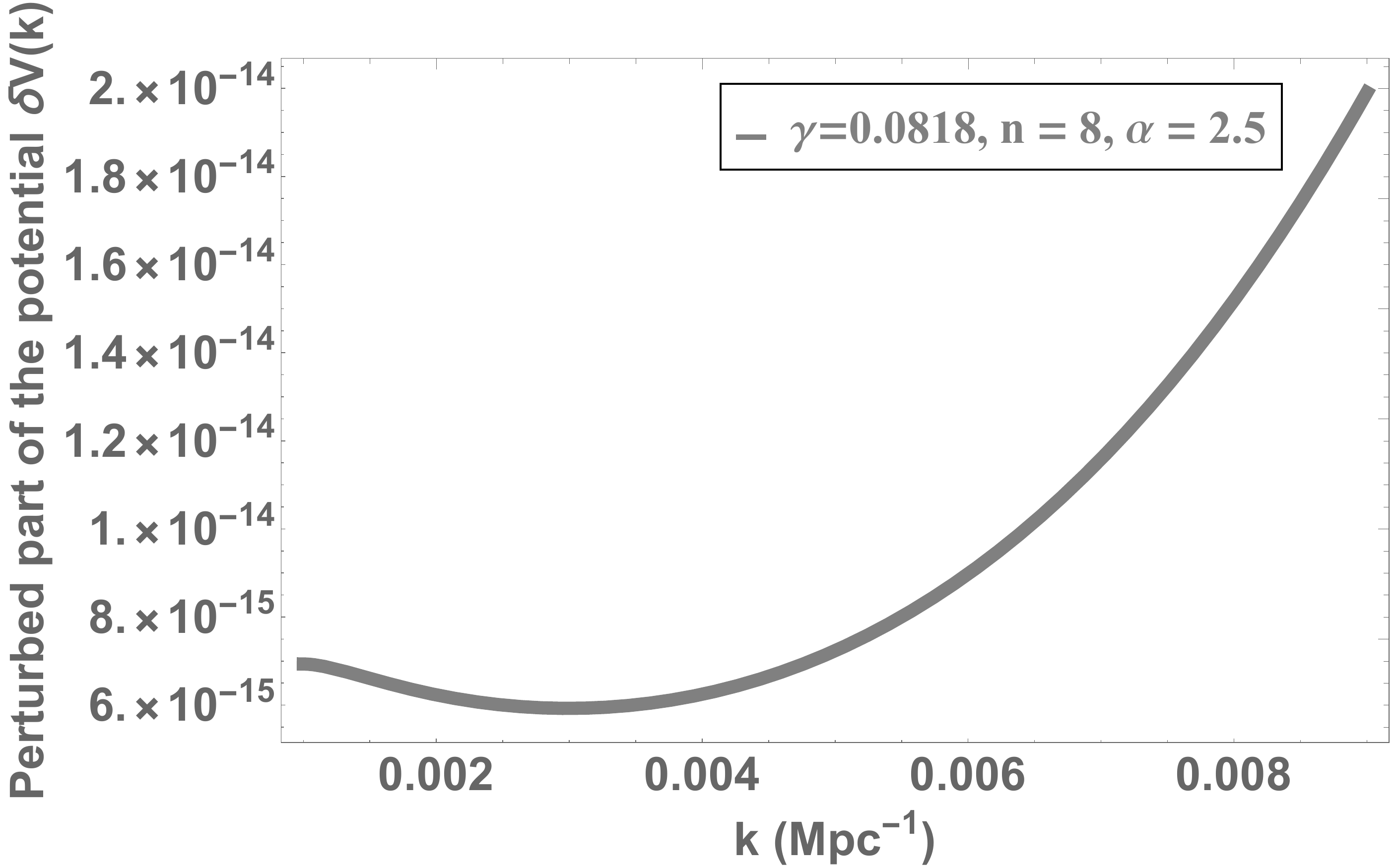}
   \subcaption{}
    \label{fig:perturbedPOT_8}
\end{subfigure}%
\begin{subfigure}{0.33\linewidth}
  \centering
   \includegraphics[width=46mm,height=40mm]{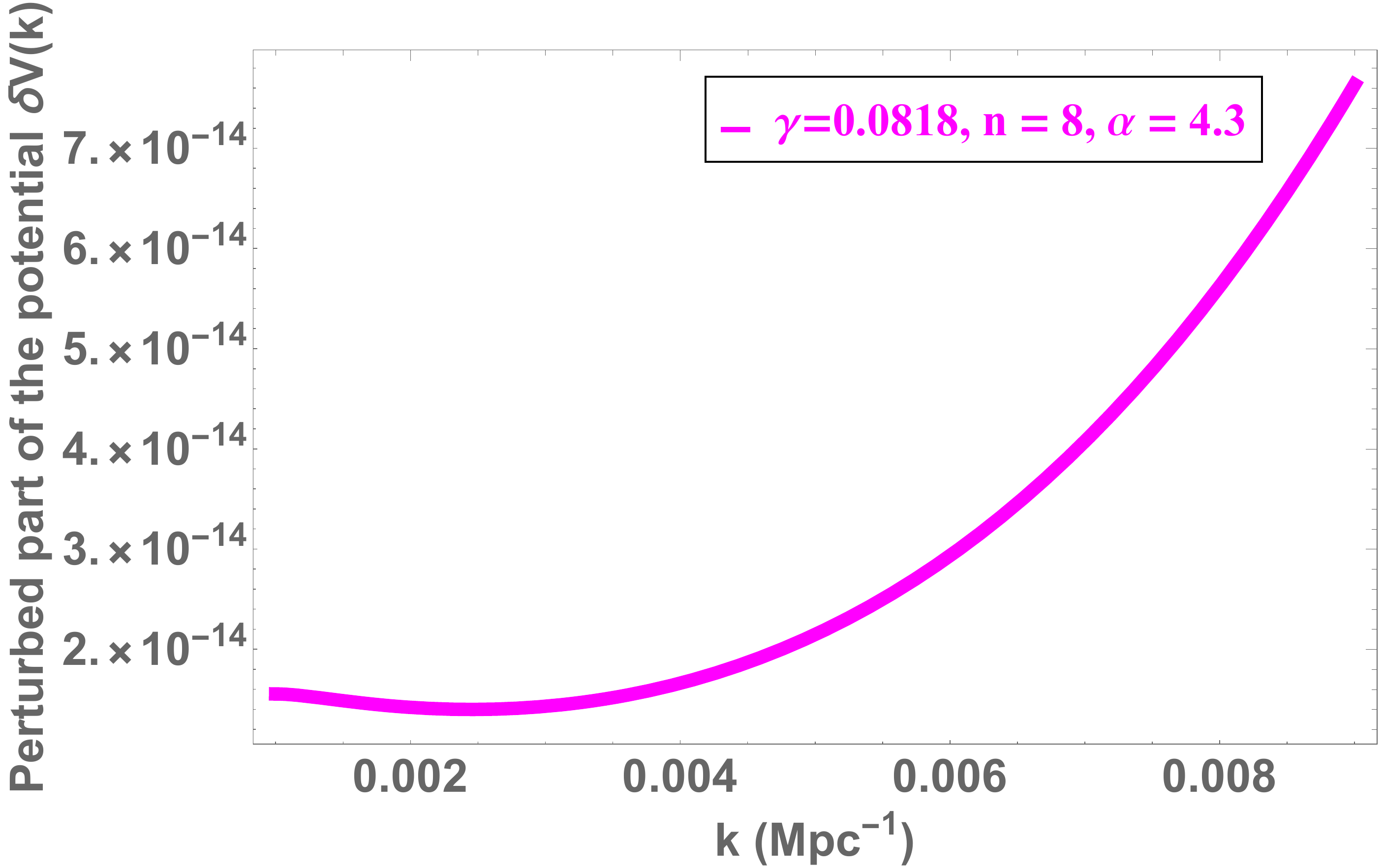}
   \subcaption{}
    \label{fig:perturbedPOT_9}
\end{subfigure}
\caption{First order perturbed parts of the potential of Eq. (\ref{eq:final_inf_pot}) for nine values of $\alpha$ for $\gamma=0.0818$ and $n=8$. The values of $\delta V(k)$ tend to increase with the increase in $\alpha$ at a particular $k$-value. After $\alpha=0.1$ the perturbation becomes stronger for $k\geq 0.005$ Mpc$^{-1}$ than $k<0.005$ Mpc$^{-1}$.}
\label{fig:perturbedPOT}
\end{figure}

 Now, the solutions of $\xi^{(0)}(k)$ and $\delta\xi(k)$ can be suitably coupled to the inflaton potential in momentum space as
\begin{equation}
    V(\xi(k))=V(\xi^{(0)}(k))+\partial_{\xi^{(0)}(k)}V(\xi^{(0)}(k))\delta\xi(k)
    =V^{(0)}(k)+\delta V(k).
\end{equation} In this way, the quantum aspects of the inflaton perturbation are transferred microscopically into the potential in $k$-space. Therefore it is quite expected that the perturbative framework should work well with respect to the potential and its perturbation part also. That is exactly what we observe in figures \ref{fig:unperturbedPOT} and \ref{fig:perturbedPOT}. The $V^{(0)}(k)$s are found in the range $9.723\times 10^{-14}$ $-$ $3.93\times 10^{-10}$ and the corresponding $\delta V(k)$s from $3\times 10^{-20}$ to $8\times 10^{-14}$. That is, $\delta V$ is about $10^4-10^6$ times smaller than $V^{(0)}$. It is also quite clear that the orders of all $V^{(0)}(k)$ values are consistent with the COBE/Planck normalisation of Eq. (\ref{eq:COBEnormalization}) (see also table \ref{tab:Table1}) which is very important for the parameter estimation, specifically the power spectra.
\subsection{Mode responses of the parameters during EMQA-inflation}
\label{subsec:ModeBehaviours}
\begin{figure}[H]
\begin{subfigure}{0.33\linewidth}
  \centering
   \includegraphics[width=46mm,height=40mm]{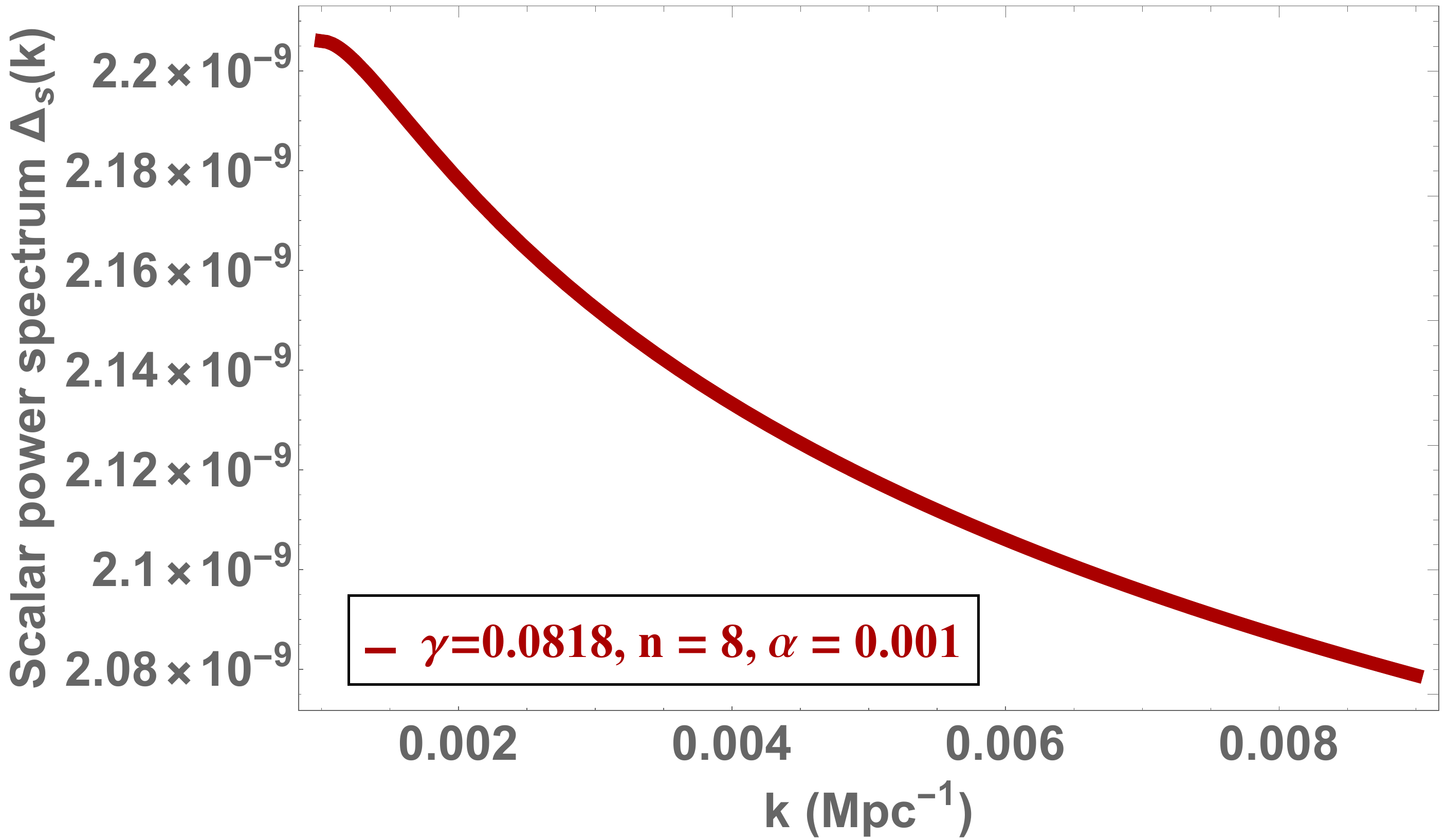} 
   \subcaption{}
   \label{fig:scalarPowerSpectrum_1}
\end{subfigure}%
\begin{subfigure}{0.33\linewidth}
  \centering
   \includegraphics[width=46mm,height=40mm]{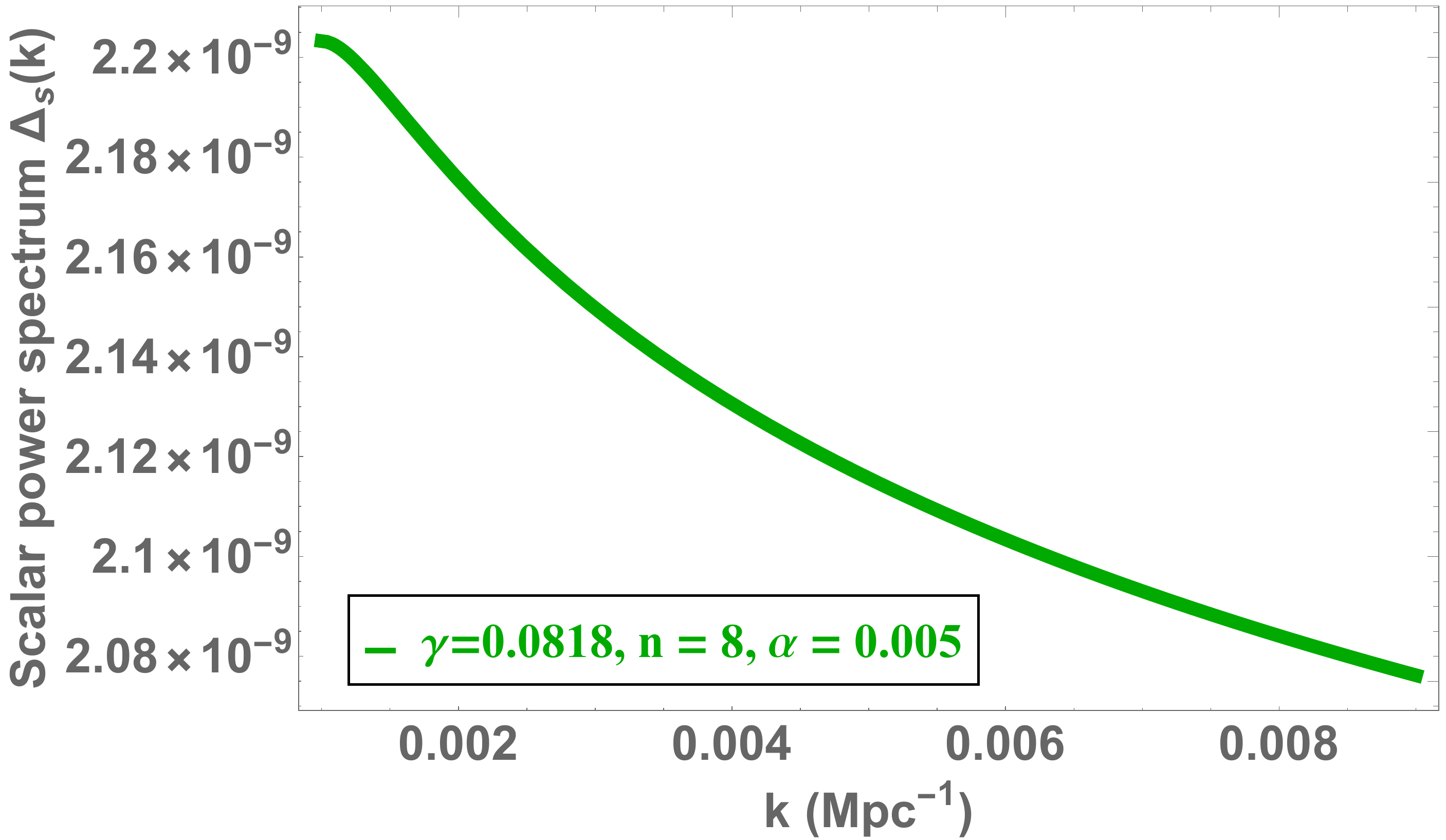}
   \subcaption{}
   \label{fig:scalarPowerSpectrum_2}
\end{subfigure}%
\begin{subfigure}{0.33\linewidth}
  \centering
   \includegraphics[width=46mm,height=40mm]{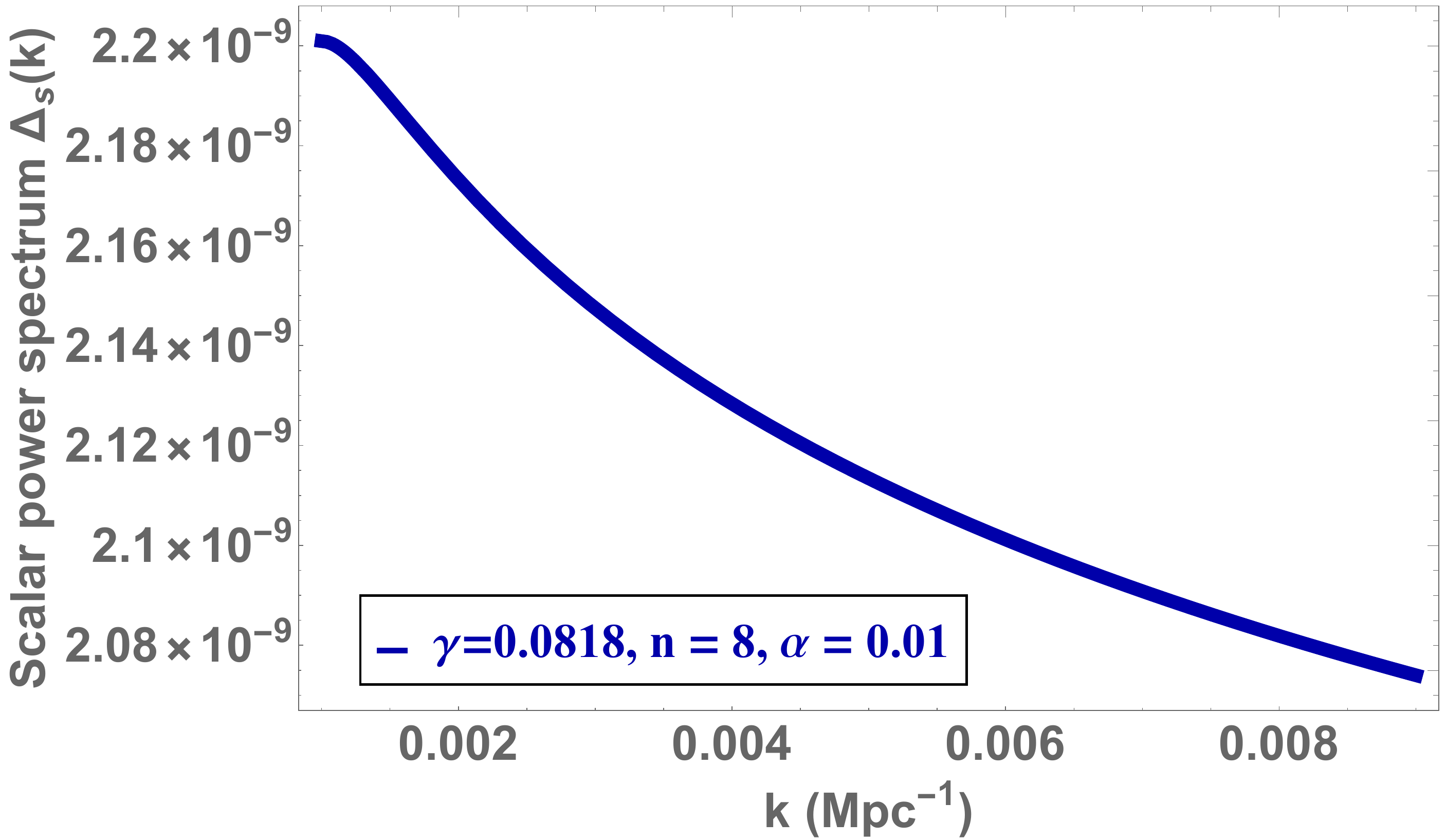}
   \subcaption{}
   \label{fig:scalarPowerSpectrum_3}
\end{subfigure}%
\vspace{0.05\linewidth}
\begin{subfigure}{0.33\linewidth}
  \centering
   \includegraphics[width=46mm,height=40mm]{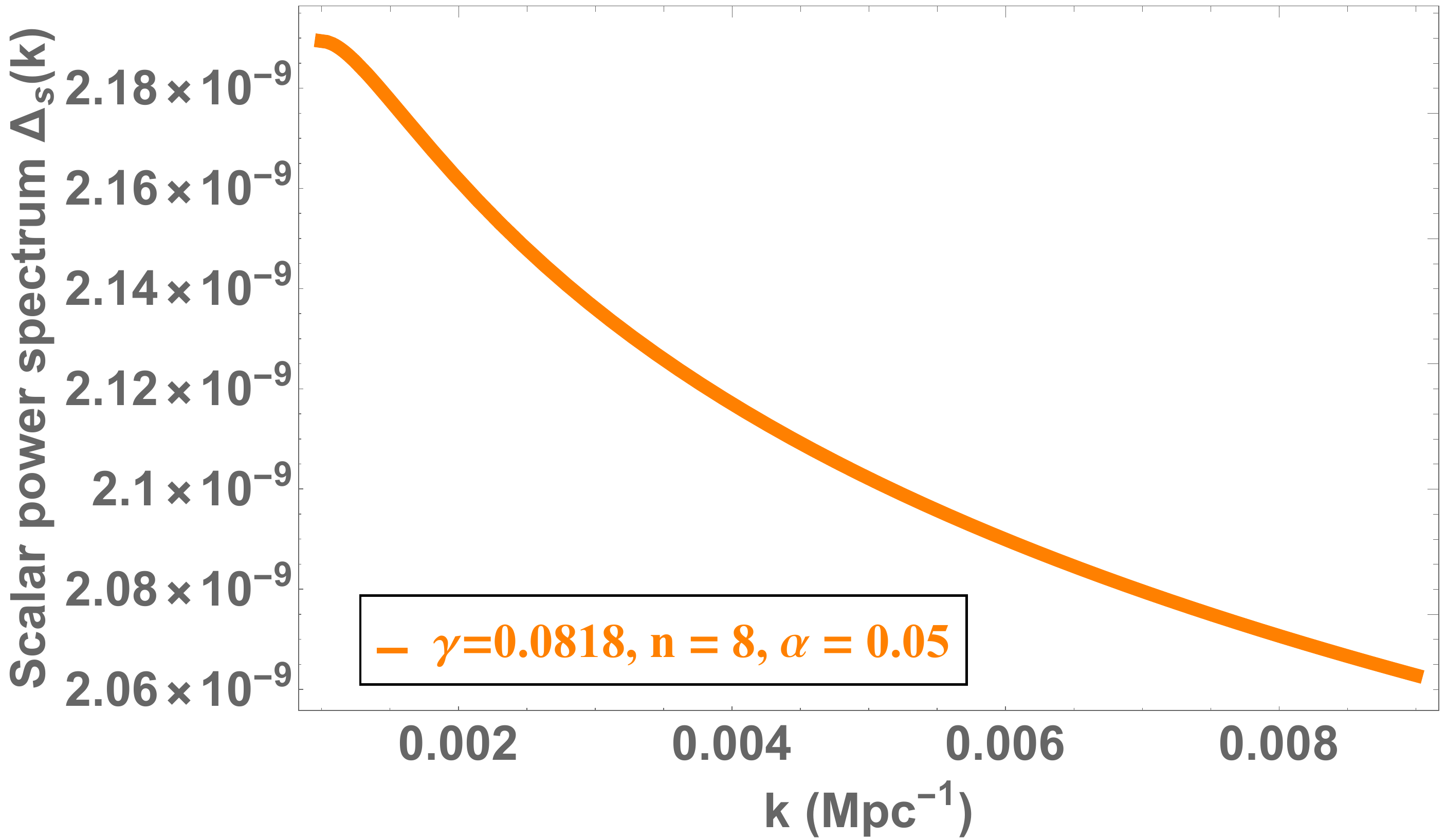}
   \subcaption{}
    \label{fig:scalarPowerSpectrum_4}
\end{subfigure}%
\begin{subfigure}{0.33\linewidth}
  \centering
   \includegraphics[width=46mm,height=40mm]{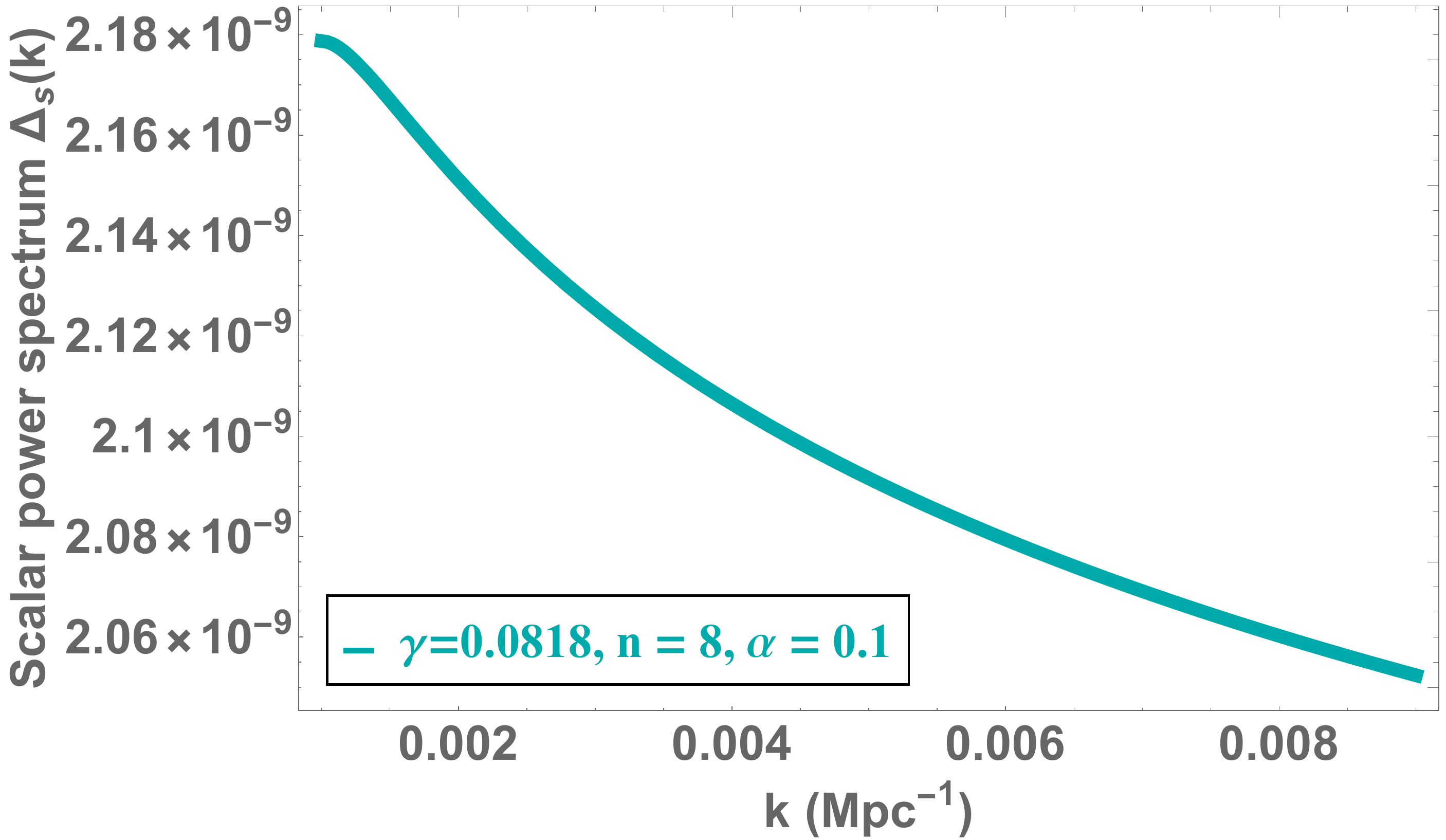}
   \subcaption{}
    \label{fig:scalarPowerSpectrum_5}
\end{subfigure}%
\begin{subfigure}{0.33\linewidth}
  \centering
   \includegraphics[width=46mm,height=40mm]{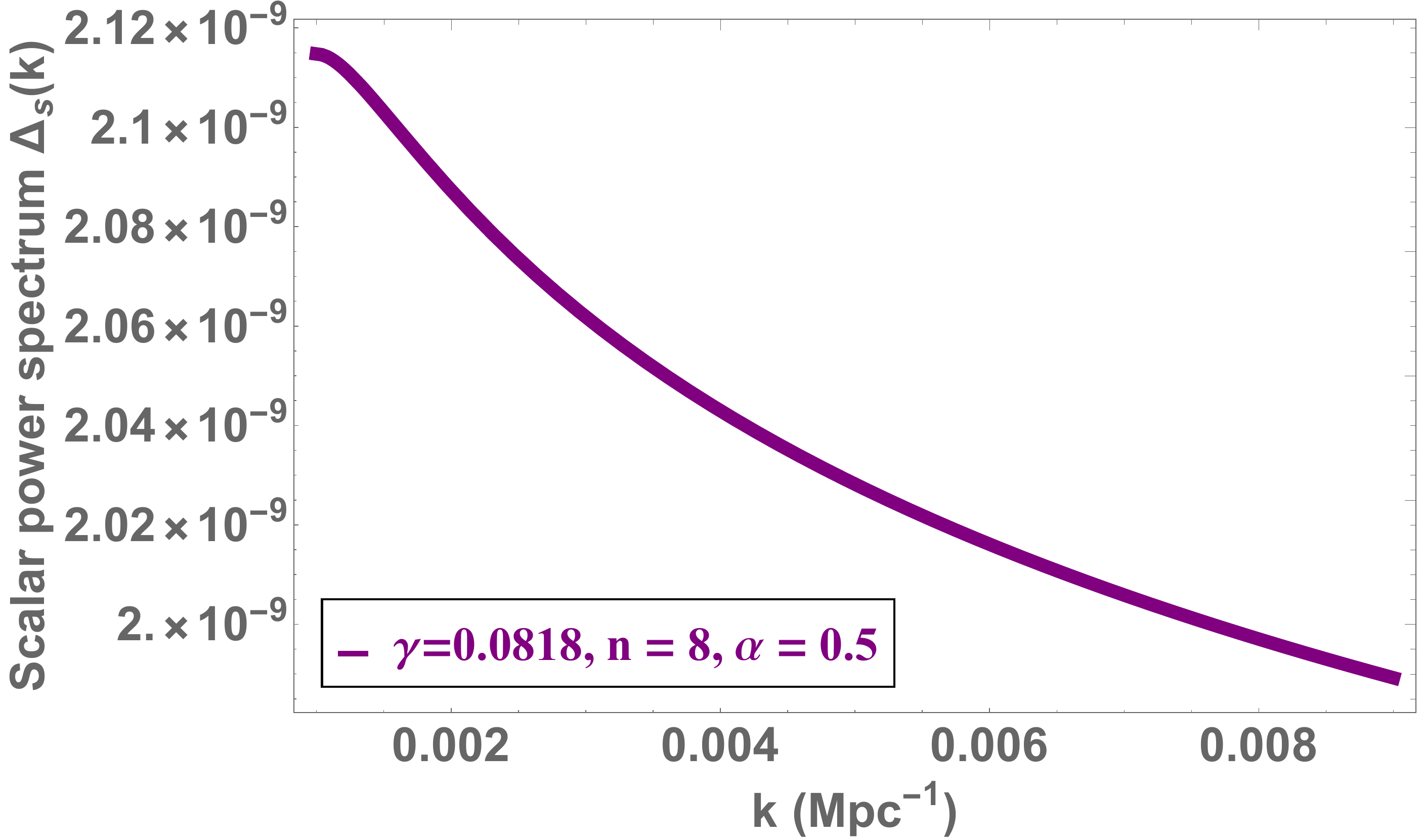}
   \subcaption{}
    \label{fig:scalarPowerSpectrum_6}
\end{subfigure}%
\vspace{0.05\linewidth}
\begin{subfigure}{0.33\linewidth}
  \centering
   \includegraphics[width=46mm,height=40mm]{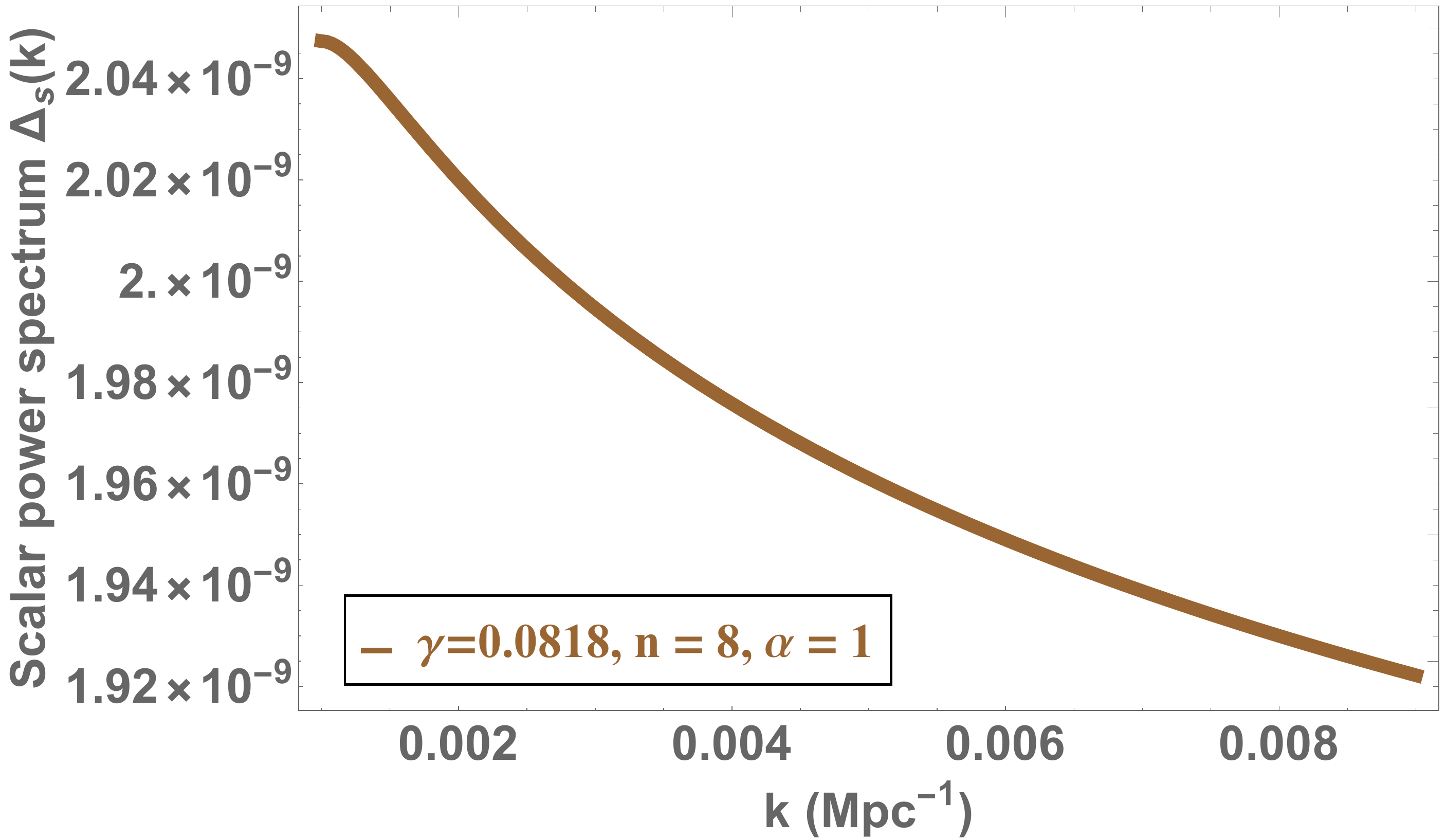}
   \subcaption{}
    \label{fig:scalarPowerSpectrum_7}
\end{subfigure}%
\begin{subfigure}{0.33\linewidth}
  \centering
   \includegraphics[width=46mm,height=40mm]{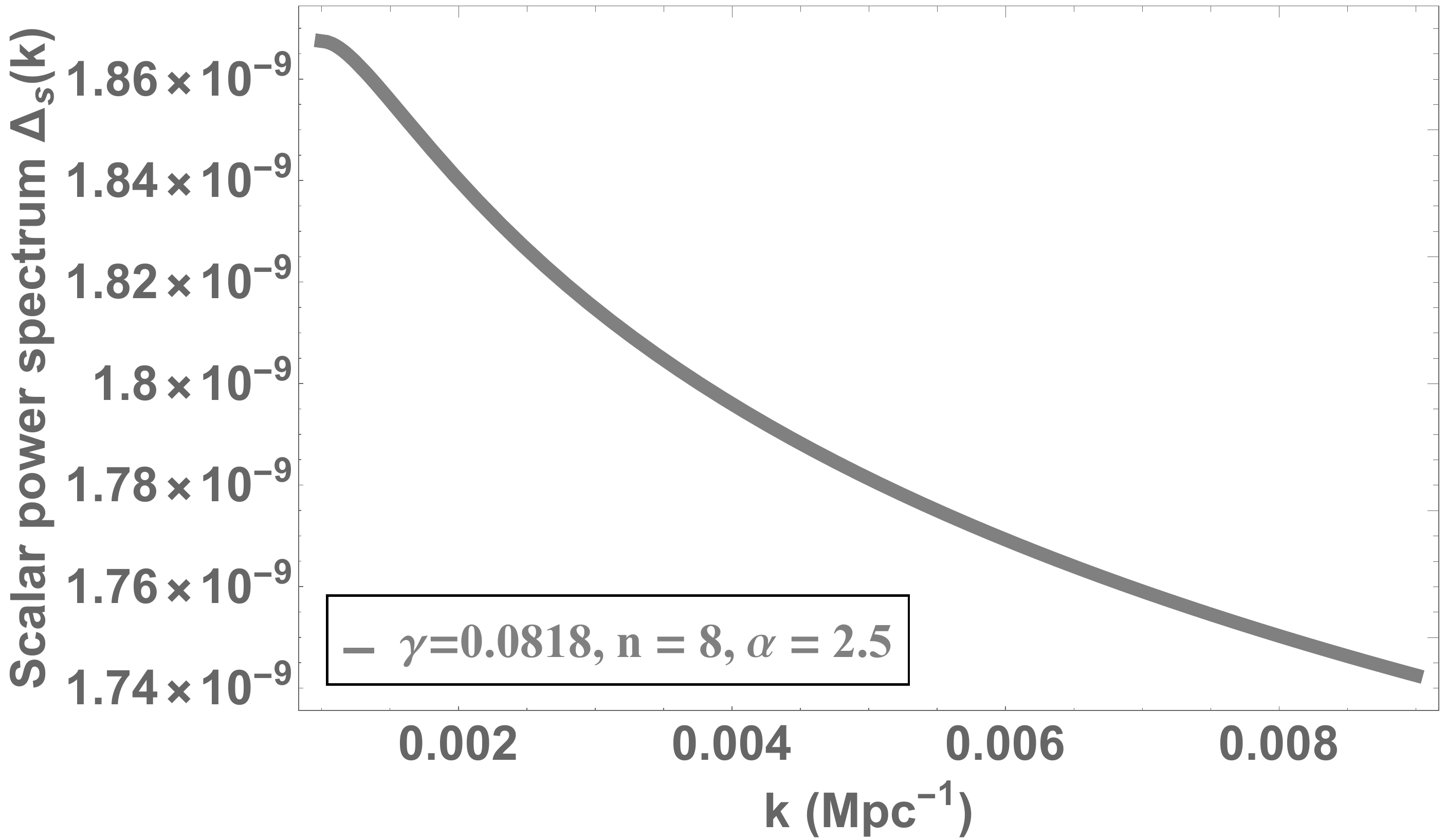}
   \subcaption{}
    \label{fig:scalarPowerSpectrum_8}
\end{subfigure}%
\begin{subfigure}{0.33\linewidth}
  \centering
   \includegraphics[width=46mm,height=40mm]{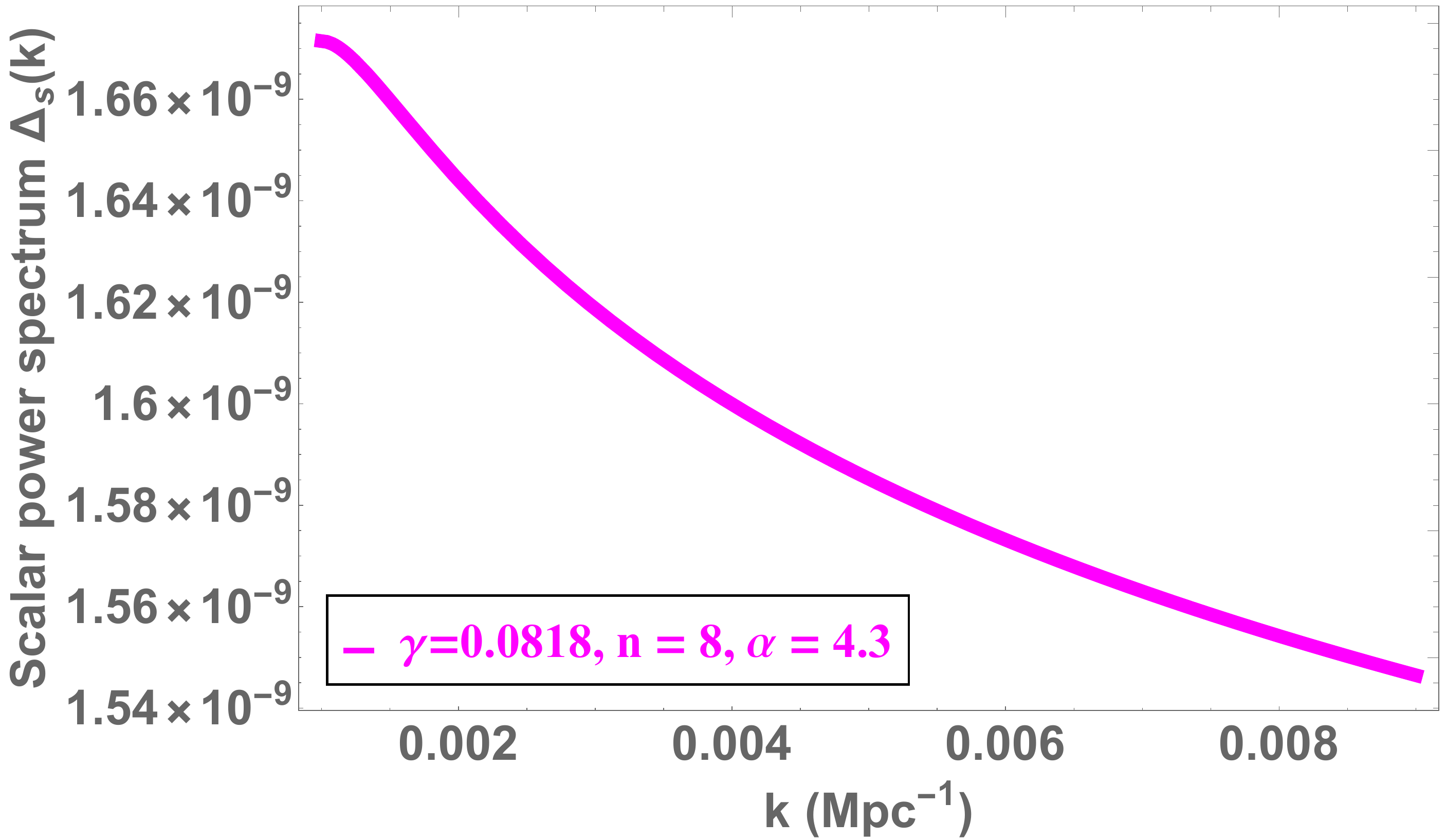}
   \subcaption{}
    \label{fig:scalarPowerSpectrum_9}
\end{subfigure}
\caption{Scalar power spectra for nine values of $\alpha$ for $\gamma=0.0818$ and $n=8$. Up to $\alpha=0.01$ the $\Delta_s(k)$ is almost insensitive to $\alpha$ and then it tends to decrease slowly with the increase in $\alpha$ at a particular $k$ value.}
\label{fig:scalarPowerSpectrum}
\end{figure}

\begin{figure}[H]
\begin{subfigure}{0.33\linewidth}
  \centering
   \includegraphics[width=46mm,height=40mm]{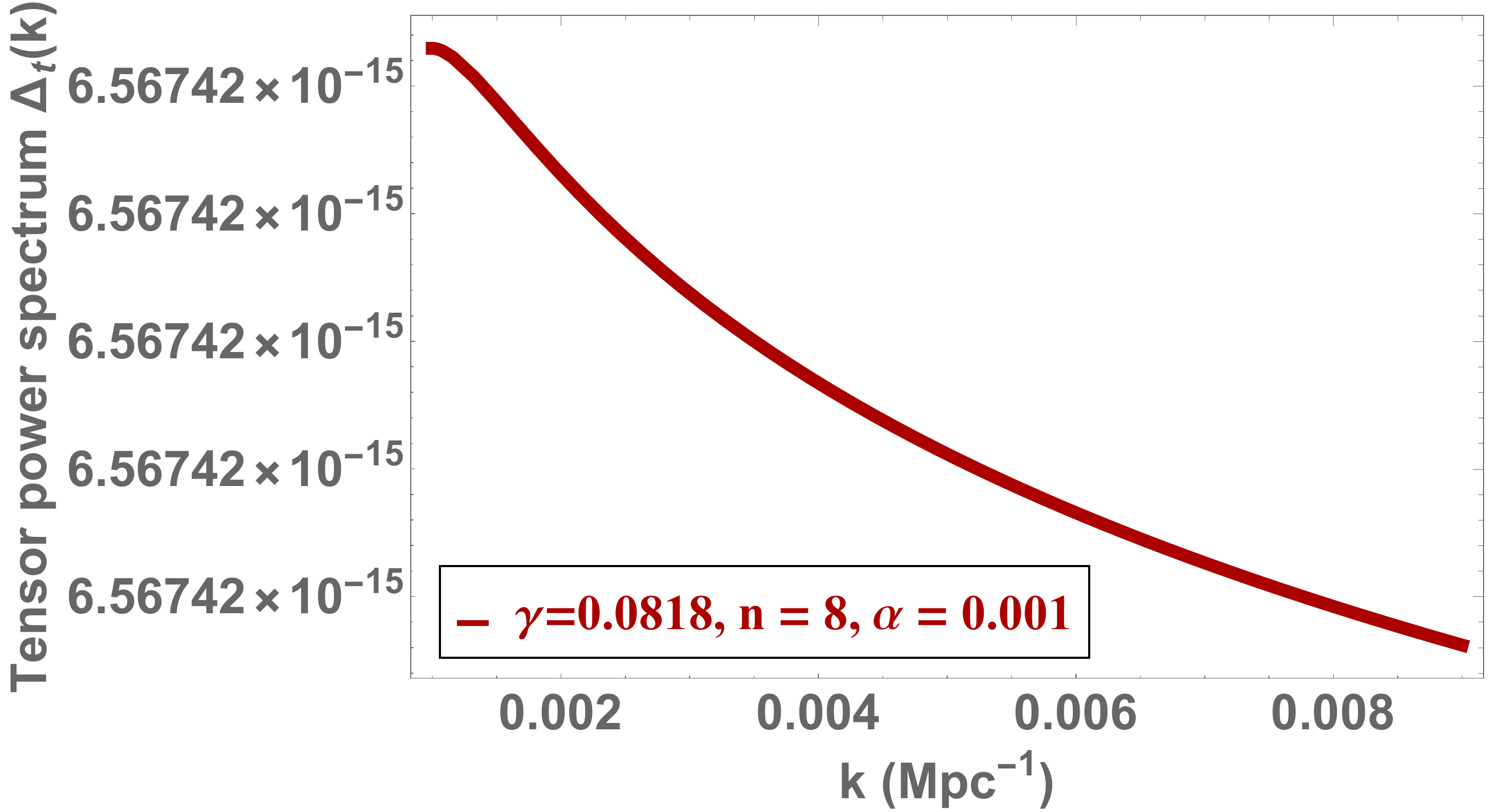} 
   \subcaption{}
   \label{fig:tensorPowerSpectrum_1}
\end{subfigure}%
\begin{subfigure}{0.33\linewidth}
  \centering
   \includegraphics[width=46mm,height=40mm]{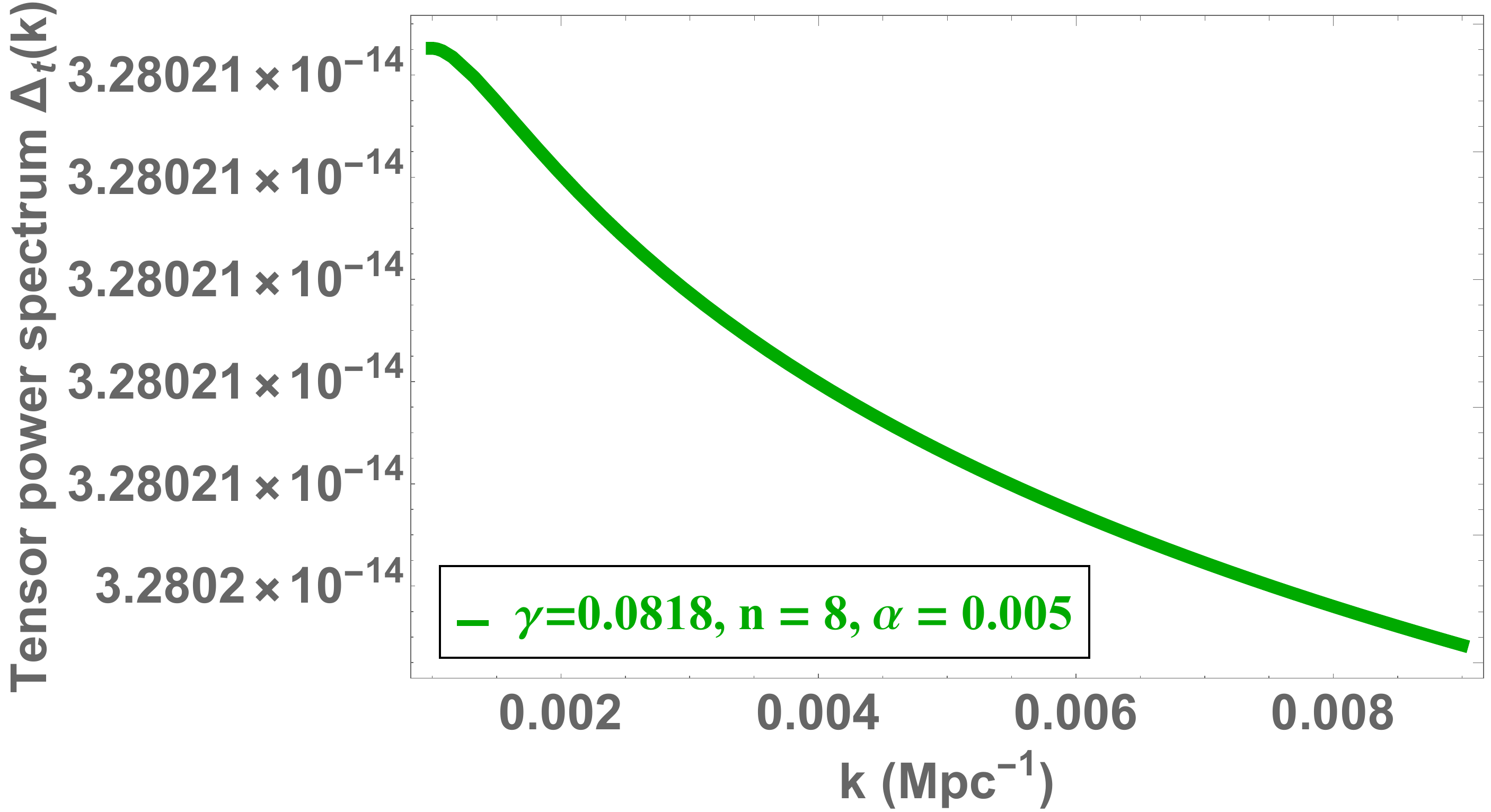}
   \subcaption{}
   \label{fig:tensorPowerSpectrum_2}
\end{subfigure}%
\begin{subfigure}{0.33\linewidth}
  \centering
   \includegraphics[width=46mm,height=40mm]{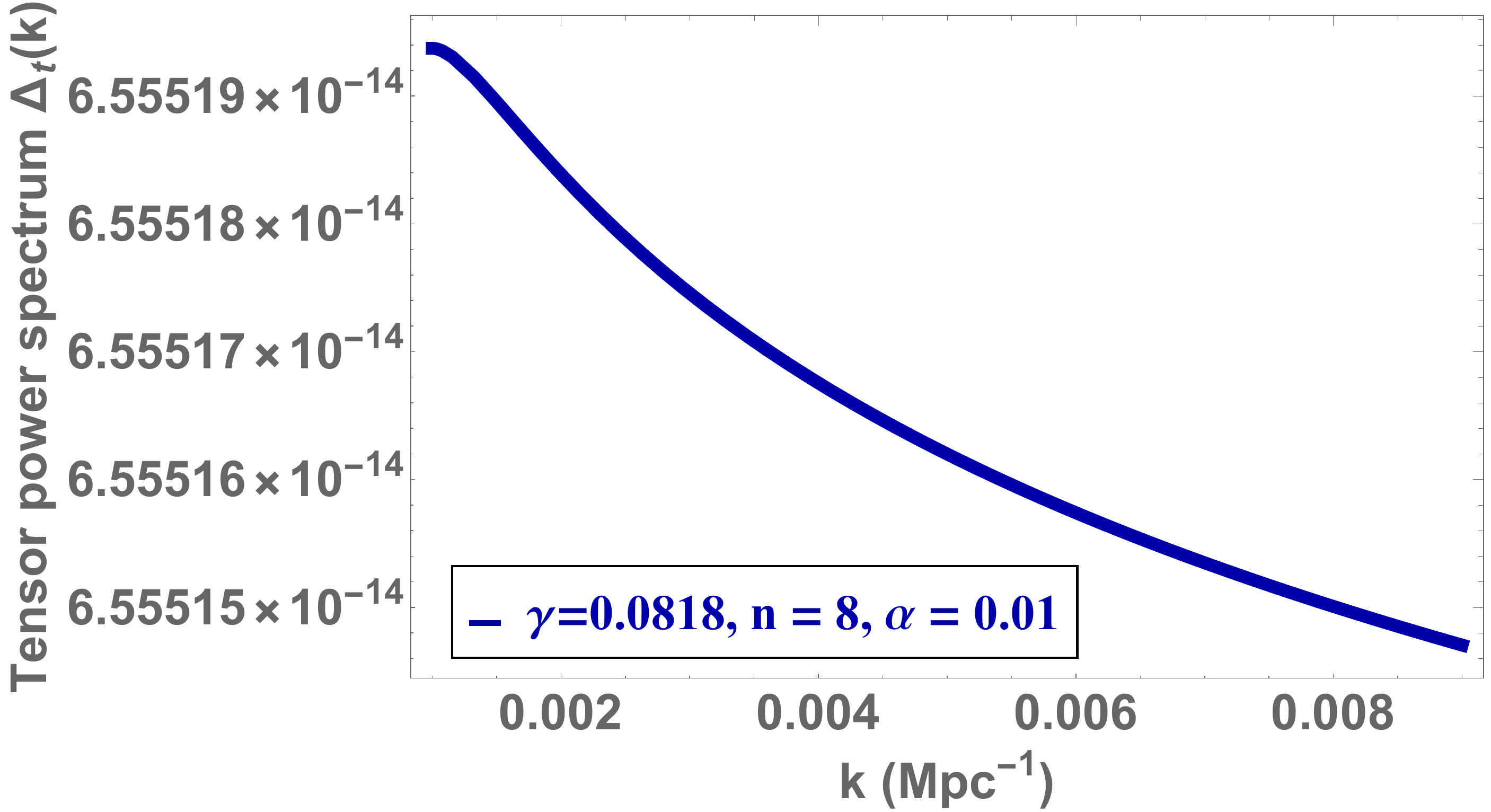}
   \subcaption{}
   \label{fig:tensorPowerSpectrum_3}
\end{subfigure}%
\vspace{0.05\linewidth}
\begin{subfigure}{0.33\linewidth}
  \centering
   \includegraphics[width=46mm,height=40mm]{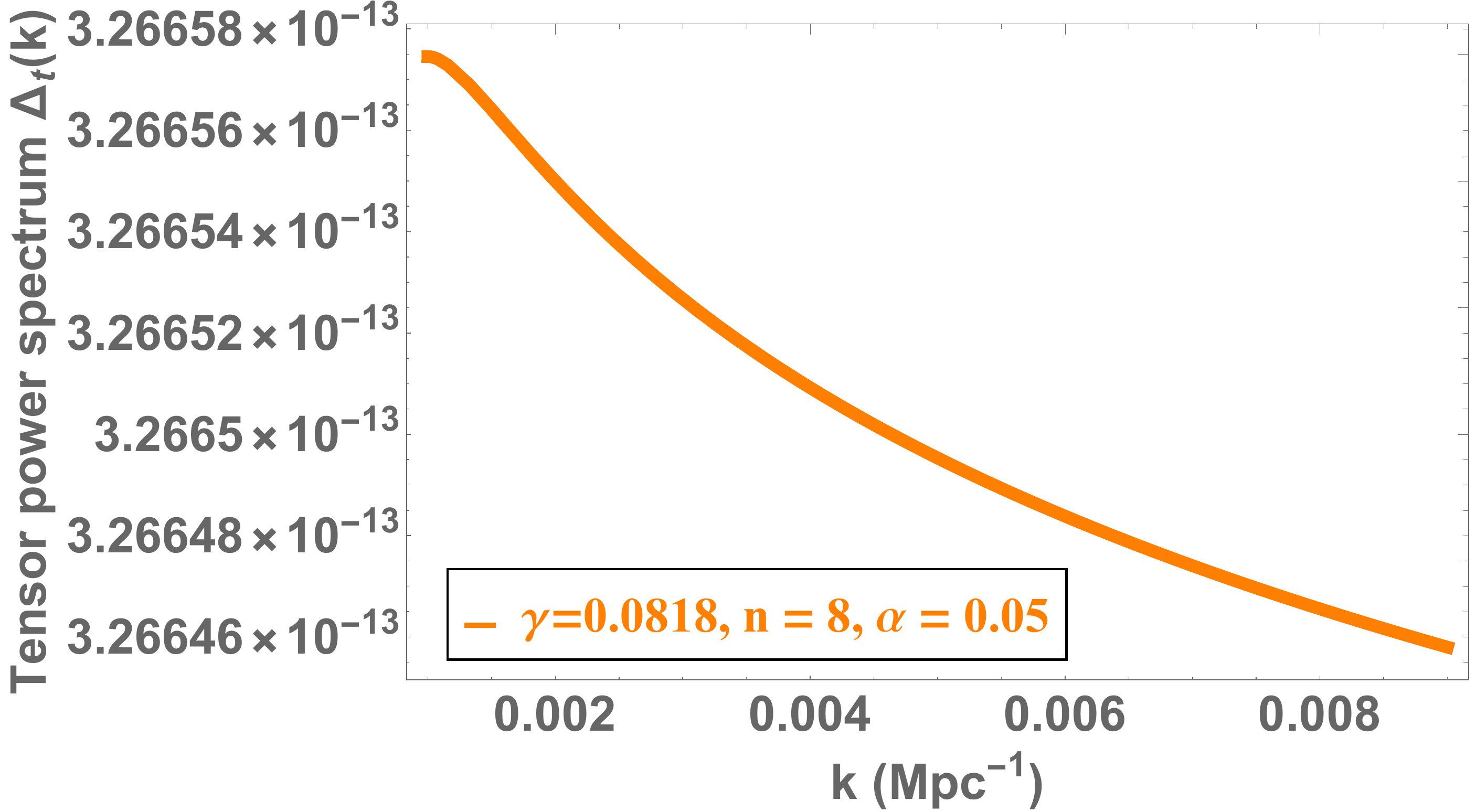}
   \subcaption{}
    \label{fig:tensorPowerSpectrum_4}
\end{subfigure}%
\begin{subfigure}{0.33\linewidth}
  \centering
   \includegraphics[width=46mm,height=40mm]{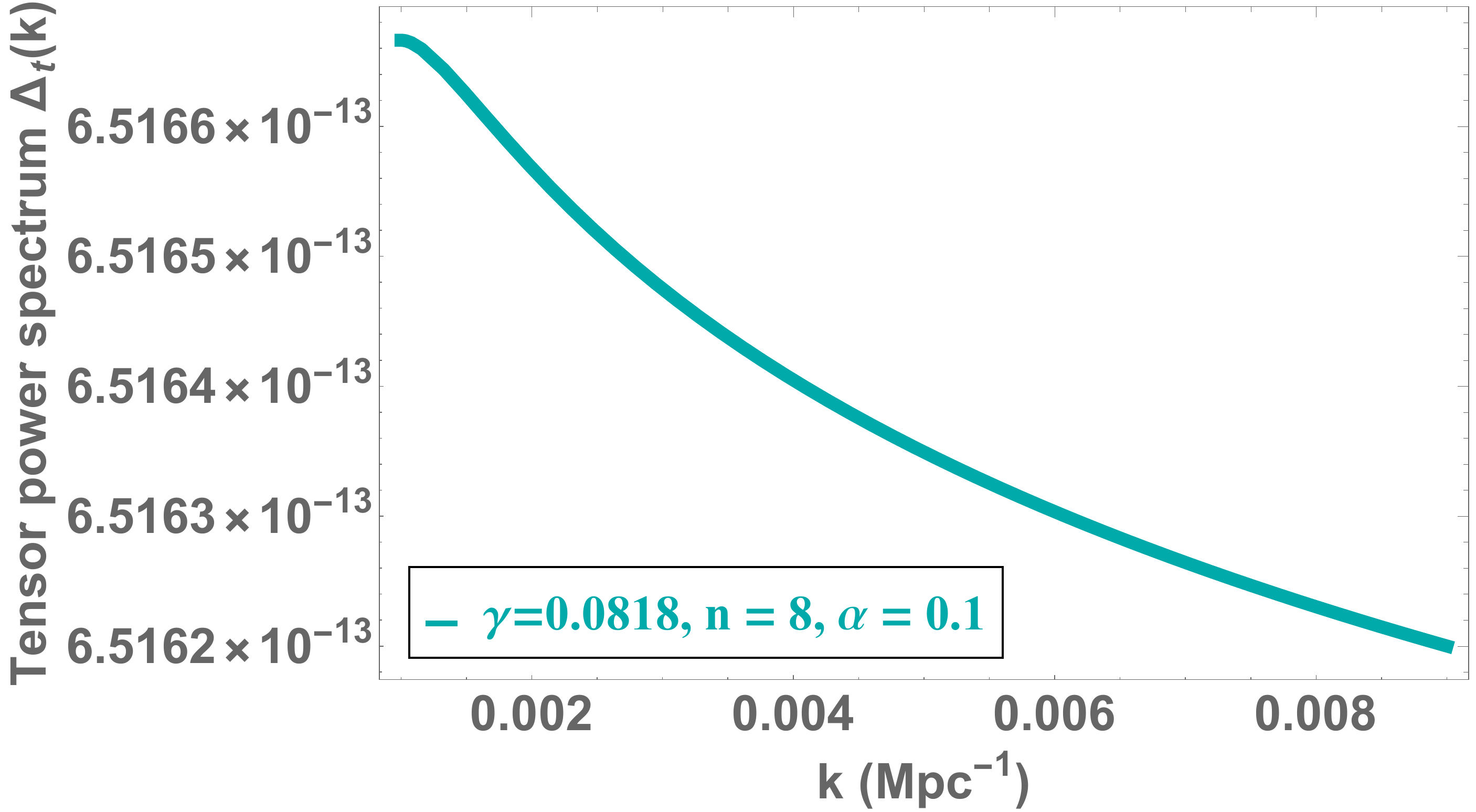}
   \subcaption{}
    \label{fig:tensorPowerSpectrum_5}
\end{subfigure}%
\begin{subfigure}{0.33\linewidth}
  \centering
   \includegraphics[width=46mm,height=40mm]{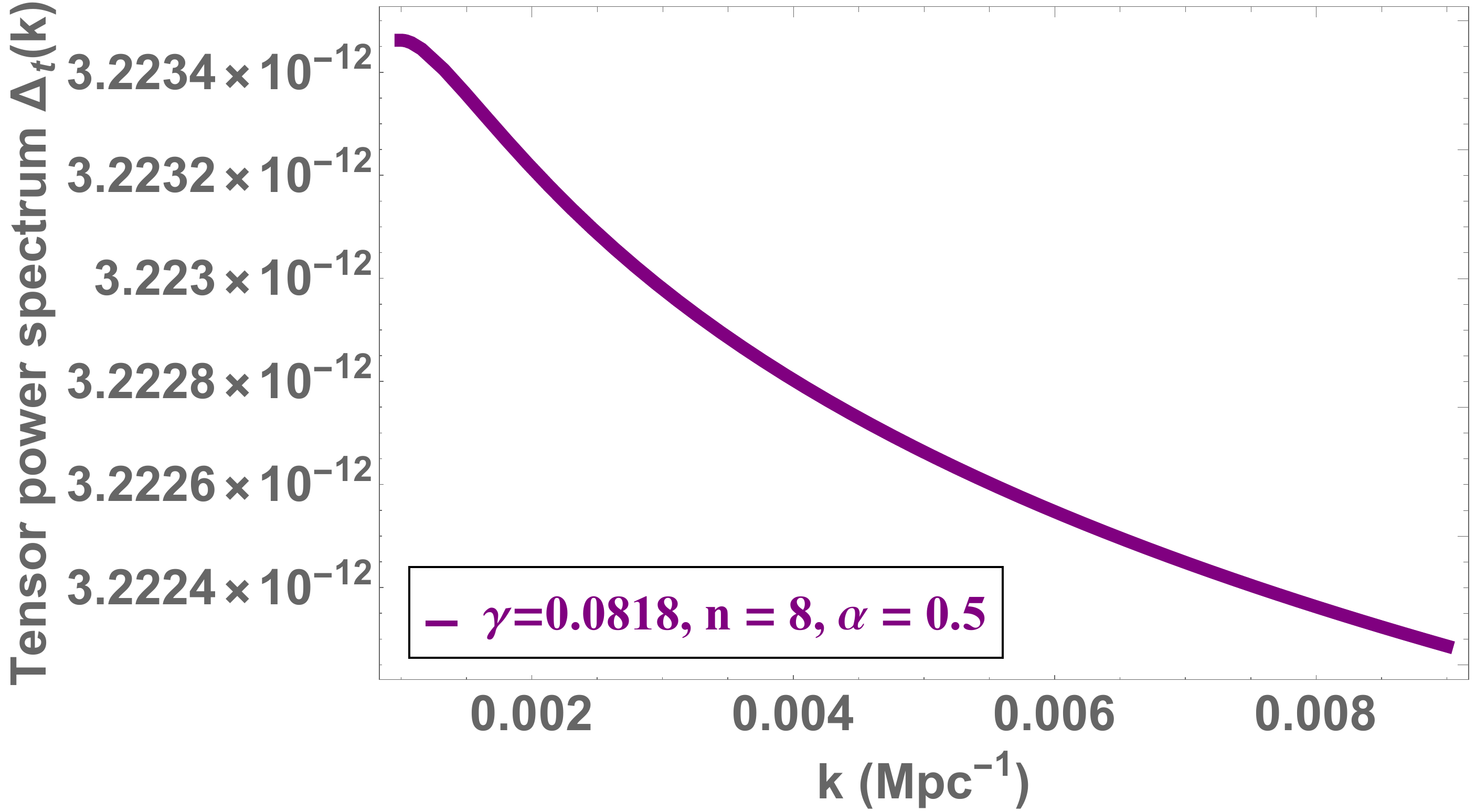}
   \subcaption{}
    \label{fig:tensorPowerSpectrum_6}
\end{subfigure}%
\vspace{0.05\linewidth}
\begin{subfigure}{0.33\linewidth}
  \centering
   \includegraphics[width=46mm,height=40mm]{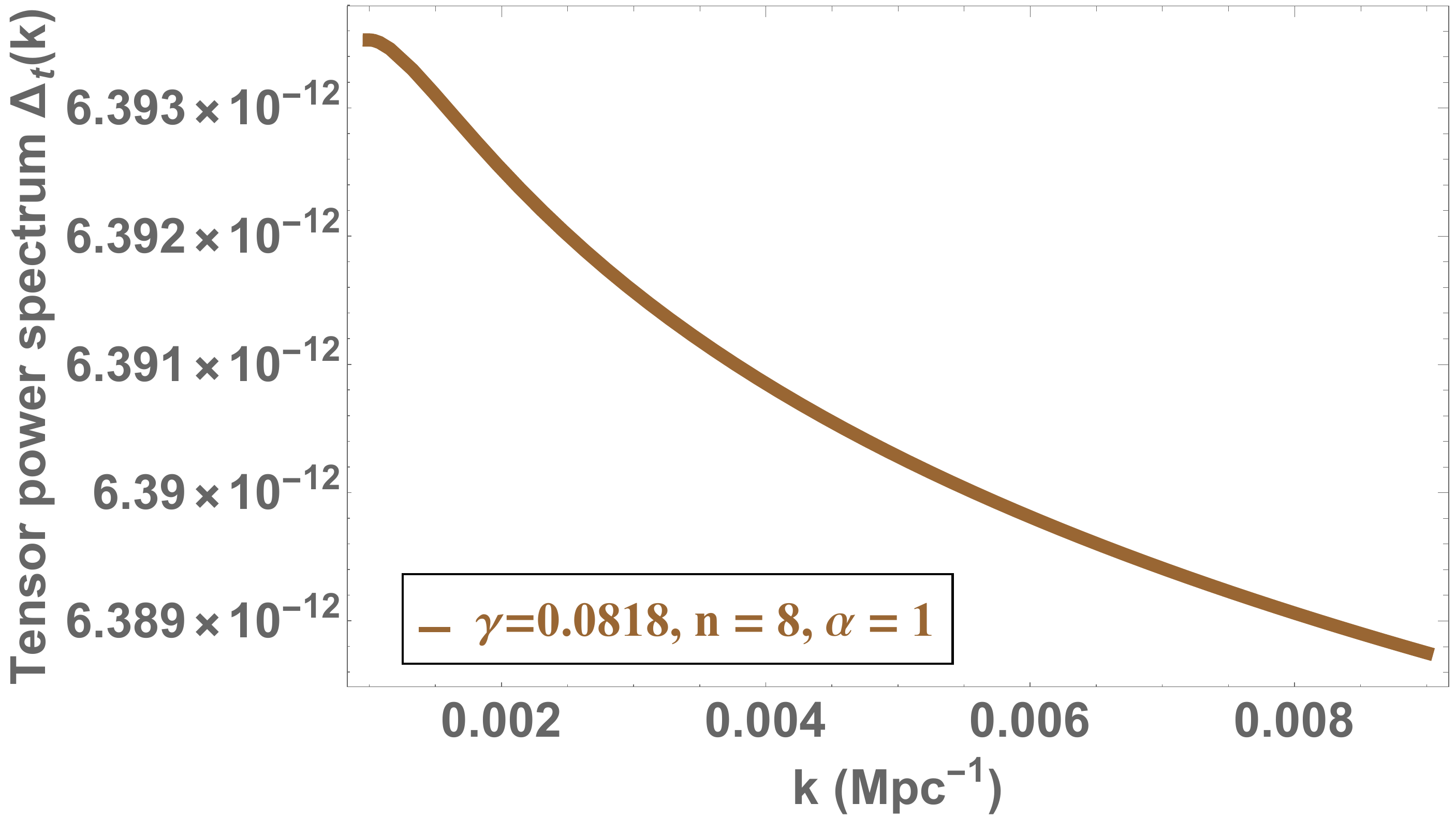}
   \subcaption{}
    \label{fig:tensorPowerSpectrum_7}
\end{subfigure}%
\begin{subfigure}{0.33\linewidth}
  \centering
   \includegraphics[width=46mm,height=40mm]{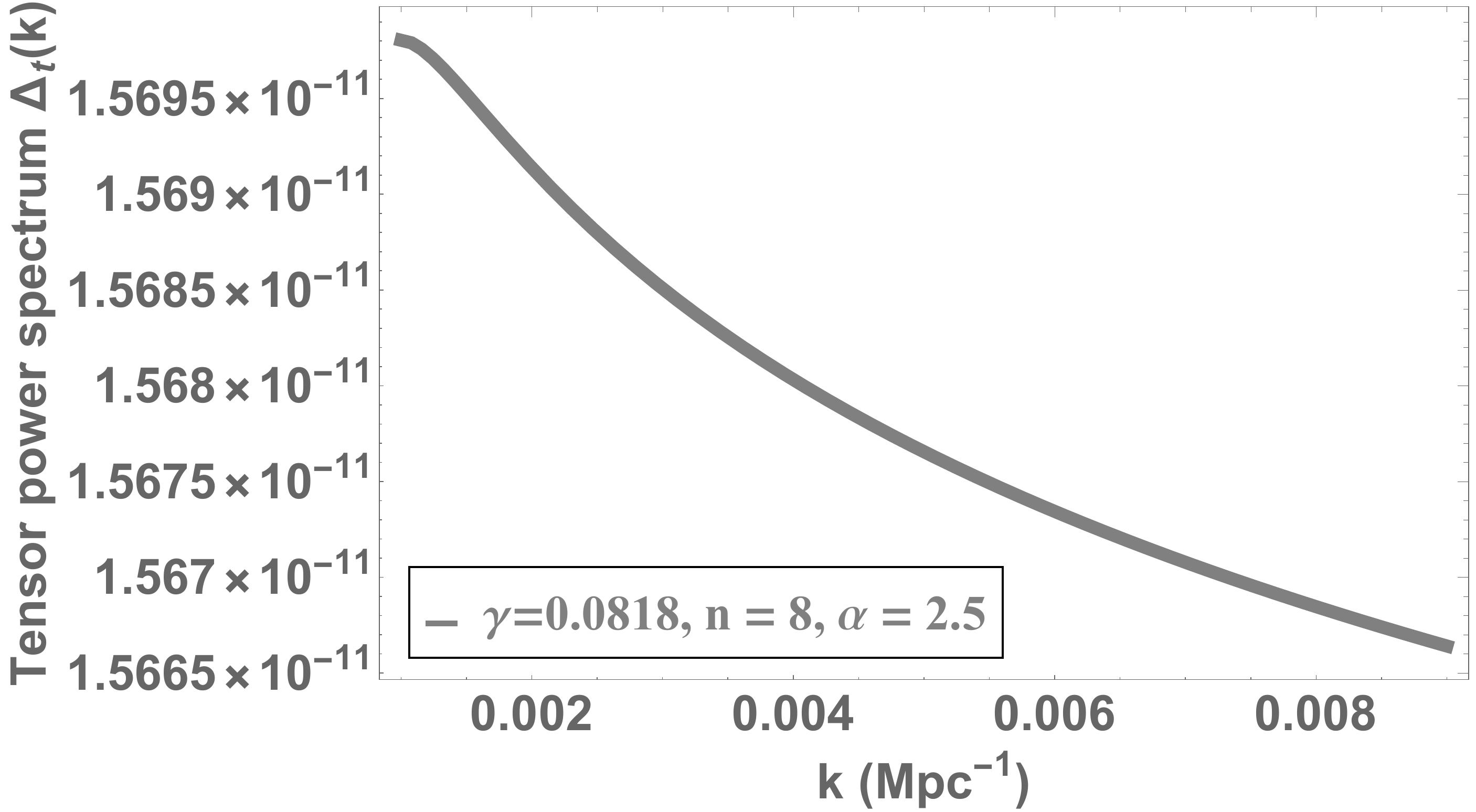}
   \subcaption{}
    \label{fig:tensorPowerSpectrum_8}
\end{subfigure}%
\begin{subfigure}{0.33\linewidth}
  \centering
   \includegraphics[width=46mm,height=40mm]{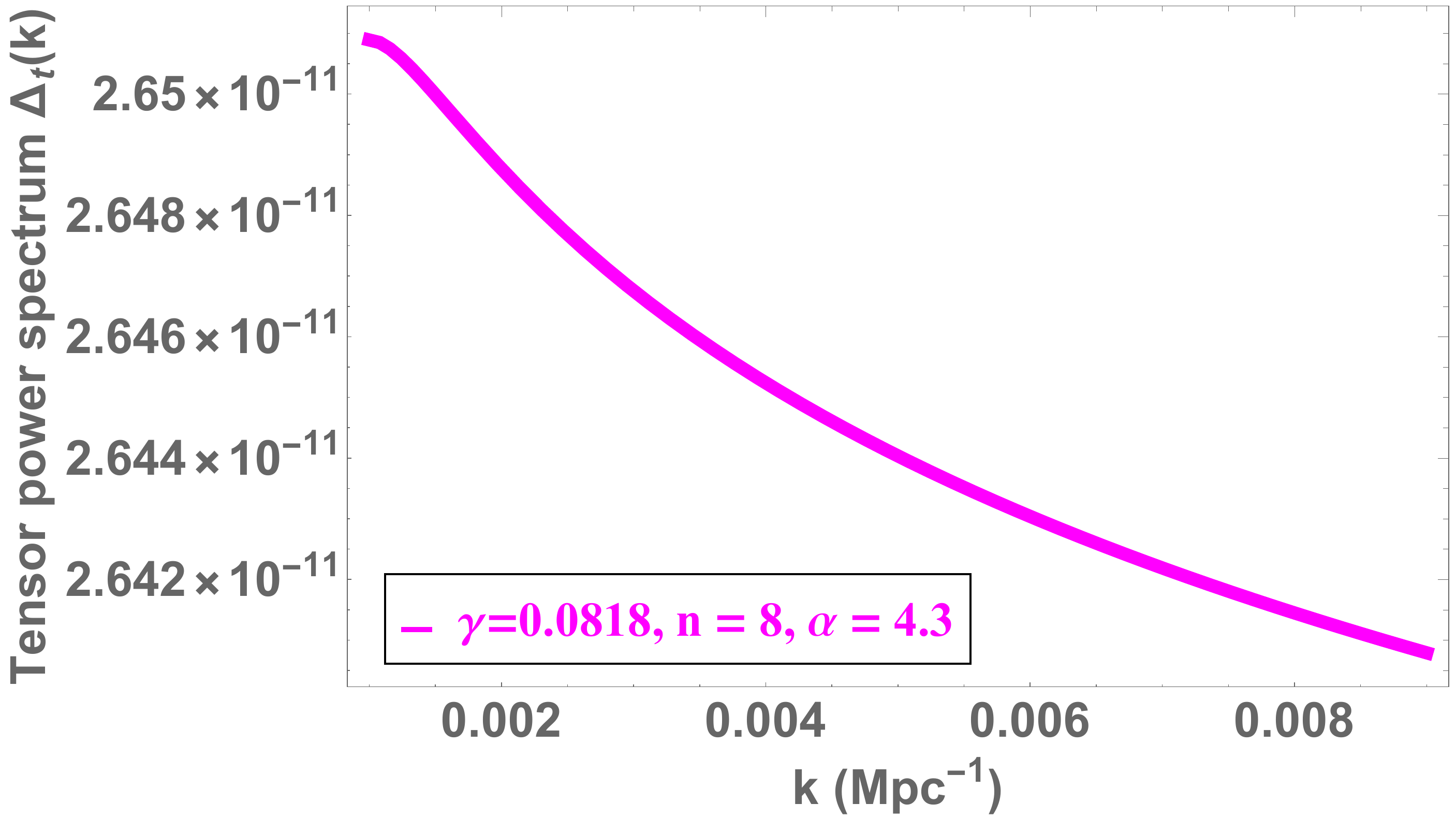}
   \subcaption{}
    \label{fig:tensorPowerSpectrum_9}
\end{subfigure}
\caption{Tensor power spectra for nine values of $\alpha$ for $\gamma=0.0818$ and $n=8$. The values of $\Delta_t(k)$ tend to increase with the increase in $\alpha$ at a particular $k$ value.}
\label{fig:tensorPowerSpectrum}
\end{figure}

Figures \ref{fig:scalarPowerSpectrum} and \ref{fig:tensorPowerSpectrum} represent the COBE/Planck normalised scalar ($\Delta_s(k)$) and tensor ($\Delta_t(k)$) power spectra calculated from Eqs. (\ref{eq:modeScalarPower}) and (\ref{eq:modeTensorPower}) respectively. As $\alpha$ increases from $\alpha=0.001$ to $\alpha=4.3$, the $\Delta_s(k)$ remains almost unaltered up to $\alpha=0.01$ and afterwords it decreases slowly, while $\Delta_t(k)$ increases monotonically at a particular value of $k$. The orders of $\Delta_s(k)$ is $ 10^{-9}$ and that for $\Delta_t(k)$ varies between $10^{-15}-10^{-11}$. These values lie within the Planck bounds (see figure \ref{fig:5}) which are consistent with the present day value of the Hubble parameter. The scalar and tensor tiny correlations are the seeds for today's observed large scale structures which are embedded in the initial conditions during inflation. Their imprints are found in the secondary anisotropies of CMB $B$-mode polarisation (which is very feeble and therefore hard to detect) and the density profile of stellar or galactic distribution (which is measured by gravitational lensing method).\par
\begin{figure}[H]
\begin{subfigure}{0.33\linewidth}
  \centering
   \includegraphics[width=46mm,height=40mm]{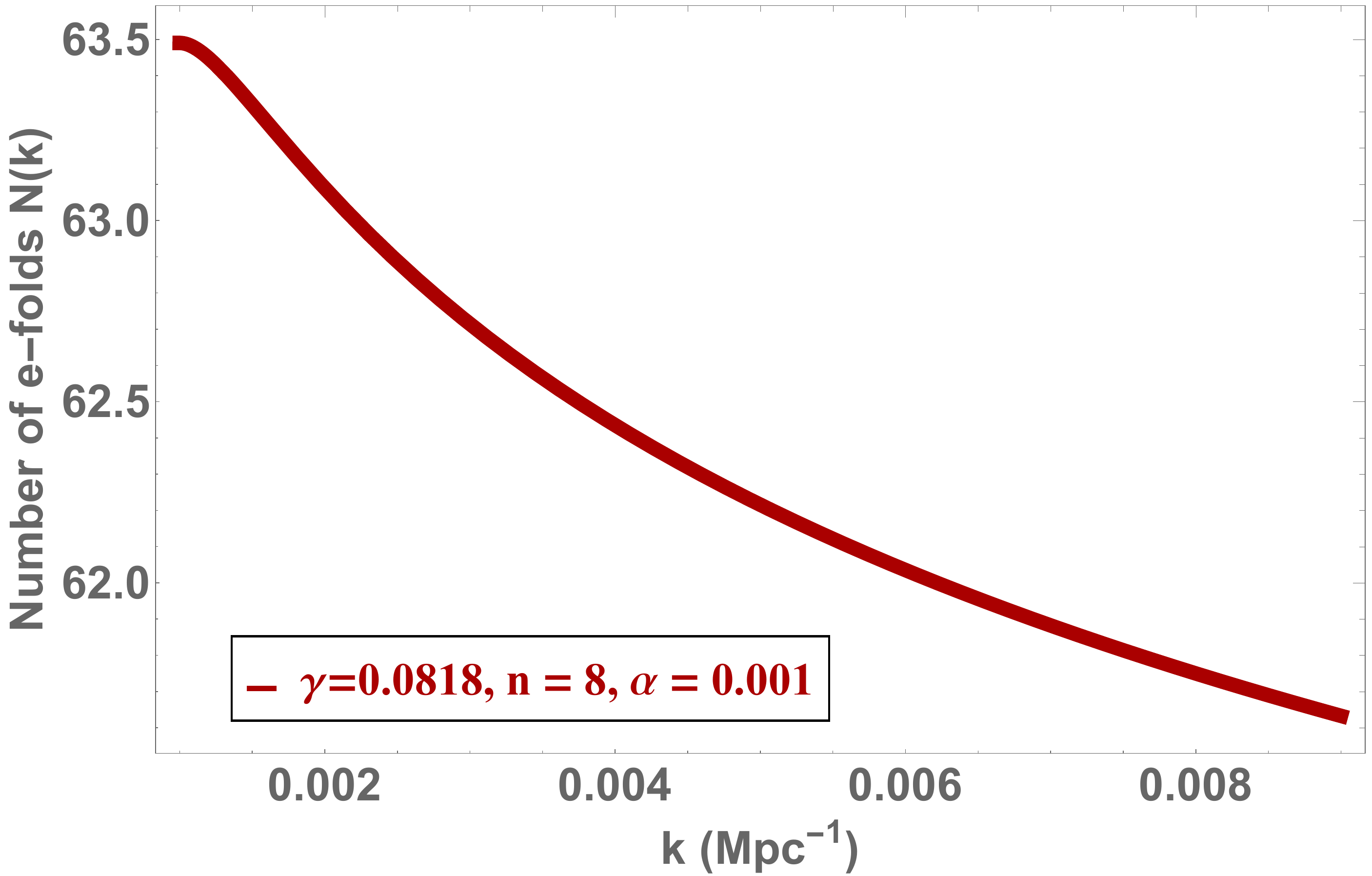} 
   \subcaption{}
   \label{fig:numberOfEFolds_1}
\end{subfigure}%
\begin{subfigure}{0.33\linewidth}
  \centering
   \includegraphics[width=46mm,height=40mm]{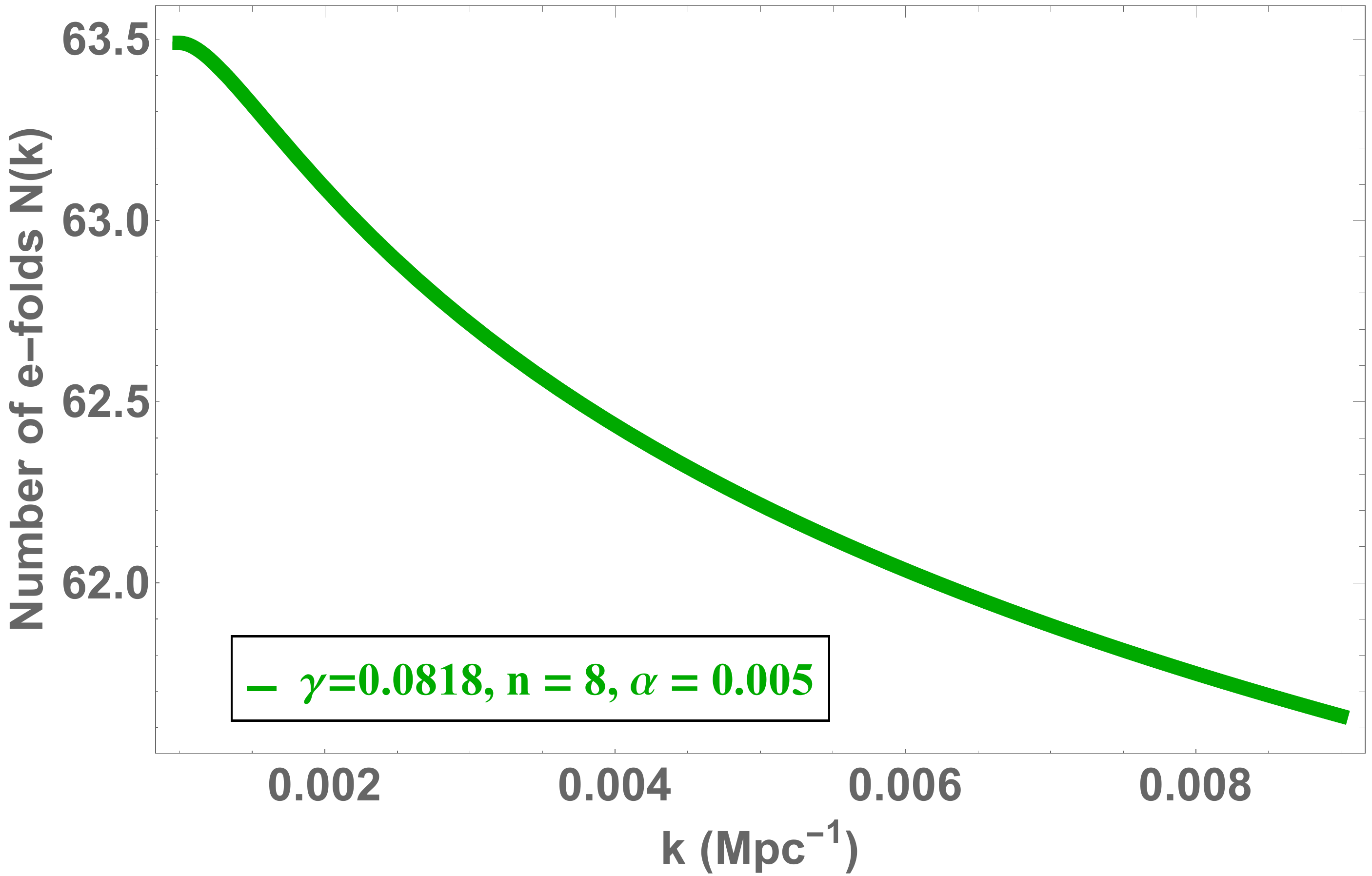}
   \subcaption{}
   \label{fig:numberOfEFolds_2}
\end{subfigure}%
\begin{subfigure}{0.33\linewidth}
  \centering
   \includegraphics[width=46mm,height=40mm]{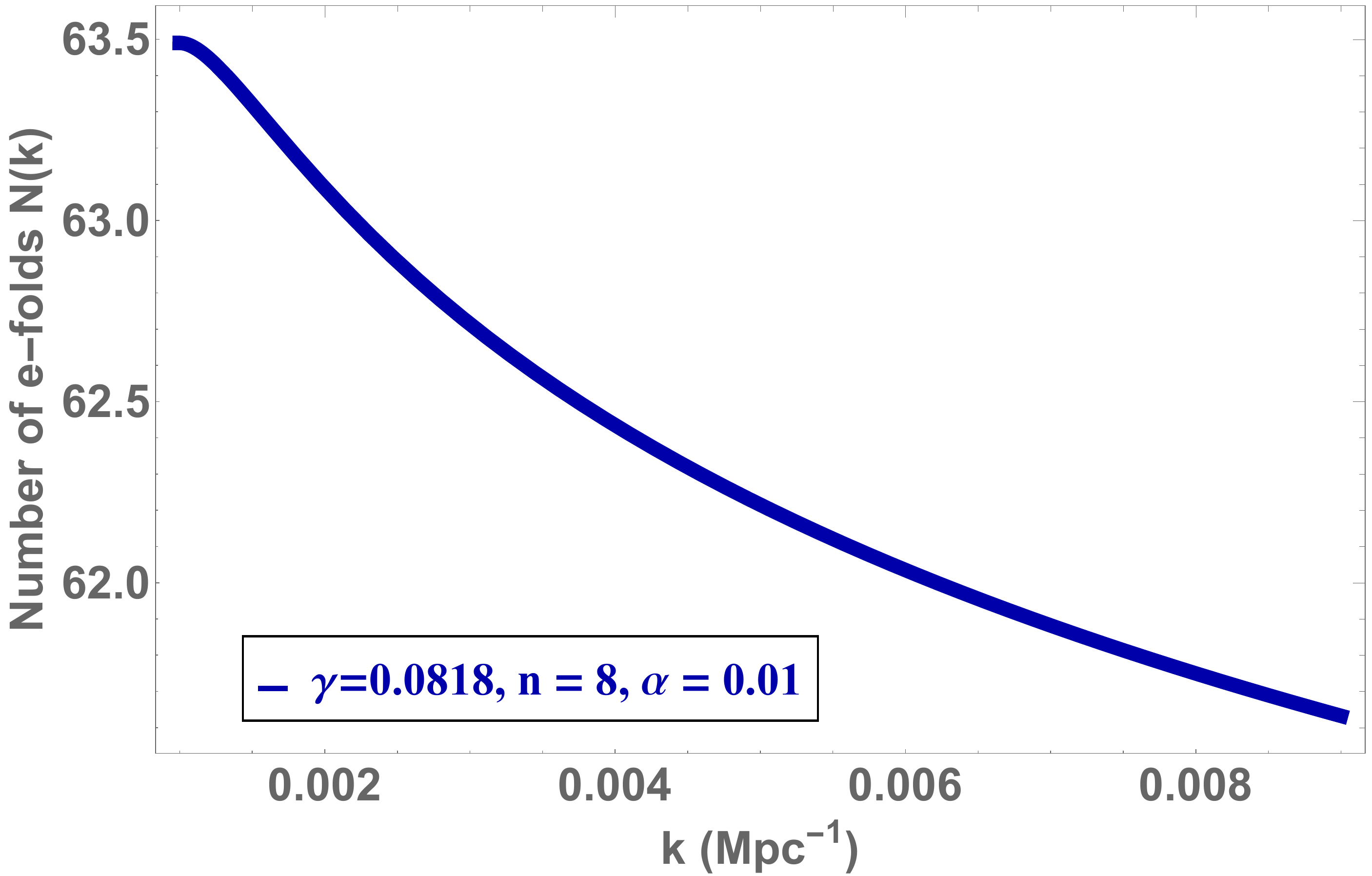}
   \subcaption{}
   \label{fig:numberOfEFolds_3}
\end{subfigure}%
\vspace{0.05\linewidth}
\begin{subfigure}{0.33\linewidth}
  \centering
   \includegraphics[width=46mm,height=40mm]{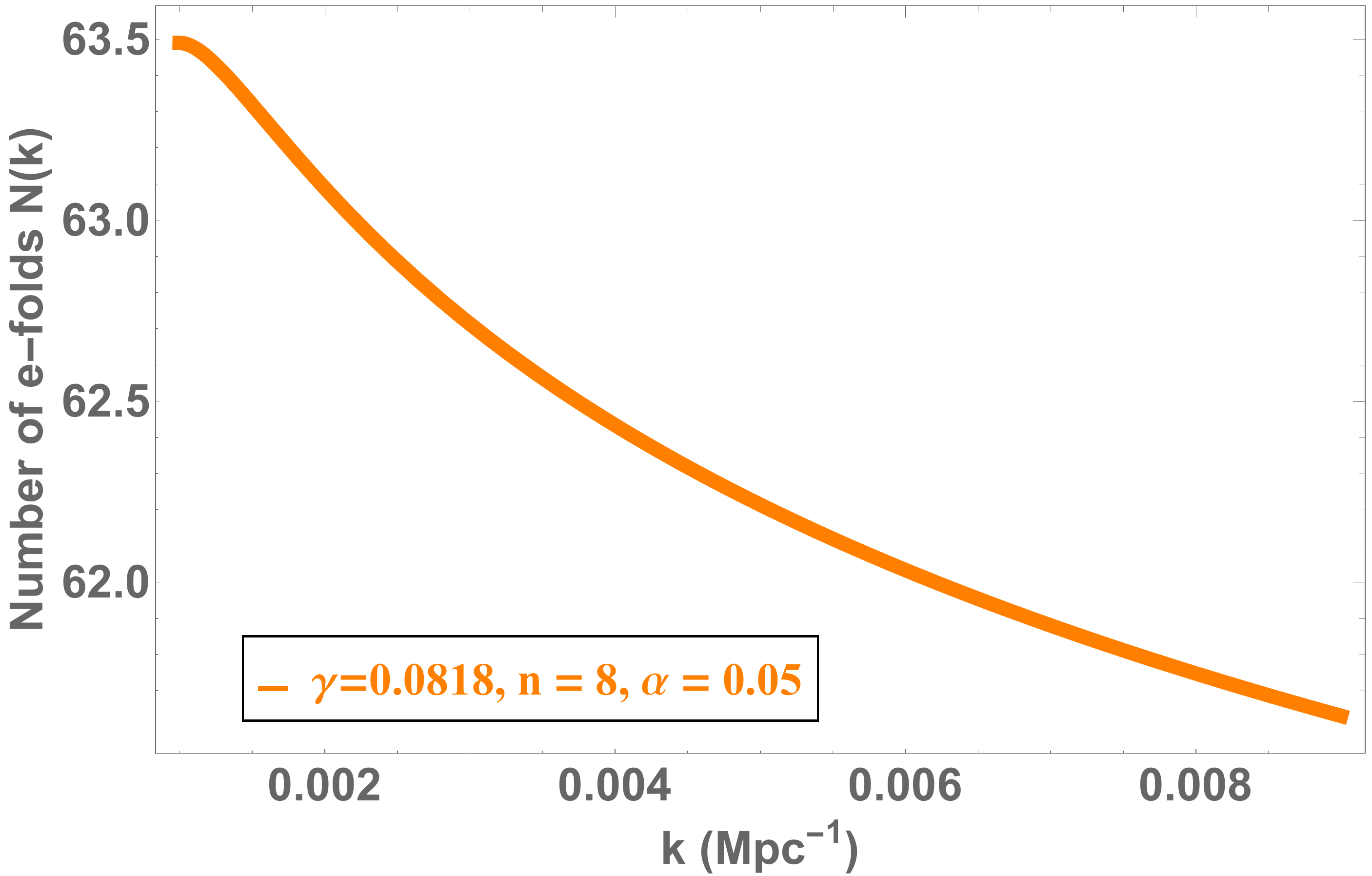}
   \subcaption{}
    \label{fig:numberOfEFolds_4}
\end{subfigure}%
\begin{subfigure}{0.33\linewidth}
  \centering
   \includegraphics[width=46mm,height=40mm]{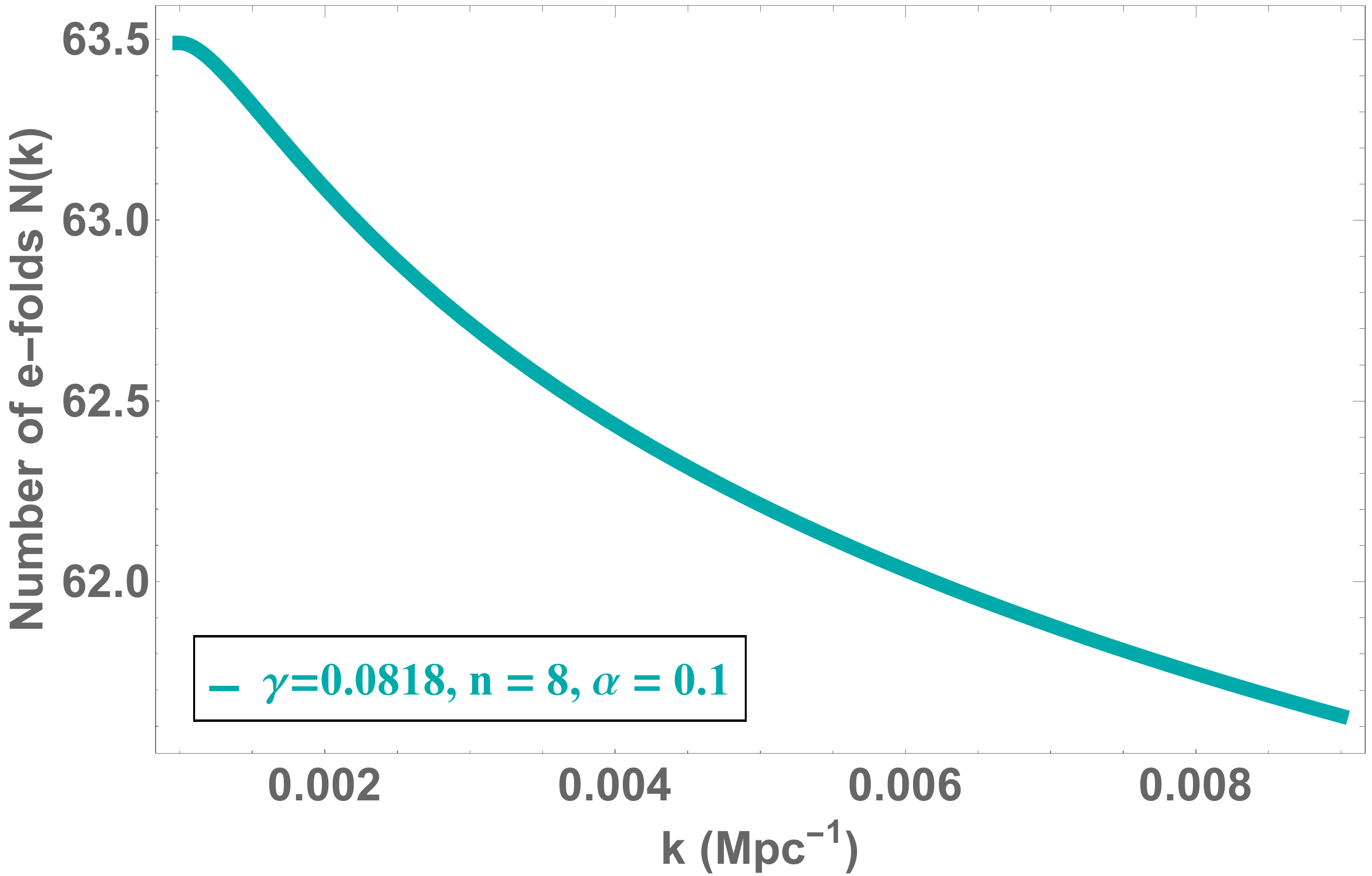}
   \subcaption{}
    \label{fig:numberOfEFolds_5}
\end{subfigure}%
\begin{subfigure}{0.33\linewidth}
  \centering
   \includegraphics[width=46mm,height=40mm]{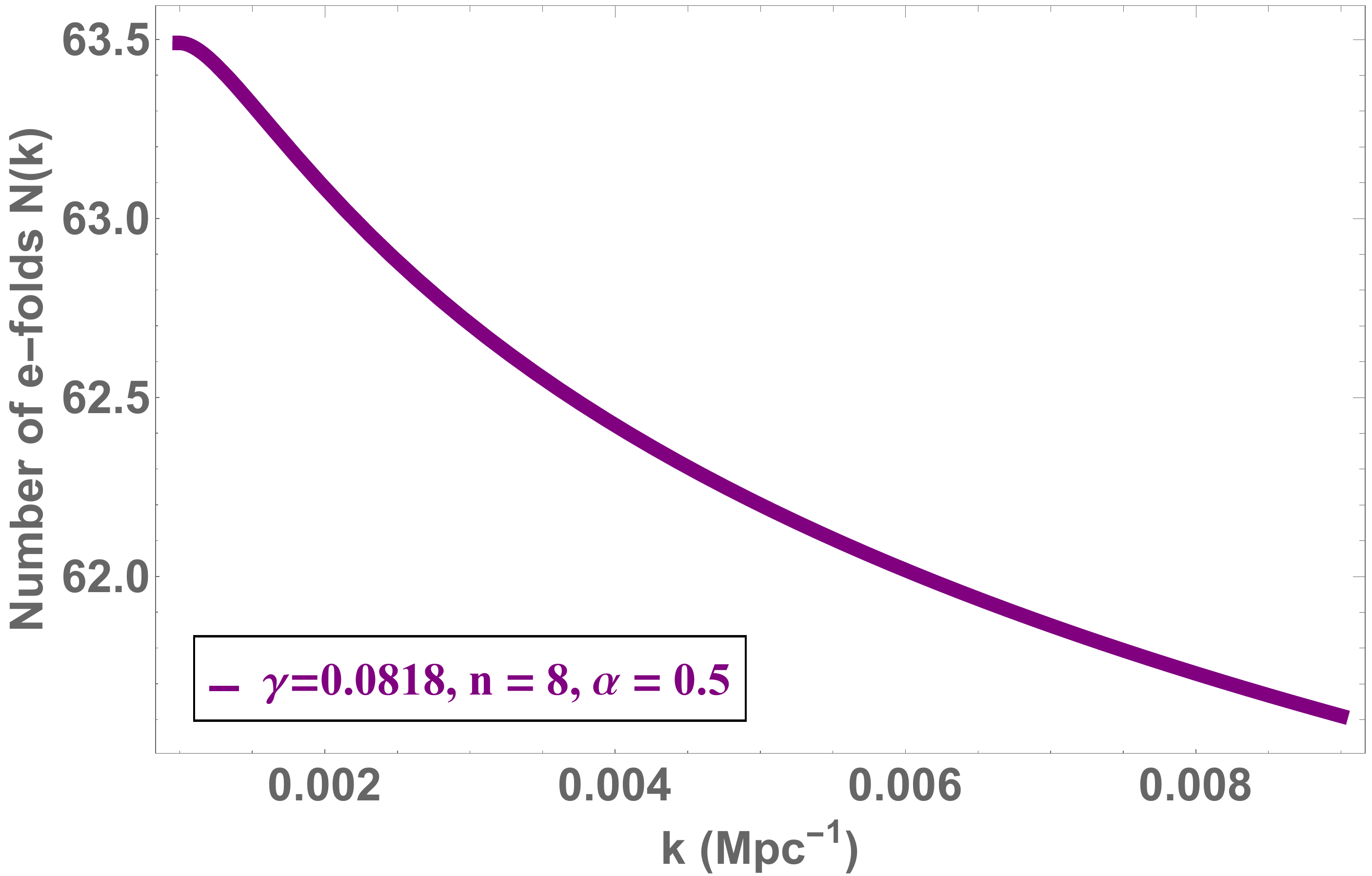}
   \subcaption{}
    \label{fig:numberOfEFolds_6}
\end{subfigure}%
\vspace{0.05\linewidth}
\begin{subfigure}{0.33\linewidth}
  \centering
   \includegraphics[width=46mm,height=40mm]{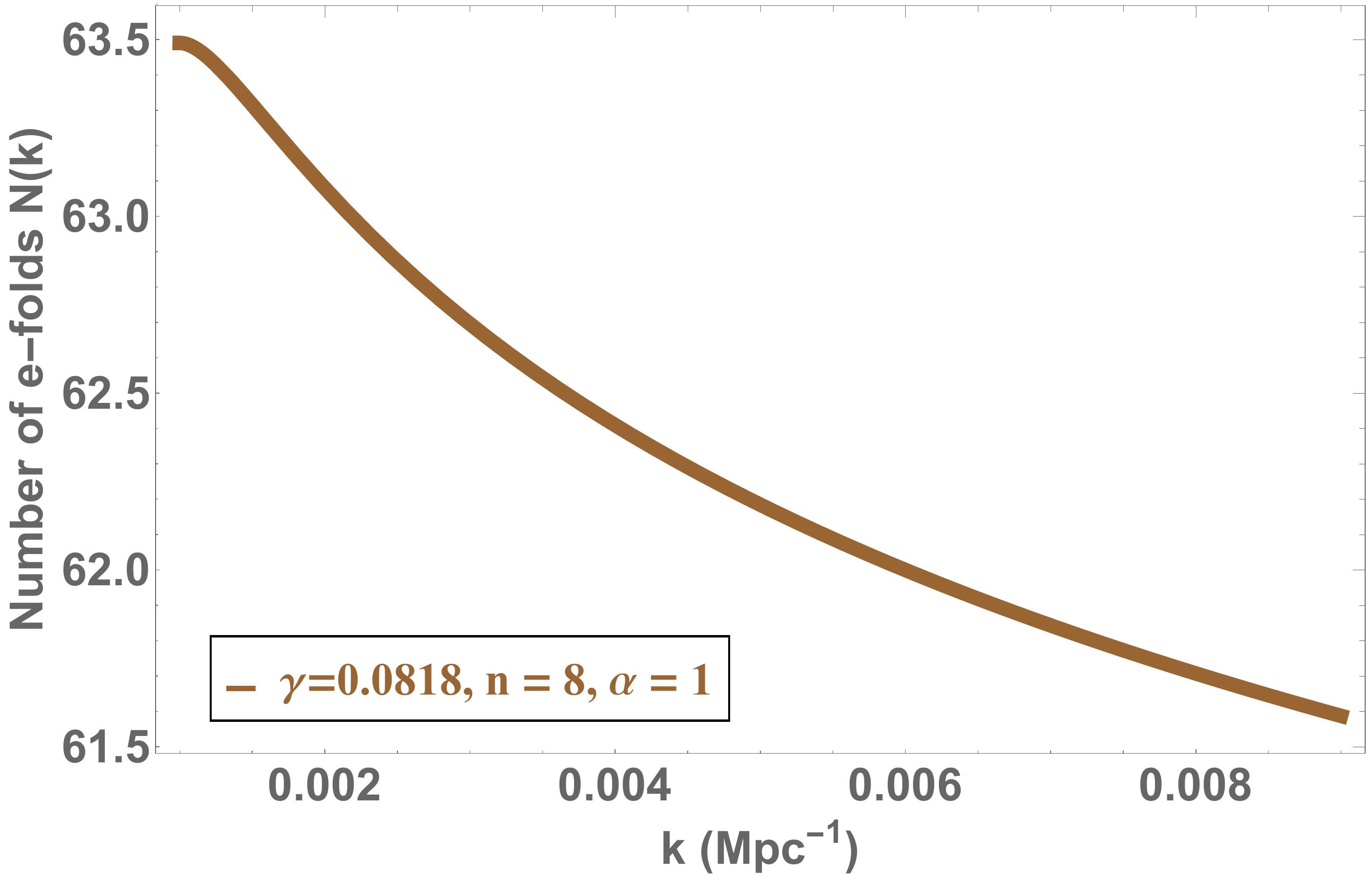}
   \subcaption{}
    \label{fig:numberOfEFolds_7}
\end{subfigure}%
\begin{subfigure}{0.33\linewidth}
  \centering
   \includegraphics[width=46mm,height=40mm]{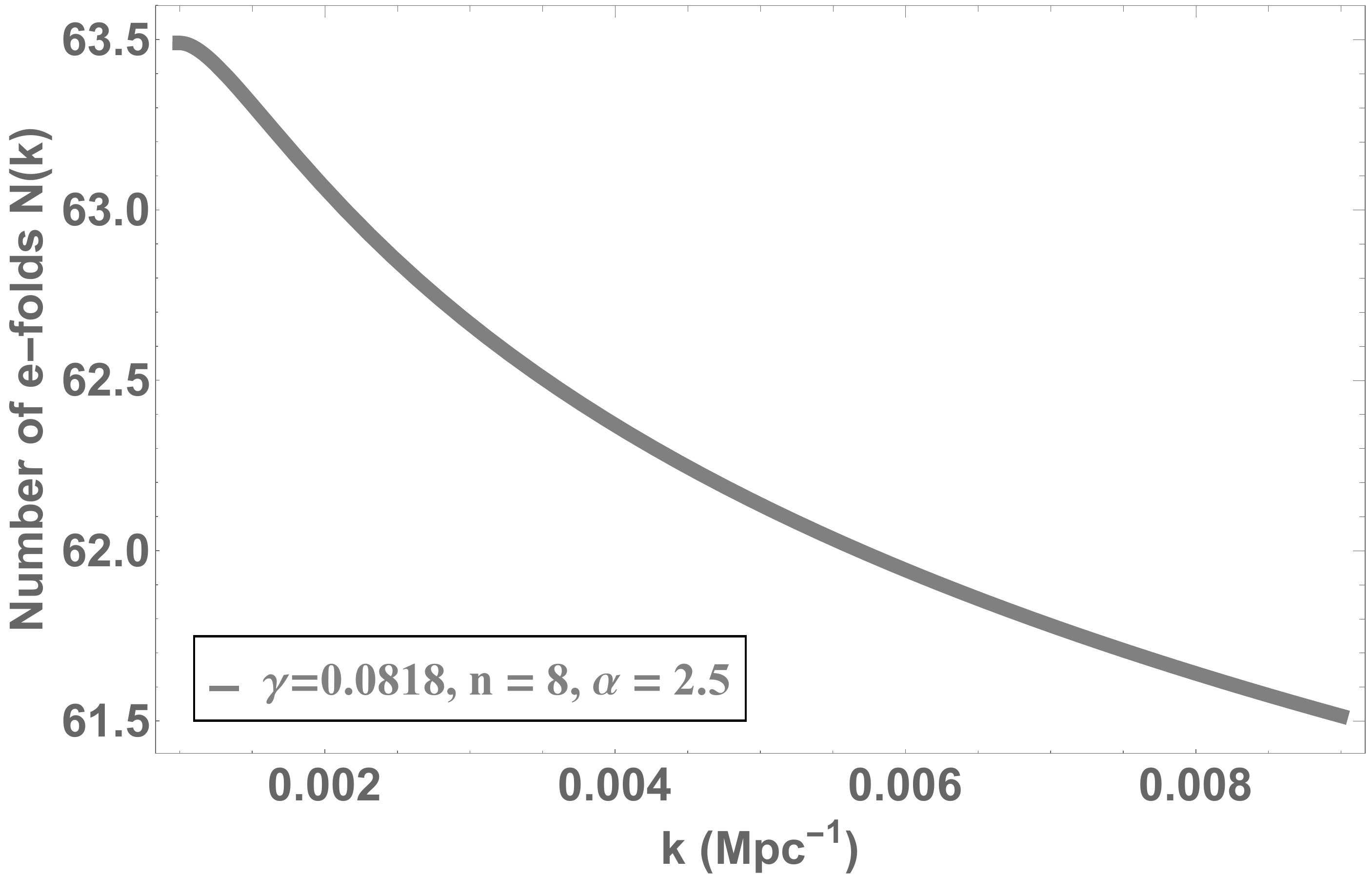}
   \subcaption{}
    \label{fig:numberOfEFolds_8}
\end{subfigure}%
\begin{subfigure}{0.33\linewidth}
  \centering
   \includegraphics[width=46mm,height=40mm]{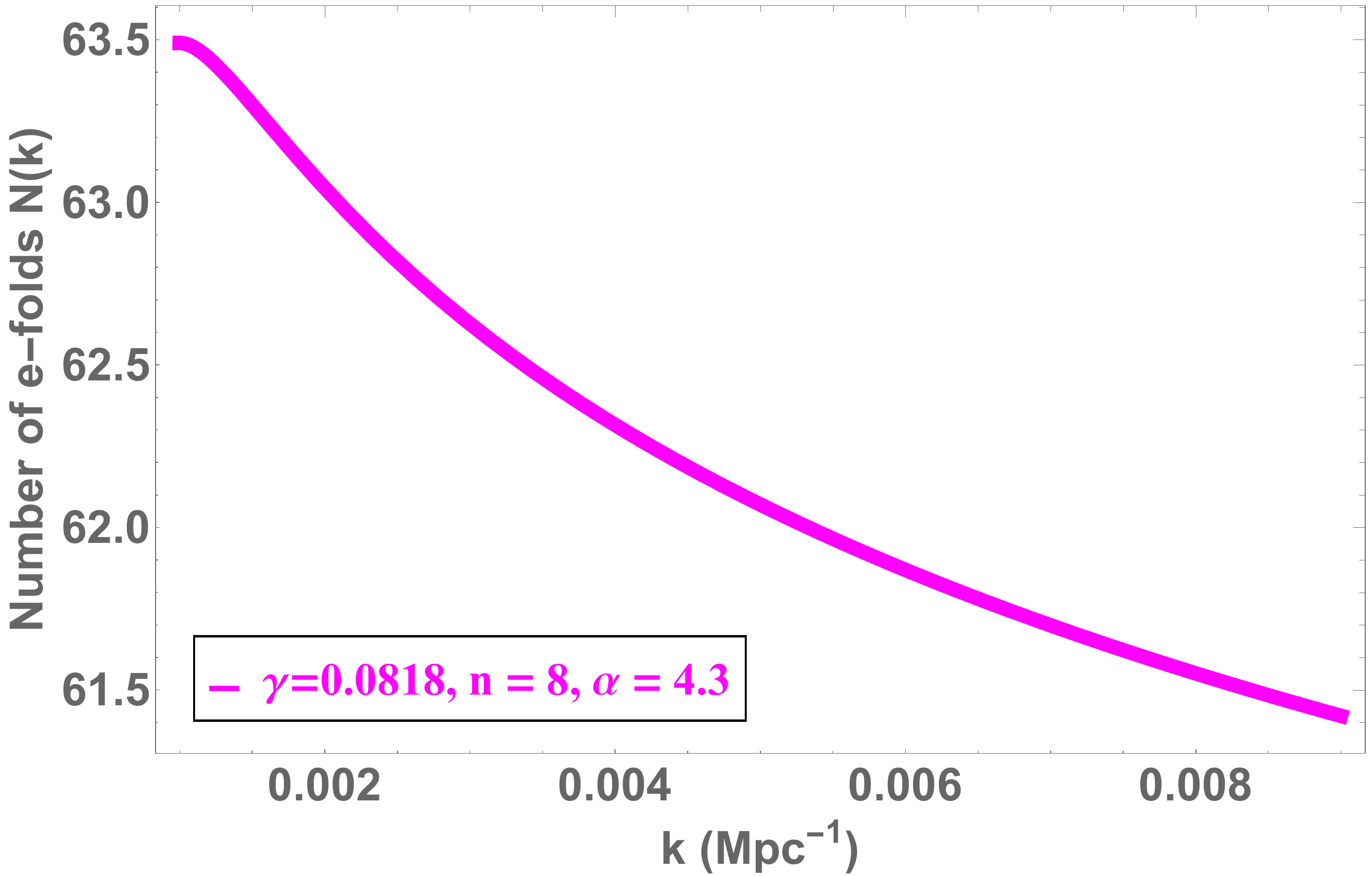}
   \subcaption{}
    \label{fig:numberOfEFolds_9}
\end{subfigure}
\caption{Number of remaining e-folds for nine values of $\alpha$ for $\gamma=0.0818$ and $n=8$. The values of $N(k)$ tend to remain same for all values of $\alpha$.}
\label{fig:numberOfEFolds}
\end{figure}

 The number of remaining e-folds $N(k)$ is plotted according to Eq. (\ref{eq:e_folds_4}) in figure \ref{fig:numberOfEFolds}. $N(k)$ varies between $61.5-63.5$ for a given value of $\alpha$, while at a fixed value of $k$ it remains almost constant. At small values of $k$ the number of remaining e-folds is approximately $63.5$ and as $k$ increases $N(k)$ decreases to around $61.5$. Such mode behaviour of $N(k)$ simulates the DHE scenario described in Section \ref{sec:formalism}, according to which small modes exit at the beginning and high modes exit at the end of Hubble sphere contraction.\par
\begin{figure}[H]
\begin{subfigure}{0.33\linewidth}
  \centering
   \includegraphics[width=46mm,height=40mm]{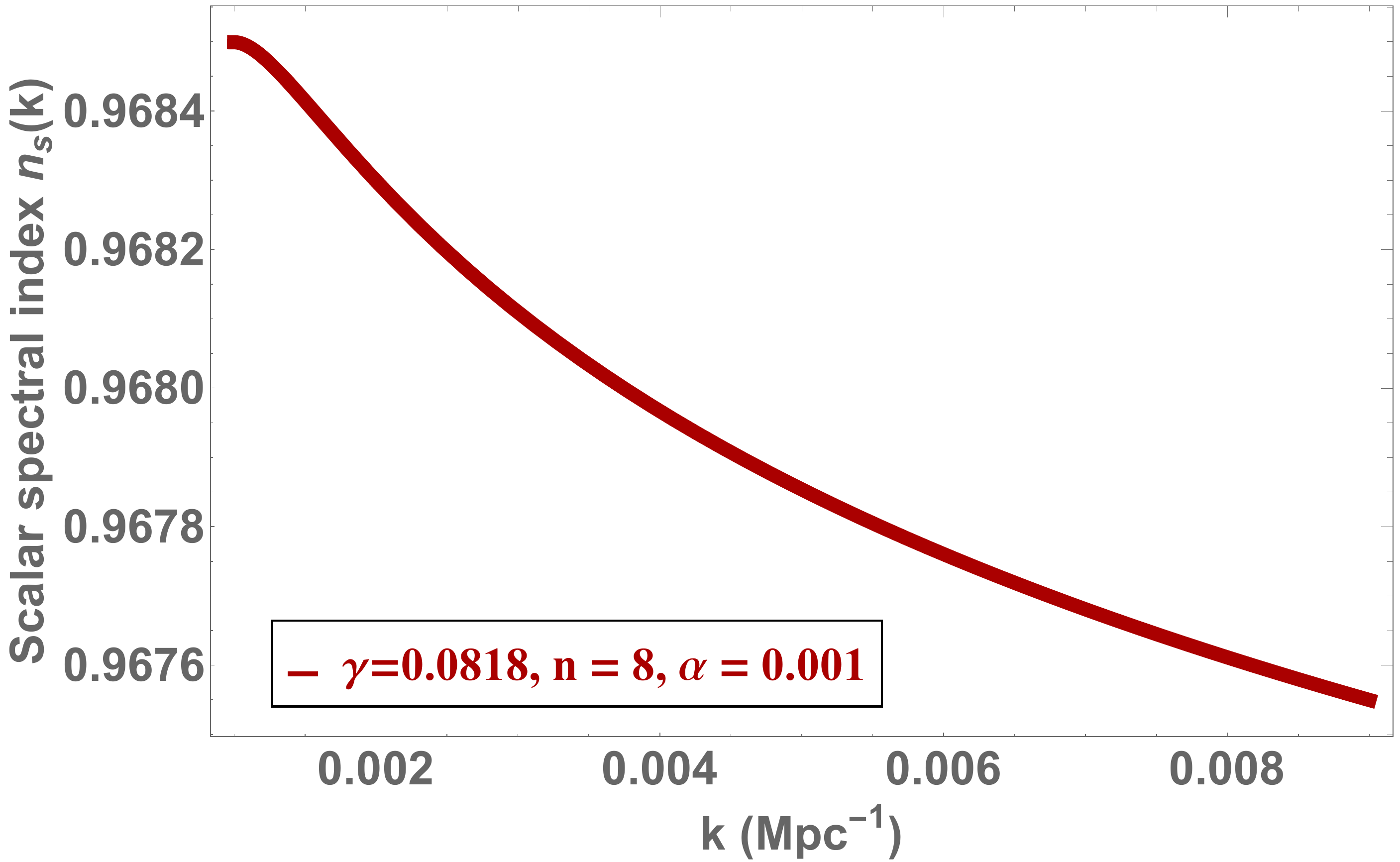}
   \subcaption{}
   \label{fig:scalarSpectralIndex_1}
\end{subfigure}%
\begin{subfigure}{0.33\linewidth}
  \centering
   \includegraphics[width=46mm,height=40mm]{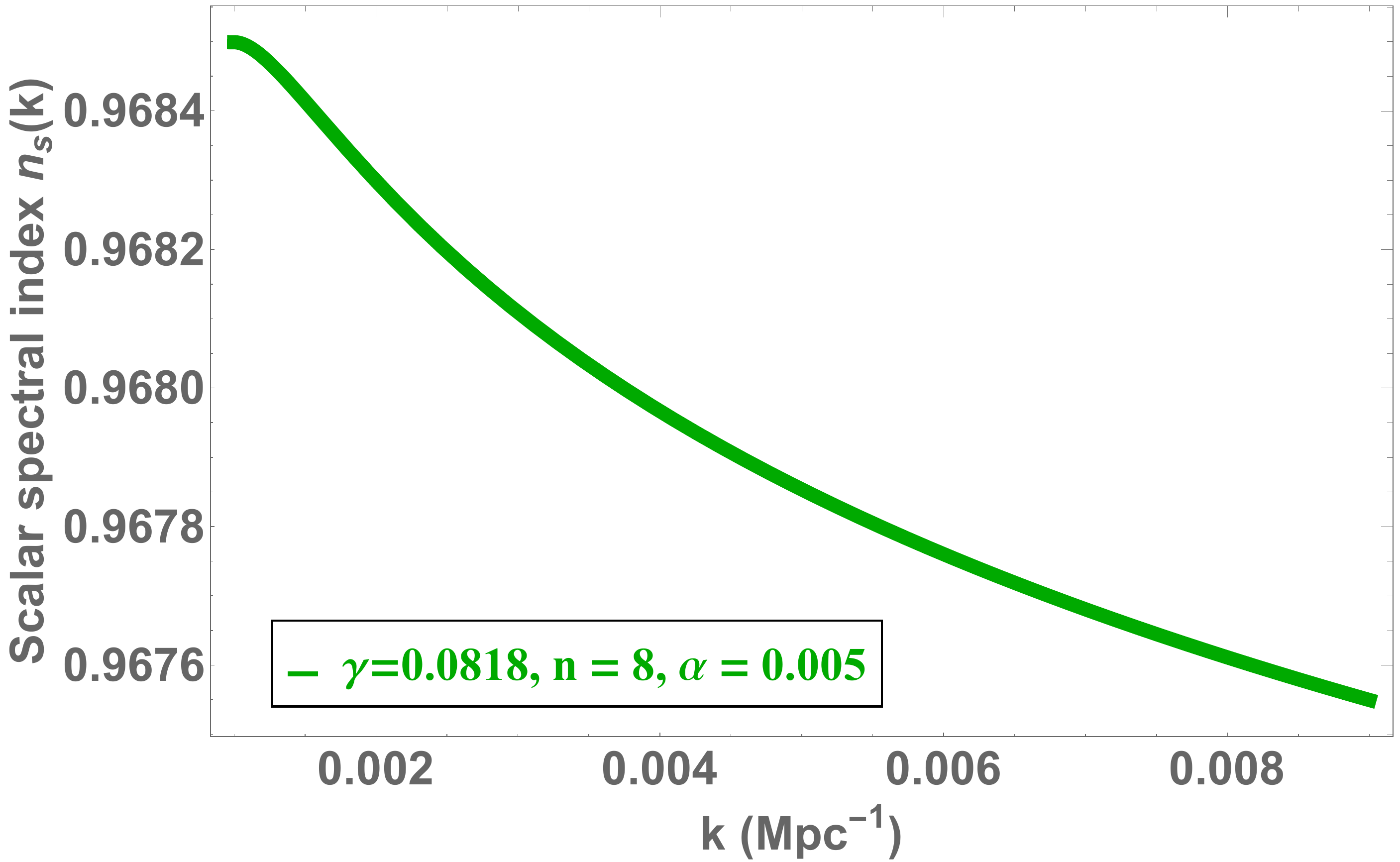}
   \subcaption{}
   \label{fig:scalarSpectralIndex_2}
\end{subfigure}%
\begin{subfigure}{0.33\linewidth}
  \centering
   \includegraphics[width=46mm,height=40mm]{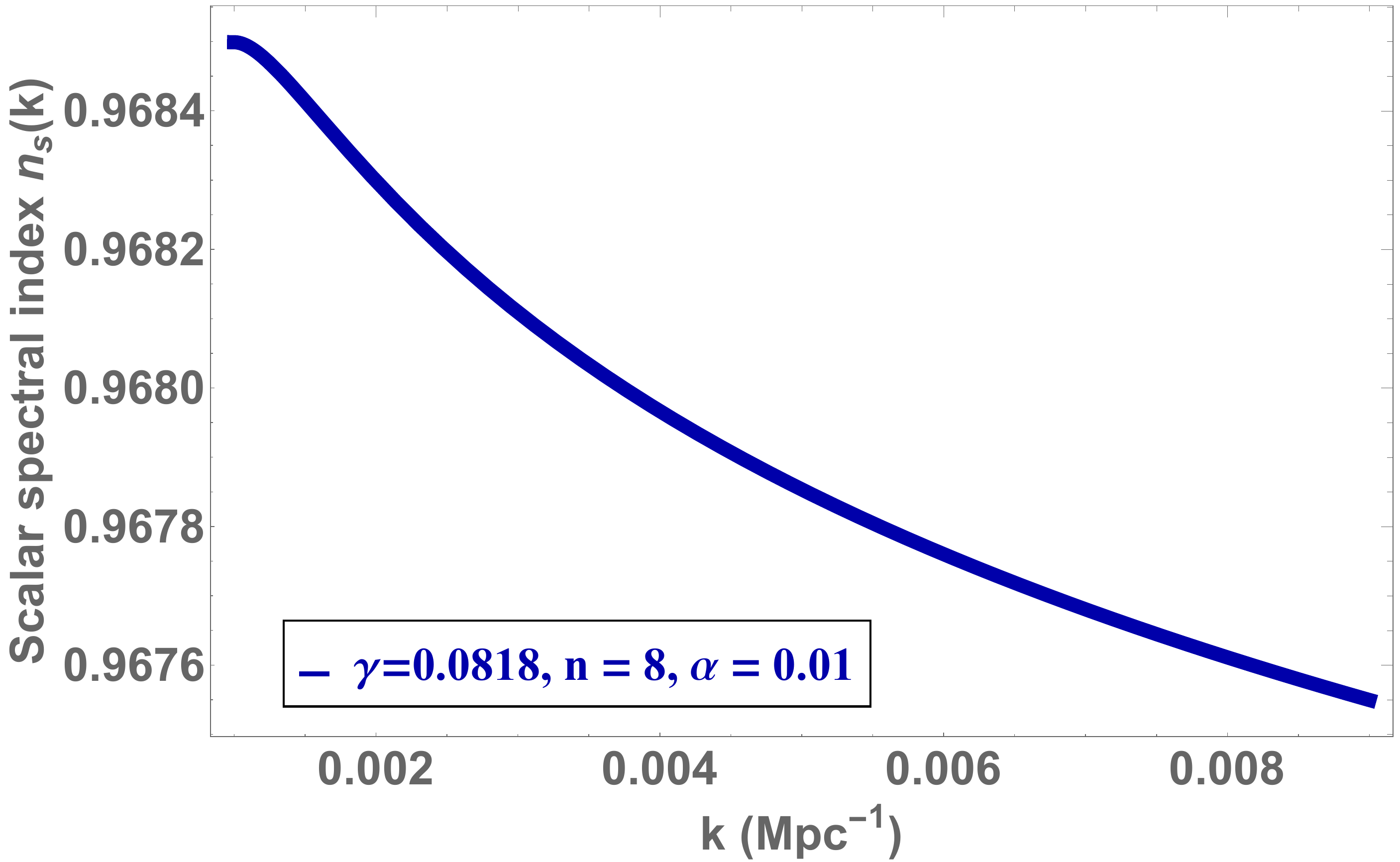}
   \subcaption{}
   \label{fig:scalarSpectralIndex_3}
\end{subfigure}%
\vspace{0.05\linewidth}
\begin{subfigure}{0.33\linewidth}
  \centering
   \includegraphics[width=46mm,height=40mm]{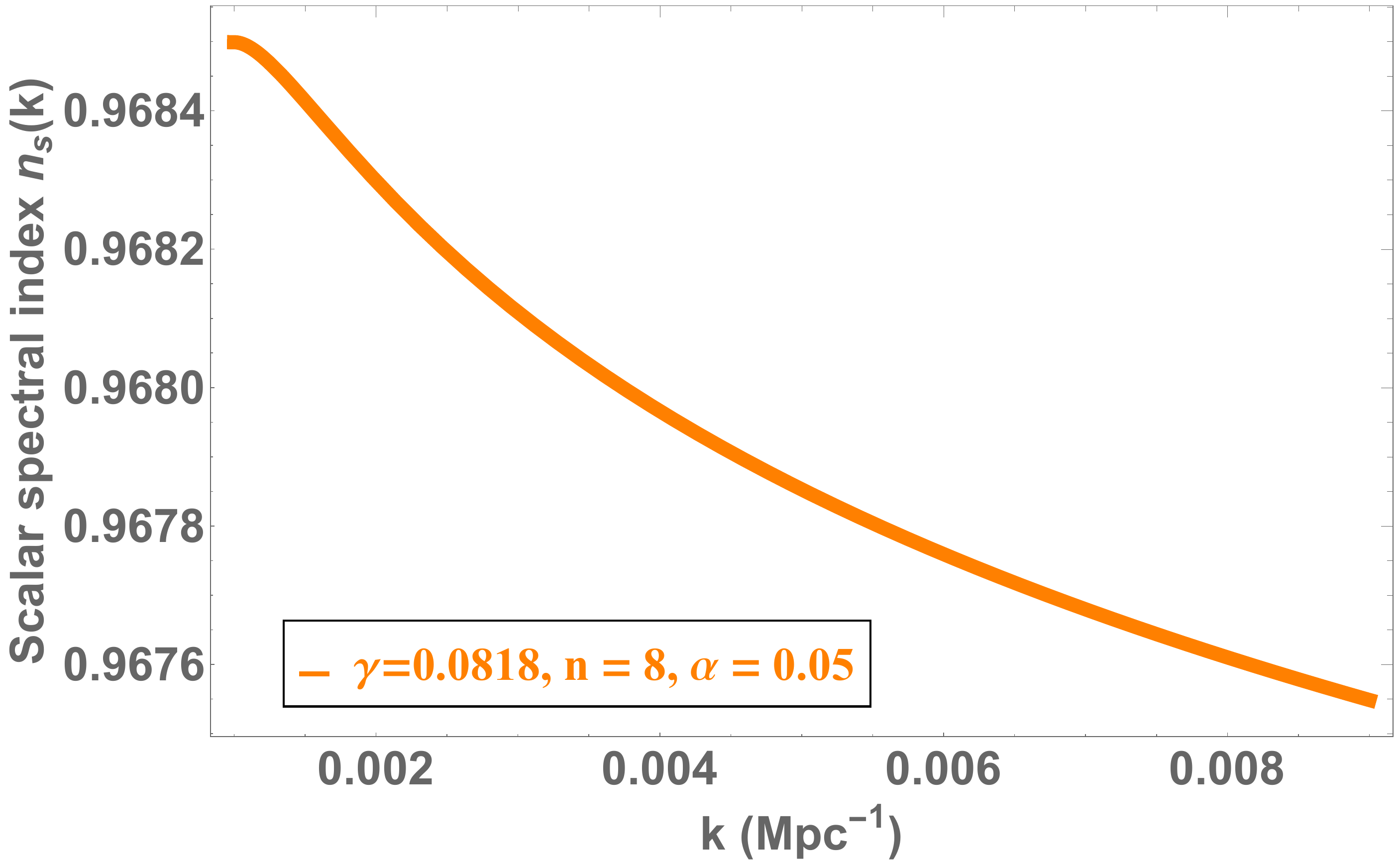}
   \subcaption{}
    \label{fig:scalarSpectralIndex_4}
\end{subfigure}%
\begin{subfigure}{0.33\linewidth}
  \centering
   \includegraphics[width=46mm,height=40mm]{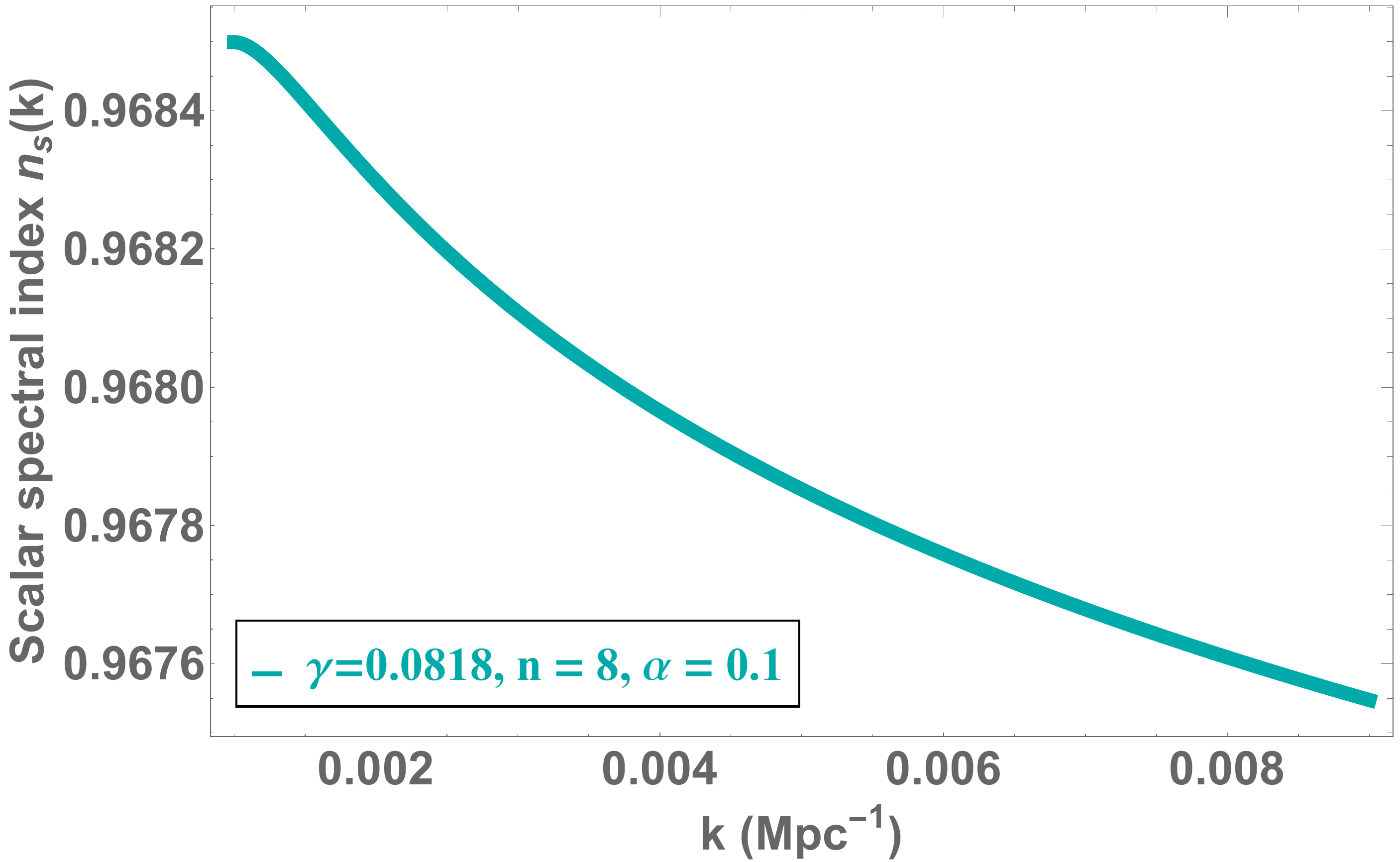}
   \subcaption{}
    \label{fig:scalarSpectralIndex_5}
\end{subfigure}%
\begin{subfigure}{0.33\linewidth}
  \centering
   \includegraphics[width=46mm,height=40mm]{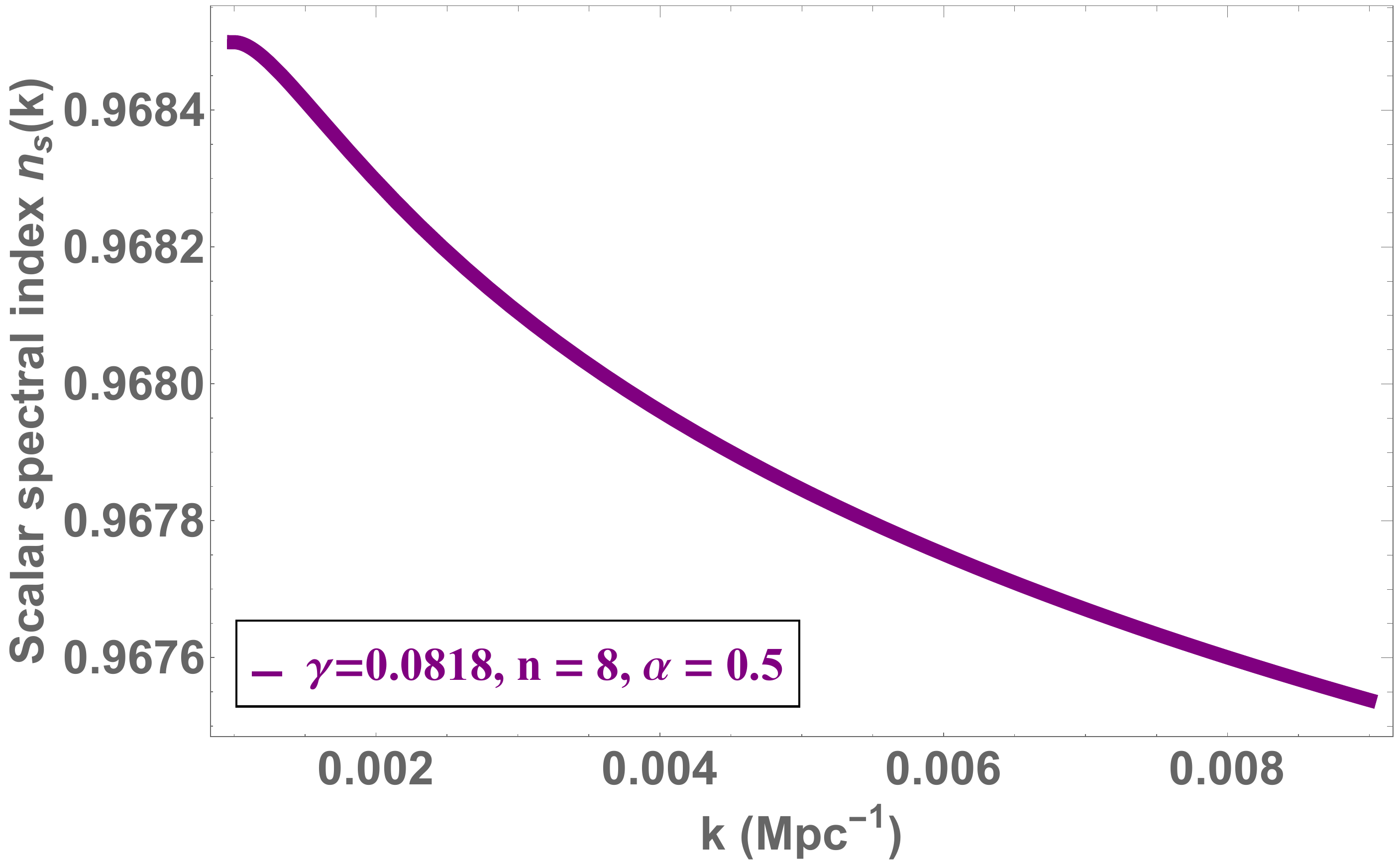}
   \subcaption{}
    \label{fig:scalarSpectralIndex_6}
\end{subfigure}%
\vspace{0.05\linewidth}
\begin{subfigure}{0.33\linewidth}
  \centering
   \includegraphics[width=46mm,height=40mm]{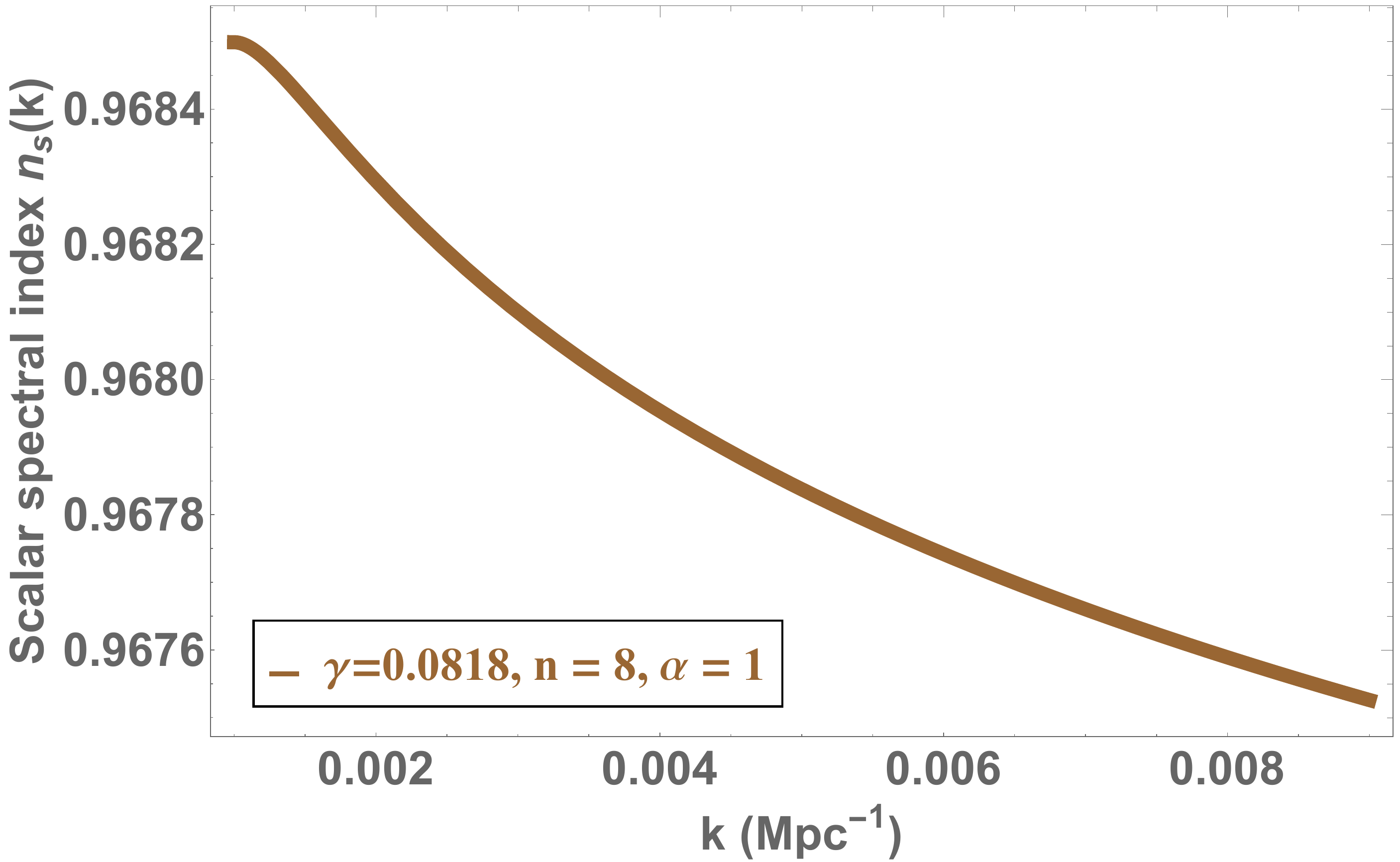}
   \subcaption{}
    \label{fig:scalarSpectralIndex_7}
\end{subfigure}%
\begin{subfigure}{0.33\linewidth}
  \centering
   \includegraphics[width=46mm,height=40mm]{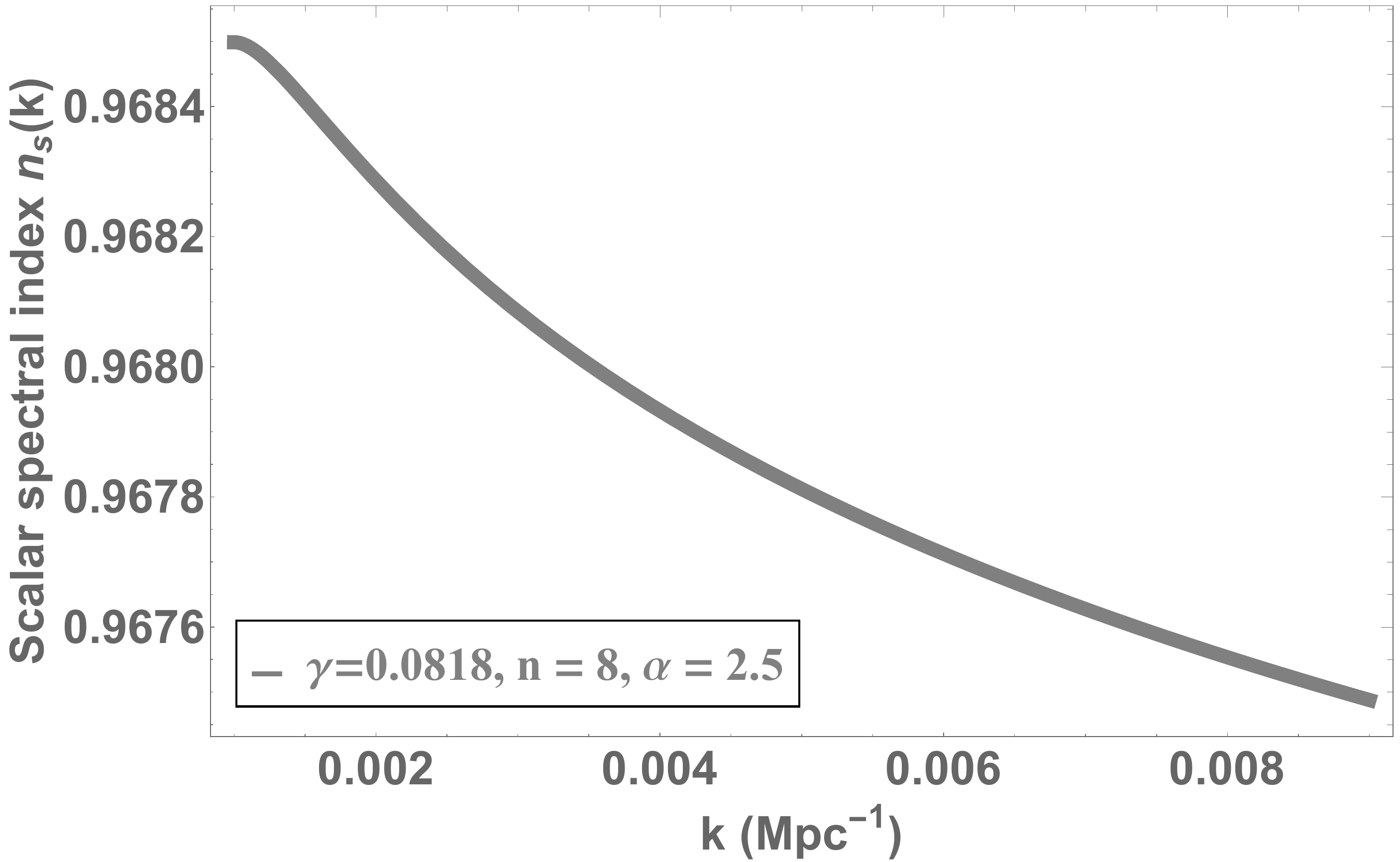}
   \subcaption{}
    \label{fig:scalarSpectralIndex_8}
\end{subfigure}%
\begin{subfigure}{0.33\linewidth}
  \centering
   \includegraphics[width=46mm,height=40mm]{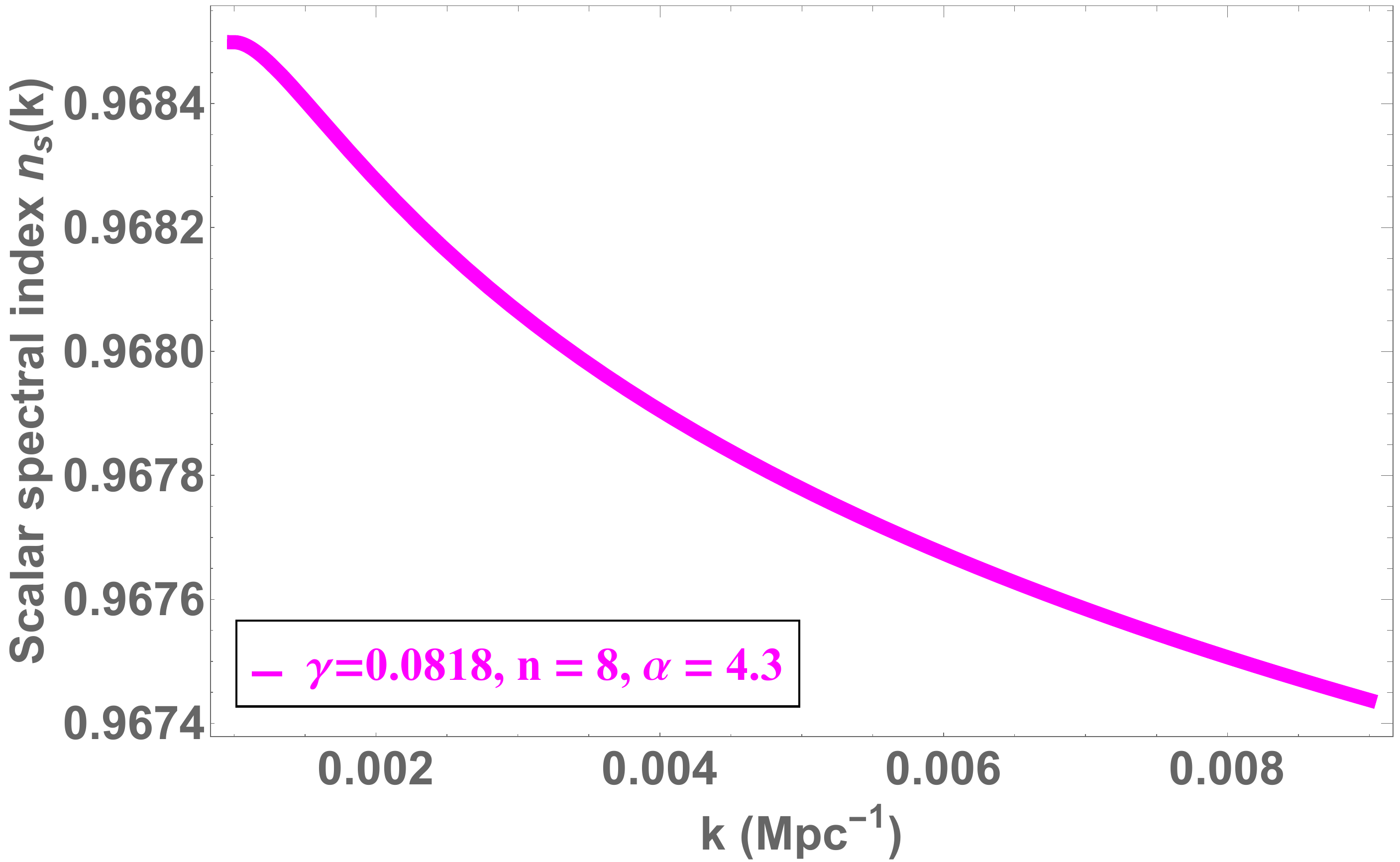}
   \subcaption{}
    \label{fig:scalarSpectralIndex_9}
\end{subfigure}
\caption{Scalar spectral indices for nine values of $\alpha$ for $\gamma=0.0818$ and $n=8$. The values of $n_s(k)$ tend to remain same for all values of $\alpha$.}
\label{fig:scalarSpectralIndex}
\end{figure}

\begin{figure}[H]
\begin{subfigure}{0.33\linewidth}
  \centering
   \includegraphics[width=46mm,height=40mm]{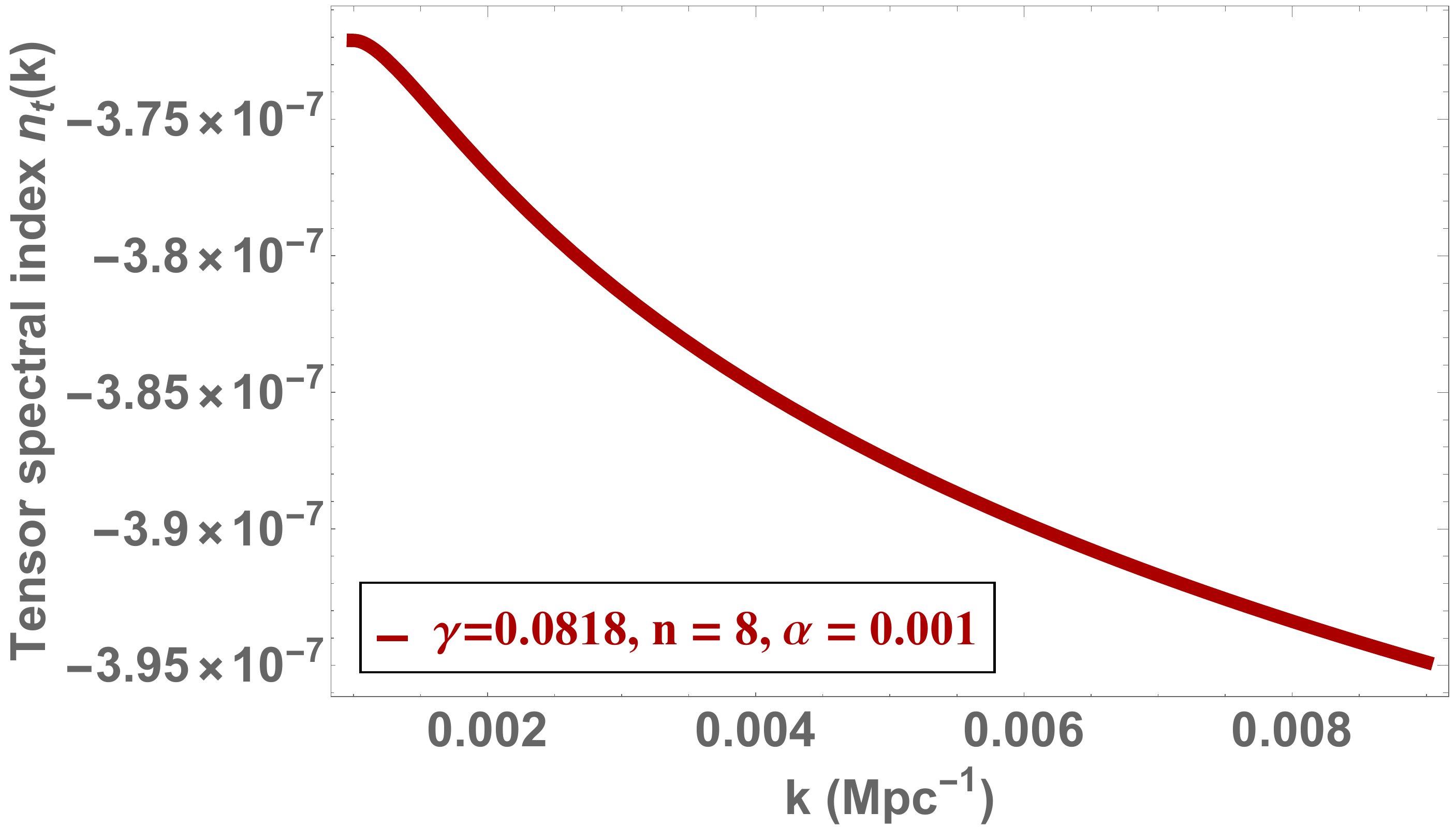}
   \subcaption{}
   \label{fig:tensorSpectralIndex_1}
\end{subfigure}%
\begin{subfigure}{0.33\linewidth}
  \centering
   \includegraphics[width=46mm,height=40mm]{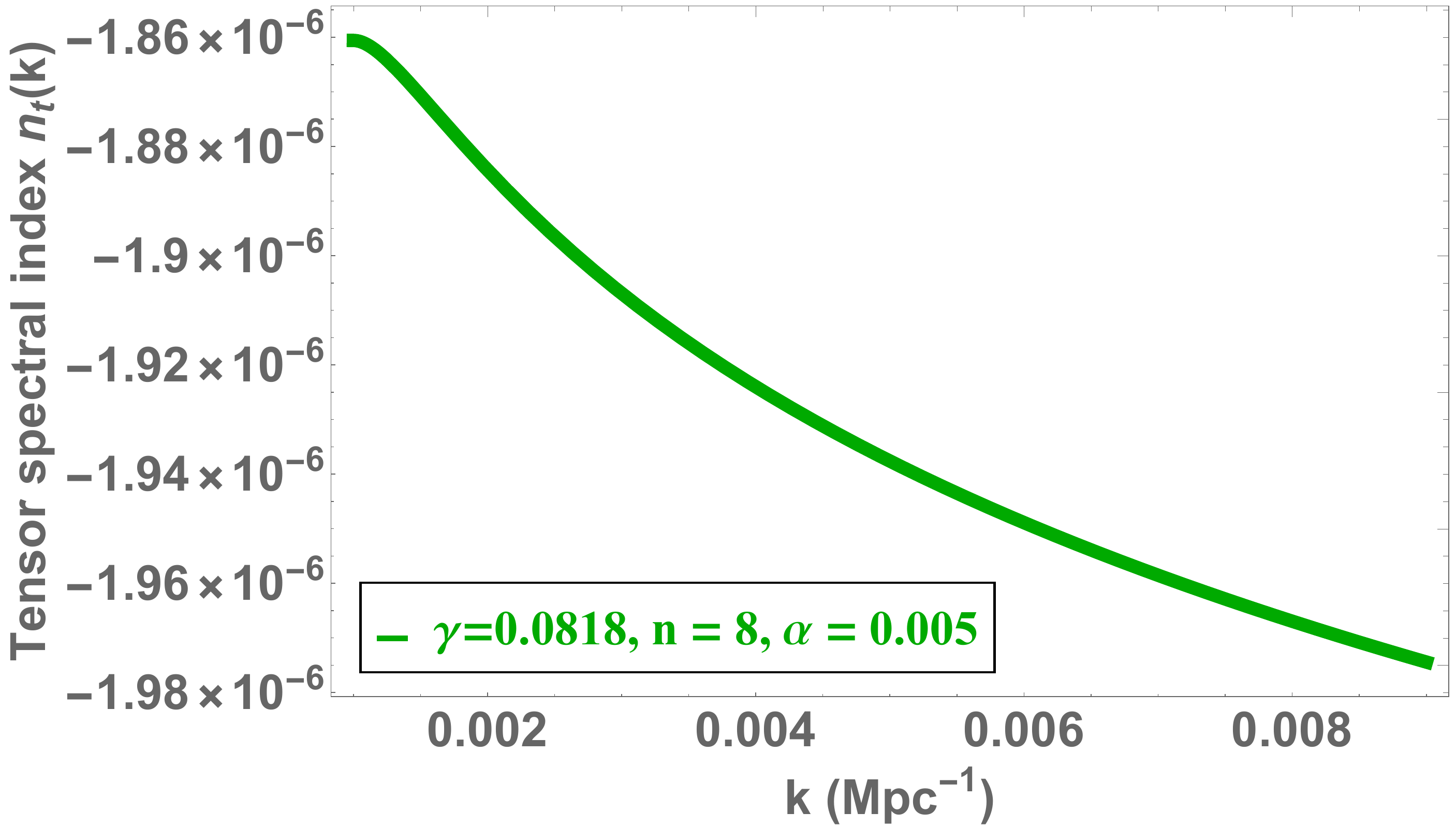}
   \subcaption{}
   \label{fig:tensorSpectralIndex_2}
\end{subfigure}%
\begin{subfigure}{0.33\linewidth}
  \centering
   \includegraphics[width=46mm,height=40mm]{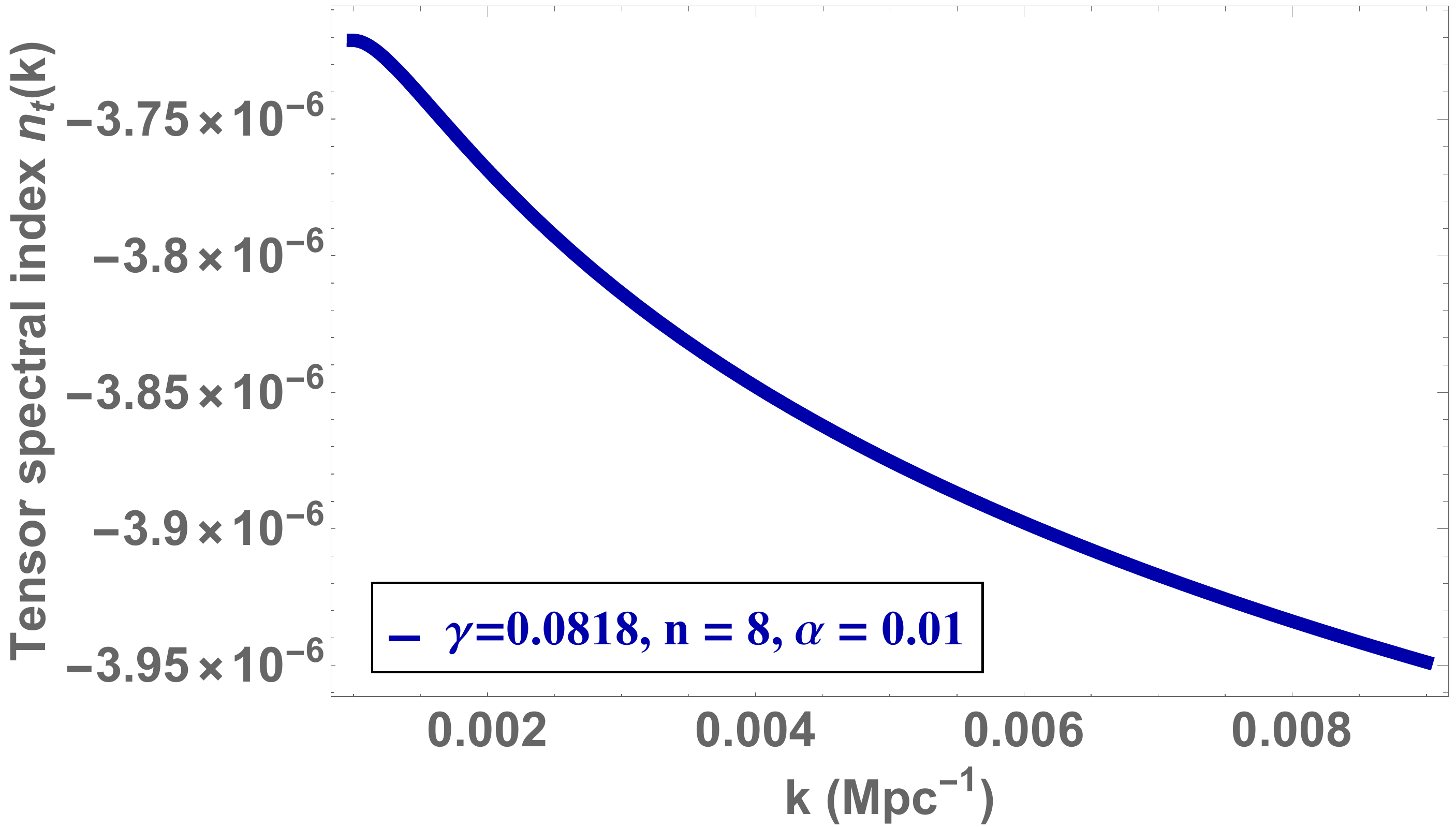}
   \subcaption{}
   \label{fig:tensorSpectralIndex_3}
\end{subfigure}%
\vspace{0.05\linewidth}
\begin{subfigure}{0.33\linewidth}
  \centering
   \includegraphics[width=46mm,height=40mm]{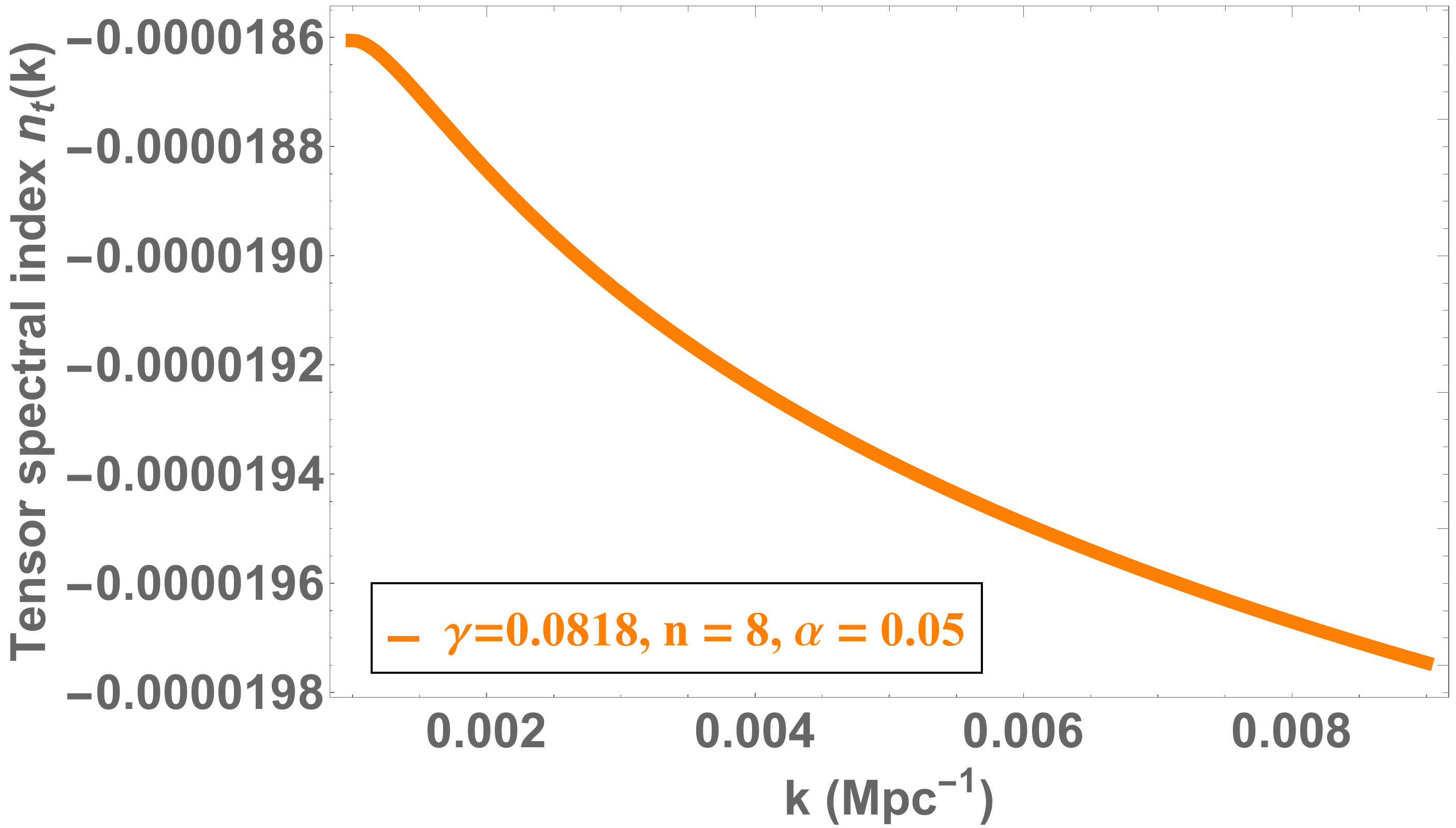}
   \subcaption{}
    \label{fig:tensorSpectralIndex_4}
\end{subfigure}%
\begin{subfigure}{0.33\linewidth}
  \centering
   \includegraphics[width=46mm,height=40mm]{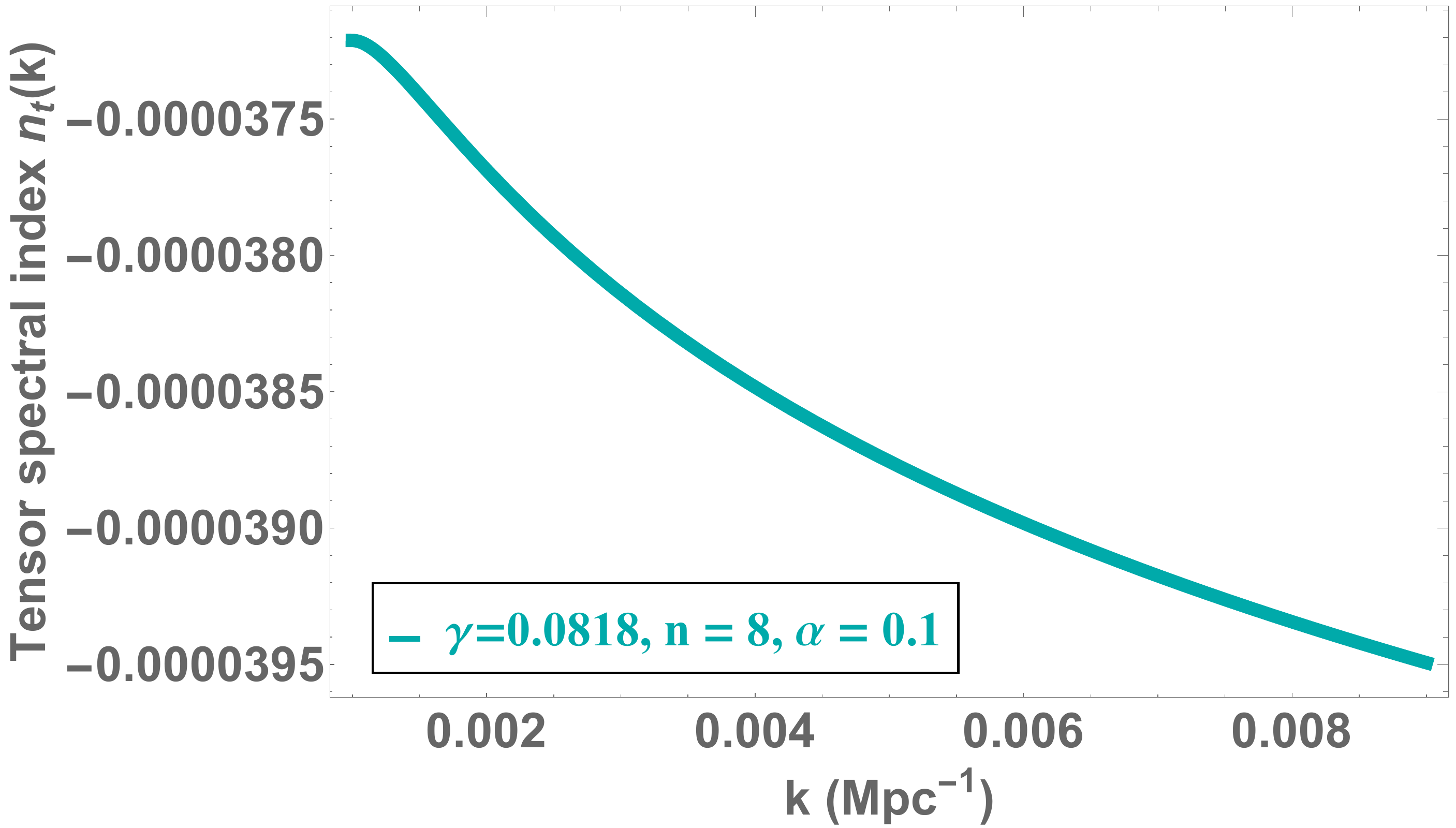}
   \subcaption{}
    \label{fig:tensorSpectralIndex_5}
\end{subfigure}%
\begin{subfigure}{0.33\linewidth}
  \centering
   \includegraphics[width=46mm,height=40mm]{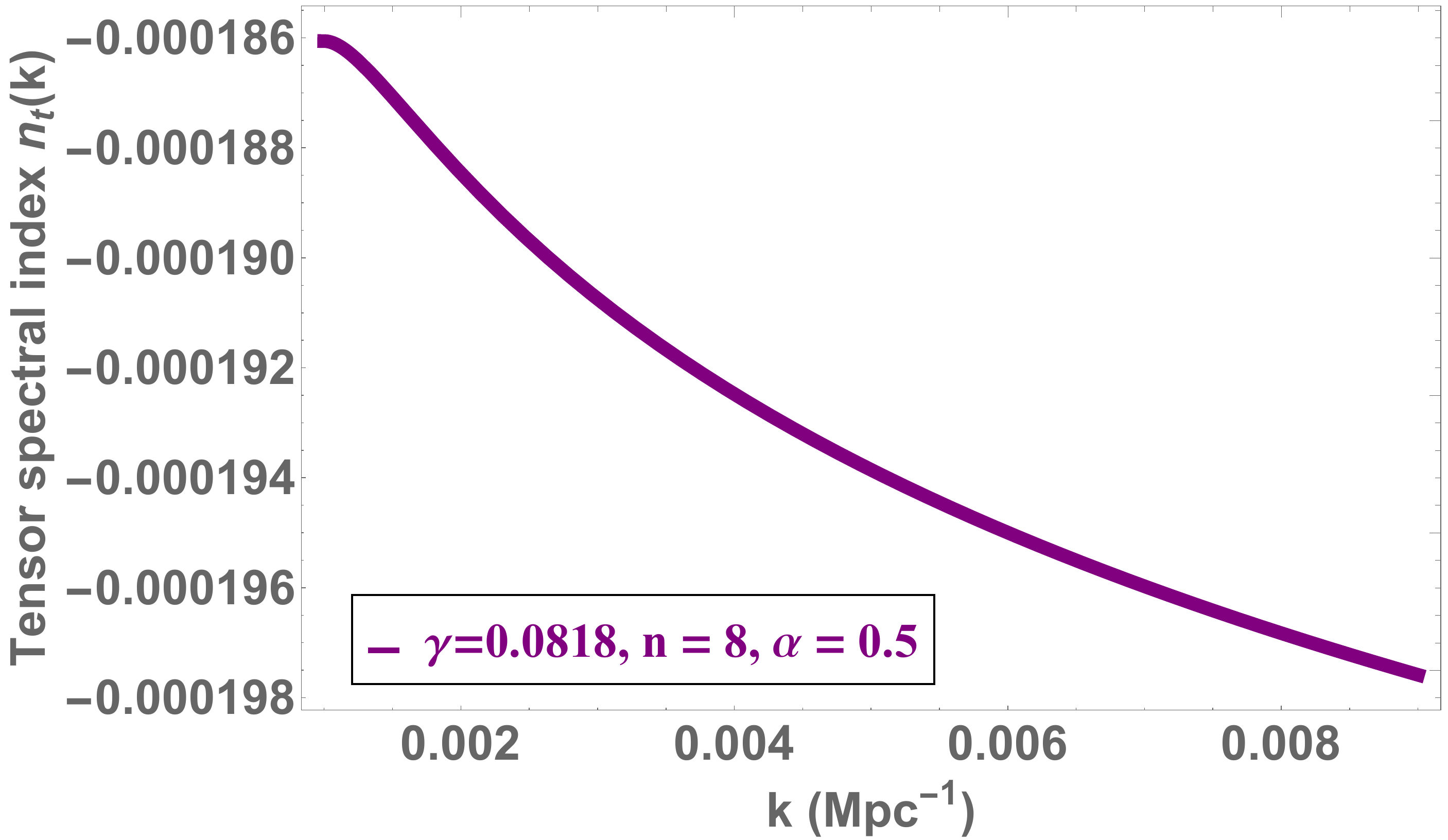}
   \subcaption{}
    \label{fig:tensorSpectralIndex_6}
\end{subfigure}%
\vspace{0.05\linewidth}
\begin{subfigure}{0.33\linewidth}
  \centering
   \includegraphics[width=46mm,height=40mm]{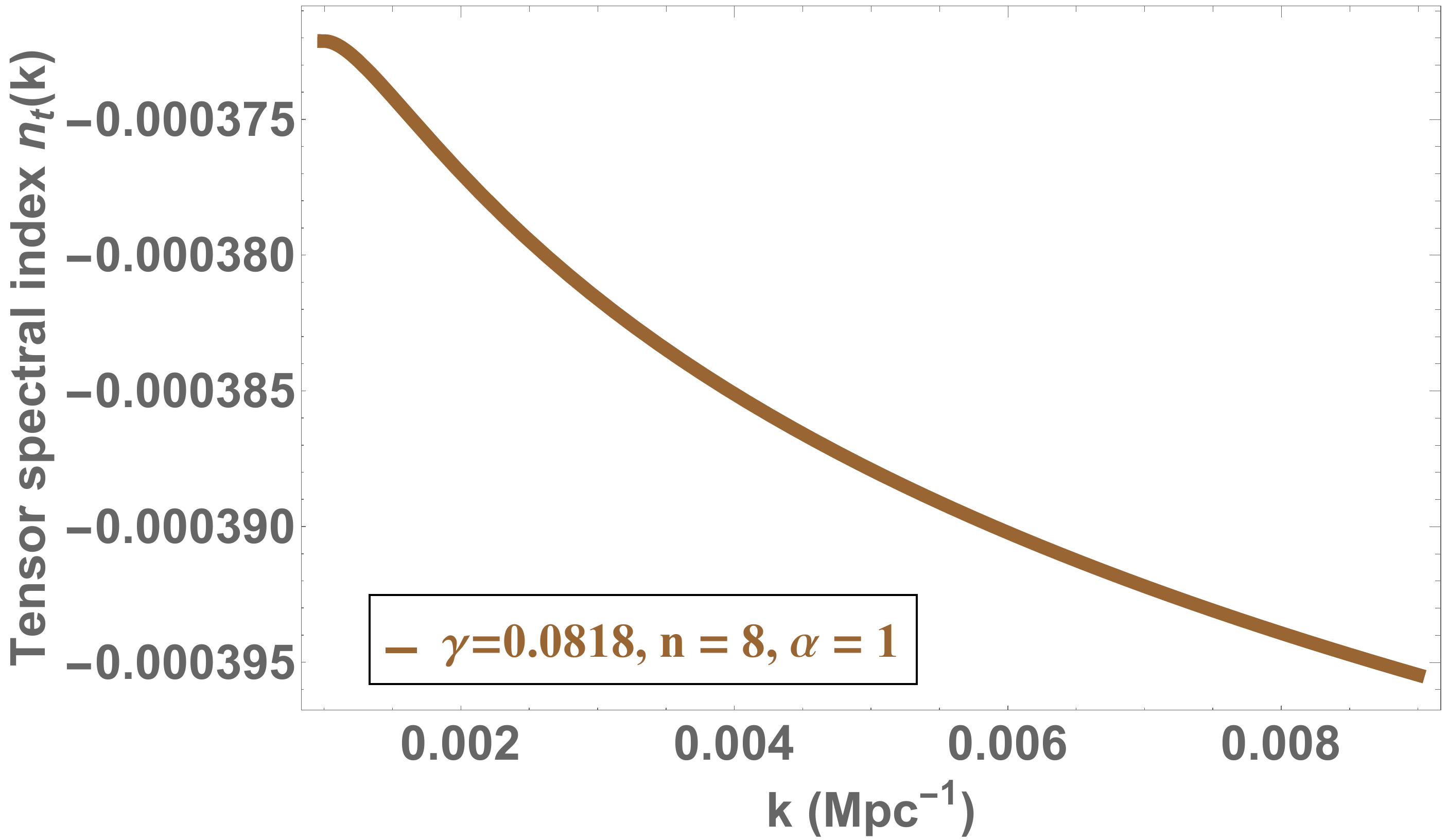}
   \subcaption{}
    \label{fig:tensorSpectralIndex_7}
\end{subfigure}%
\begin{subfigure}{0.33\linewidth}
  \centering
   \includegraphics[width=46mm,height=40mm]{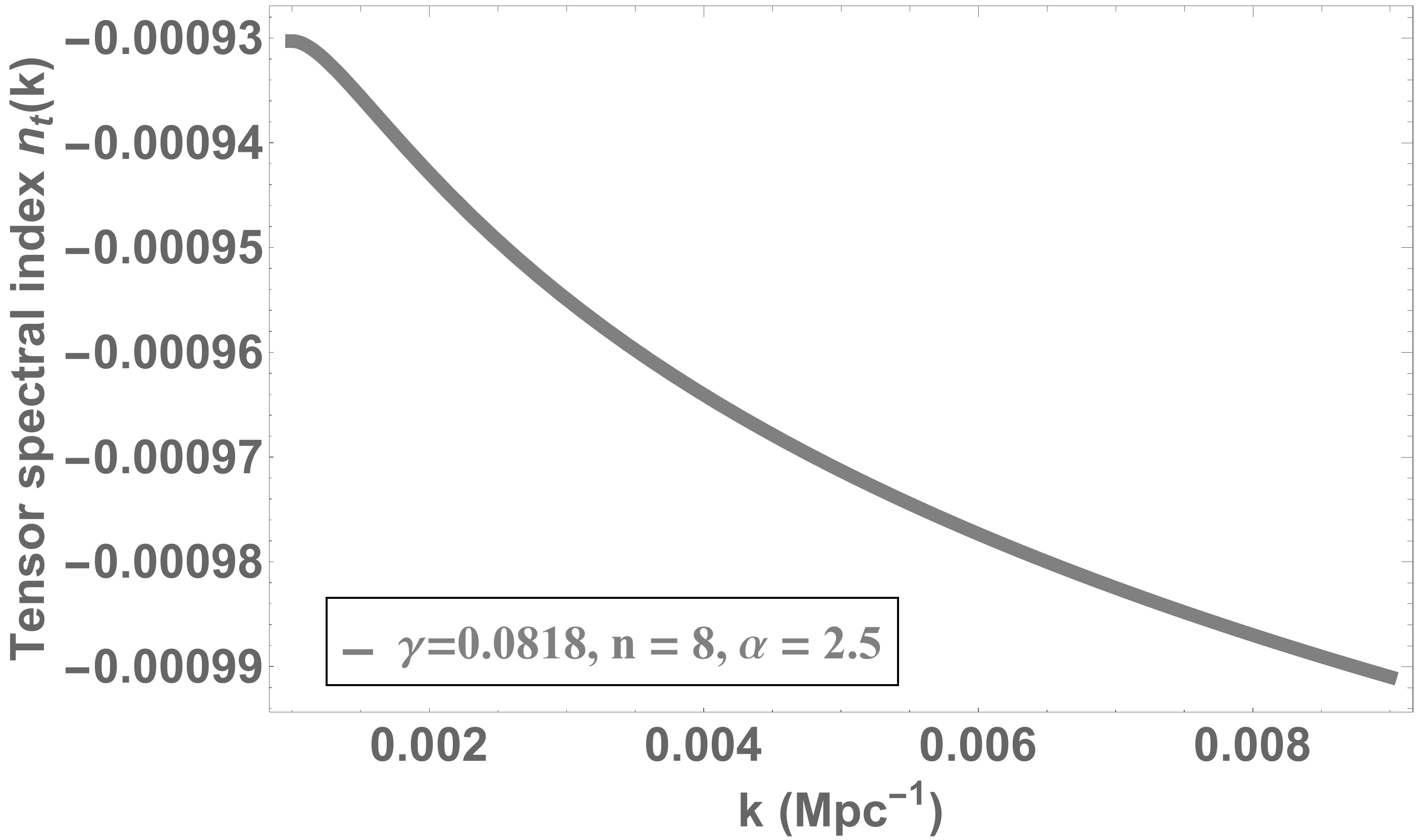}
   \subcaption{}
    \label{fig:tensorSpectralIndex_8}
\end{subfigure}%
\begin{subfigure}{0.33\linewidth}
  \centering
   \includegraphics[width=46mm,height=40mm]{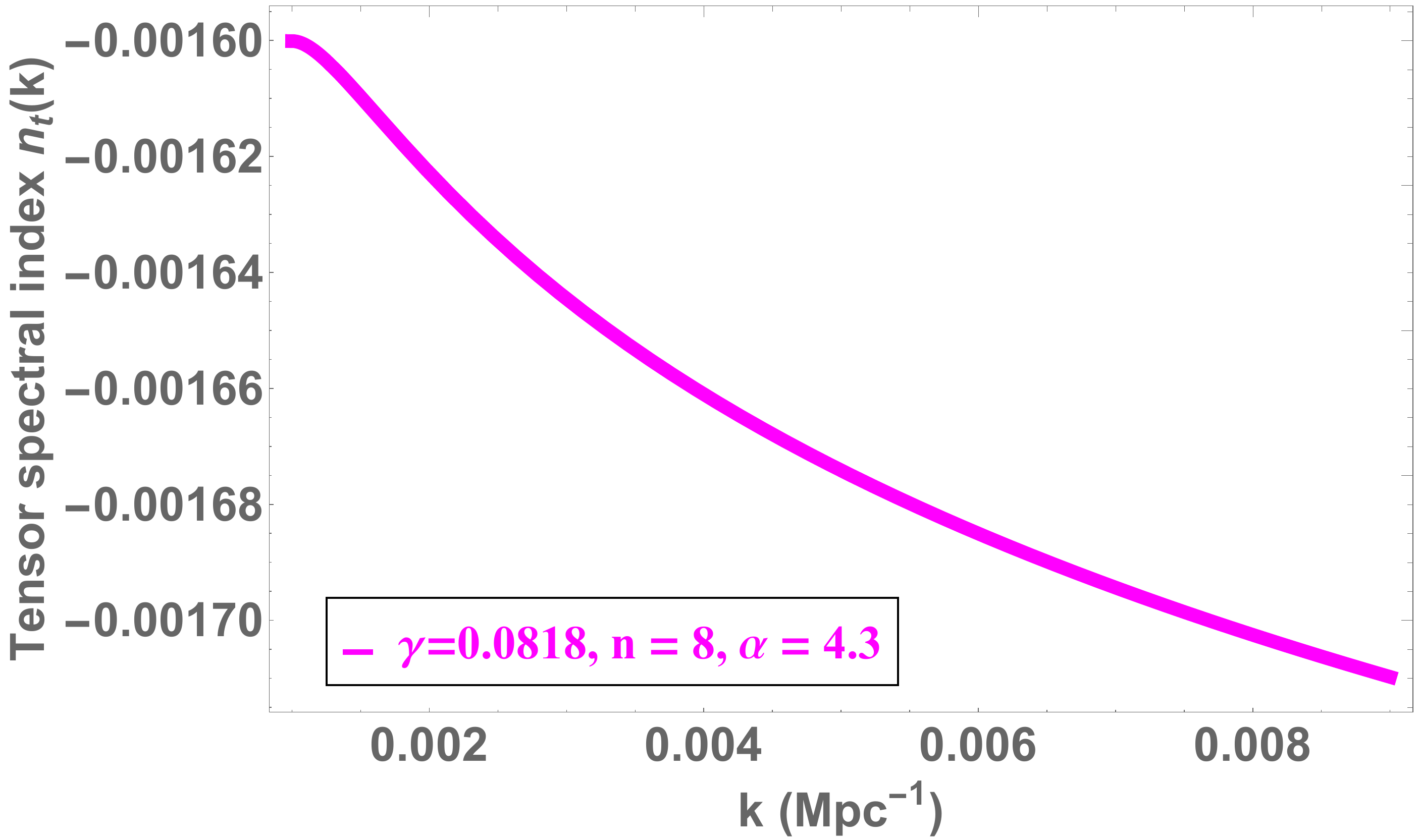}
   \subcaption{}
    \label{fig:tensorSpectralIndex_9}
\end{subfigure}
\caption{Tensor spectral indices for nine values of $\alpha$ for $\gamma=0.0818$ and $n=8$. The values of $|n_t(k)|$ tend to increase with the increase in $\alpha$ at a particular $k$ value.}
\label{fig:tensorSpectralIndex}
\end{figure}

\begin{figure}[H]
\begin{subfigure}{0.33\linewidth}
  \centering
   \includegraphics[width=46mm,height=40mm]{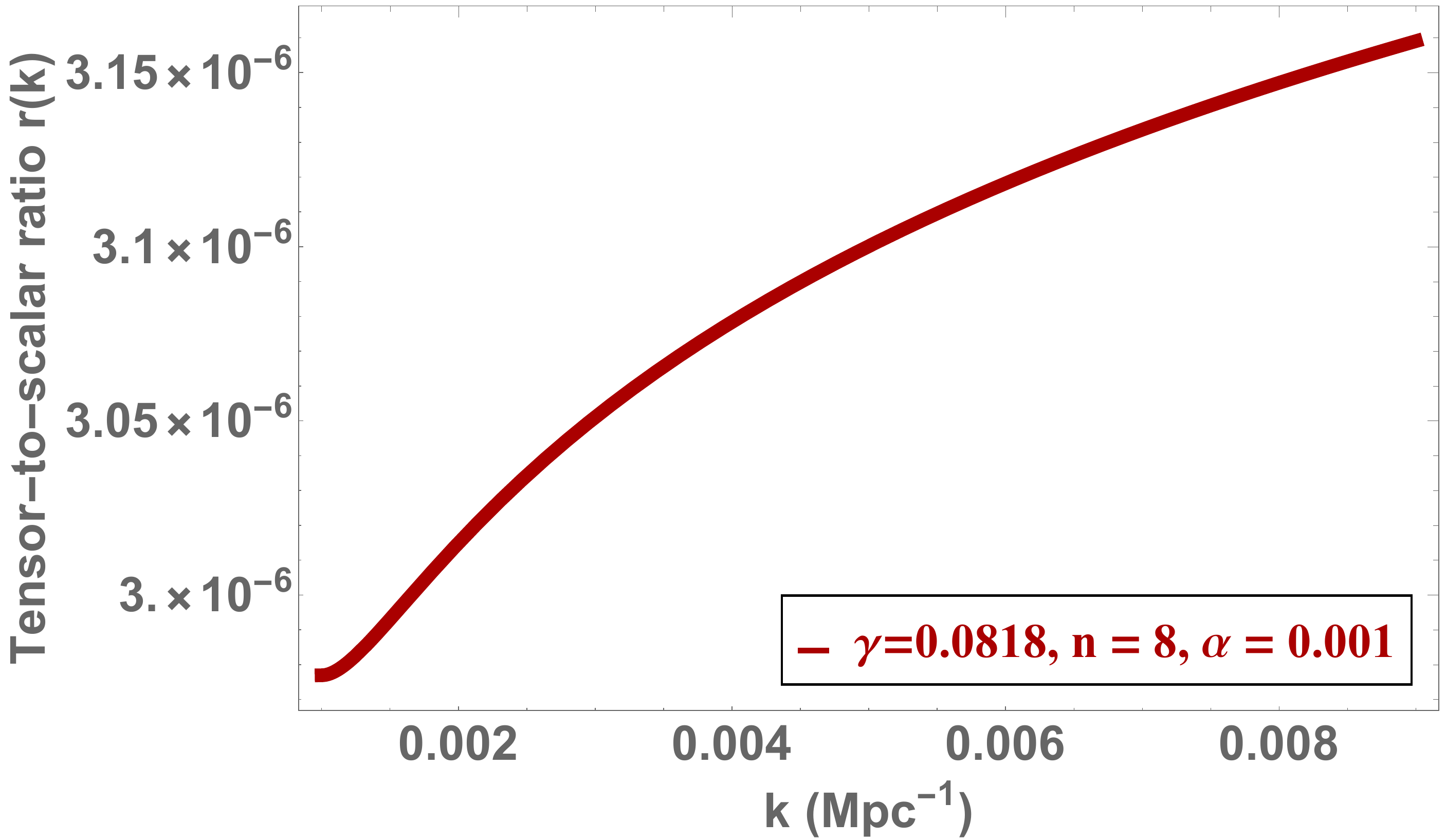}
   \subcaption{}
   \label{fig:tensorToScalarRatio_1}
\end{subfigure}%
\begin{subfigure}{0.33\linewidth}
  \centering
   \includegraphics[width=46mm,height=40mm]{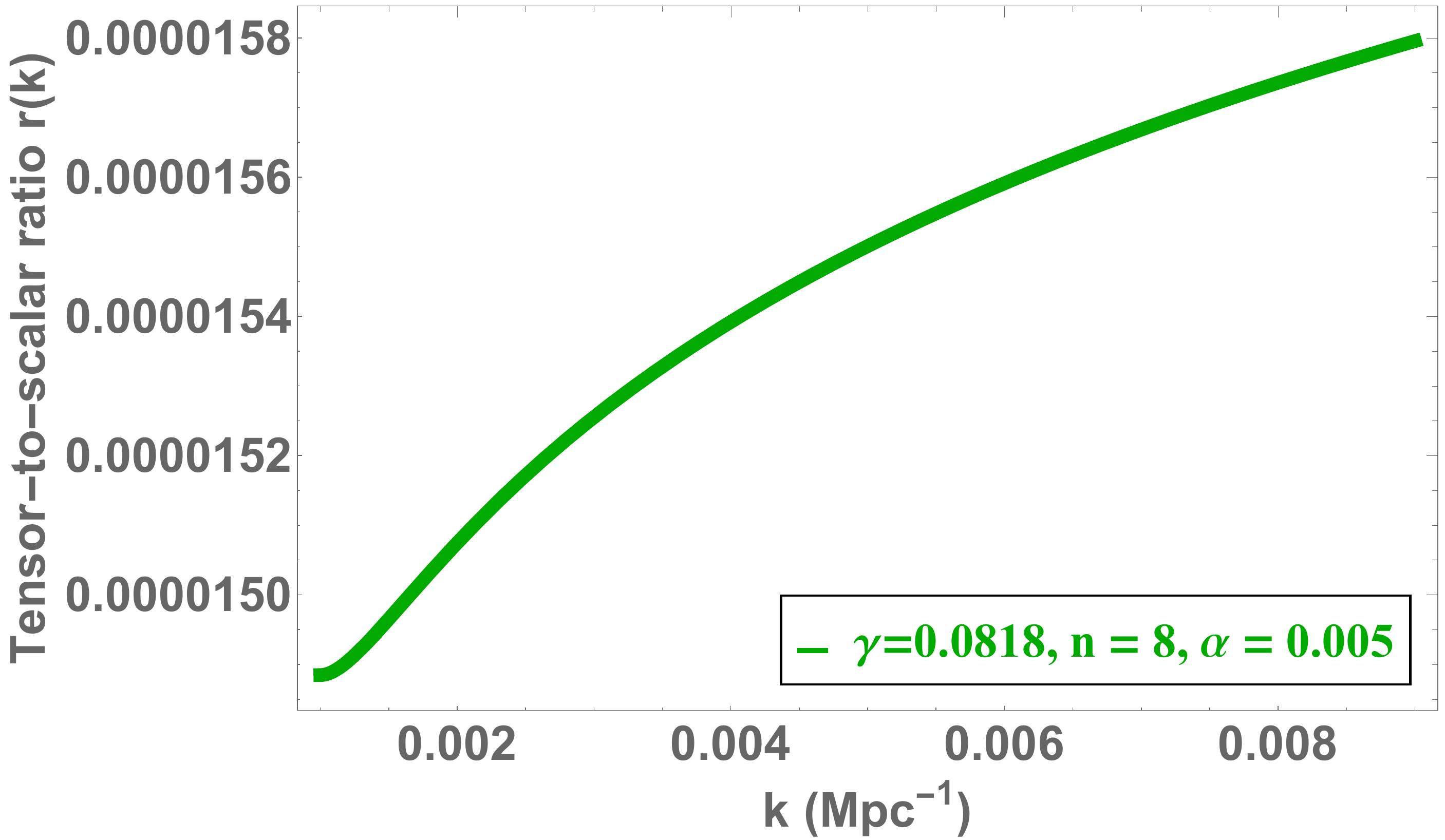}
   \subcaption{}
   \label{fig:tensorToScalarRatio_2}
\end{subfigure}%
\begin{subfigure}{0.33\linewidth}
  \centering
   \includegraphics[width=46mm,height=40mm]{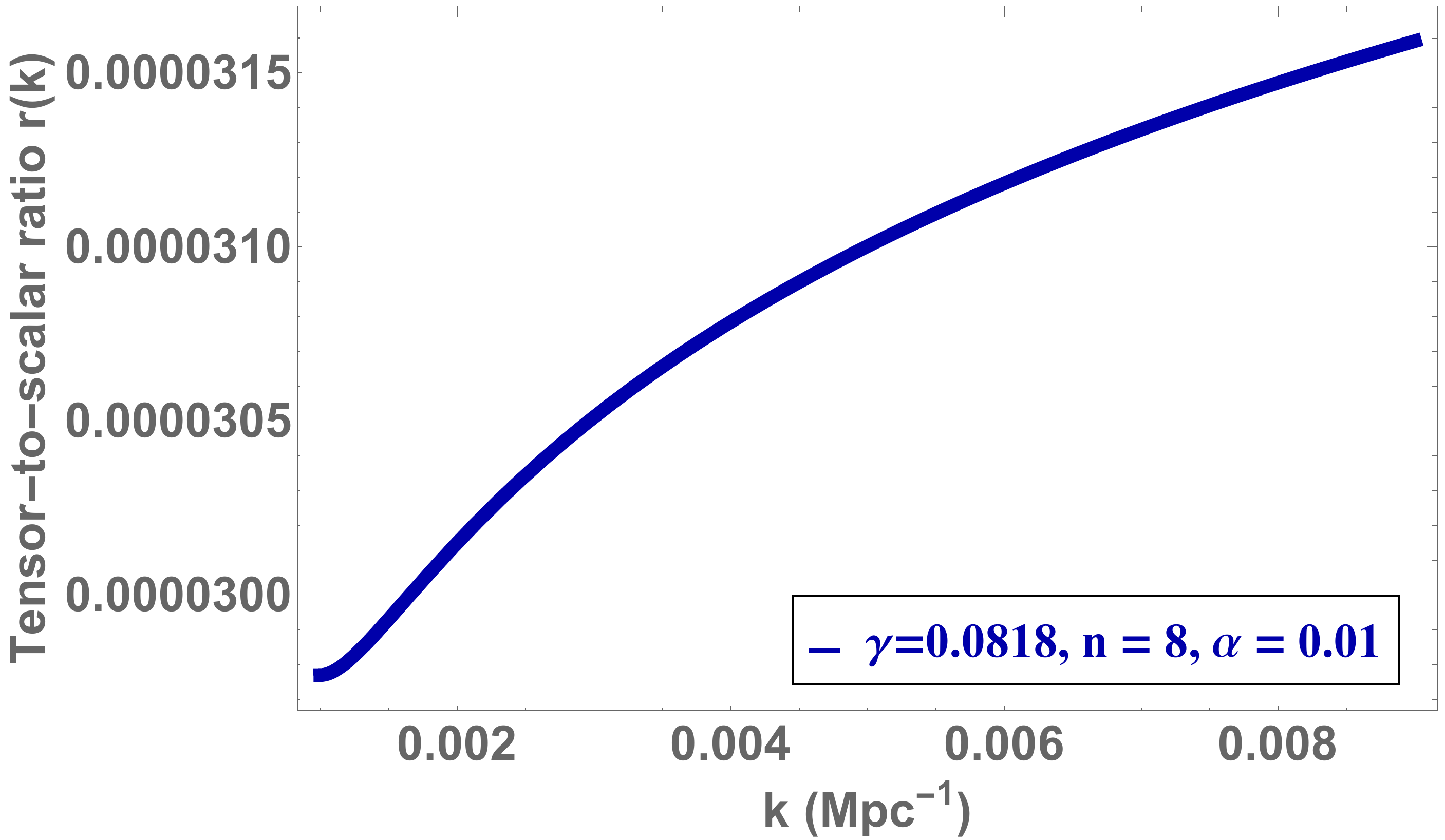}
   \subcaption{}
   \label{fig:tensorToScalarRatio_3}
\end{subfigure}%
\vspace{0.05\linewidth}
\begin{subfigure}{0.33\linewidth}
  \centering
   \includegraphics[width=46mm,height=40mm]{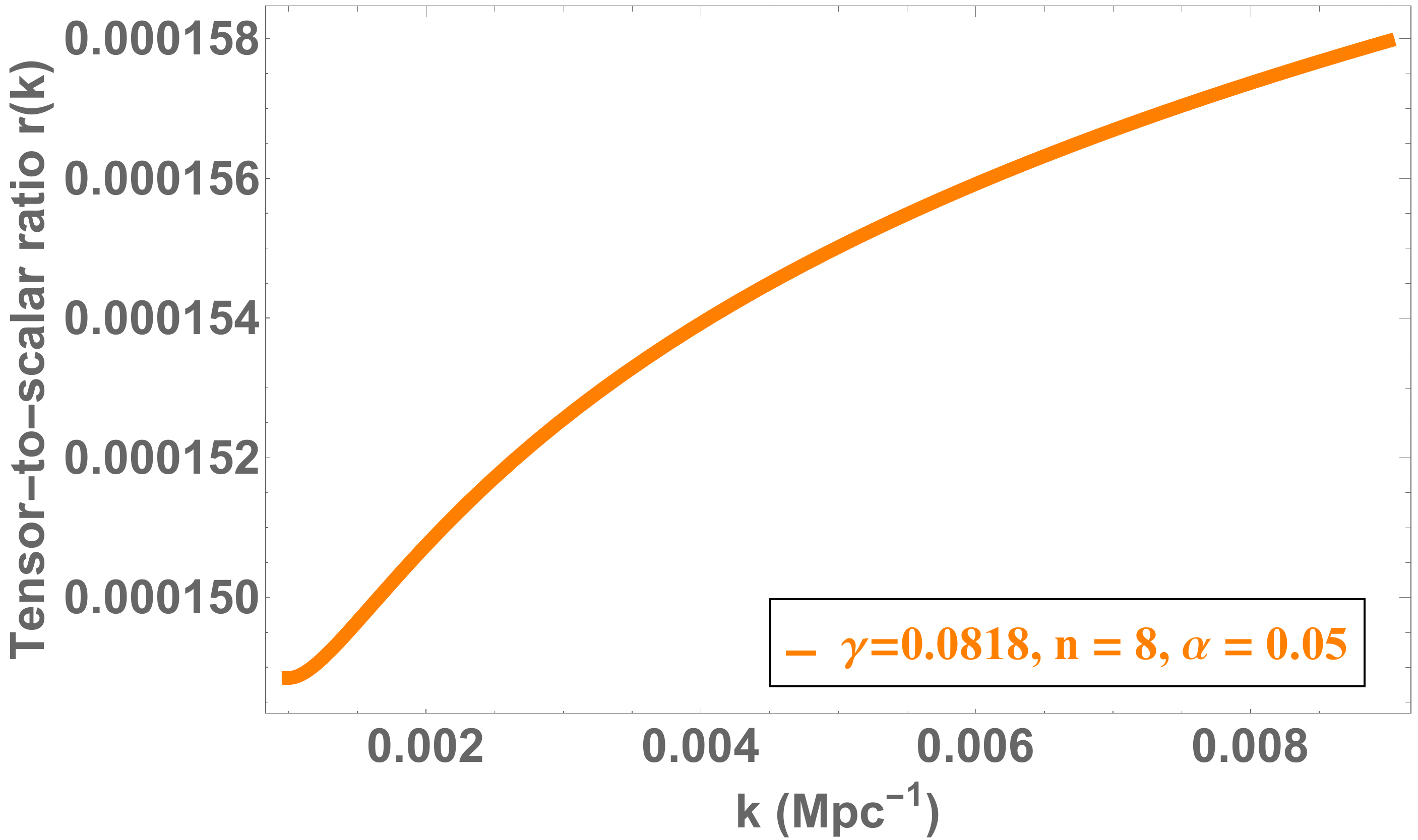}
   \subcaption{}
    \label{fig:tensorToScalarRatio_4}
\end{subfigure}%
\begin{subfigure}{0.33\linewidth}
  \centering
   \includegraphics[width=46mm,height=40mm]{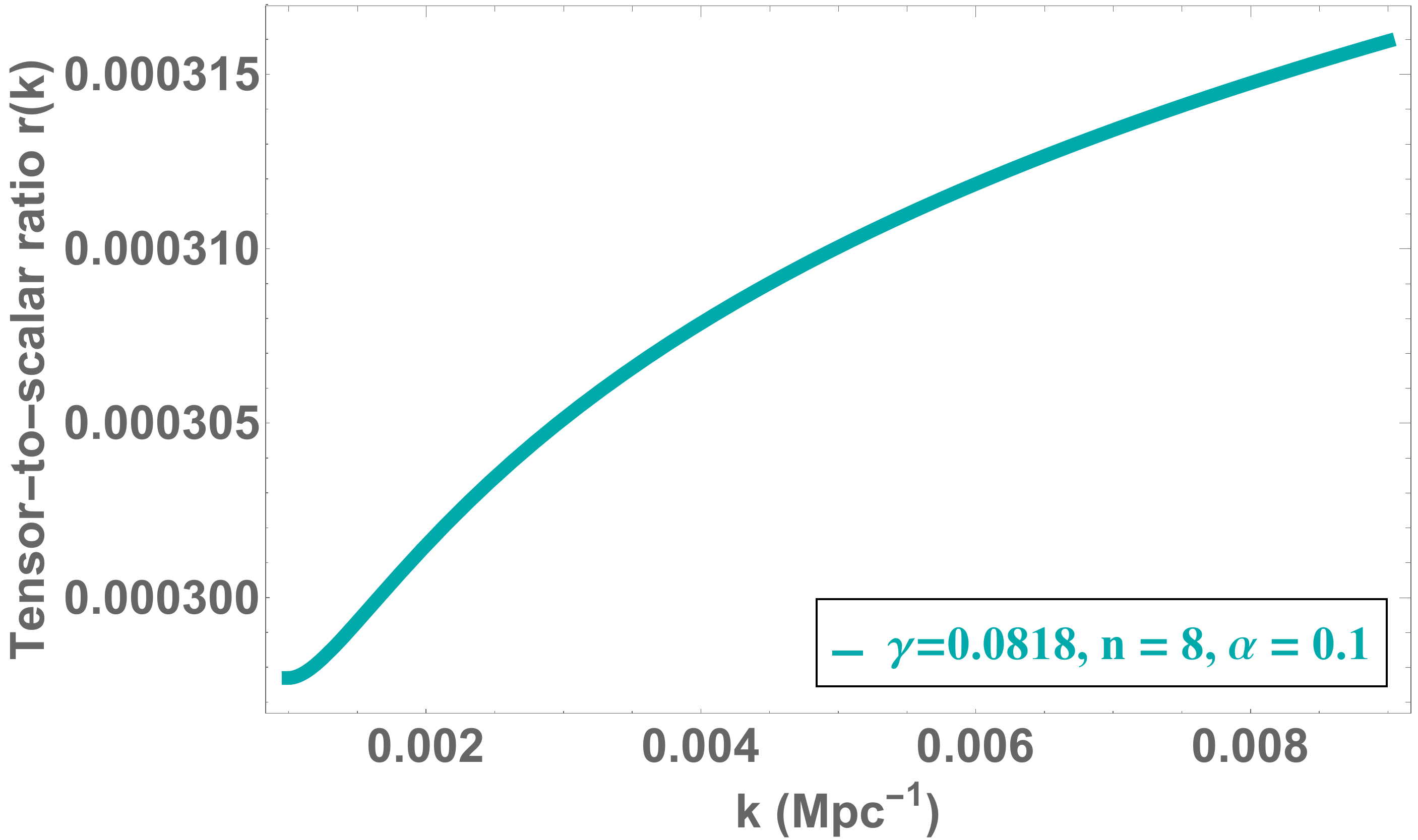}
   \subcaption{}
    \label{fig:tensorToScalarRatio_5}
\end{subfigure}%
\begin{subfigure}{0.33\linewidth}
  \centering
   \includegraphics[width=46mm,height=40mm]{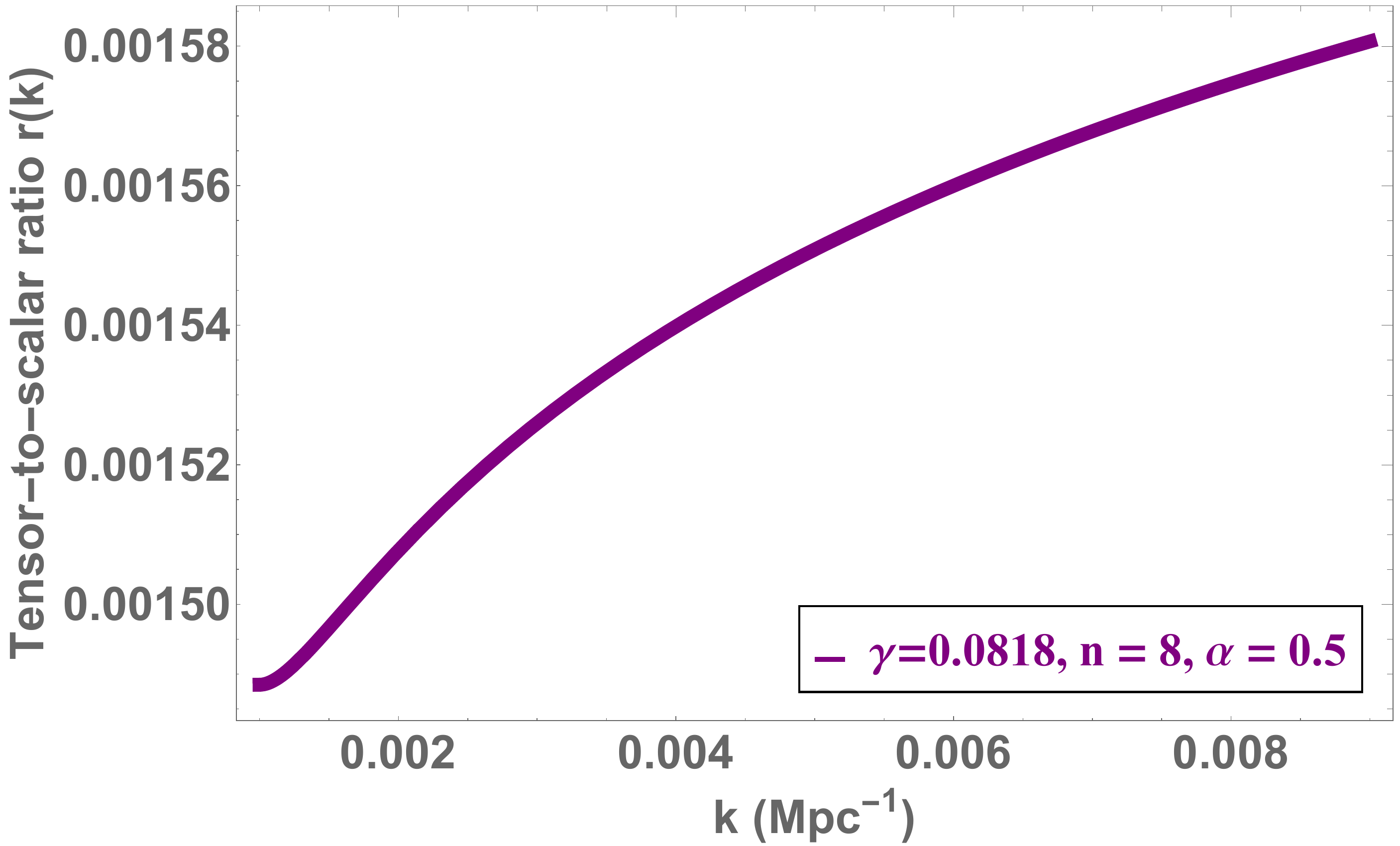}
   \subcaption{}
    \label{fig:tensorToScalarRatio_6}
\end{subfigure}%
\vspace{0.05\linewidth}
\begin{subfigure}{0.33\linewidth}
  \centering
   \includegraphics[width=46mm,height=40mm]{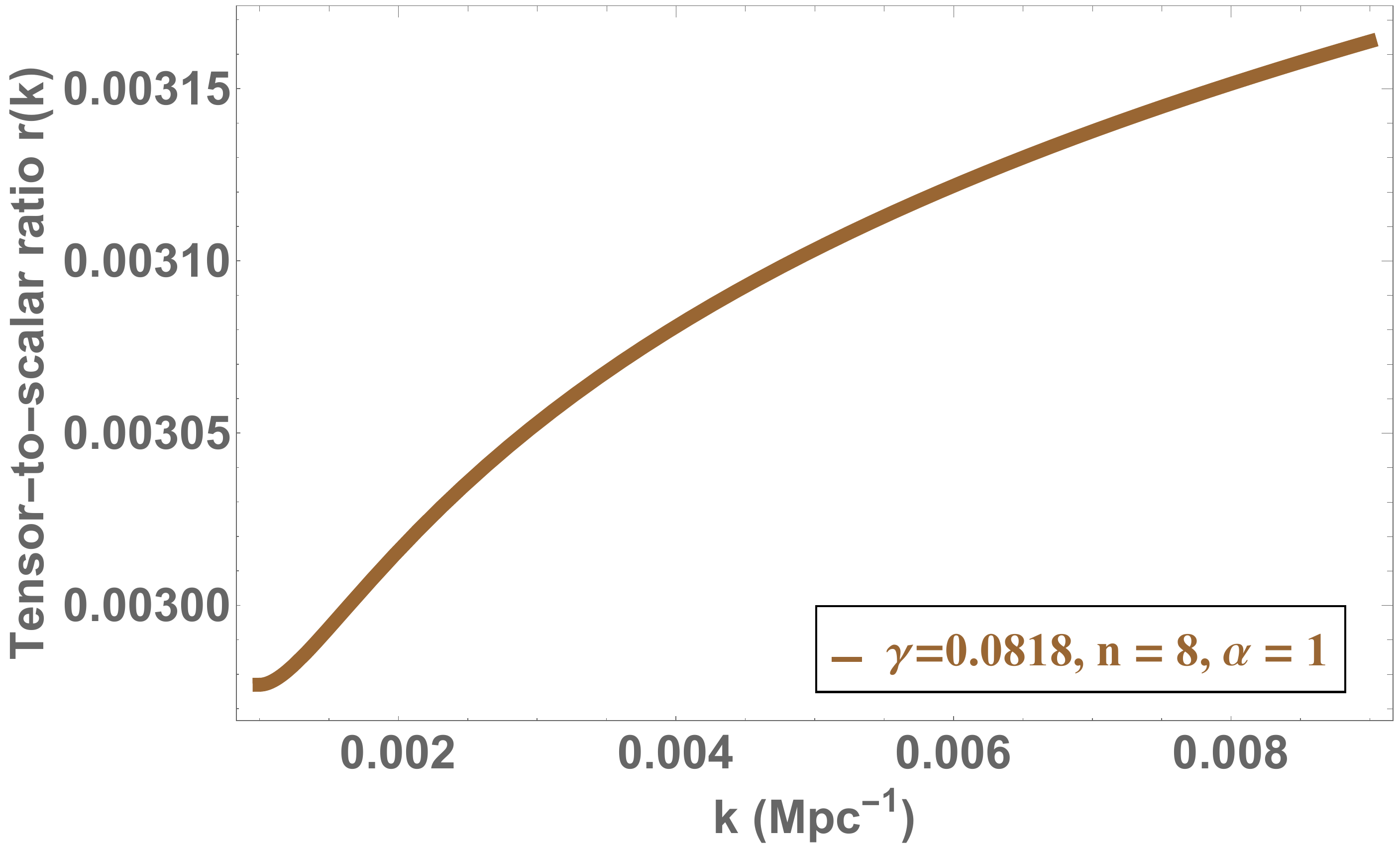}
   \subcaption{}
    \label{fig:tensorToScalarRatio_7}
\end{subfigure}%
\begin{subfigure}{0.33\linewidth}
  \centering
   \includegraphics[width=46mm,height=40mm]{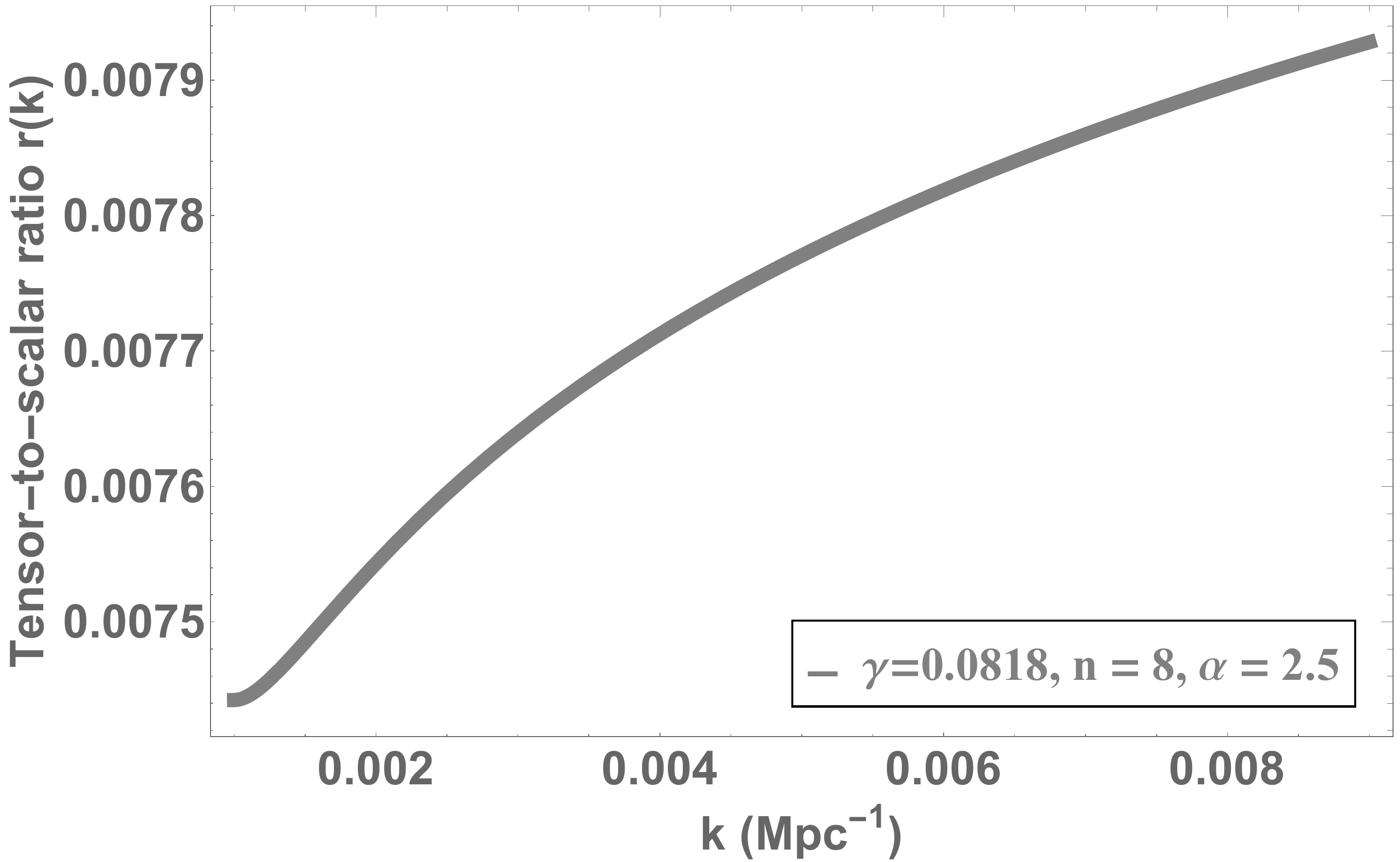}
   \subcaption{}
    \label{fig:tensorToScalarRatio_8}
\end{subfigure}%
\begin{subfigure}{0.33\linewidth}
  \centering
   \includegraphics[width=46mm,height=40mm]{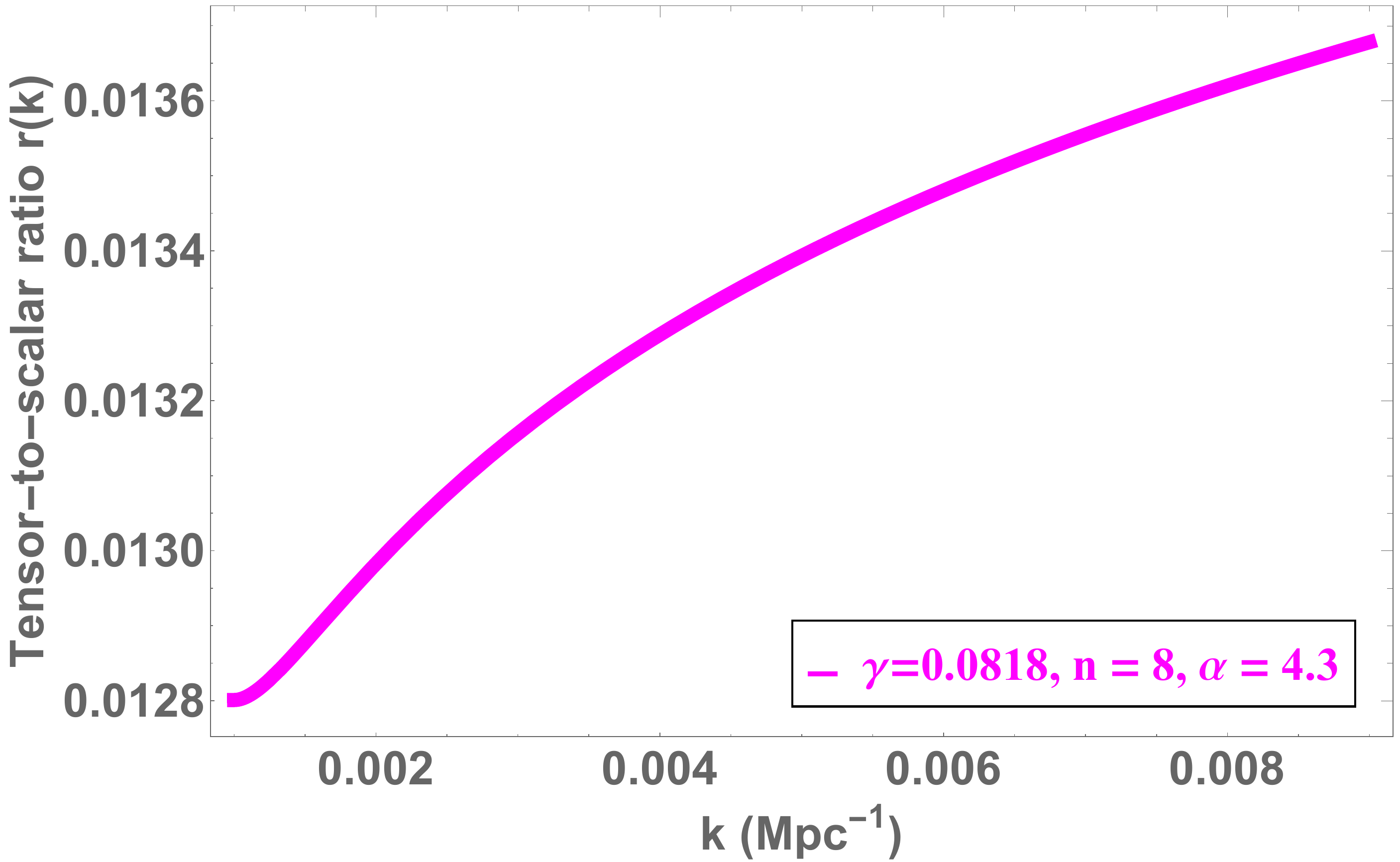}
   \subcaption{}
    \label{fig:tensorToScalarRatio_9}
\end{subfigure}
\caption{Tensor-to-scalar ratios for nine values of $\alpha$ for $\gamma=0.0818$ and $n=8$. The values of $r(k)$ tend to increase with the increase in $\alpha$ at a particular $k$ value.}
\label{fig:tensorToScalarRatio}
\end{figure}

 We shall now discuss about the spectral indices \textit{i.e.} the scalar spectral index $n_s(k)$, tensor spectral index $n_t(k)$ and tensor-to-scalar ratio $r(k)$ which are related to the momentum dependencies of the power spectra $\Delta_s(k)$ and $\Delta_t(k)$ and their relative variations. For pole inflation (as in the present case), they obey simple expressions called the \textit{attractor equations} which are dependent on the number of e-folds. We derived such equations (Eqs. (\ref{eq:final_ns})-(\ref{eq:final_r})) in Section \ref{sec:formalism}. These equations show that $n_s$ is independent of $\alpha$ while $|n_t|$ and $r$ are directly proportional to $\alpha$. In figure \ref{fig:numberOfEFolds} we have seen that $N(k)$ remains within $61.5 - 63.5$, irrespective of $\alpha$. Thus it can be anticipated that $n_s(k)$ should follow the same pattern as that of $N(k)$, while $|n_t(k)|$ and $r(k)$ should increase with $\alpha$. Figures \ref{fig:scalarSpectralIndex}, \ref{fig:tensorSpectralIndex} and \ref{fig:tensorToScalarRatio} demonstrate these facts, as expected. Now, the reason behind the monotonic increment of $r(k)$ with $k$ for a specific value of $\alpha$ lies in the $k$-space behaviour of the two power spectra. At a particular value of $\alpha$, the scalar power spectrum decreases with faster rate than that of tensor power spectrum. As a result, their ratio \textit{i.e.} $r(k)$ increases with $k$. All the values of $n_s(k)$ ($0.9685-0.9674$), $n_t(k)$ ($(-3.95\times 10^{-7})$-$(-1.72\times 10^{-3})$) and $r(k)$ ($3.15\times 10^{-6}-1.28\times 10^{-2}$) satisfy the Planck constraints (see figure \ref{fig:5}) quite well. Detecting such small orders of $r(k)$ ($\lesssim 10^{-3}$) is the target of many ongoing and forthcoming $B$-mode surveys \cite{Tristram:2021tvh}. \par
\begin{figure}[H]
\begin{subfigure}{0.33\linewidth}
  \centering
   \includegraphics[width=46mm,height=40mm]{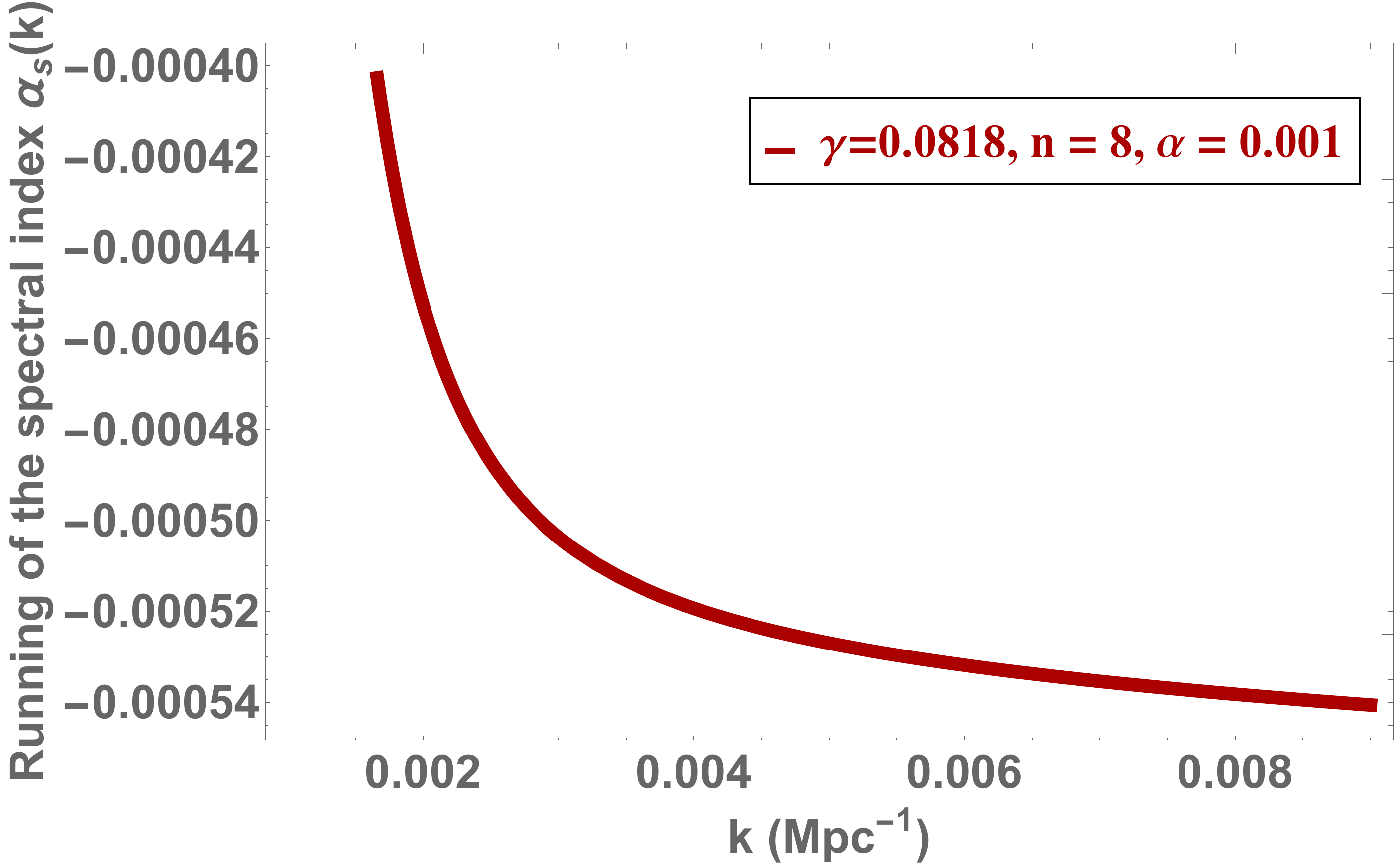}
   \subcaption{}
   \label{fig:RunningSpectralIndex_1}
\end{subfigure}%
\begin{subfigure}{0.33\linewidth}
  \centering
   \includegraphics[width=46mm,height=40mm]{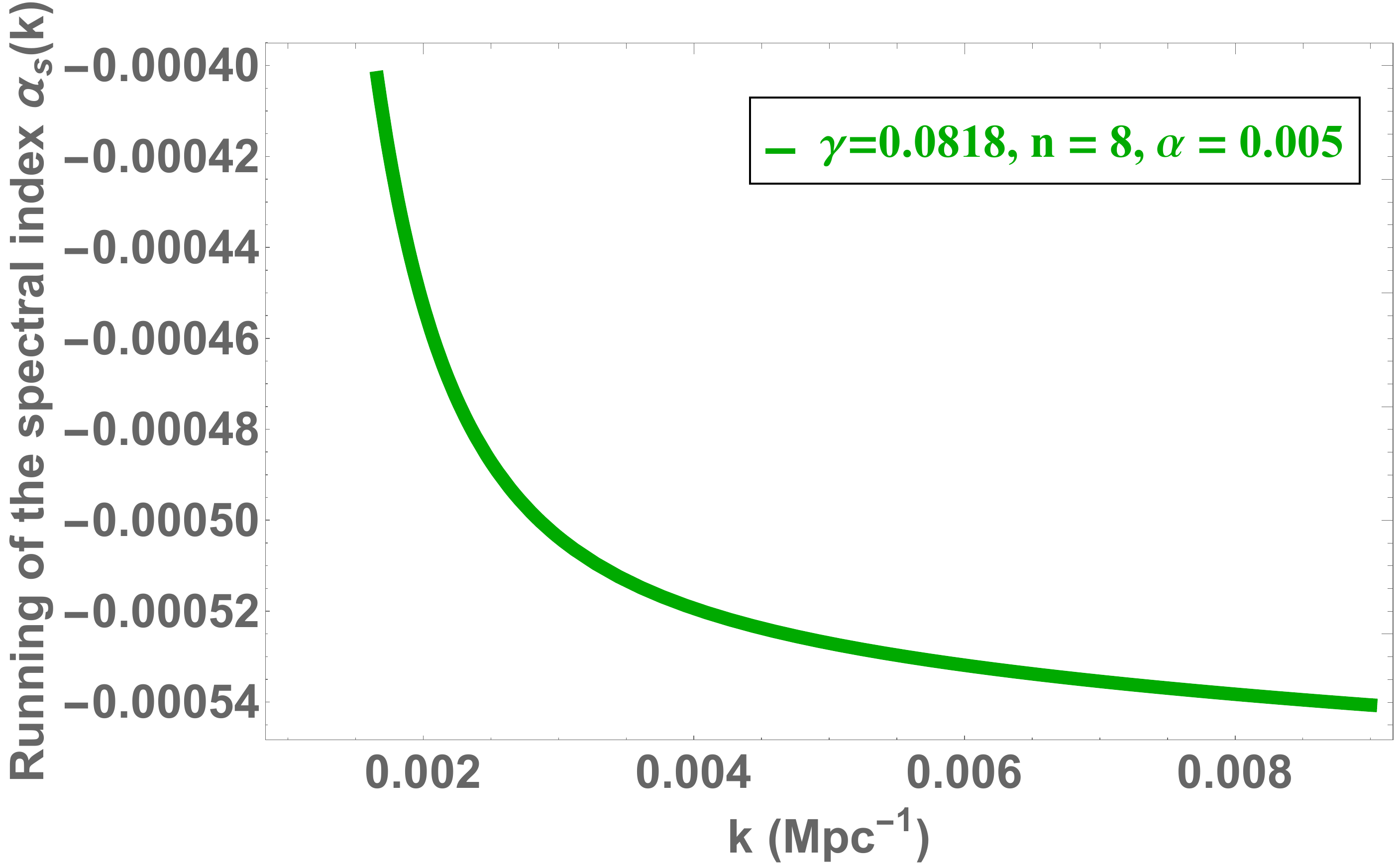}
   \subcaption{}
   \label{fig:RunningSpectralIndex_2}
\end{subfigure}%
\begin{subfigure}{0.33\linewidth}
  \centering
   \includegraphics[width=46mm,height=40mm]{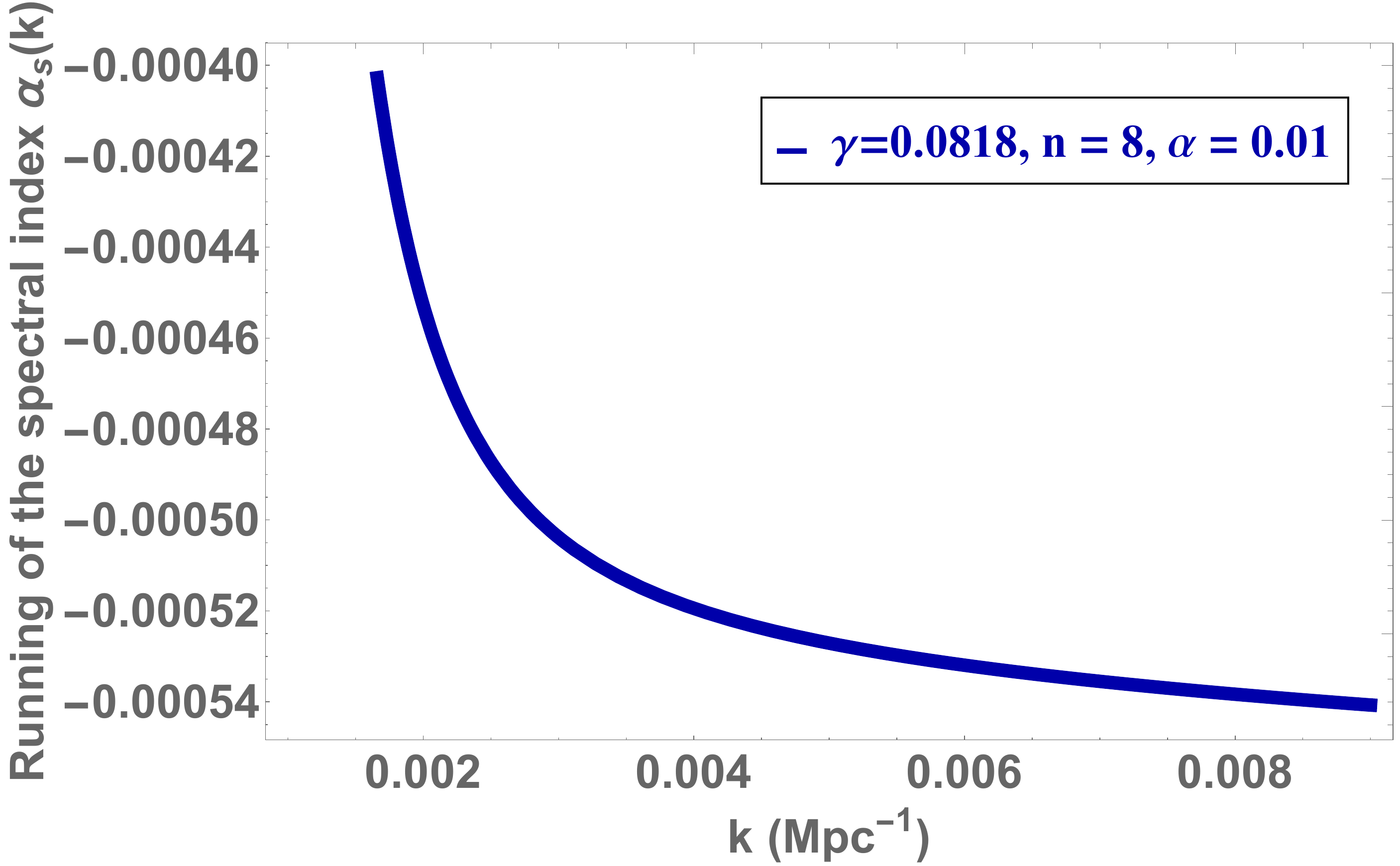}
   \subcaption{}
   \label{fig:RunningSpectralIndex_3}
\end{subfigure}%
\vspace{0.05\linewidth}
\begin{subfigure}{0.33\linewidth}
  \centering
   \includegraphics[width=46mm,height=40mm]{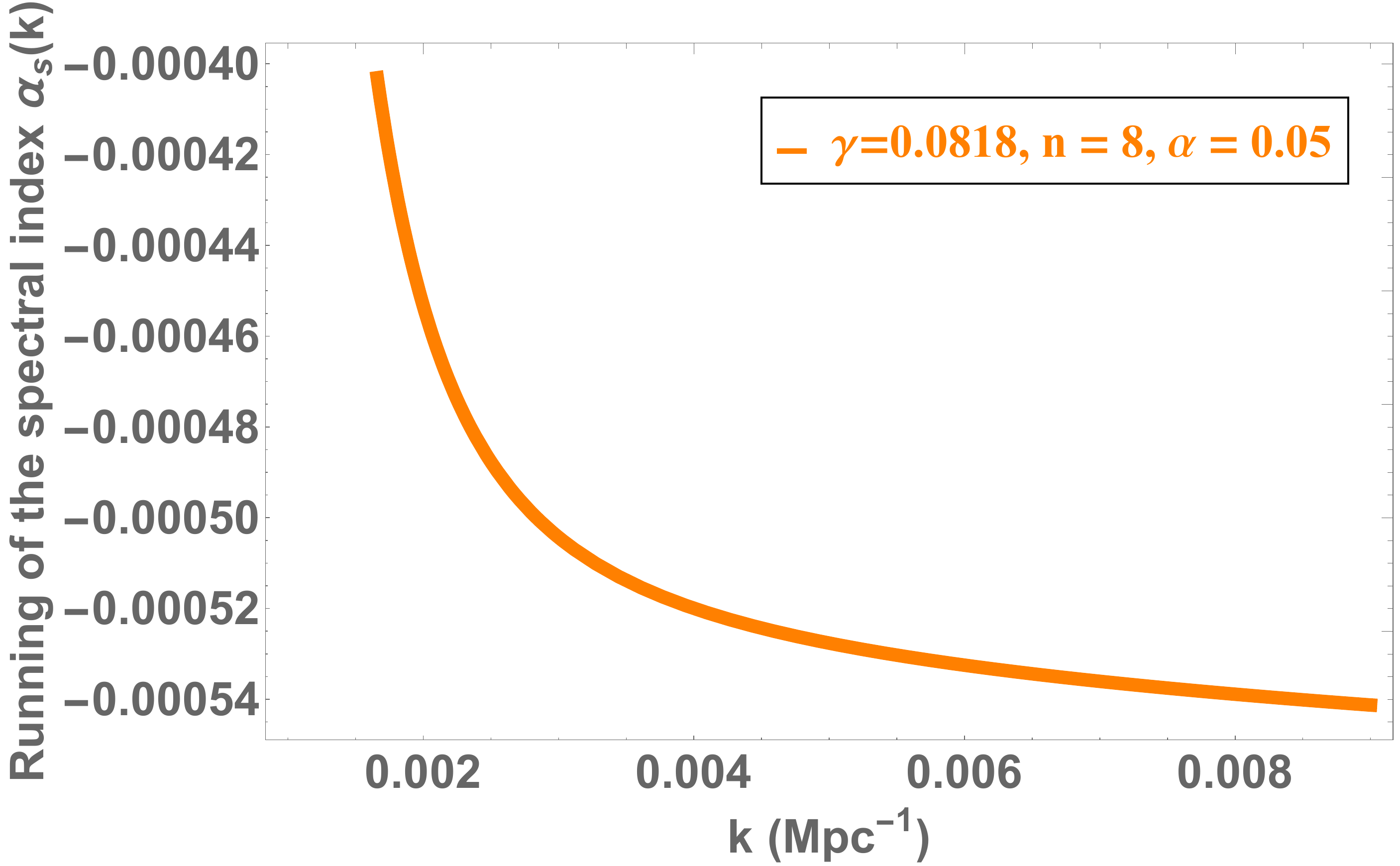}
   \subcaption{}
    \label{fig:RunningSpectralIndex_4}
\end{subfigure}%
\begin{subfigure}{0.33\linewidth}
  \centering
   \includegraphics[width=46mm,height=40mm]{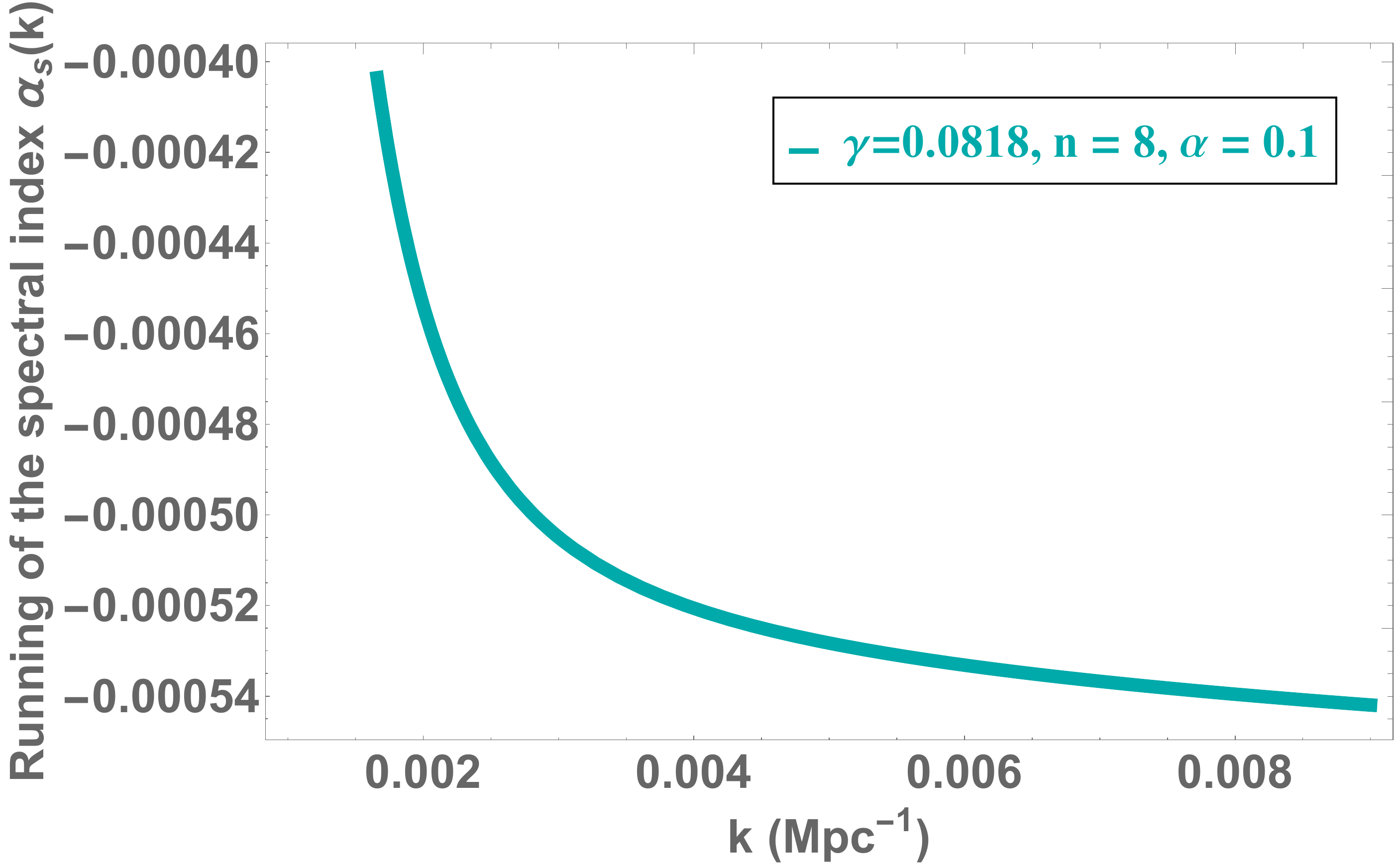}
   \subcaption{}
    \label{fig:RunningSpectralIndex_5}
\end{subfigure}%
\begin{subfigure}{0.33\linewidth}
  \centering
   \includegraphics[width=46mm,height=40mm]{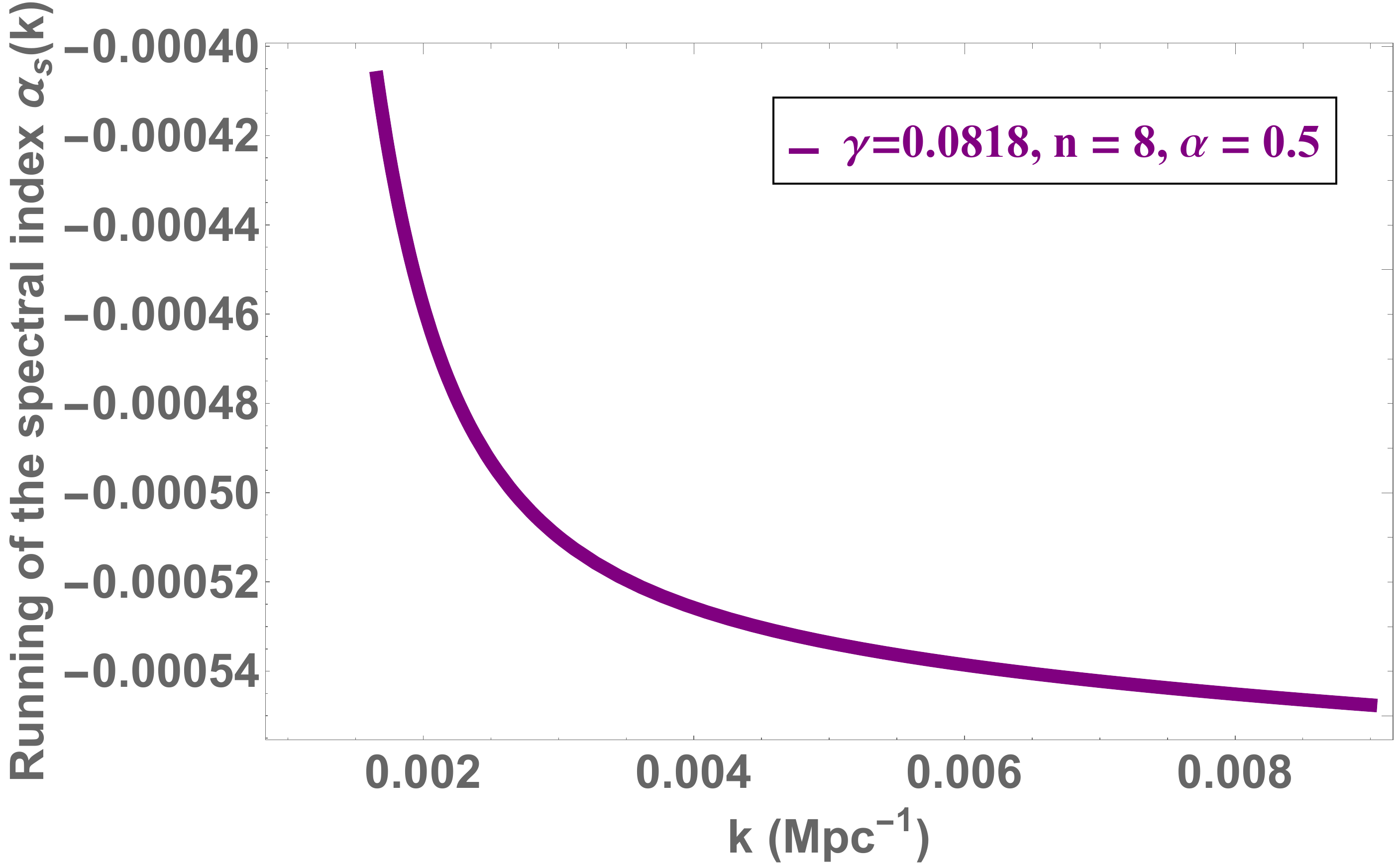}
   \subcaption{}
    \label{fig:RunningSpectralIndex_6}
\end{subfigure}%
\vspace{0.05\linewidth}
\begin{subfigure}{0.33\linewidth}
  \centering
   \includegraphics[width=46mm,height=40mm]{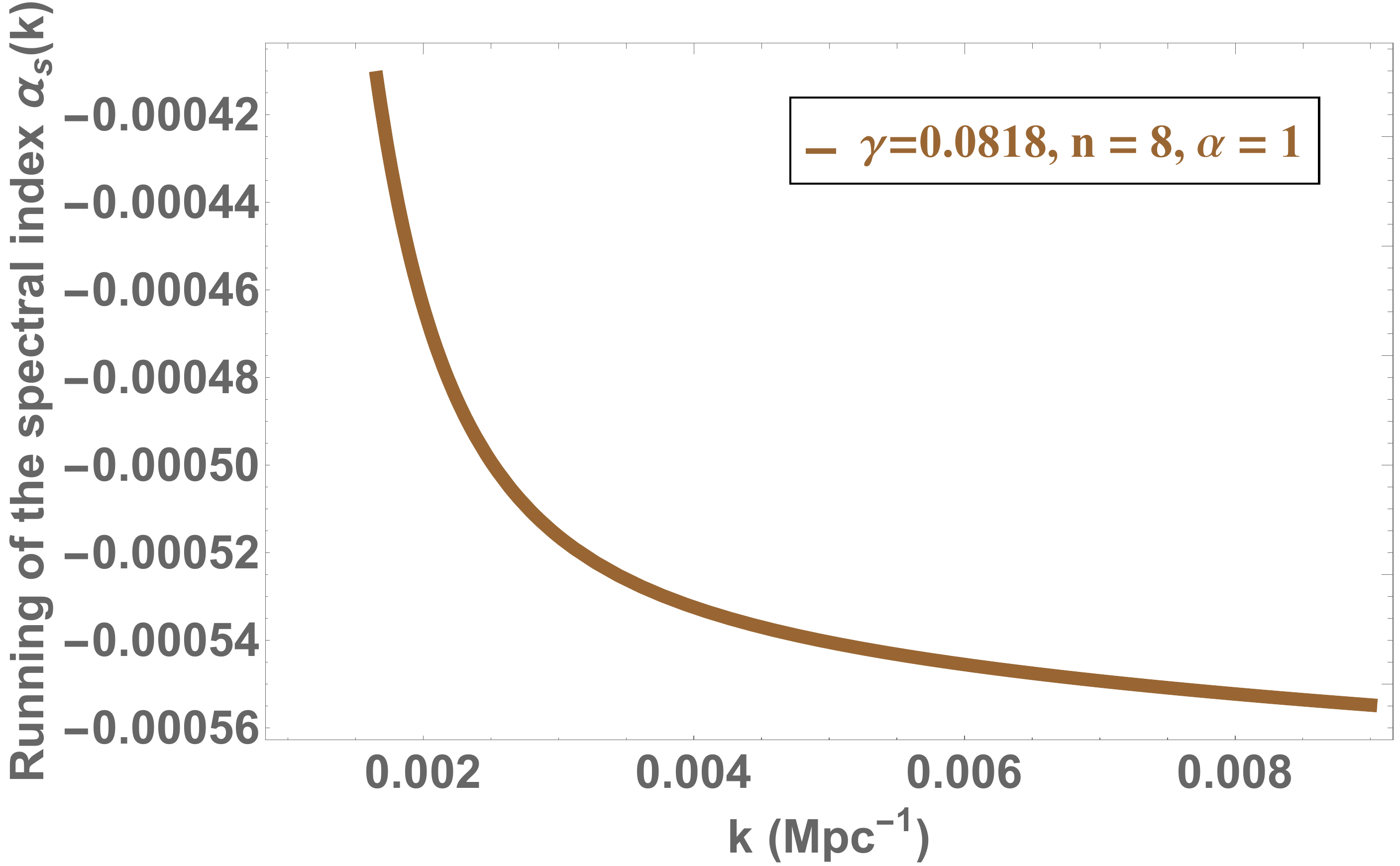}
   \subcaption{}
    \label{fig:RunningSpectralIndex_7}
\end{subfigure}%
\begin{subfigure}{0.33\linewidth}
  \centering
   \includegraphics[width=46mm,height=40mm]{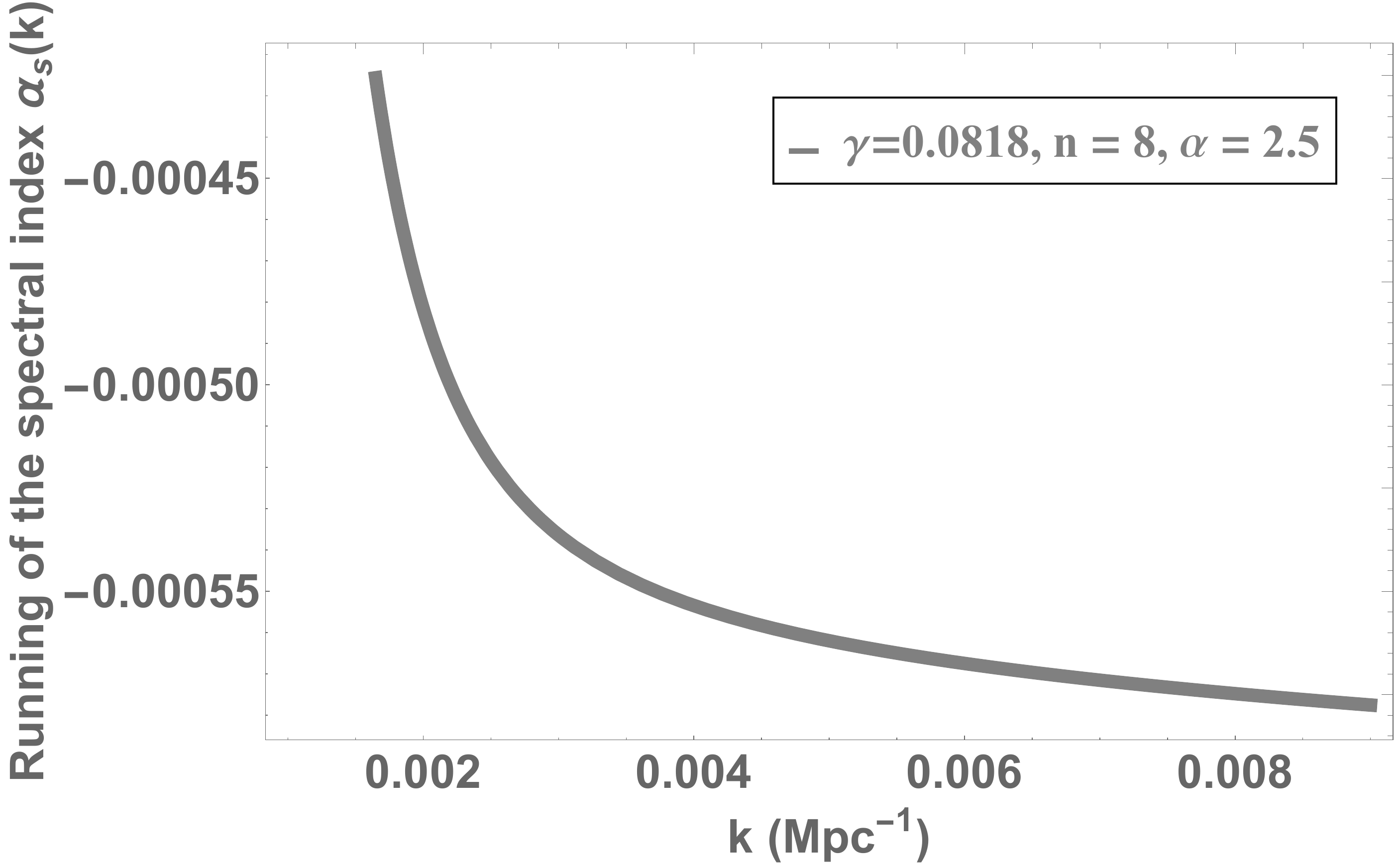}
   \subcaption{}
    \label{fig:RunningSpectralIndex_8}
\end{subfigure}%
\begin{subfigure}{0.33\linewidth}
  \centering
   \includegraphics[width=46mm,height=40mm]{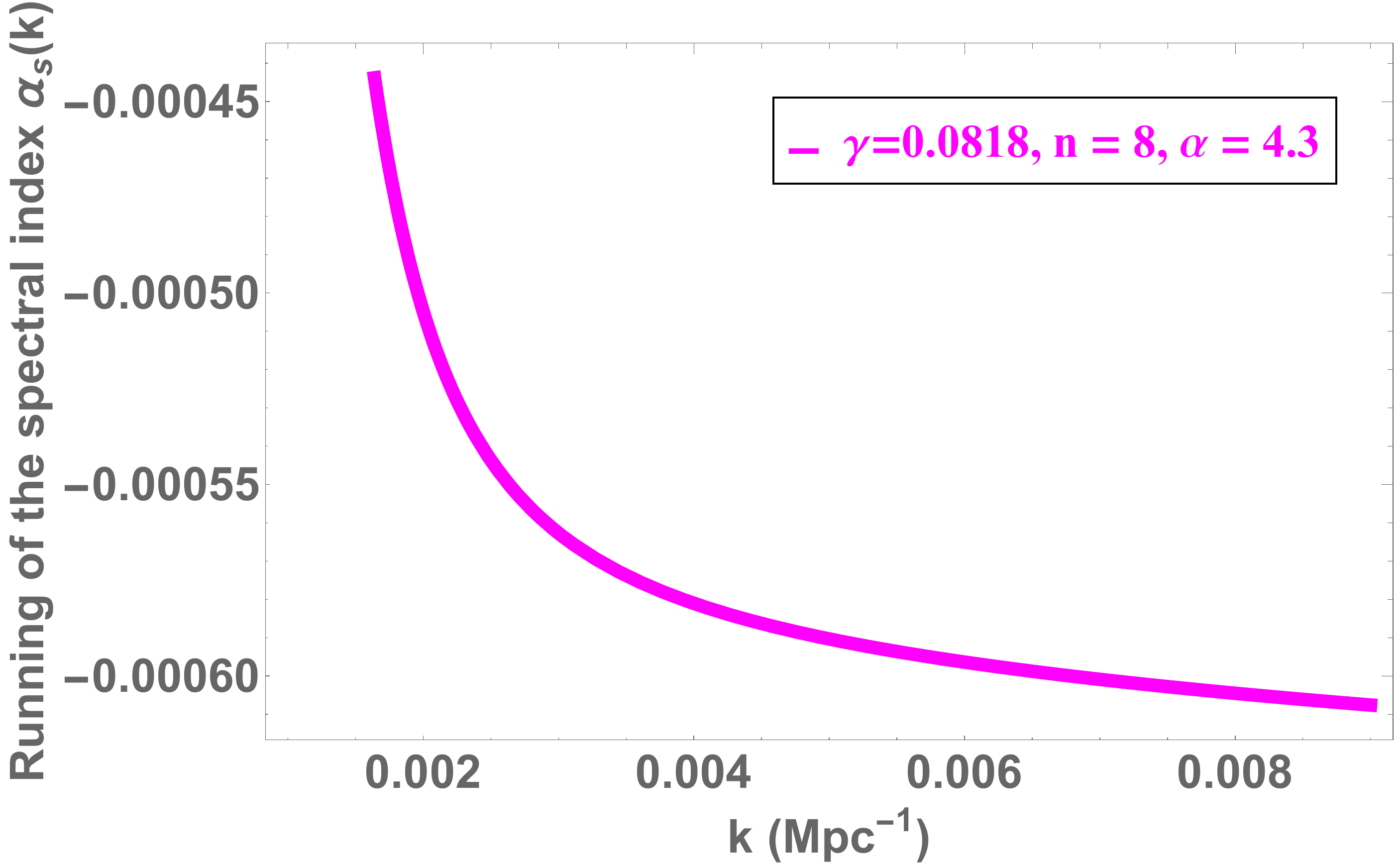}
   \subcaption{}
    \label{fig:RunningSpectralIndex_9}
\end{subfigure}
\caption{Running of spectral index for nine values of $\alpha$ for $\gamma=0.0818$ and $n=8$. The values of $\alpha_s(k)$ do not vary significantly with the increase in $\alpha$ at a particular $k$ value.}
\label{fig:RunningSpectralIndex}
\end{figure}

 The little scale dependency of $n_s(k)$ is usually visualized by the running of spectral index $\alpha_s(k)$ which we demonstrate by the figure \ref{fig:RunningSpectralIndex}. $\alpha_s(k)$ varries from $(-0.00040)$-$(-0.00060)$ for all values of $\alpha$ in the entire $k$ range which match Planck data satisfactorily. The negative sign signifies the decreasing mode behaviour and the small values indicate approximate scale invariance of $n_s(k)$ with $k$.\par
\begin{figure}[H]
\begin{subfigure}{0.33\linewidth}
  \centering
   \includegraphics[width=46mm,height=40mm]{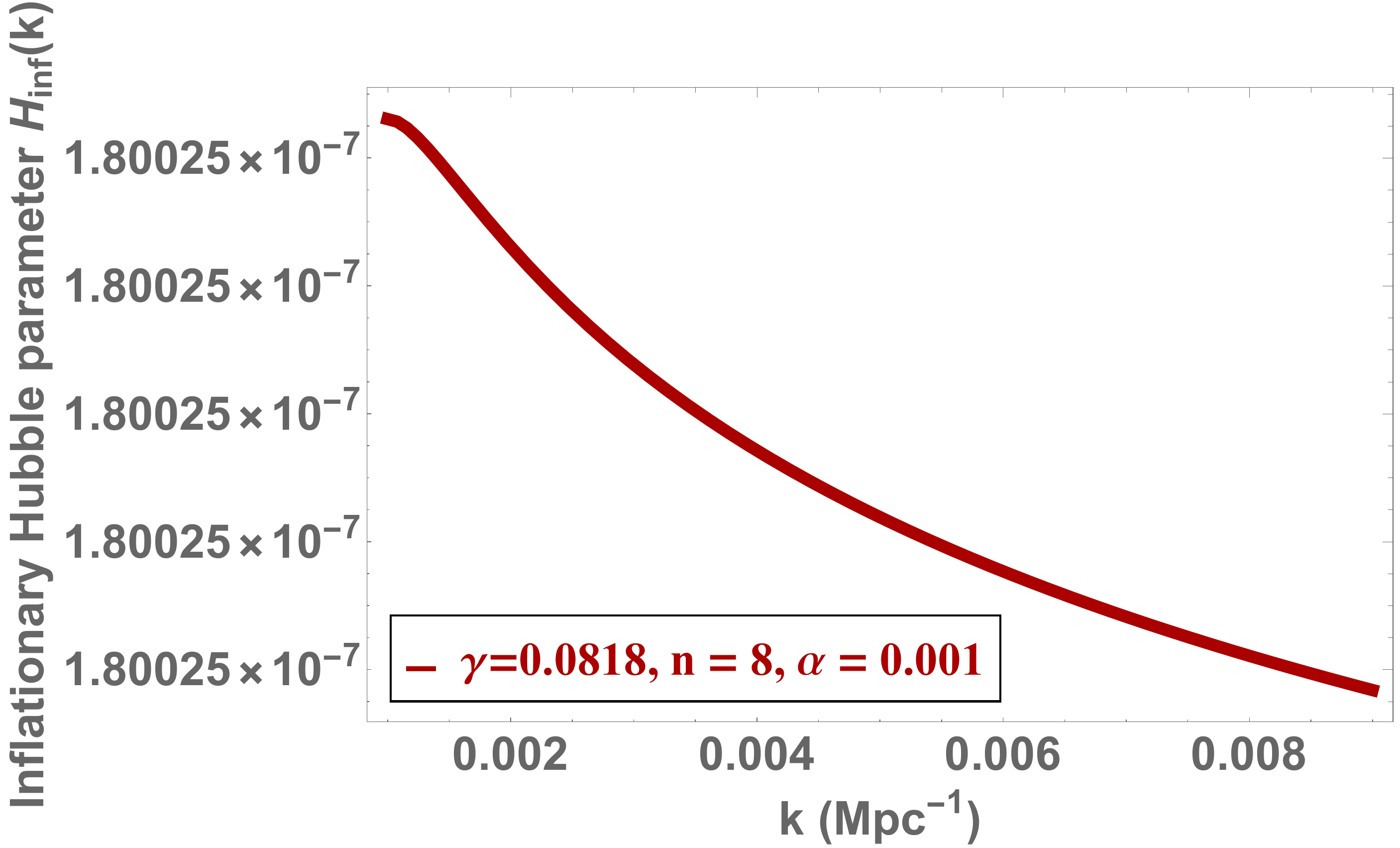}
   \subcaption{}
   \label{fig:InfHubbleParameter_1}
\end{subfigure}%
\begin{subfigure}{0.33\linewidth}
  \centering
   \includegraphics[width=46mm,height=40mm]{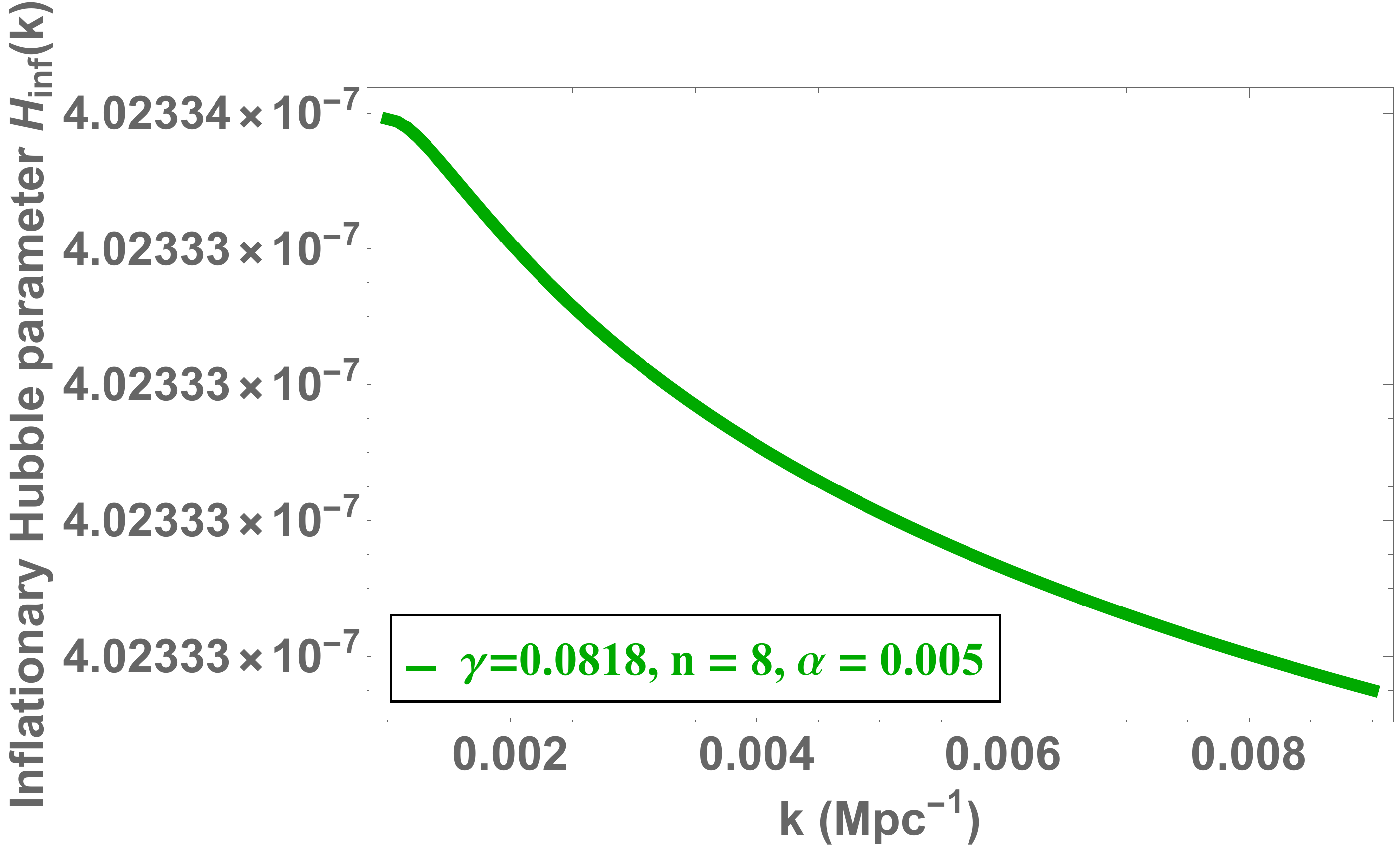}
   \subcaption{}
   \label{fig:InfHubbleParameter_2}
\end{subfigure}%
\begin{subfigure}{0.33\linewidth}
  \centering
   \includegraphics[width=46mm,height=40mm]{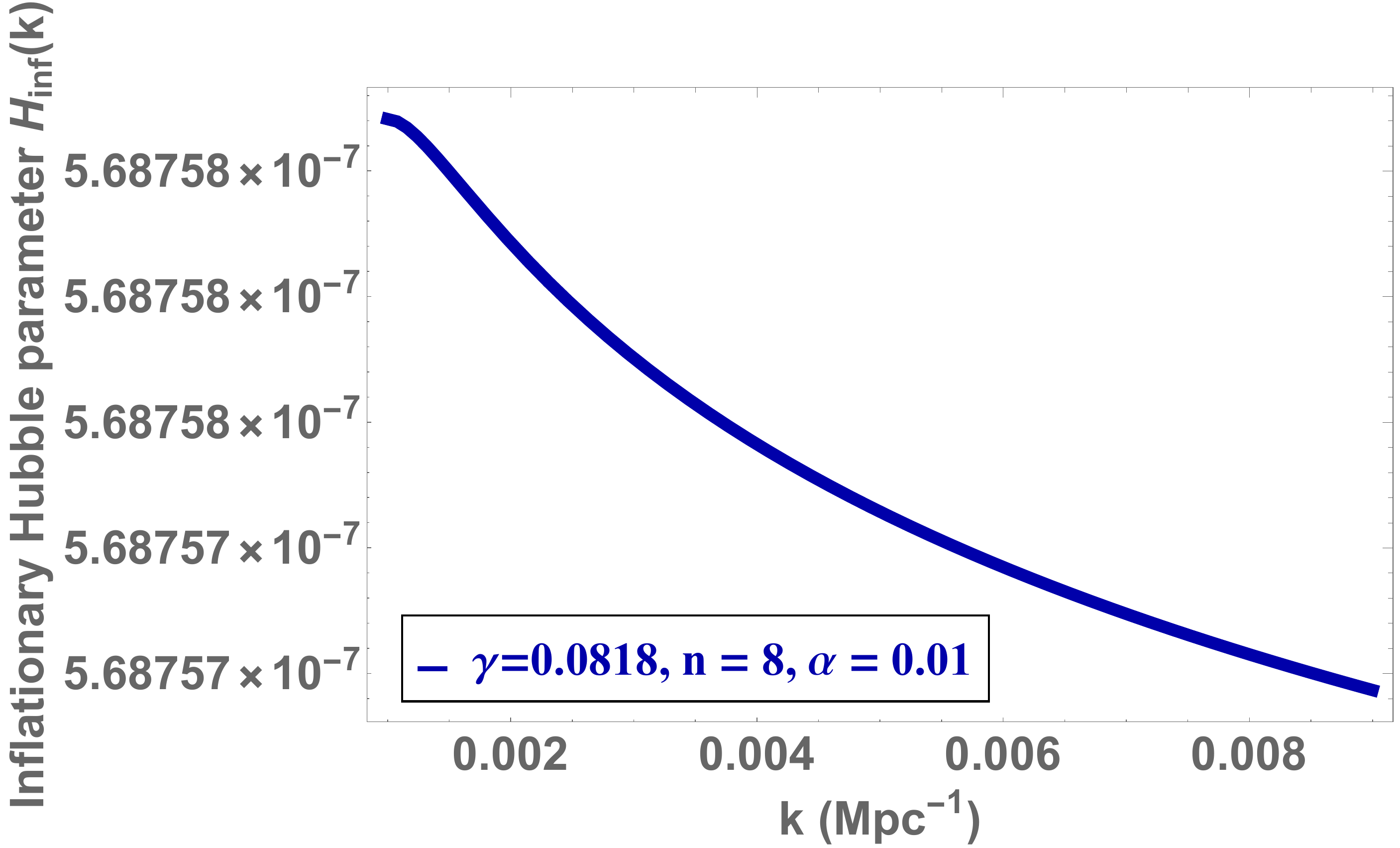}
   \subcaption{}
   \label{fig:InfHubbleParameter_3}
\end{subfigure}%
\vspace{0.05\linewidth}
\begin{subfigure}{0.33\linewidth}
  \centering
   \includegraphics[width=46mm,height=40mm]{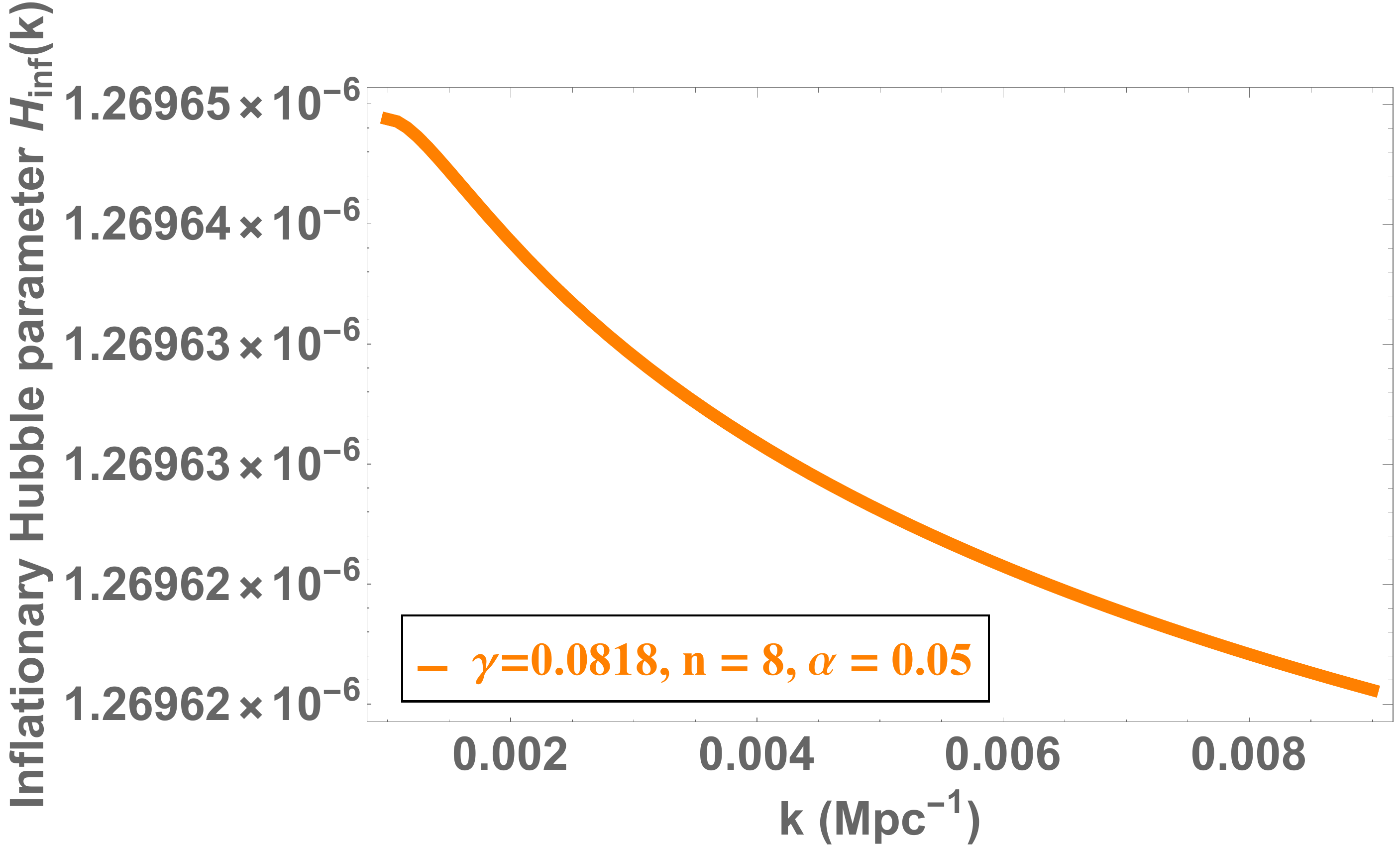}
   \subcaption{}
    \label{fig:InfHubbleParameter_4}
\end{subfigure}%
\begin{subfigure}{0.33\linewidth}
  \centering
   \includegraphics[width=46mm,height=40mm]{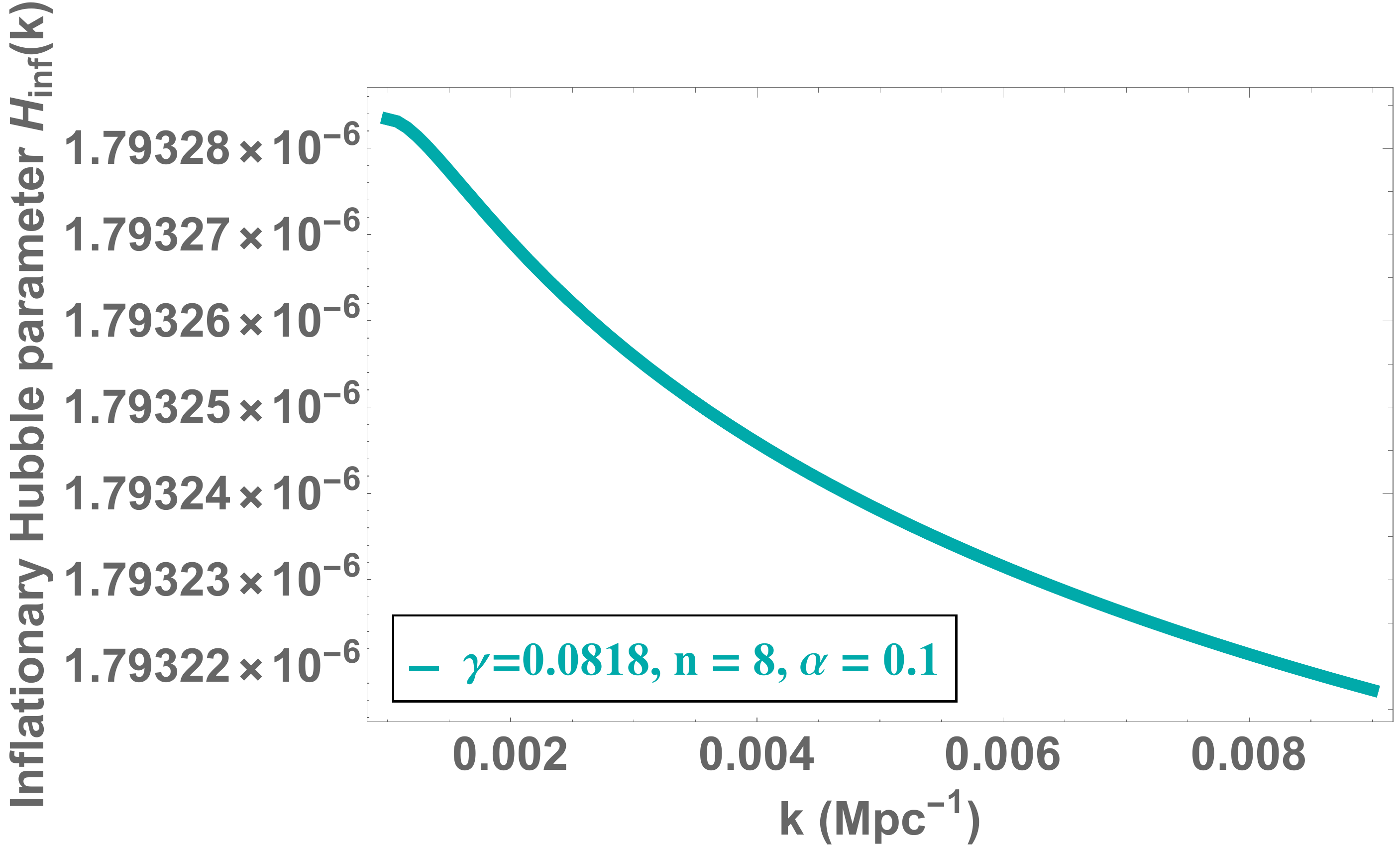}
   \subcaption{}
    \label{fig:InfHubbleParameter_5}
\end{subfigure}%
\begin{subfigure}{0.33\linewidth}
  \centering
   \includegraphics[width=46mm,height=40mm]{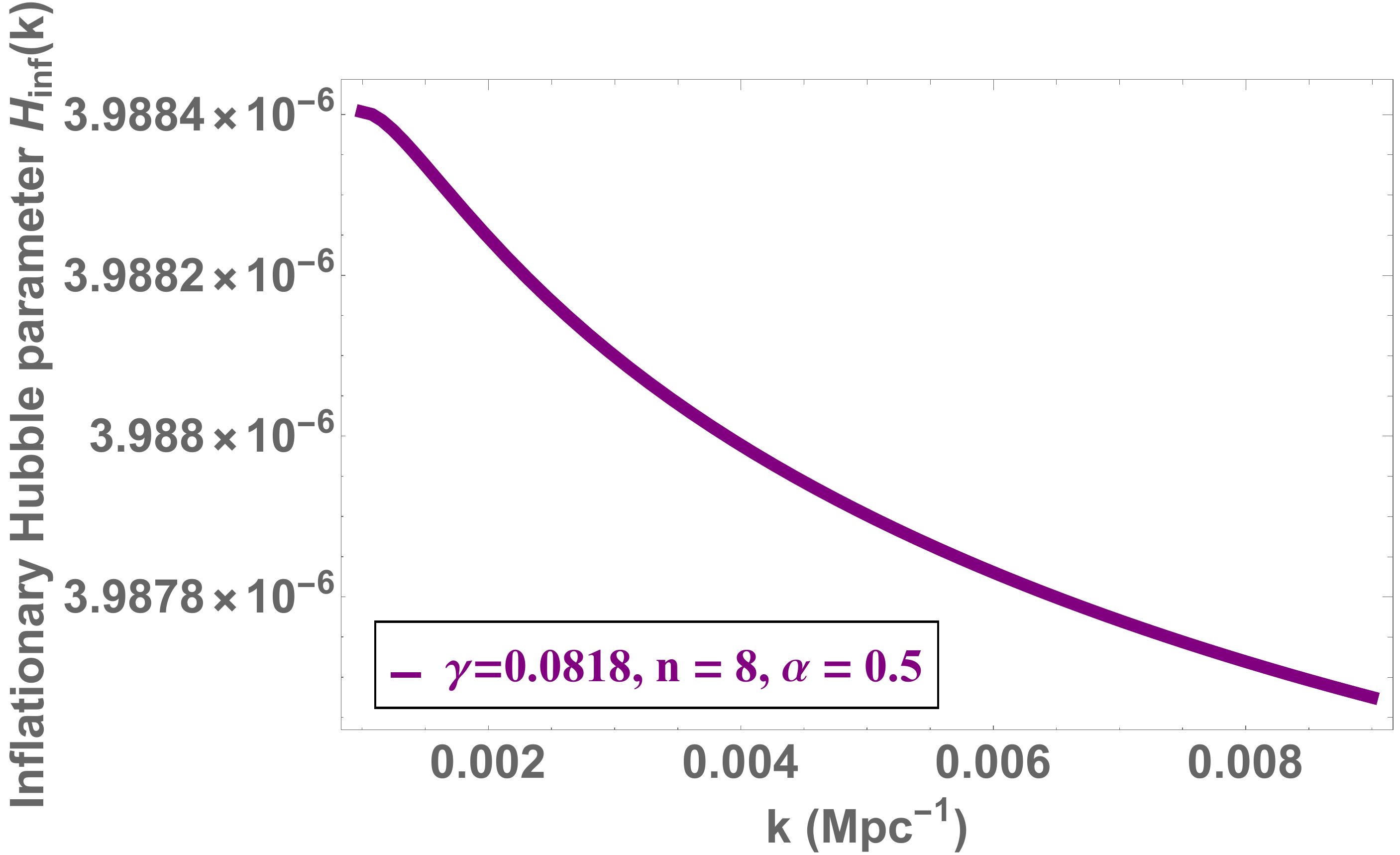}
   \subcaption{}
    \label{fig:InfHubbleParameter_6}
\end{subfigure}%
\vspace{0.05\linewidth}
\begin{subfigure}{0.33\linewidth}
  \centering
   \includegraphics[width=46mm,height=40mm]{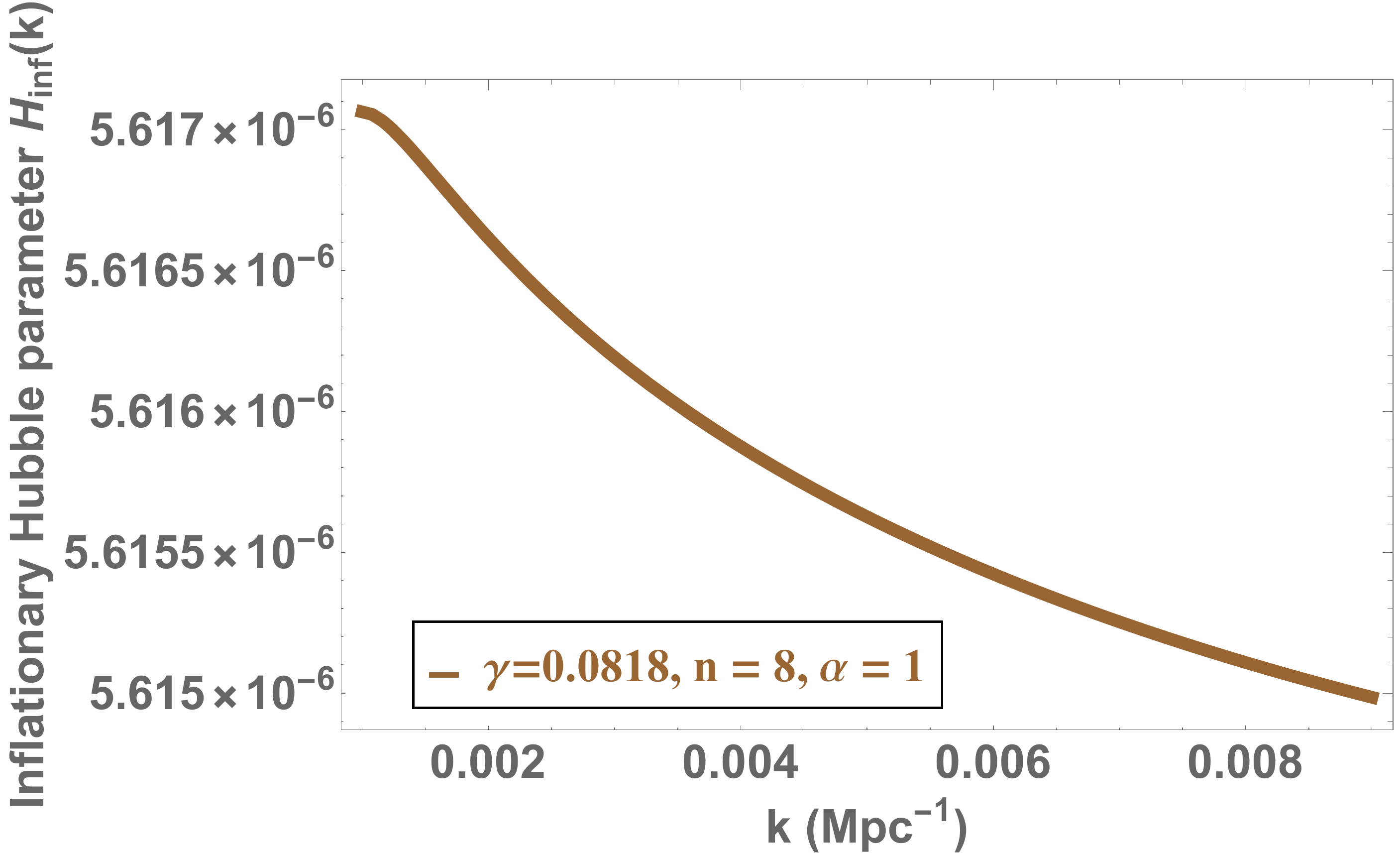}
   \subcaption{}
    \label{fig:InfHubbleParameter_7}
\end{subfigure}%
\begin{subfigure}{0.33\linewidth}
  \centering
   \includegraphics[width=46mm,height=40mm]{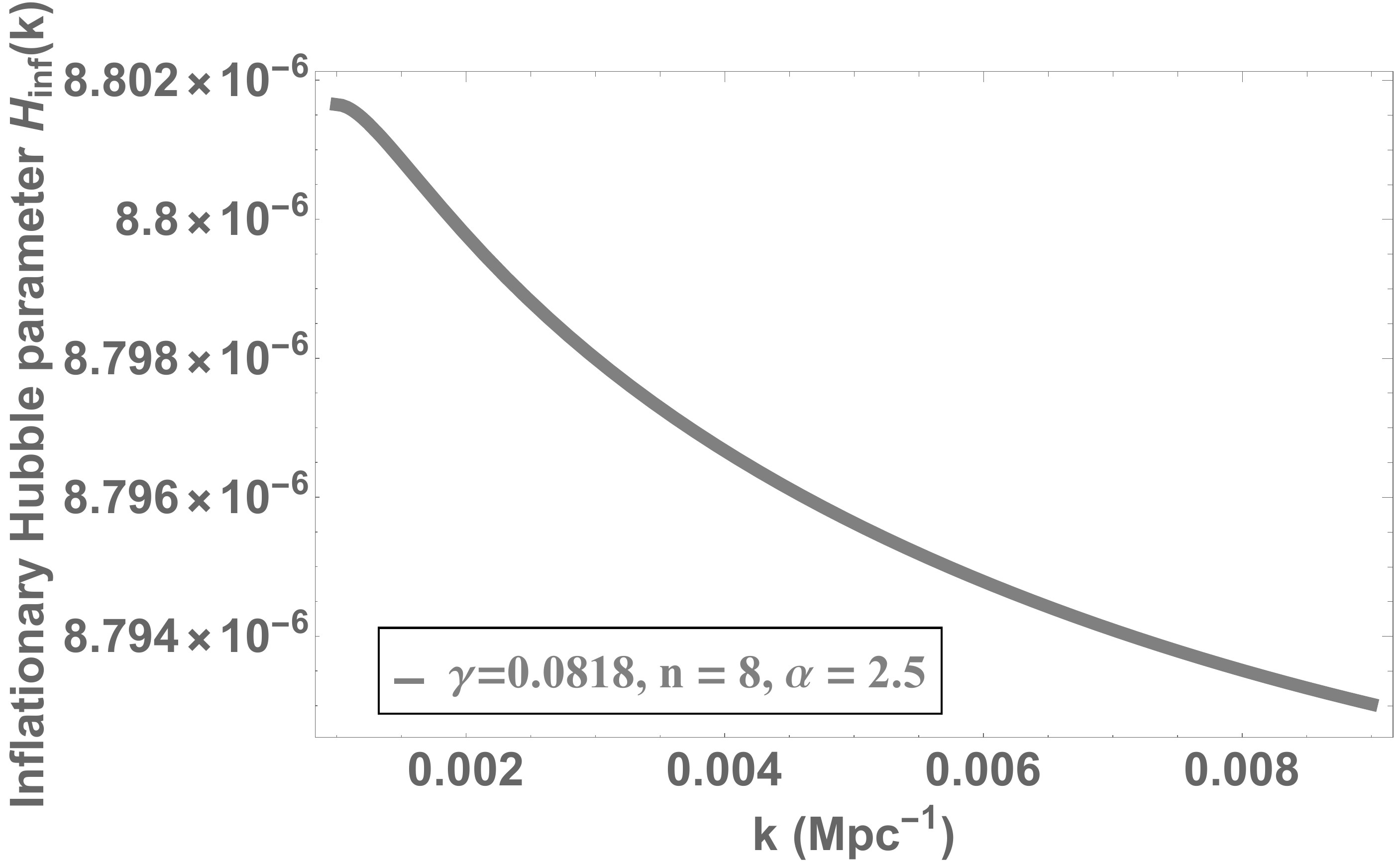}
   \subcaption{}
    \label{fig:InfHubbleParameter_8}
\end{subfigure}%
\begin{subfigure}{0.33\linewidth}
  \centering
   \includegraphics[width=46mm,height=40mm]{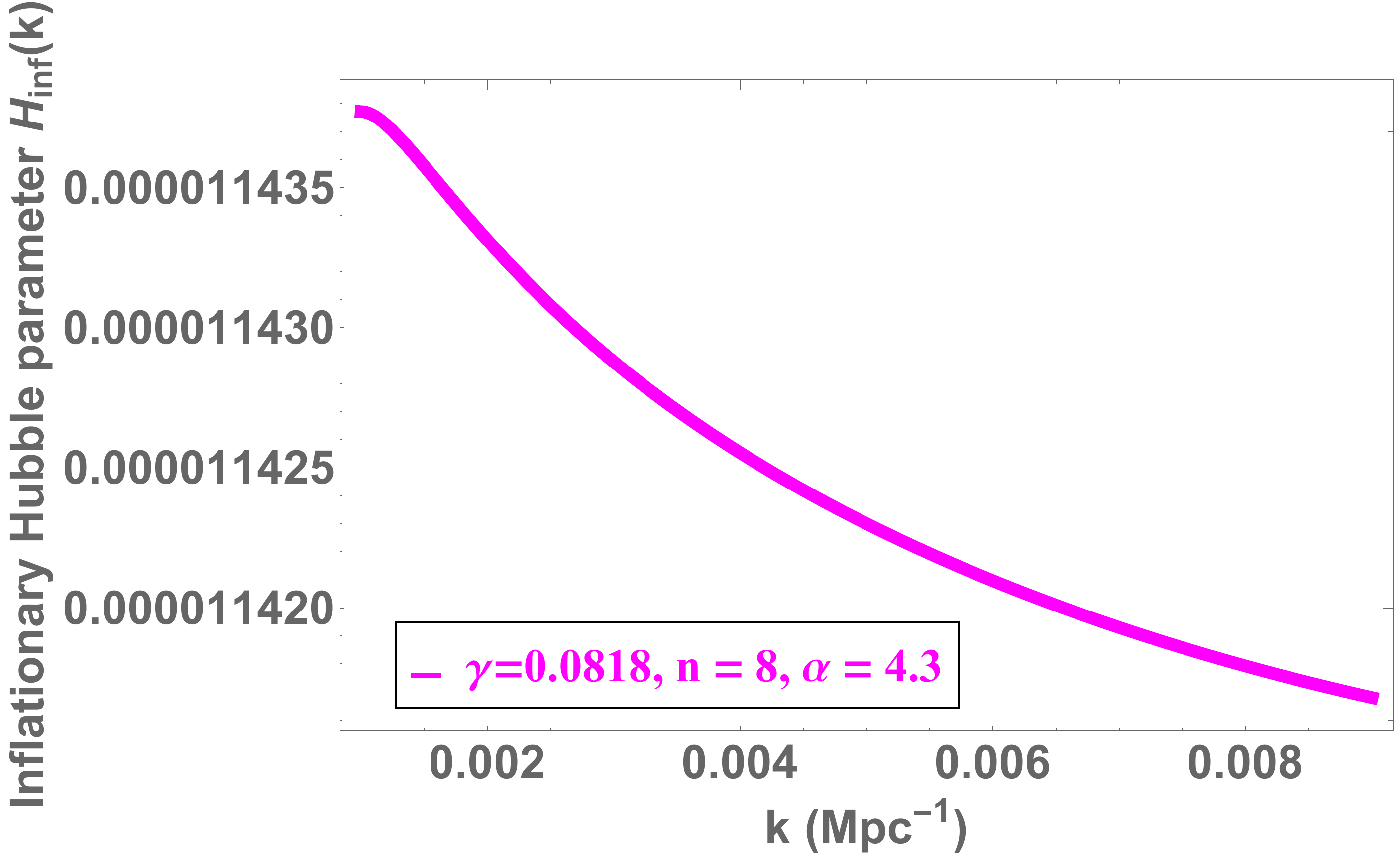}
   \subcaption{}
    \label{fig:InfHubbleParameter_9}
\end{subfigure}
\caption{Inflationary Hubble parameters for nine values of $\alpha$ for $\gamma=0.0818$ and $n=8$. The values of $H(k)$ tend to increase with the increase in $\alpha$ at a particular $k$-value.}
\label{fig:InfHubbleParameter}
\end{figure}

 Let us end our numerical and graphical analysis part by showing the inflationary Hubble parameter $H_{\mathrm{inf}}(k)$ in figure \ref{fig:InfHubbleParameter} according to Eq. (\ref{eq:Inf_Hubble_parameter}). The calculated values are in the range, $1.8\times 10^{-7} - 1.14\times 10^{-5}$, which are supported by Planck data (see section \ref{sec:formalism}). The orders of the $H_{\mathrm{inf}}(k)$ values suggest that the inflation takes place in GUT scale ($\sim 10^{15}$ GeV).\par In the course of dynamical mode analysis, we find that the obtained results for all the mode-dependent cosmological parameters are within the allowed ranges of Planck-2018 and do not deviate significantly. In fact, the graphical results also show that the cosmological parameters in quintessential EDE model are not very different in comparison to those in the model without EDE \cite{Sarkar:2023cpd}, specifically for $\alpha\geq 0.1$. Thus, we have verified a characteristic property of EDE, that, it does not considerably affect the parameters of the $\Lambda$CDM model. However it can have influence on $V_{\Lambda}$ (see table \ref{tab:Table2}) and also on $\alpha$, as discussed in the next subsection.

\subsection{Role of ESP in constraining the parameter \texorpdfstring{$\alpha$}{a}}
\label{subsec:Role_of_ESP}
Now, we decode the unusual behaviour of $\xi^{(0)}(k)$ with $k$ within $0.05\leq\alpha\leq 4.3$. Let us begin with the expressions of number of remaining e-folds of Eqs. (\ref{eq:Preefolds}) and (\ref{eq:e_folds_4}). It is enough to satisfy the inflationary data of Planck around $60$-efolds and it is sufficient to consider a single field slow-roll inflation.\par Here we denote $\xi$ for $N=63.49$ at $k=0.001$ Mpc$^{-1}$ as $\xi_{*}$, which depends on three model parameters \textit{viz.,} $n$, $\gamma$ and $\alpha$. For particular values of $n$ and $\gamma$, $\xi_{*}$ is a complicated function of $\alpha$,
\begin{equation}
    \xi_{*}(\alpha)=\sqrt{\frac{3\alpha}{2}}\ln{\left(\frac{4n\gamma N e^{-n}}{3\alpha}\right)}.
    \label{eq:xi_star}
\end{equation}
\begin{figure}[H]
    \centering
    \includegraphics[width=100mm, height=70mm]{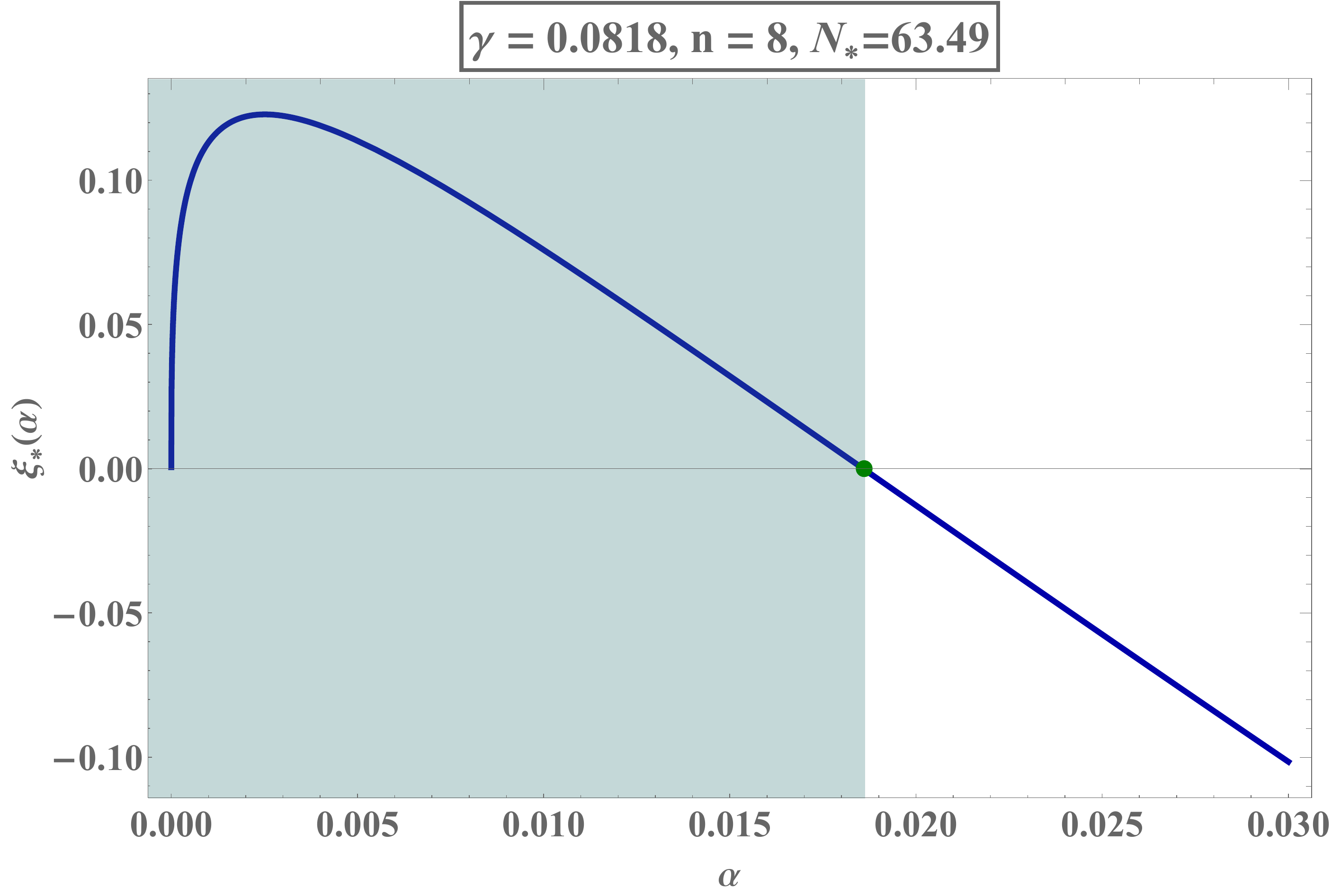}
    \caption{Variation of $\xi_{*}$ with $\alpha$ for a fixed set of model parameters. The shaded region describes the allowed range of $\alpha$ for which $\xi_{*}>0$.}
    \label{fig:ESPconstrain}
\end{figure}
In figure \ref{fig:ESPconstrain}, $\xi_{*}(\alpha)$ is plotted against $\alpha$ for the chosen model parameters, $\gamma=0.0818$ and $n=8$. As $\alpha$ increases, $\xi_{*}$ increases, reaches a maximum value and then monotonically decreases. After a certain value of $\alpha$ (denoted by a red dot) \textit{viz.,} $\alpha=0.0186$, it becomes negative and does not return to positive value. Such kind of behaviour we have noticed in figures \ref{fig:unperturbedINF_1} - \ref{fig:unperturbedINF_9}. Now, as discussed in Section \ref{sec:our model}, the inflation takes place for $\xi>0$ region above the Enhanced Symmetry Point (ESP) at $\xi=0$ (see figure \ref{fig:Fig1}). The inflaton field stops rolling after hitting the ESP, remains frozen there for a while behaving like EDE and then freely falls rapidly through the steepest portion of the potential during kination until it reaches the throat of the potential tail to become DE in the present universe. Therefore, $\xi>0$ is an indispensable criterion for quintessential inflation. So, from figure \ref{fig:ESPconstrain} and Eq. (\ref{eq:xi_star}) we can infer that the ESP at $\xi=0$ sets an upper cut-off of $\alpha$,
\begin{equation}
    \alpha_{\mathrm{max}}=\frac{4n\gamma N e^{-n}}{3}
    \label{eq:alpha_cutoff}
\end{equation} beyond which $\xi_{*}<0$. In our case, $\alpha_{\mathrm{max}}=0.0186$. That is why, we obtain negative values of $\xi_{*}$ from $\alpha=0.05$ onward in figure \ref{fig:unperturbedINF}. Therefore the presence of ESP \textit{vis-\`{a}-vis} the EDE restricts $\alpha$ as $\alpha<\alpha_{\mathrm{max}}$. In non-EDE version of quintessential inflation \cite{Sarkar:2023cpd}, $\xi_{*}$ is always positive, no matter what value of $\alpha$ is chosen. But in the EDE-version the presence of the factor $\gamma e^{-n}$ bears the signature of EDE, which confines $\alpha$ below $\alpha_{\mathrm{max}}$. Evidently, this upper limit of $\alpha$ depends on $n$ and $\gamma$, which are again constrained by the present-day vacuum density as shown in Eq. (\ref{eq:final_V_lambda_5}). Thus, apart from setting the correct energy scale for quintessence, the EDE plays a crucial role (albeit, secondary\footnote{Its primary role is to resolve the Hubble tension, staying at the post inflationary regime.}) in providing a consistent inflaton potential by constraining the parameter $\alpha$.
\begin{figure}[H]
    \begin{subfigure}{0.5\linewidth}
  \centering
   \includegraphics[width=70mm,height=60mm]{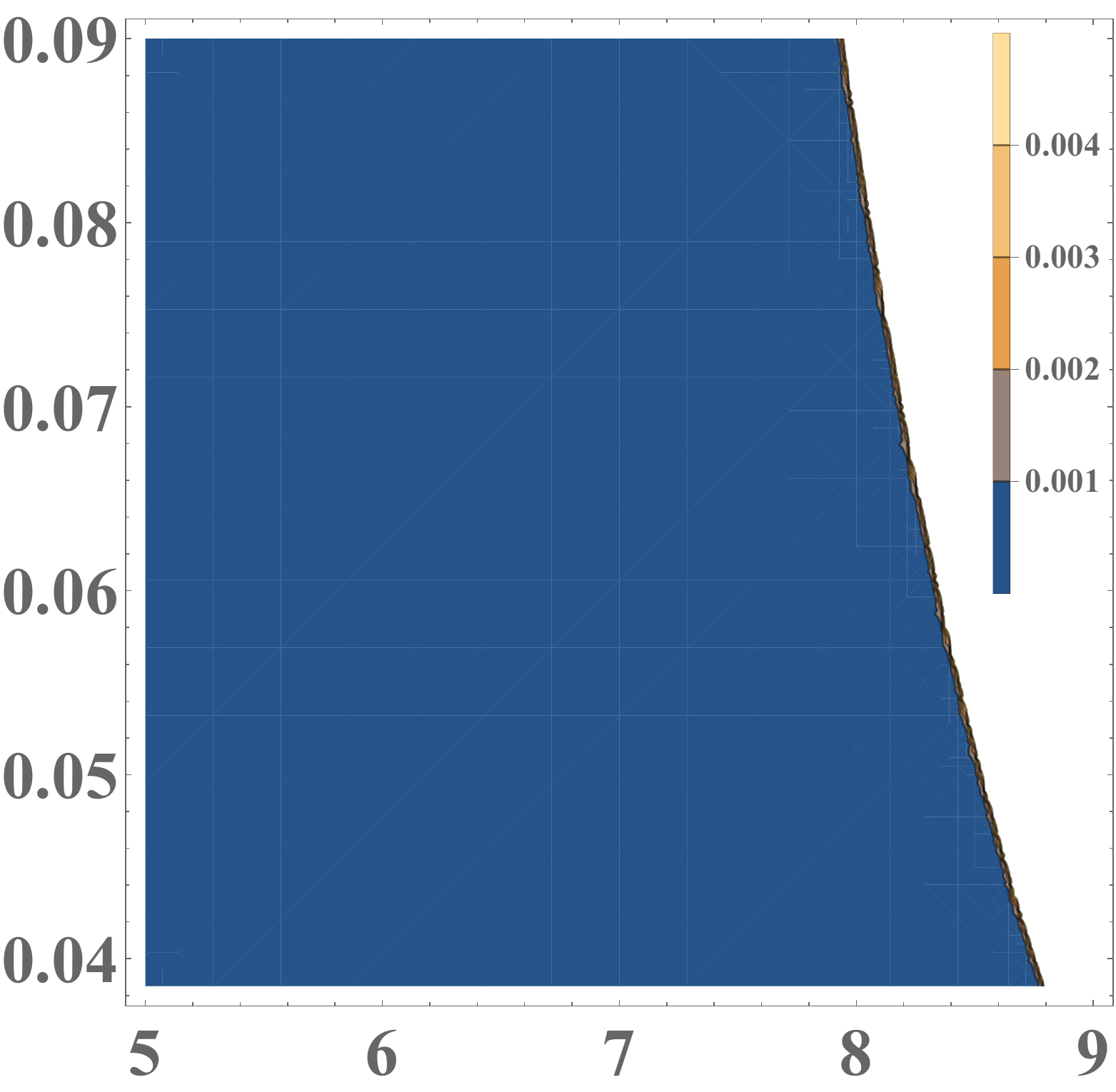}
   \subcaption{}
   \label{fig:theoreticalcontourplot_1}
\end{subfigure}
\begin{subfigure}{0.5\linewidth}
  \centering
   \includegraphics[width=70mm,height=60mm]{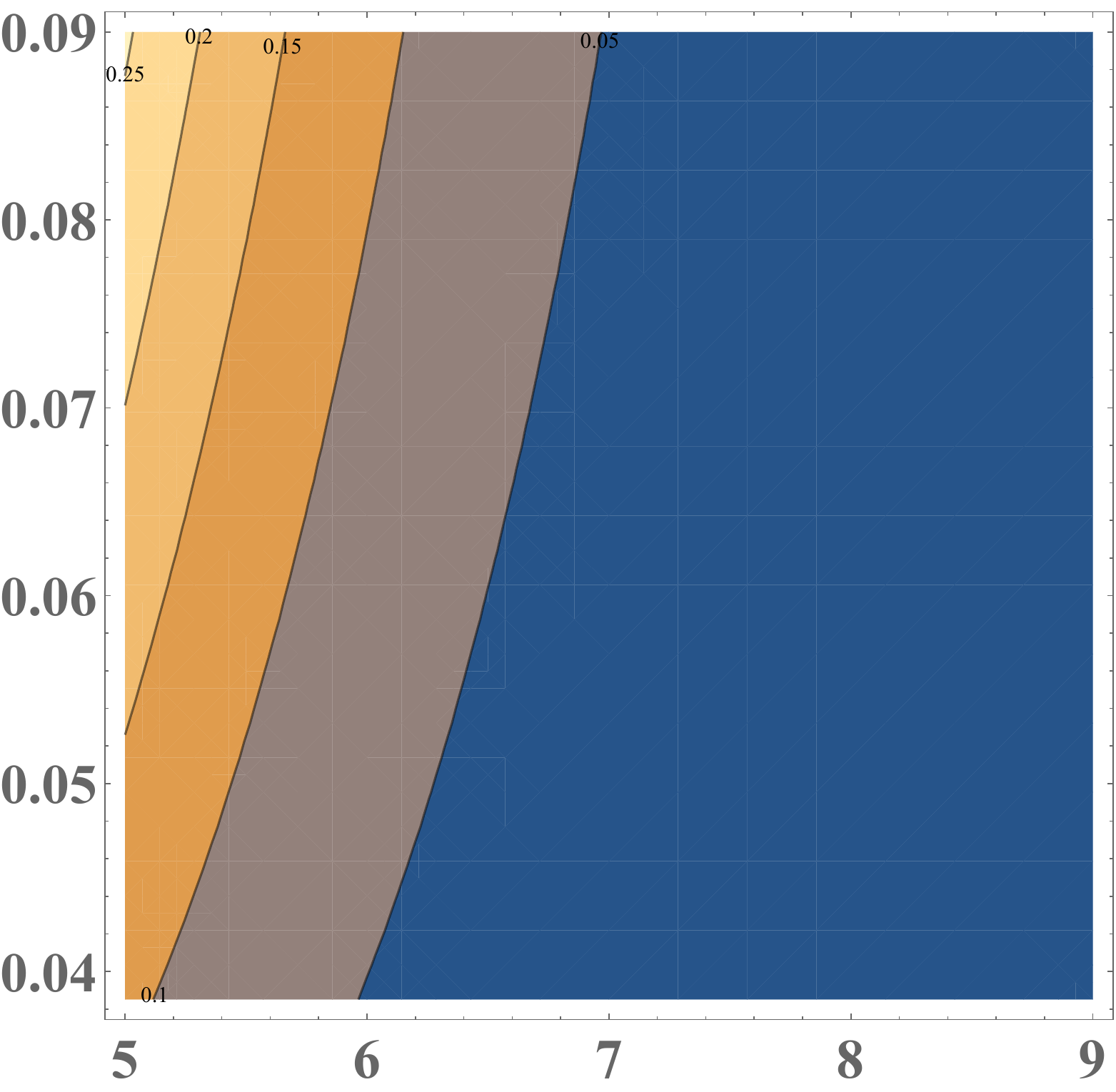}
   \subcaption{}
    \label{fig:theoreticalcontourplot_2}
\end{subfigure}
     \caption{Contour plots of $\gamma$ vs. $n$ with respect to $M$ (left) according to Eq. (\ref{eq:final_V_lambda_5}) and $\alpha_{\mathrm{max}}$ (right) according to Eq. (\ref{eq:alpha_cutoff}) around the chosen values of $n$ and $\gamma$. The vertical bar in figure \ref{fig:theoreticalcontourplot_1} shows the scale for the obtained values of $M$ (see table \ref{tab:Table1}). In this figure the $\gamma - n$ boundary signifies that the favourable values lie within a region restricted by $n\in [8,9)$ and $\gamma\in [0.035,0.09]$ and below these ranges $\gamma$ and $n$ are independent of each other. The coloured regions in figure \ref{fig:theoreticalcontourplot_2} represent the corresponding values of $\alpha_{\mathrm{max}}$, indicated at the end of the respective boundaries.}
    \label{fig:contourplot_1}
\end{figure}
\begin{figure}[H]
    \begin{subfigure}{0.5\linewidth}
  \centering
   \includegraphics[width=70mm,height=60mm]{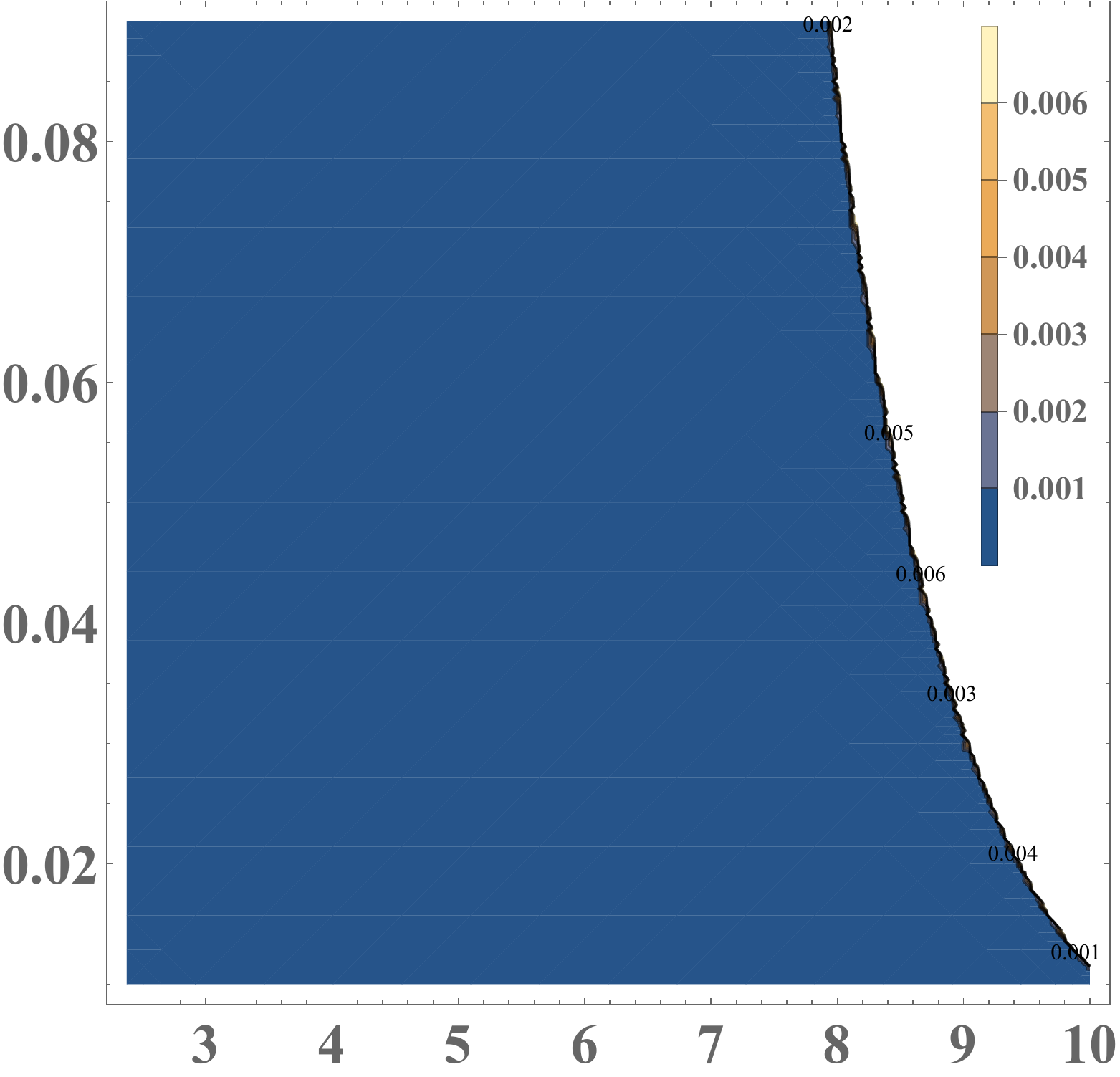}
   \subcaption{}
   \label{fig:experimentalcontourplot_1}
\end{subfigure}
\begin{subfigure}{0.5\linewidth}
  \centering
   \includegraphics[width=70mm,height=60mm]{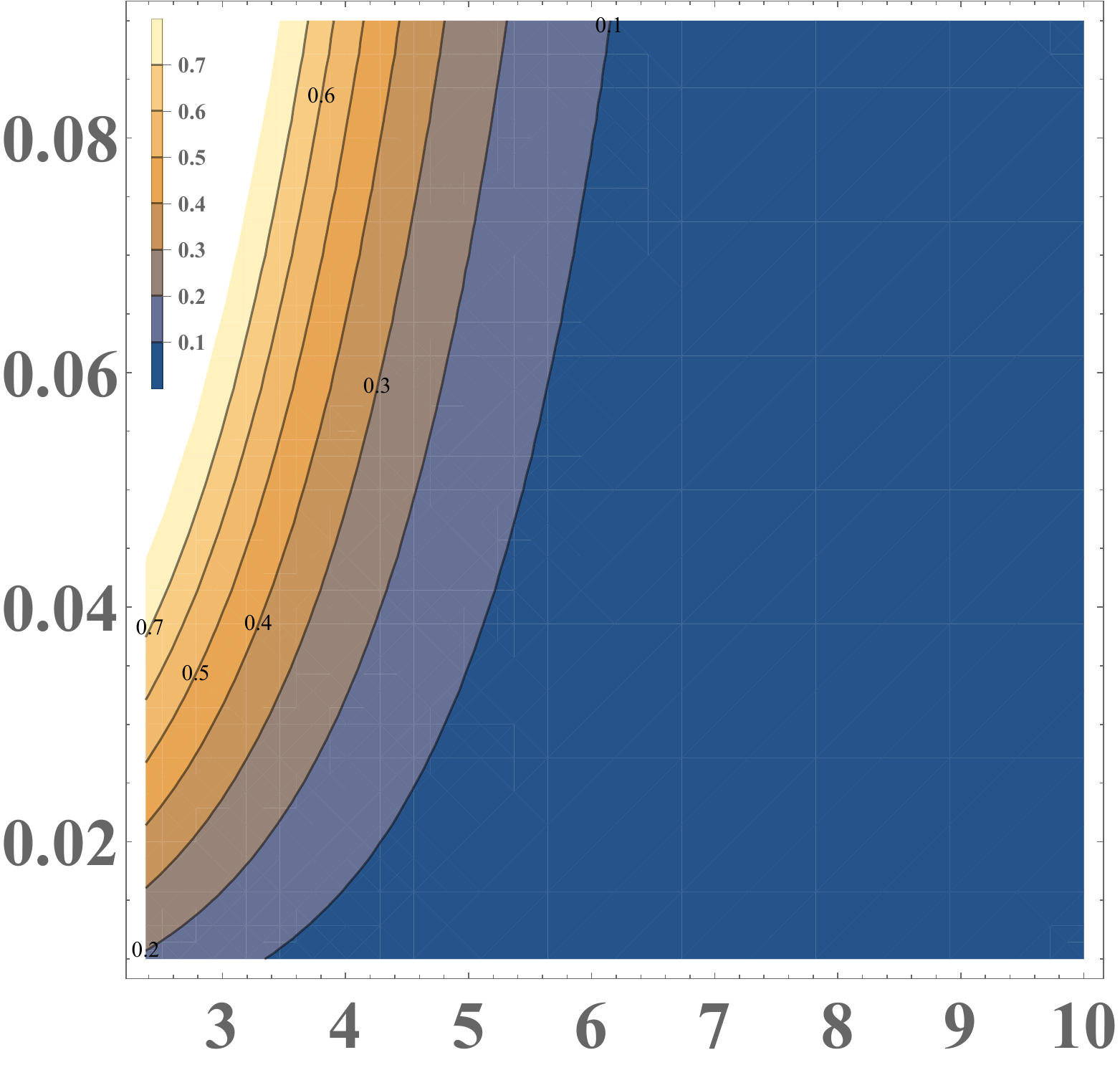}
   \subcaption{}
    \label{fig:experimentalcontourplot_2}
\end{subfigure}
     \caption{Contour plots of  $\gamma$ vs. $n$ with respect to $M$ (left) according to Eq. (\ref{eq:final_V_lambda_5}) and $\alpha_{\mathrm{max}}$ (right) according to Eq. (\ref{eq:alpha_cutoff}) about broad ranges of values of $n$ and $\gamma$. The vertical bar in figure \ref{fig:experimentalcontourplot_1} shows the scale for the Planck-supported values of $M$  (see table \ref{tab:Table1}). In this figure the $\gamma - n$ boundary signifies that the favourable values lie within a region restricted by $n\in [8,10]$ and $\gamma\in [0.01,0.09]$ and below these ranges $\gamma$ and $n$ are independent of each other. The coloured regions in figure \ref{fig:experimentalcontourplot_2} represent the corresponding values of $\alpha_{\mathrm{max}}$, indicated at the end of the respective boundaries.}
    \label{fig:contourplot_2}
\end{figure}
In figures \ref{fig:contourplot_1} and \ref{fig:contourplot_2} we show the $\gamma - n$ contour plots with respect to the obtained and Planck-allowed values of $M$  following Eq. (\ref{eq:final_V_lambda_5}); and with respect to $\alpha_{\mathrm{max}}$ following Eq. (\ref{eq:alpha_cutoff}). It is clear from these plots that $2.4\leq n\leq 10$ and $0.01\leq \gamma\leq 0.09$ can be considered as legitimate ranges for $n$ and $\gamma$, respectively, in order to support the Planck data. For all cases the maximum allowed value of $\alpha_{\mathrm{max}}$ is $\sim 10^{-1}$. Figures \ref{fig:theoreticalcontourplot_1} and \ref{fig:experimentalcontourplot_1} show that most of the $M$-values lie on the $n-\gamma$ boundaries characterised by $n\in [8,10]$, $\gamma\in [0.01,0.09]$ and the associated upper limit of $\alpha_{\mathrm{max}}$ is constrained by $\alpha_{\mathrm{max}}< 0.1$. So, we can consider that $8\leq n\leq 10$, $0.01\leq\gamma\leq 0.09$ and $\alpha_{\mathrm{max}}<0.1$ (exact value can be estimated for a specific choices of $\gamma$ and $n$ from Eq. (\ref{eq:alpha_cutoff})) are the final ranges of the model parameters for the EDE-motivated quintessential $\alpha$-attractor potential.\par So far as the lower limit of $\alpha$ is concerned, it depends on the power of convergence into self-consistent solutions of the mode equations for the assigned value of $\alpha$ as boundary condition. In our case, we find that below $\alpha=0.001$ no solution is obtained for all possible boundary conditions. Therefore we set the lower limit of $\alpha$ as $\alpha_{\mathrm{min}}=0.001$. However, it may be different for different quintessential $\alpha$-attractor models. In Ref. \cite{Brissenden:2023yko} it is found that $\alpha\sim 10^{-3}$ which is the same as the lower limit in our case. So, we can say that $0.001\leq \alpha< 0.1$ is the final range of $\alpha$ for $8\leq n\leq 10$ and $0.01\leq\gamma\leq 0.09$. For the values of $n$ and $\gamma$ used for mode analysis (see figures \ref{fig:unperturbedINF} - \ref{fig:InfHubbleParameter}) \textit{viz.,} $n=8$ and $\gamma=0.0818$, $\alpha$ lies within the range $0.001\leq \alpha\leq 0.0186$.\par Thus, we have constrained the parameter $\alpha$ by two different ways. The lower limit is determined by the consistency checking of the first order perturbative mode equations of the inflaton field and the upper limit is fixed by the ESP in the EDE-motivated quintessential $\alpha$-attractor model. Within this range of $\alpha$, cosmological parameters do not get affected appreciably \textit{vis-\`{a}-vis} the Planck data. This is one kind of verification of the EDE characteristics \cite{Brissenden:2023yko} by $k$-space analysis. 
\subsection{Obtained results in light of PLANCK-2018}
\label{subsec:PLACKcomp}
We explicitly compare our results for the newly obtained range of $\alpha$ with Planck-2018 data. We take help of the joint marginalised contour plots of $n_s$ versus $r$ generated from the simulation data available in Planck Legacy Archive (\url{https://pla.esac.esa.int/}) by running in GetDist (\url{https://getdist.readthedocs.io/en/latest/}) plotting utility and Python Jupyter notebook environment (\url{https://jupyter.org/}). All these data include the effects of CMB $E$ mode polarisation, lensing and BAO. The estimated values of $n_s$ and $r$ described in subsection \ref{subsec:ModeBehaviours} are used to compare with the Planck data.\par
\begin{figure}[H]
\begin{subfigure}{0.33\linewidth}
  \centering
   \includegraphics[width=50mm,height=45mm]{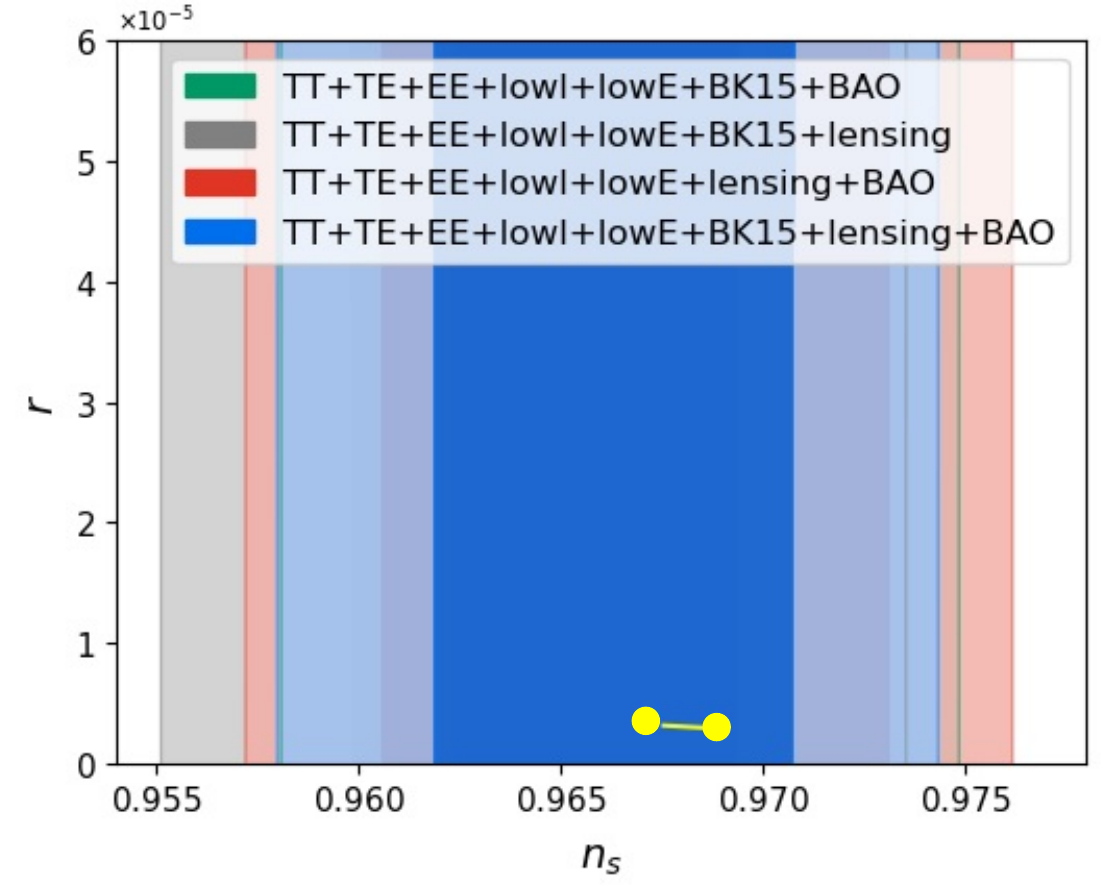}
   \subcaption{$\alpha = 0.001$}
   \label{fig:PL_1}
\end{subfigure}%
\begin{subfigure}{0.33\linewidth}
  \centering
   \includegraphics[width=50mm,height=45mm]{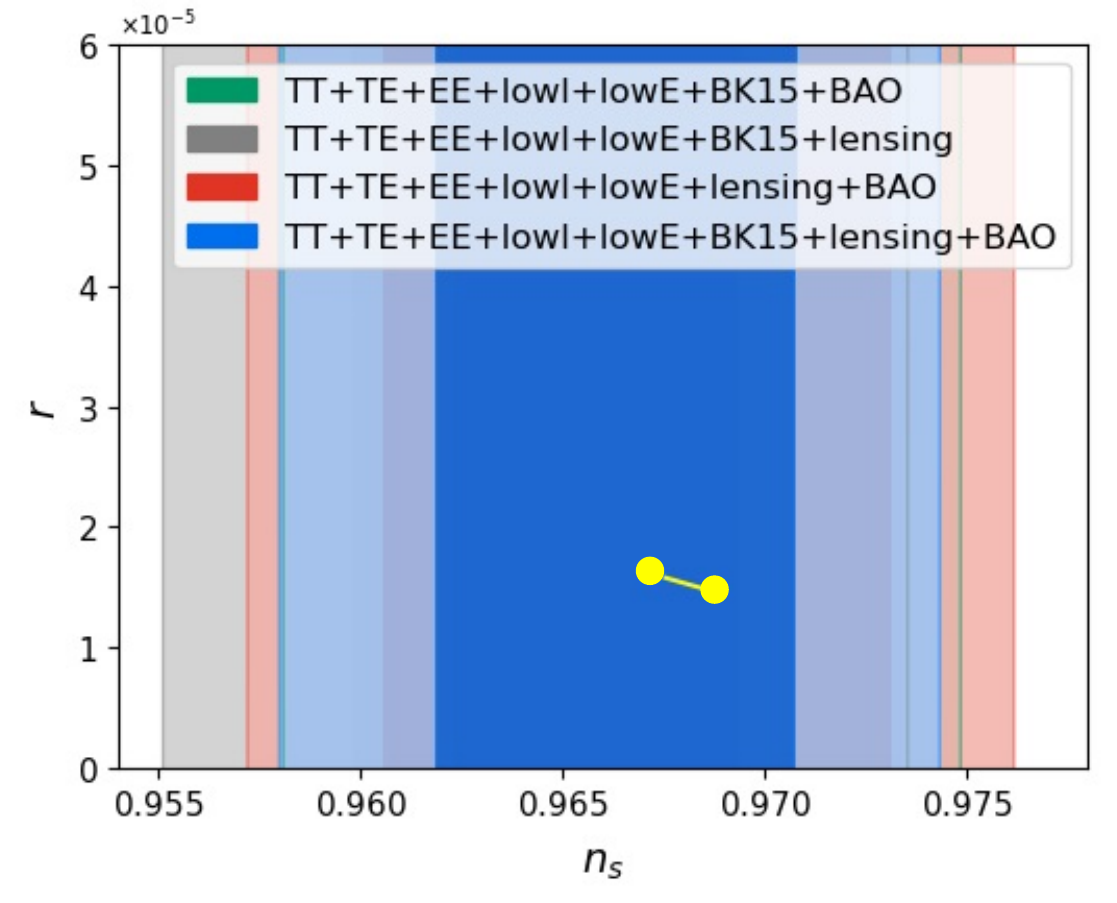}
   \subcaption{$\alpha = 0.005$}
   \label{fig:PL_2}
\end{subfigure}%
\begin{subfigure}{0.33\linewidth}
  \centering
   \includegraphics[width=50mm,height=45mm]{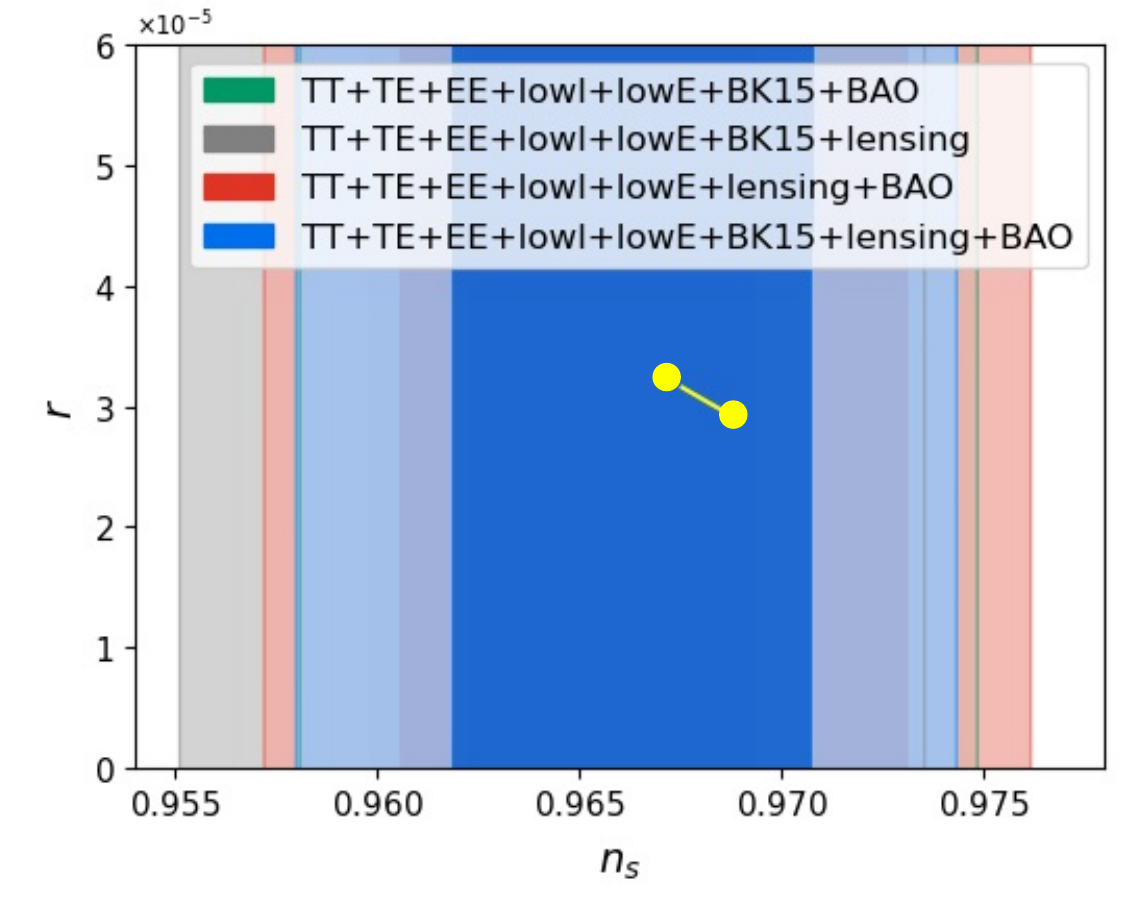}
   \subcaption{$\alpha = 0.01$}
   \label{fig:PL_3}
\end{subfigure}%
\caption{Comparisons of all the calculated values of $n_s$ and $r$ with the Planck constraints for first three values of $\alpha$. As $\alpha$ increases the parametric line is shifted upwards but does not cross the boundary of $68\%$ CL. The yellow bubbles at the ends represent that all the values in the graphs are computed from the number of remaining e-folds $N=63.49$.}
\label{fig:PlanckComparison_1}
\end{figure}

 Figure \ref{fig:PlanckComparison_1} shows the results of the comparison between the parameters calculated from the dynamical mode analysis and that of Planck data for three values of $\alpha$ \textit{viz.,} $\alpha=0.001,~0.005$, $0.01$ in the $k$-modes ranging from $0.001-0.009$ Mpc$^{-1}$. Axes of the graphs are reasonably extended to fit the data of $r$ in the $y$ direction. The cumulative mode responses of $n_s$ and $r$ are demonstrated by the yellow lines with two identical dots at the ends. These dots signify that the outputs are derived from a single value of number of remaining e-folds \textit{i.e.} $N=63.49$ as initial condition in the entire $k$-space. Blue and red areas in the graphs indicate the $68\%$ and $95\%$ CL zones respectively. All the yellow lines lie within $68\%$ CL for the given values of $\alpha$ and no results are found to exist beyond that. As $\alpha$ increases from $0.001$ up to $0.01$, the line segments are lifted towards higher and higher values of $r$ which is a direct verification of the double pole behaviour of the concerned model.\par In Ref. \cite{Sarkar:2021ird} we found that, in ordinary $\alpha$-attractor $E$ and $T$ models $\alpha$ is restricted between $\alpha=1$ and $\alpha=15$ in discrete manner. $\alpha\leq 10$ and $\alpha=15$ results lie within $68\%$ and $95\%$ CL zones respectively. This range of $\alpha$ becomes more stringent on coupling the quintessence in the inflationary framework in order to explain both the early and late-time expansions of the universe. In Ref. \cite{Sarkar:2023cpd} we found that the upper and lower bounds of $\alpha$ are confined between $\alpha=4.3$ and $0.1$ in continuous fashion. It may be noted that the fractional values of $\alpha$ are important requirements for the $\alpha$-attractor formulation in minimal $\mathcal{N}=1$ \cite{Kallosh:2013yoa} and maximal $\mathcal{N}=8$ \cite{Kallosh:2017ced} supergravity theories.\par The present work shows that, $\alpha$ is restricted in the closed interval $[0.001,0.0186]$. The lower bound of $\alpha$, which was set to $\alpha=0.1$ in earlier case, now becomes $\alpha=0.001$. Therefore around hundred order of magnitude is dropped by the attachment of EDE with quintessence. Such a tiny values of $\alpha$ are currently reported in \cite{Brissenden:2023yko} to resolve the Hubble tension. Also, these extremely low limits of $\alpha$ are the new $B$-mode targets of ongoing and upcoming LSS surveys \cite{QUaD:2009gka,QUIET:2012szu,SPT:2019nip,ACTPol:2014pbf,SPT:2015htm,Henderson:2015nzj,SPT-3G:2014dbx,LiteBIRD:2020khw,Heymans:2012gg,Hildebrandt:2016iqg,Kohlinger:2017sxk,Dawson:2015wdb,DES:2015gax,DESI:2016fyo,DESI:2016igz,LSSTScience:2009jmu,LSST:2017ags,Spergel:2015sza,Hounsell:2017ejq,Amendola:2016saw}.\par
\begin{figure}[H]
	\centering
	\includegraphics[width=0.8\linewidth]{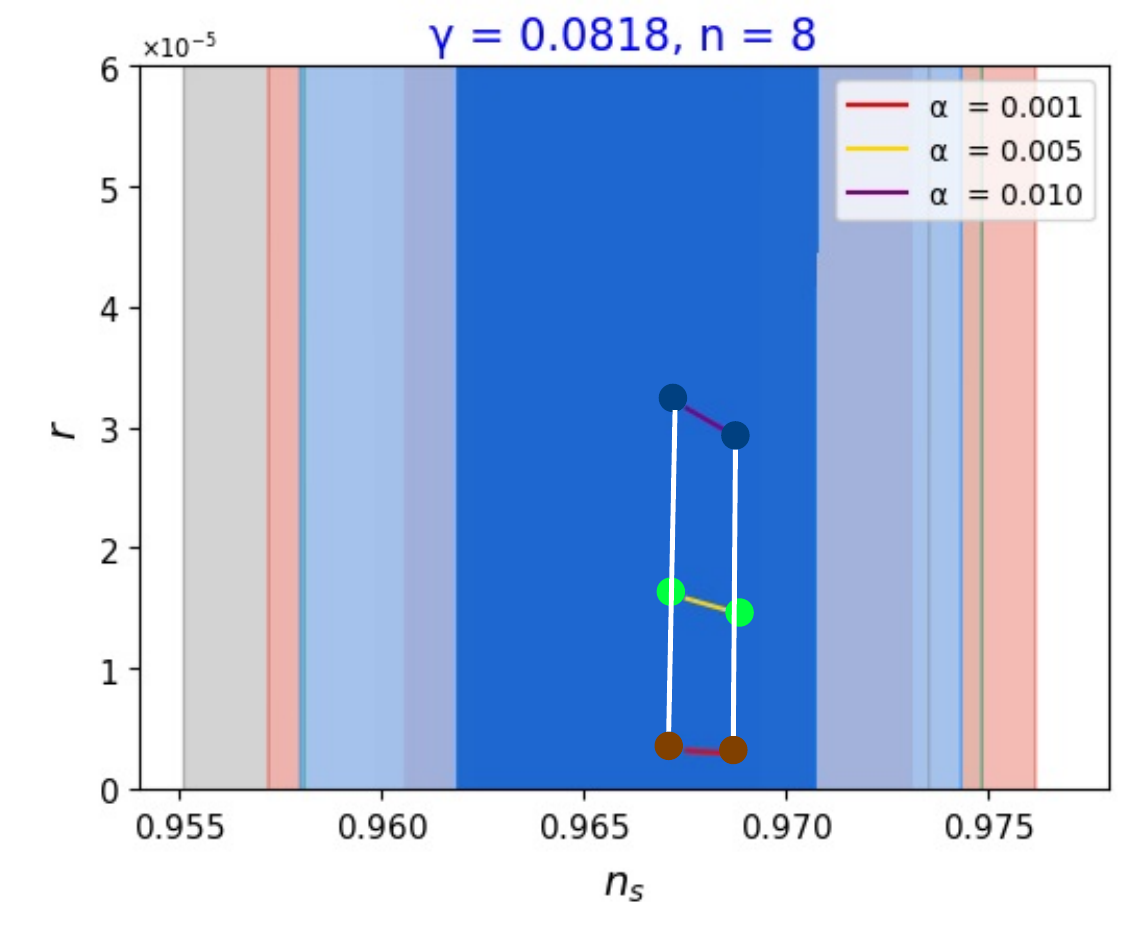}
	\caption{Combined results of figure \ref{fig:PlanckComparison_1} for all three values of $\alpha$. White line signifies that all possible values of $\alpha$ are permitted between two extreme limits. The slopes of the lines indicate that the potential is always concave in nature.}
	\label{fig:PlanckComparison_2}
\end{figure}

 In the last figure \ref{fig:PlanckComparison_2} we merge all the plots of figure
\ref{fig:PlanckComparison_1}. Each coloured line represents the results for a particular value of $\alpha$. The spectral tilts of all the lines are described by the negative and positive mode variations of $n_s$ and $r$, respectively. The tilts are such that the results always lie within the regime of $68\%$ CL. This reveals that the potential considered in this paper is a single field concave type having a long slow-roll plateau. The increasing behaviour of the parametric lines stops within the blue zone signifying the fact that the model is always a slow-roll one \textit{i.e.} Planck supported for all values of the model parameters.\par On the whole, we believe that the model, considered here, has the efficacy to explain inflation, quintessence and the EDE in unified way.
\subsection{A remark on the resolution of the Hubble tension}
In the present paper, we do not intend to solve the Hubble tension, rather our aim is to constrain the model parameters ($n,~\gamma,~\alpha$), specifically $\alpha$, from the aspects of ESP \textit{vis-\`{a}-vis} the EDE in the inflationary dynamics in the light of Planck+BICEP2/Keck results. However, we can get a very rough idea about the estimations of the model parameters, which can be helpful in the resolution of the Hubble tension.\par
\begin{figure}[H]
	\centering
	\includegraphics[width=90mm,height=75mm]{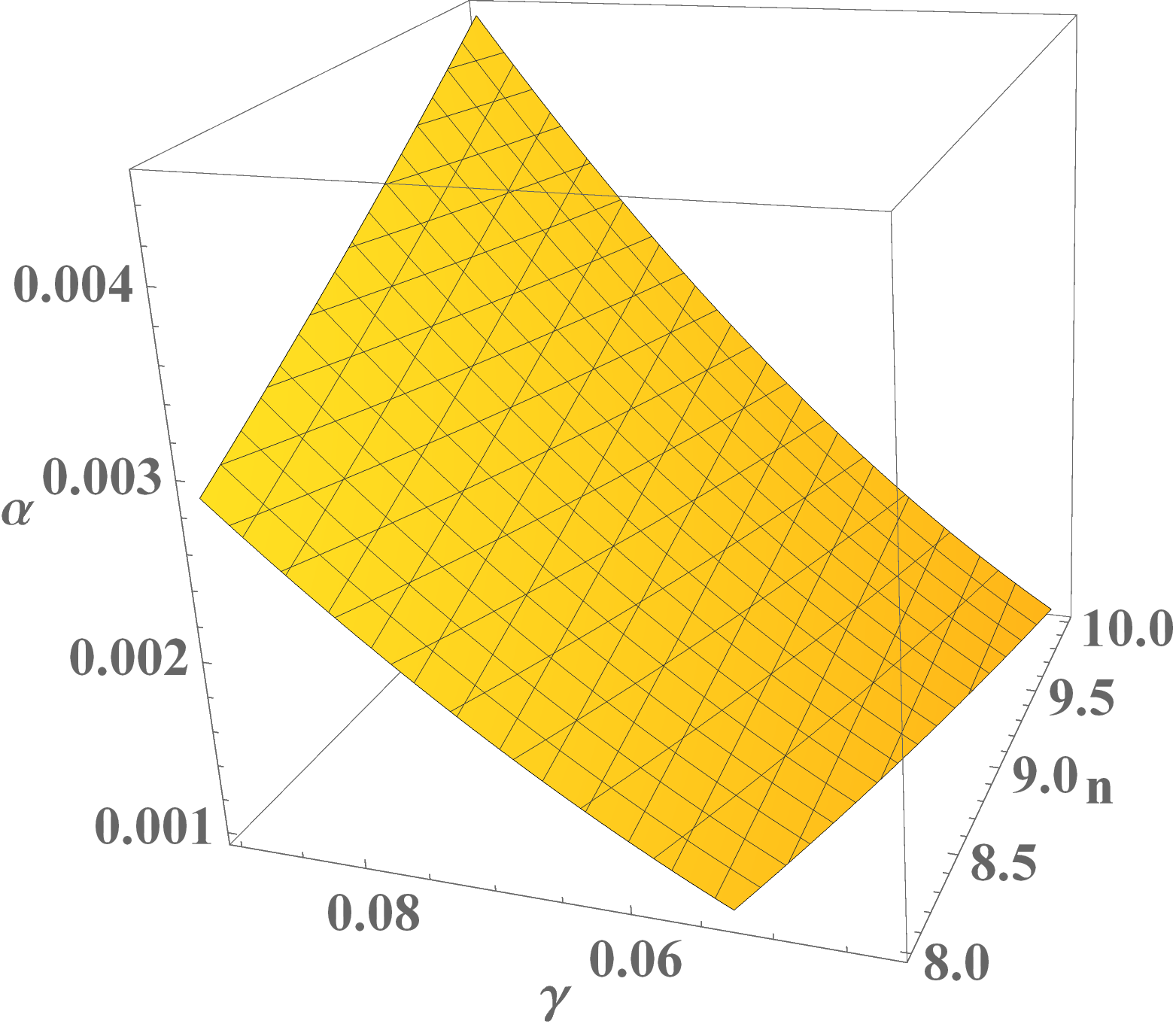}
	\caption{Three dimensional contour plot of $n$, $\gamma$ and $\alpha$ for Eq. (\ref{eq:EDE_omega_2}). The plot shows the necessary ranges of the model parameters for having the required value of $\Omega_{\mathrm{equal}}$ ($\approx0.1$) for the resolution of $H_0$ tension.}
	\label{fig:hubble_3D}
\end{figure}
 As discussed in Section \ref{sec:our model}, the $\xi$-field behaves as the EDE as long as it is trapped at the ESP during a time, which includes starting of formation of radiation and matter and then matter-radiation equality, before CMB decoupling. Once the density parameter reaches its maximum value $\Omega_{\mathrm{equal}}$, the field unfreezes and undergoes a free-fall in kination period to appear at the quintessential runaway to become (late) dark energy at present time. Now, as indicated in Refs. \cite{Smith:2020rxx,Brissenden:2023yko,Brissenden:2023cne}, in order to resolve the Hubble tension we should have $\Omega_{\mathrm{equal}}=0.10\pm0.02\simeq 0.1$ \textit{i.e.} $\rho_{\xi}(0)$ should be about $10\%$ of the total density of the universe during equality at redshift $4070_{-840}^{+400}$ \cite{Brissenden:2023cne}. Following this idea we can further constrain the model parameters using Eq. (\ref{eq:EDEomega}) as follows. The three dimensional contour plot of the equation
 \begin{equation}
     \Omega_{\mathrm{equal}}=\frac{18\alpha}{(n\gamma)^2}=0.1
     \label{eq:EDE_omega_2}
 \end{equation} in figure \ref{fig:hubble_3D} shows that, in order to have the required value of density parameter at equality, the model parameters should be within the following range: $8\leq n\leq 10$, $0.04\leq \gamma \leq 0.09$ and $0.001\leq \alpha \leq 0.0045$. Therefore, for the EDE motivated quintessential $\alpha$-attractor inflaton potential considered here, the ranges of $\gamma$ and $\alpha$ are further tightened over the ranges found by the mode analysis and the $\alpha$-cut-off, described in previous subsection, to solve the $H_0$ tension. However, a full simulation, based on statistical analysis (like MCMC), is required for actual results using post inflationary attributes. See \cite{Brissenden:2023yko} as an example. It is at least clear, here, that the parameter $\alpha$ should be $\sim 10^{-3}$ for building up a successful model for the EDE-motivated quintessential $\alpha$-attractor inflation in conjunction with the resolution of the Hubble tension. 
\section{Conclusions}
\label{sec:conclusion}
In conclusion, we have,
\begin{enumerate}
    \item constructed a new version of quintessential $\alpha$-attractor potential with EDE of non-oscillating type. The potential comprises two asymptotic poles corresponding to the inflationary slow-roll plateau and the quintessential tail, a steep slope for kinetically driven free fall and an ESP at the origin,
    \item studied in details the scalar field dynamics of inflation near pole boundaries and the geometric structure of the inflaton potential, featuring the aspects of EDE,
    \item extracted two regimes of operation by successive approximations in field space corresponding to inflation and quintessence,
    \item performed a first-order quantum mode analysis of the inflaton perturbation by DHE method in a perturbed metric background,
    \item examined the mode responses of the cosmological parameters for $k=0.001-0.009$ Mpc$^{-1}$ for $\gamma=0.0818$, $n=8$ and nine values of $\alpha$ within $\alpha=0.001-4.3$. The estimated values of the parameters satisfy Planck-2018$+$BICEP$2$/Keck-2015 data at $68\%$ CL,
    \item employed the variations of $n_s$ and $r$ in $k$-space in order to probe their cumulative mode behaviour in $n_s-r$ parametric space of Planck. The resulting line segments mimic the double-pole feature of the model and the associated spectral tilts show that the obtained results lie always within $68\%$ CL zone for all values of the model parameters,
    \item verified that from observational viewpoint, EDE has no effect on the cosmological parameters, which is one of the rudimentary property of EDE. But its presence in the model manifests in two ways. One is in the improved values of some parameters compared to the earlier ones \cite{Sarkar:2023cpd,Sarkar:2021ird}, specifically the energy scales of inflation $M$ and the EDE-modified present day vacuum density $V^{\mathrm{exact}}_{\Lambda}$. Their improved values are found to be $M=5.58\times 10^{-4}-4.57\times 10^{-3} M_P$ and $V_{\Lambda}^{\mathrm{exact}}=1.042\times 10^{-119}-4.688\times 10^{-116} M_P^4$. These values satisfy the required COBE/Planck normalization and the experimental value of vacuum density $V_{\Lambda}^{\mathrm{Planck}}\sim 10^{-120} M_P^4$. The second one is in the presence of ESP at the origin in the potential and thereby making the inflaton field positive ($\xi>0$), and
    \item finally, constrained the crucial parameter $\alpha$ by two different methods. The upper limit is fixed by a new probe \textit{viz.,} the ESP of the potential and the lower limit is determined by the consistent initial conditions of first order perturbative mode equations of the inflaton field. We find that $\alpha$ should be within $0.001\leq\alpha< 0.1$ continuously for $\gamma$ and $n$ lying within $0.01\leq\gamma\leq 0.09$ and $8\leq n \leq 10$ respectively. For the chosen values of the model parameters \textit{i.e.} $\gamma=0.0818$ and $n=8$, the range of $\alpha$ is obtained as $0.001\leq \alpha\leq 0.0186$. The lower and upper limits are substantially diminished from $0.1$ to $0.001$ and $4.3$ to $0.0186$, respectively, in comparison to the earlier model in Ref. \cite{Sarkar:2023cpd}, by the incorporation of the EDE in quintessential inflaton field. We have also found by an approximate calculations that, in order to resolve the Hubble tension the ranges of $\gamma$ and $\alpha$ are further restricted to $0.04\leq \gamma \leq 0.09$ and $0.001\leq \alpha \leq 0.0045$. Therefore, lower end of $\alpha$ is capable of resolving the $H_0$ discrepancy, as also described in Ref. \cite{Brissenden:2023yko}. However, the entire space of $\alpha$ is essential for studying the early and late-time signatures of spacetime expansion of the universe.
    \end{enumerate} 


\section*{Acknowledgments}
The authors acknowledge the University Grants Commission, The Government of India for the CAS-II program in the Department of Physics, The University of Burdwan. AS acknowledges The Government of West Bengal for granting him the Swami Vivekananda fellowship.





\bibliographystyle{utcaps}
\bibliography{biblio}
\end{document}